\documentclass[10pt,journal,compsoc]{IEEEtran}
\ifCLASSOPTIONcompsoc
\usepackage{cite}

\ifCLASSINFOpdf
\else
\fi

\usepackage{amsmath}
\usepackage{amssymb}
\usepackage{enumitem}
\usepackage{mathtools}
\usepackage{amsthm}
\usepackage{mathrsfs}
\usepackage{multirow}
\usepackage{subfigure}
\usepackage{titletoc}
\usepackage{bm}
\usepackage{colortbl}
\usepackage{wrapfig}
\usepackage{tabularx}
\usepackage{caption}
\usepackage{multirow}
\usepackage{booktabs}
\usepackage{url} 
\usepackage{ragged2e}
\usepackage[ruled]{algorithm2e}
\usepackage{tcolorbox}     
\usepackage[framemethod=tikz]{mdframed}
\setlist {nolistsep} 

\usepackage[colorlinks,
linkcolor=blue,
anchorcolor=pink,
citecolor=orange
]{hyperref}

\theoremstyle{plain}
\newtheorem{theorem}{Theorem}
\newtheorem{proposition}{Proposition}
\newtheorem{lem}{Lemma}
\newtheorem{coro}{Corollary}
\theoremstyle{definition}
\newtheorem{defi}{Definition}
\newtheorem{assu}{Assumption}
\newtheorem{rem}{Remark}
\newtheorem{property}{Property}

\newtheorem*{rthm1}{\textbf{Restate of Theorem} \ref{them1}}
\newtheorem*{rthm2}{\textbf{Restate of Theorem} \ref{pami:thm2}}
\newtheorem*{rthm3}{\textbf{Restate of Theorem} \ref{pami:thm3}}
\newtheorem*{rlem1}{\textbf{Restate of Lemma} \ref{covering_lem}}
\newtheorem*{rlem2}{\textbf{Restate of Lemma} \ref{pami:lem2}}
\newtheorem*{rlem3}{\textbf{Restate of Lemma} \ref{pami:lem3}}
\newtheorem*{rdef2}{\textbf{Restate of Definition} \ref{def2}}
\newtheorem*{rdef3}{\textbf{Restate of Definition} \ref{def3}}
\newtheorem*{rassu1}{\textbf{Restate of Assumption} \ref{assu1}}
\newtheorem*{rcor1}{\textbf{Restate of Corollary} \ref{cor1}}

\DeclareMathOperator*{\expe}{\mathbb{E}} 
\DeclareMathOperator*{\mcu}{\mathcal{U}} 
\DeclareMathOperator*{\mci}{\mathcal{I}}
\DeclareMathOperator*{\mcd}{\mathcal{D}}
\DeclareMathOperator*{\mcl}{\mathcal{L}}
\DeclareMathOperator*{\mcr}{\mathcal{R}}
\def \bmg{\boldsymbol{g}}
\def \bmga{\boldsymbol{\gamma}}

\usepackage{xcolor}
\definecolor{ballblue}{HTML}{338EA7}
\definecolor{lightseagreen}{HTML}{759D39}
\definecolor{lightred}{HTML}{DD7769}
\definecolor{org}{HTML}{F8A145}
\definecolor{blu}{HTML}{63ACE5}
\definecolor{lightseagreen}{HTML}{759D39}

\usepackage[textsize=tiny]{todonotes}

\tcbuselibrary{skins}
\newenvironment{qbox}
{\begin{tcolorbox}[enhanced jigsaw, drop shadow=black!50!white,colback=white, width=0.95\linewidth, center, left=2pt,right=2pt,top=1pt,bottom=1pt]}	
	{\end{tcolorbox}}

\usepackage[framemethod=tikz]{mdframed}
\usepackage{lipsum}

\begin{document}
	\title{Improved Diversity-Promoting Collaborative Metric Learning for Recommendation}
	\author{Shilong~Bao,
		Qianqian~Xu*,~\IEEEmembership{Senior~Member,~IEEE},
		Zhiyong Yang, 
		Yuan He, \\
		Xiaochun~Cao,~\IEEEmembership{Senior~Member,~IEEE},
		and~Qingming Huang*,~\IEEEmembership{Fellow,~IEEE}
		\IEEEcompsocitemizethanks{\IEEEcompsocthanksitem Shilong Bao is with State Key Laboratory of Information Security (SKLOIS), Institute of Information Engineering, Chinese Academy of Sciences, Beijing 100093, China, and also with School of Cyber Security, University of Chinese Academy of Sciences, Beijing 100049, China (e-mail: \texttt{baoshilong@iie.ac.cn}).}
		\IEEEcompsocitemizethanks{\IEEEcompsocthanksitem Qianqian Xu is with the Key Laboratory of Intelligent Information Processing, Institute of Computing Technology, Chinese Academy of Sciences, Beijing 100190, China (e-mail: \texttt{xuqianqian@ict.ac.cn}).}
		\IEEEcompsocitemizethanks{\IEEEcompsocthanksitem Zhiyong Yang is with the School of Computer Science and Technology, University of Chinese Academy of Sciences, Beijing 101408, China (e-mail: \texttt{yangzhiyong21@ucas.ac.cn}).}
		\IEEEcompsocitemizethanks{\IEEEcompsocthanksitem Yuan He is with the Security Department of Alibaba Group, Hangzhou 311121, China (e-mail : \texttt{heyuan.hy@alibaba-inc.com}).}
		\IEEEcompsocitemizethanks{\IEEEcompsocthanksitem Xiaochun Cao is with School of Cyber Science and Technology, Shenzhen Campus of Sun Yat-sen University, Shenzhen 518107, China (e-mail: \texttt{caoxiaochun@mail.sysu.edu.cn}).}
		\IEEEcompsocitemizethanks{\IEEEcompsocthanksitem Qingming Huang is with the School of Computer Science and Technology, University of Chinese Academy of Sciences, Beijing 101408, China, also with the Key Laboratory of Big Data Mining and Knowledge Management (BDKM), University of Chinese Academy of Sciences, Beijing 101408, China, also with the Key Laboratory of Intelligent Information Processing, Institute of Computing Technology, Chinese Academy of Sciences, Beijing 100190, China, and also with Peng Cheng Laboratory, Shenzhen 518055, China (e-mail: \texttt{qmhuang@ucas.ac.cn}).}
		\IEEEcompsocitemizethanks{\IEEEcompsocthanksitem * Corresponding authors}
	}

	\markboth{IEEE TRANSACTIONS ON PATTERN ANALYSIS AND MACHINE INTELLIGENCE}%
	{Shell \MakeLowercase{\textit{et al.}}: Bare Demo of IEEEtran.cls for Computer Society Journals}
	
	\IEEEtitleabstractindextext{%
		\begin{abstract}
			\justifying
			Collaborative Metric Learning (CML) has recently emerged as a popular method in recommendation systems (RS), closing the gap between metric learning and collaborative filtering. Following the convention of RS, existing practices exploit unique user representation in their model design. This paper focuses on a challenging scenario where a user has multiple categories of interests. Under this setting, the unique user representation might induce preference bias, especially when the item category distribution is imbalanced. To address this issue, we propose a novel method called \textit{Diversity-Promoting Collaborative Metric Learning} (DPCML), with the hope of considering the commonly ignored minority interest of the user. The key idea behind DPCML is to introduce a set of multiple representations for each user in the system where users' preference toward an item is aggregated by taking the minimum item-user distance among their embedding set. Specifically, we instantiate two effective assignment strategies to explore a proper quantity of vectors for each user. Meanwhile, a \textit{Diversity Control Regularization Scheme} (DCRS) is developed to accommodate the multi-vector representation strategy better. Theoretically, we show that DPCML could induce a smaller generalization error than traditional CML. Furthermore, we notice that CML-based approaches usually require \textit{negative sampling} to reduce the heavy computational burden caused by the pairwise objective therein. In this paper, we reveal the fundamental limitation of the widely adopted hard-aware sampling from the One-Way Partial AUC (OPAUC) perspective and then develop an effective sampling alternative for the CML-based paradigm. Finally, comprehensive experiments over a range of benchmark datasets speak to the efficacy of DPCML. Code are available at \url{https://github.com/statusrank/LibCML}.
		\end{abstract}



		}

	\maketitle
	
	\IEEEdisplaynontitleabstractindextext
	\IEEEpeerreviewmaketitle
	
	\section{Introduction}\label{intro}
	Recommender system (RS) is a well-known building block in eCommerce, which can assist buyers to find products they wish to purchase by giving them the relevant recommendations. The key recipe behind RS is to learn from user-item interaction records \cite{DBLP:conf/pakdd/WangZ0HC20, DBLP:conf/nips/MaZ0Y019, DBLP:conf/kdd/MaZYCW020, DBLP:journals/aei/LvZWWW20,DBLP:journals/tkde/JiangCCW0Y15}. In practice, since user preferences are hard to collect, such records often exist as implicit feedback \cite{DBLP:conf/nips/WangGZZ18,DBLP:conf/sigir/AskariSS21,DBLP:conf/www/TogashiKOS21} where only indirect actions are provided (say clicks, collections, reposts, etc.). Such a property of implicit feedback raises a great challenge for RS-targeted machine learning methods and thus stimulates a wave of relevant studies along this course \cite{DBLP:conf/icml/XuRKKA21,DBLP:conf/icml/ZhengTDZ16,DBLP:conf/aaai/WangWSSL20}. 
	
	
	Over the past two decades, most literature follows a typical paradigm known as One-Class Collaborative Filtering (OCCF) \cite{DBLP:conf/icdm/PanZCLLSY08}, where the items not being observed are usually assumed to be of less interest to the user and labeled as negative instances. In the early days, the vast majority of studies in the OCCF community focus on Matrix Factorization (MF) based algorithms, where the inner product between their embeddings conveys the preference of a specific user toward an item \cite{DBLP:journals/ijon/ZhangR21, DBLP:conf/aaai/ChenL019}. Recently, a milestone study \cite{hsieh2017collaborative} points out that the inner product violates the triangle inequality, resulting in a sub-optimal topological embedding space. Inspired by the strength of metric learning \cite{DBLP:conf/iccv/KumarTZ07}, a novel framework called \textit{Collaborative Metric Learning} (CML) \cite{hsieh2017collaborative} is proposed, achieving promising performance in practice. Hereafter, many efforts have been made along the research direction to improve CML
	\cite{tran2019improving,DBLP:conf/www/TayTH18, DBLP:conf/icdm/ParkKXY18, DBLP:conf/mm/BaoXMYCH19,DBLP:journals/nn/WuZNC20,wang2019group, DBLP:conf/ijcai/ZhouLL019,DBLP:conf/dasfaa/ZhangZLXF0SC19, DBLP:conf/recsys/TranSHM21}. 
	
	\begin{figure}
		\begin{center}
			\includegraphics[width=1.03\columnwidth]{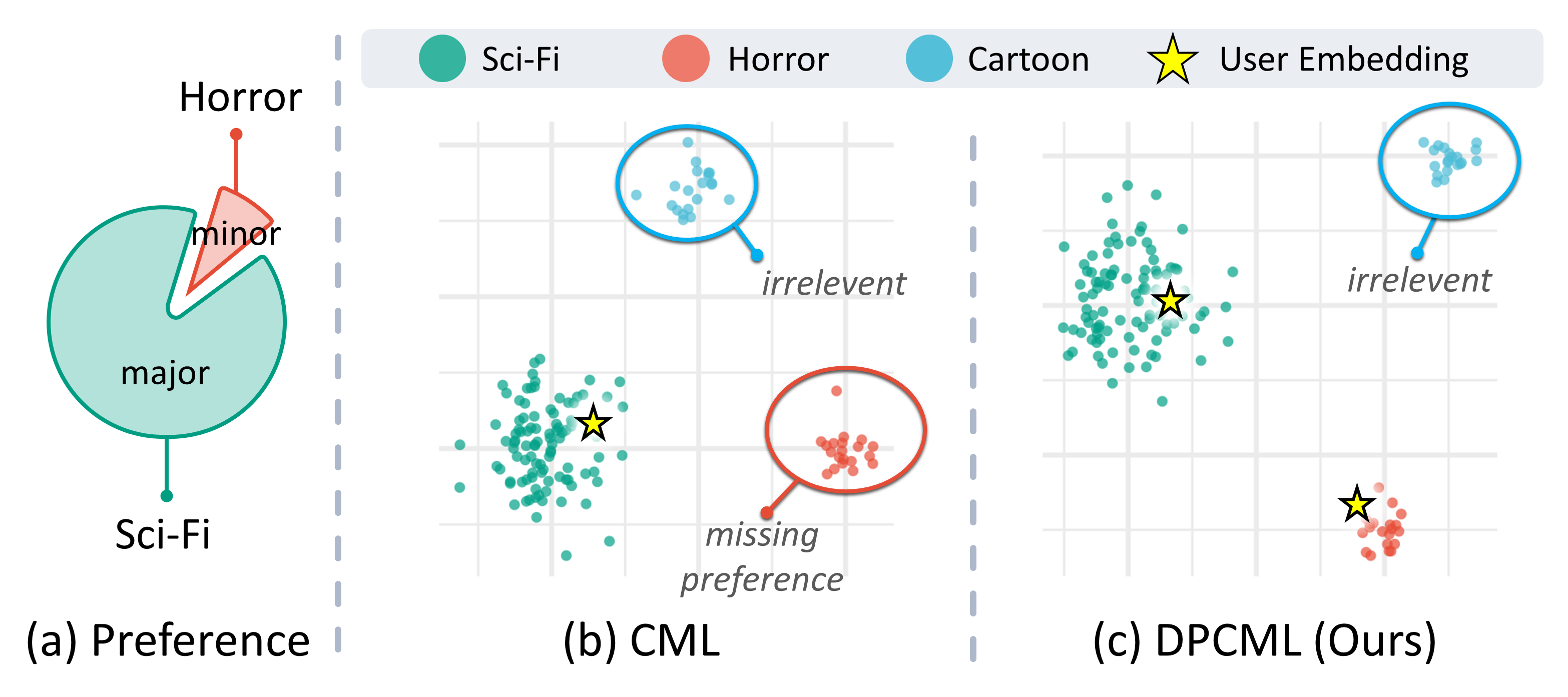}
		\end{center}
		\caption{An illustration shows the benefit of DPCML when a user has multiple diverse preferences. Taking movies as an example, we assume that Sci-Fi/Horror is the majority/minority interest of the user while Cartoon is an irrelevant movie type. It is easy to see that if the item embeddings are distributed like depicted in the figure, we can hardly find a single user embedding to capture both interests simultaneously.}
		\label{motivation}
	\end{figure}
	
	Despite great success, we observe that users usually have multiple categories of preferences, as evidenced by a critical example in Sec.\ref{Sec3.2}. Moreover, such interest groups are often not equally distributed, where the amount of some groups dominates the others. Under this case, as shown in Fig.\ref{motivation}, the existing studies might induce preference bias since they tend to meet the majority interest while missing the other potential preference. Therefore, in this paper, we are interested in the following problem:
	
	\begin{qbox}
		\begin{center}
			\textit{How to develop an effective CML-based algorithm to accommodate the diversity of user preferences?}
		\end{center}
	\end{qbox}
	\noindent\textbf{Contributions.} In search of an answer, we propose a novel algorithm called \textit{Diversity-Promoting Collaborative Metric Learning (DPCML)}. The key recipe is introducing a set of $C_{u_i}$ embeddings for each user $u_i$ to span multiple interest groups of items. In this sense, $u_i$'s preference toward a given item embedding $\boldsymbol{g}_{v_j}$ is defined by the minimum distance between $u_i$'s embedding set $\{\boldsymbol{g}^c_{u_i}\}_{c=1}^{C_{u_i}}$ and $\boldsymbol{g}_{v_j}$, i.e., $s(u_i,v_j) =\min\limits_{c} \|\boldsymbol{g}^c_{u_i} - \boldsymbol{g}_{v_j}\|^2$. Thereafter, the model could exploit different user vectors to fit diverse interest groups such that all potential preferences would be captured, as shown in Fig.\ref{motivation}-(c). A central challenge here is how to determine the number of embeddings for each user. To this end, we instantiate two assignment rules called \textit{Basic Preference Assignment} (BPA) strategy and \textit{Adaptive Preference Assignment} (APA) strategy, respectively. Generally speaking, BPA assumes all users have the same number of $C$ interest clusters, while APA could adaptively determine a proper value of $C_{u_i}$ for each user from their historical records. Meanwhile, we observe that the diversity of the embeddings among the same user representation set also plays a vital role in the model. Therefore, we further present a novel \textit{Diversity Control Regularization Scheme} (DCRS) to accommodate the multi-vector representation strategy better.
	
	To show the effectiveness of DPCML, we continue to investigate the generalization performance of the CML paradigm from a theoretical point of view. To the best of our knowledge, \textbf{such a problem remains barely explored in the existing literature.} Here the major challenges fall into two aspects: 1) The pairwise risk of CML-based algorithms could not be expressed as a sum of independently and identically distributed (i.i.d.) loss terms, making the standard Rademacher Complexity-based \cite{DBLP:conf/colt/BartlettM01, DBLP:books/daglib/0034861} theoretical arguments unavailable; 2) The annoying minimum operation involved in DPCML is not continuous, which cannot be analyzed easily in the Rademacher complexity framework. To address these challenges, we employ the covering number and $\epsilon$-net arguments to derive the generalization upper bound, which only requires a weaker Lipschitz continuous property over the hypothesis space instead of the i.i.d. condition. Meanwhile, this approach also helps manage the annoying minimum operation. The generalization bound (Thm.\ref{them1}) shows that DPCML could induce a smaller generalization error than traditional CML with a high probability. 

	Taking a step further, we notice that the optimization objective of CML is usually expressed in a pairwise learning manner, leading to unaffordable computational burdens. To ease this, CML-based approaches usually require \textit{negative sampling}, where only a few items (denoted as $U$) sampled from each user's unobserved candidates would be regarded as irrelevant samples during training. At present, hard negative sampling (HarS) \cite{DBLP:conf/iccv/VasudevaDB0C21,hsieh2017collaborative,DBLP:conf/mm/BaoXMYCH19}, merely leveraging the ``hardest'' negative sample (i.e., $U \equiv 1$), has become the most effective way in the CML community. 
	
	However, we argue that the current HarS is insufficient for the top-$N$ recommendation. We start from the equivalent reformulation between the generalized HarS and the OPAUC optimization (Prop.\ref{pro1}), where \textbf{the sampling number $U$ particularly corresponds to the FPR range} in the OPAUC optimization \cite{DBLP:journals/corr/abs-2210-03967,DBLP:conf/icml/0001XBHCH21}. Then, through theoretical analysis (Thm.\ref{pami:thm2}) of the performance guarantee between the top-$N$ recommendation and OPAUC, we show that the sampling parameters should \textbf{positively correlate} with the value of $N$ to pursue promising performance. Unfortunately, the existing HarS-based CML merely considers the fixed number of items (i.e., $U \equiv 1$) no matter how $N$ is. In light of this, we propose a novel OPAUC-oriented Differentiable HarS-based algorithm (DiHarS), which can include a proper number of ``hardest'' examples via maximizing the OPAUC performance. 

	Finally, we conduct comprehensive empirical studies over $6$ widely used benchmarks to show the superiority of DPCML, including recommendation performance/diversity comparisons, qualitative analysis, how to leverage side information and solve cold-start problems. The results consistently speak to the efficacy of DPCML.  

	
	%
	
		This work extends our NeurIPS 2022 Oral paper \cite{DPCML}, where we advanced a diversity-promoting CML-based algorithm to accommodate the diverse preferences of users. In this version, we rethink the design of DPCML carefully and make a series of substantial ameliorations in methodologies and experiments. The novelty of the extended version is summarized as follows:
	\begin{itemize}[leftmargin=*]
		\item \textbf{A New Representation Assignment Strategy.} The original DPCML follows the BPA scheme, i.e., simply assigning $C$ representation vectors for each user in the system. This might fail to capture all users' diverse preferences accurately, leading to limited performance gain. To alleviate this, we explore an APA strategy to accommodate the diversity of user preferences better.
		\item \textbf{A Novel OPAUC-driven Efficient Optimization.} The conference version of DPCML adopts two off-the-shelf sampling strategies, i.e., uniform \cite{hsieh2017collaborative,DBLP:conf/kdd/YangDZYZT20,DBLP:conf/wsdm/RendleF14} and hard \cite{DBLP:conf/iccv/VasudevaDB0C21,hsieh2017collaborative,DBLP:conf/mm/BaoXMYCH19} to ease its heavy optimization burdens. This paper reveals the fundamental limitation of HarS and proposes a novel OPAUC-oriented Differentiable HarS-based algorithm (DiHarS), which can achieve promising performance with a theoretical guarantee.
		\item \textbf{Enhancing the Applicability of DPCML.} Limited by the CML paradigm, the original DPCML cannot exploit other semantic information in the system and will lose efficacy for cold-start scenarios. Motivated by the idea of DropoutNet (DN) \cite{DBLP:conf/nips/VolkovsYP17}, this paper also presents an extended DPCML with DN (Sec.\ref{major:Sec7.5}) to enhance the applicability of DPCML in practice. 
		\item \textbf{New Experiments}. We conduct a wide range of new empirical studies, including $3$ new collaborative filtering-based competitors, $2$ new benchmarks, $9$ new diversity-promoting competitors, $2$ new diversification metrics, a series of quantitive studies and fine-grained analysis.
		\item \textbf{Miscellaneous Contents.} We also improve some existing contents to make the work more complete, including the abstract, introduction, review of prior arts (Sec.\ref{pami:sec2.2}, Sec.\ref{sec2.4} and Sec.\ref{pami:sec2.5}), preliminary (Sec.\ref{rw:opauc}), methodology (Sec.\ref{pami:sec4} and Sec.\ref{pami:sec6.2}), and experiments (Sec.\ref{exp}).
	\end{itemize}
	\section{Prior Arts} \label{rel_work}
	In this section, we briefly review the closely related studies along with our main topic.
	
	\subsection{One-Class Collaborative Filtering}
	
	In many real-world applications, the vast majority of interactions are implicitly expressed by users' behaviors, e.g., downloads of movies, clicks of products, and browses of news. In this sense, we can only know the users' interest in the observed records, while their preferences for the rest are usually not available. Therefore, in order to develop RS from such implicit feedback, researchers usually formulate the recommendation task as the \textit{One-Class Collaborative Filtering} (OCCF) problem \cite{DBLP:conf/icdm/PanZCLLSY08, DBLP:journals/www/YaoTYXZSL19, DBLP:conf/icml/HeckelR17,DBLP:journals/kbs/ZhangR21,DBLP:conf/sigir/LeeKJPY21, DBLP:journals/ijon/ZhangR21}. Generally speaking, the critical assumption of OCCF is that users' preferences toward items not being observed are less than those known interacted ones. In what follows, we will briefly review two simple but effective OCCF frameworks, i.e., Matrix Factorization (MF) and Collaborative Metric Learning (CML). 
	
	
	\noindent \textbf{Matrix Factorization (MF) based Algorithm}. Over the past decades, the Matrix Factorization (MF)-based algorithms are one of the most classical OCCF solutions \cite{DBLP:conf/ijcai/0001DWTTC18,DBLP:conf/icml/ZhengTDZ16,DBLP:conf/uai/RendleFGS09, DBLP:conf/aaai/ChenL019}. The key idea of MF is to express each user/item in RS as a latent vector such that the user-item interaction could be recovered by a product between their corresponding latent embeddings. Many successful studies have been made to build practical MF-based approaches in the OCCF community. For instance, \cite{DBLP:conf/sigir/HeZKC16} proposes an item-oriented MF method with implicit feedback, which employs an element-wise alternating least squares strategy to optimize the MF model with variably-weighted missing data. Besides, Neural Collaborative Filtering (NCF) \cite{DBLP:conf/www/HeLZNHC17} regards the recommendation task as a regression problem and then develops a general framework unifying the advantages of MF and neural networks together. Despite the effectiveness of MF-based approaches, recent studies argue that the inner product of MF might fail to the triangle inequality property, leading to sub-optimal performance.
	

	
	\noindent \textbf{Collaborative Metric Learning (CML) based Algorithm}. To mitigate the fundamental limitation of the MF-based framework, \cite{hsieh2017collaborative} proposes the \textit{Collaborative Metric Learning} (CML) paradigm, which has demonstrated significant performance gain. Generally speaking, the idea of CML is to learn a joint user-item metric space to reflect the users' preferences, which is highly inspired by the success of metric learning \cite{DBLP:journals/symmetry/KayaB19, DBLP:journals/pami/MilbichRBO22, DBLP:journals/pami/EleziSWVTPL23}. At present, the advances of CML have attracted great research attention in the RS community, giving birth to many competitive recommendation methods. To name a few, inspired by the knowledge translation mechanism in the knowledge graph, \cite{DBLP:conf/icdm/ParkKXY18} proposes a collaborative translation metric learning (short for TransCF) method, which aims to learn an exclusive latent relation vector for each user-item interaction to model the users' interests precisely. Similar to TransCF, \cite{DBLP:conf/www/TayTH18} designs a latent relational metric learning (LRML) framework, which adopts an attention-based memory-based framework to obtain the translation vector for each user-item interaction. Besides, to deal with sparse and insufficient interest records, \cite{DBLP:conf/mm/BaoXMYCH19} proposes a collaborative preference embedding (CPE) technique. \cite{DBLP:conf/dasfaa/ZhangZLXF0SC19} proposes a memory component and an attention mechanism to integrate the item-side representation interacted by the user as the adaptive interest for the user. \cite{DBLP:conf/recsys/TranSHM21} employs the memory-based attention networks to hierarchically capture users' preferences from both latent user-item and item-item relations. Different from the existing literature, this paper targets a challenging scenario where a user has multiple categories of interests. Unfortunately, in this case, the current literature equipped with unique user representation might induce preference bias, especially when the item category distribution is imbalanced. 
	\subsection{Learning with Negative Sampling \label{pami:sec2.2}}
	Apart from the reasonable regard of user-item interaction records in the system, another primary concern is how to efficiently optimize a model built on implicit signals, because the large space of unobserved items usually brings about heavy optimization burdens. Most current studies resort to a so-called \textit{negative sampling} technique to improve efficiency, where merely a few items would be selected from unknown interest items as negative items for optimization \cite{DBLP:journals/nn/WuZNC20,DBLP:conf/www/HeLZNHC17,DBLP:conf/sigir/Wang0WFC19,DBLP:conf/kdd/Wang00LC19,DBLP:journals/corr/abs-2302-03472}. Note that,  negative sampling has been employed in various machine learning tasks to boost the model performance while reducing computing complexity, such as deep metric learning \cite{DBLP:conf/iccv/QianSSHTLJ19,DBLP:journals/pami/ZhengLZ21,DBLP:conf/iccv/HarwoodGCRD17,DBLP:conf/iccv/VasudevaDB0C21} and contrastive learning \cite{DBLP:conf/iclr/RobinsonCSJ21,DBLP:journals/corr/abs-2206-01197} in computer vision. In this paper, we narrow our attention to \textbf{CML-based algorithms learning with negative sampling}. Generally speaking, one of the choices is to employ a uniform sampling strategy \cite{hsieh2017collaborative,DBLP:conf/kdd/YangDZYZT20,DBLP:conf/wsdm/RendleF14}, which will uniformly construct negative user-item pairs at each mini-batch to optimize the pairwise empirical risk. In addition, popularity-based sampling \cite{DBLP:conf/sigir/WuVSSR19}, two-stage negative sampling \cite{tran2019improving} and hard negative sampling \cite{DBLP:journals/pr/GajicAG21} are also applied to the CML framework. In practice, learning with the hard negative sampling technique could induce a more promising performance than others. However, the fundamental reasons for its effectiveness are still an attractive mystery. Furthermore, the default version of hard negative sampling only considers the "hardest" (one item) achieved by a ranking selection process. This might limit its performance because: 1) Merely using the one hardest sample could not guarantee obtaining a promising Top-$N$ recommendation performance. 2) The ranking operation is non-differentiable, making optimizing it challenging. We will present elaborate discussions about this in Sec.\ref{pami:sec6.2}.
	
		\subsection{Diversity in Recommendation System}\label{sec2.4}
		\textit{Diversification}, one of the most significant measures for evaluating the quality of online user experiences, has received increasing research attention in the RS community \cite{DBLP:journals/kbs/KunaverP17, DBLP:journals/csur/ZangerleB23,DBLP:reference/sp/2022rsh, DBLP:conf/coling/RazaBN22, DBLP:journals/tbd/XieLLZ00L22}. At the early stage, most conventional methods \cite{DBLP:journals/tkde/AdomaviciusK12,vaishnavi2013ranking, DBLP:conf/ijcai/ShaWN16,DBLP:conf/ijcai/AshkanKBW15,DBLP:conf/nips/ChenZZ18,DBLP:conf/icml/GillenwaterKMV19} generally consider developing post-processing methods conducted on top of the ordered recommendation candidate predicted by relevance. Such a re-ranking strategy is independent of the underlying ``relevance model'' and can be easily applied to most recommendation systems.  For example, \cite{DBLP:conf/ah/BridgeK06} proposes a bounded greedy selection algorithm to enhance diversity for collaborative recommendations. \cite{DBLP:conf/compsac/PremchaiswadiPJP13} designs a total diversity effect ranking method to guarantee maximum diversification in the recommendations list. However, considering relevance and diversity separately is insufficient for optimal outcomes \cite{DBLP:conf/recsys/SuYCY13,DBLP:conf/www/ChengWMSX17,DBLP:conf/dasc/LiZZZL17,DBLP:conf/aaai/0020X0MZZT20,DBLP:conf/sigir/LiangQLY21} due to the trade-off between them. To address this issue, researchers attempt to regard relevance and diversity simultaneously during training. Typically, personalized ranking with diversity \cite{DBLP:conf/recsys/Hurley13} is proposed, which incorporates the diversity goal into a ranking objective for implicit feedback recommendation. \cite{DBLP:conf/uic/YangFW18} advocates boosting the recommendation diversity from the item-diversity point of view, where a variance minimization regularization term is adopted to prevent biased predictions of item potential groups. Besides, an end-to-end graph-based model is developed \cite{DBLP:conf/www/ZhengGCJL21} for diversified recommendations. To better balance accuracy and diversity, \cite{DBLP:journals/ipm/IsufiPH21} introduces graph convolutions to diversify user-item similarities and item-item dissimilarities based on a neighbor graph conveyed by historical interactions. To summarize, existing solutions along this direction either \textbf{(1)} are built on simple rank-based frameworks (say MF-based) with an extra regularization term, leading to limited performance, or \textbf{(2)} depend on external side information (e.g., tag and category), which might be challenging to collect in practice sufficiently. Unlike the existing literature, in this paper, we propose a diversity-promoting framework from the CML-based perspective due to its simplicity and efficacy in the RS community. Our proposed DPCML method does not simply introduce the diversity goal by regularization. Instead, we develop a novel multiple representation strategy and design an effective diversity control regularization scheme to serve our purpose better. By doing so, our proposed method can pursue a win-win situation for relevance and diversity \textbf{using collaborative data only without any side information.}
	\subsection{Recommendation against Joint Accessibility} \label{Sec.2.2}
	Recently, some studies \cite{DBLP:conf/eaamo/GuoKJG21, DBLP:conf/icml/CurmeiDR21, DBLP:conf/fat/DeanRR20} have pointed out a \textit{joint accessibility} problem in the recommendation, which determines the opportunities for users to discover interesting content. More precisely, joint accessibility measures whether an item candidate with size $K$ could be jointly accessed by a user in a Top-$K$ recommendation \cite{DBLP:conf/eaamo/GuoKJG21}. In other words, joint accessibility also somewhat captures a fundamental requirement of content diversity. If there are sufficient preference records of a target user, he/she should be able to be recommended any combination of $K$ items that he/she may be interested in. In this direction, noteworthy is the work present in \cite{DBLP:conf/eaamo/GuoKJG21}, which provides the theoretically necessary and sufficient conditions to meet joint accessibility. Subsequently, \cite{DBLP:conf/eaamo/GuoKJG21} proposes an alternative MF-based model (M2F) to improve joint accessibility. Formally, with respect to each user, it assigns $m$ feature vectors to users, and thus the predicted score of each item is defined as $
	s(j) = \max\limits_{i \in [m]} \boldsymbol{u}_i^{\top} \boldsymbol{v}_j$, where $\boldsymbol{u}_i, i \in [m]$ is the $i$-th user latent vector; $\boldsymbol{v}_j$ is the item feature and $[m] = \{1, \dots, m\}$. 
	Finally, M2F adopts the least square \cite{DBLP:conf/recsys/TakacsT12} loss to recover the missing values in the user-item matrix. The existing line of such work merely focuses on the MF-based algorithms, while we take a further step to explore the problem under the context of CML. It is also interesting to note that, under mild conditions, we could see that M2F is a particular case of our method (shown in Sec.\ref{sec.3.6}). In this sense, we generalize the original idea of joint accessibility.
	
	\subsection{General Metric Learning}\label{pami:sec2.5}
	Metric learning aims to learn a distance metric that can establish or reflect the similarities between all data points, where similar samples will be assigned smaller distances and dissimilar ones induce larger values \cite{DBLP:journals/symmetry/KayaB19,DBLP:journals/pami/MilbichRBO22}. Over the past two decades, metric learning has attracted significant research attention \cite{DBLP:conf/iccv/QianSSHTLJ19,DBLP:journals/pami/ZhengLZ21,DBLP:conf/iccv/HarwoodGCRD17,DBLP:conf/iccv/VasudevaDB0C21} in the machine learning community due to its promising performance over a wide range of downstream tasks. One of the most successful applications is the image retrieval and classifications \cite{DBLP:conf/eccv/MensinkVPC12, DBLP:journals/tmm/YaoSZYCW21, DBLP:conf/cvpr/WangSLRWPCW14,DBLP:conf/iccv/WangZWLL17}. Typically, \cite{DBLP:conf/cvpr/KarlinskySHSAFG19} proposes an end-to-end representative-based metric learning framework for image classifications and few-shot object detections. \cite{DBLP:journals/tmm/YaoSZYCW21} develops an adaptive metric learning method and proposes a unified multi-task optimization to serve the purpose of affective image retrieval and classification simultaneously. Apart from computer visions, a simple but effective RS framework called Collaborative Metric Learning (CML) \cite{hsieh2017collaborative,DBLP:conf/ijcai/FengLZCCY15, DBLP:conf/mm/BaoXMYCH19,DBLP:journals/pami/BaoX0CH23} is proposed inspired by the idea of the largest margin nearest neighbor algorithm (LMNN) \cite{DBLP:journals/jmlr/WeinbergerS09}. Generally speaking, following the principles of metric learning, the fundamental mechanism of current methods applied in various downstream tasks is very similar. Nonetheless, there are still a few technical differences when dealing with different tasks. Take CML and general metric learning for image retrieval and classifications as an example: (1) The primary concern is different. In terms of image retrieval, the goal is to determine the visual similarity between any two images in a unified space and then respond to the candidates given a query image. By contrast, CML cares about the similarities between users and items in the space, while the relationships between items are not explicitly considered. Meanwhile, different tasks usually require distinct metric spaces for accurate measurements. (2) The accessibility of data is also different. Under the context of implicit feedback, CML could only know a few positive user-item interactions, which belongs to the so-called one-class classification problem \cite{DBLP:conf/icdm/PanZCLLSY08}. Besides, for general metric learning, we can usually determine the exact ground-truth label for each sample or similarity for each pair in a supervised manner. The above two-fold factors motivate us to explore CML model designs, sampling and optimization strategies to unleash the power of metric learning in recommendations as much as possible. 
	\section{Preliminary \label{pami:pre}}
	Before presenting the diversity-promoting CML framework, we first make some brief reviews of the top-$N$ recommendation with implicit feedback and the One-way partial AUC (OPAUC) optimization problem.
	\subsection{Top-N Recommendation with Implicit Feedback} \label{major:sec3.1}
	In this paper, we focus on how to develop an effective CML-based recommendation system on top of the implicit feedback signals (say clicks, browses, and bookmarks).
	Assume there is a pool of users and items in the system, denoted by $\mathcal{U}=\{u_1, u_2, \dots, u_{|\mathcal{U}|}\}$ and $\mathcal{I} = \{v_1, v_2\, \dots, v_{|\mathcal{I}|}\}$, respectively. Here $|\cdot|$ denotes the cardinality of the set. For each user $u_i \in \mathcal{U}, i=1, 2, \dots, |\mathcal{U}|$, let $\mathcal{D}^+_{u_i} = \{v_1^+, v_2^+, \dots, v_{n_{i}^+}^+\}$ denote the set of items that user $u_i$ has interacted with (i.e., observed user-item interactions) and the rest of the items (i.e., unobserved interactions) are denoted by $\mathcal{D}^{-}_{u_i} = \{v_1^-, v_2^-, \dots, v_{n_{i}^-}^-\}$, where $n_i^+,n_i^-$ are the number of observed/unobserved interactions of user $u_i$. We have $\mathcal{I} = \mathcal{D}_{u_i} = \mathcal{D}^+_{u_i} \cup \mathcal{D}^-_{u_i}$ and $|\mathcal{I}| = n_i^+ + n_i^-$. In the standard settings of OCCF, one usually assumes that users tend to have a higher preference for the items contained in $\mathcal{D}^+_{u_i}$ than the items in $\mathcal{D}^-_{u_i}$. Therefore, given a target user $u_i \in \mathcal{U}$ and his/her historical interaction records, the goal of RS is to discover the most interested $N$ items by a score function $f_{\Theta}(v_j|u_i), v_j \in \mathcal{D}_{u_i}^-$, $\Theta$ is the corresponding learnable parameters, and then recommends the items with the top-$N$ (bottom-$N$) score. The top-$N$ item list for user $u_i$ is denoted as $\mathcal{I}_N^{u_i}$.
	
	\subsection{One-way Partial AUC Learning \label{rw:opauc}}
	Without loss of generality, we discuss the AUC learning problem for a specific target user $u_i$ throughout this section. To this end, we abbreviate the score function $f_{\Theta}(v_*|u_i)$ as $f_{\Theta}(v_*), v_* \in \mci$ for the sake of expressions. Note that, similar conclusions could be easily extended to all users. 
	
\noindent\textbf{AUC Learning.} The standard AUC is defined as the entire \textit{Area Under the ROC Curve (AUC)} obtained by plotting the True Positive Rate (TPR) against the False Positive Rate (FPR) of a given classifier with all possible thresholds \cite{DBLP:conf/nips/CortesM03,DBLP:journals/tnn/GultekinSRP20,DBLP:journals/corr/abs-2206-00260}. Mathematically, AUC could be expressed as follows: 
		\[
		\texttt{AUC}(f_\Theta) = \int_{0}^{1} \text{TPR}_{f_\Theta}\left(\text{FPR}_{f_\Theta}^{-1}(s)\right) ds,
		\]
		where $f$ represents the predictor, $\Theta$ is its parameters.

	Additionally, we have the following definitions of TPR and FPR:
		\begin{equation}\label{pami:eq1}
			\begin{aligned}
				\text{TPR}_{f_\Theta}(t) =& \mathbb{P}[{f_\Theta}(v_*) > t | v_* \in \mathcal{D}_*^+],\\
				\text{FPR}_{f_\Theta}(t) =& \mathbb{P}[{f_\Theta}(v_*) > t | v_* \in \mathcal{D}_*^-], \\
			\end{aligned}
		\end{equation}
		where $\mathcal{D}_*^+, \mathcal{D}_*^-$ denote the sets of positive and negative instances, respectively; $f_{\Theta}(v_*)$ is the probability that a sample is inferred as a positive one \cite{DBLP:conf/uai/LyuY18}. 

	Then, for a given $s \in [0, 1]$, we have
	\[
	\text{FPR}_{f_\Theta}^{-1}(s) = \inf \{t \in \mathbb{R}: \text{FPR}(t) \le s\}.
	\]
	
	Practically, if we assume that there are no tied scores between positive and negative samples, AUC is equivalent to the probability of a positive sample ranking higher than a negative one, which could be formulated as \cite{DBLP:journals/pami/YangXBCH22}:
		\[
		\texttt{AUC}(f_\Theta) = \mathbb{P}[f_{\Theta}(v_j^+) > f_{\Theta}(v_k^-)|v_j^+ \in \mathcal{D}_*^+, v_k^- \in \mathcal{D}_*^-].
		\]
	
	Because it is challenging to know the exact distributions of $\mathcal{D}_*^+, \mathcal{D}_*^-$, we usually consider the unbiased estimation of AUC as follows:
		\begin{equation}
			\hat{\texttt{AUC}}(f_\Theta) = 1 - \sum_{j=1}^{|\hat{\mathcal{D}}_*^+|}\sum_{k=1}^{|\hat{\mathcal{D}}_*^-|} \frac{\ell_{0-1}(f_{\Theta}(v_j^+) - f_{\Theta}(v_k^-))}{|\hat{\mathcal{D}}_*^+||\hat{\mathcal{D}}_*^-|},
		\end{equation}
		where $\hat{\mathcal{D}}_*^+, \hat{\mathcal{D}}_*^-$ represent the empirical data of positive and negative instances, respectively; $\ell_{0-1}(\cdot)$ is the $0-1$ loss with $\ell_{0-1}(z)=1$ if $z < 0$ and $\ell_{0-1}(z)=0$ otherwise.
	
	\noindent\textbf{One-way Partial AUC (OPAUC) Learning.} Unlike standard AUC measure, OPAUC merely pays attention to the performance within a specific region of FPR interval $s \in [\alpha, \beta]$, which is more practical in some real-world applications \cite{DBLP:conf/nips/WenXYHH21,DBLP:conf/iccvw/LiuHQGLJ19} such as recommendation and medical diagnosis. Without loss of generality, in this work, we care about a special case of OPAUC with $\alpha \equiv 0$ defined by:
		\[
		\texttt{OPAUC}(f_\Theta, \beta) = \int_{0}^{\beta} \text{TPR}_{f_\Theta}\left(\text{FPR}_{f_\Theta}^{-1}(s)\right) ds.
		\]
	
	Similar to standard AUC, as shown in \cite{DBLP:journals/corr/abs-2210-03967,DBLP:conf/icml/0001XBHCH21,DBLP:conf/icml/ZhuLWWY22}, OPAUC could be expressed as the possibility that a positive sample enjoys a higher score than a negative example within a specific range, i.e., 
		\begin{equation}
			\begin{aligned}
				\texttt{OPAUC}(f_\Theta, \beta) &= \\
				\mathbb{P}[f_{\Theta}(v_j^+) &> f_{\Theta}(v_k^-)|v_j^+ \in \mathcal{D}_*^+, v_k^- \in \mathcal{D}_*^-(\beta)], \\
			\end{aligned}
		\end{equation}
		where $\mathcal{D}_*^-(\beta)$ denotes the set of negative samples whose scores belong to $[s_{\beta}(f_{\Theta}), 1]$, i.e., $f_{\Theta}(v_k^-) \in [s_{\beta}(f_{\Theta}), 1], \ \ s.t. \ \ \mathbb{P}[f_{\Theta}(v_k^-) \ge s_{\beta}|v_k^- \in \mathcal{D}_*^-] = \beta.$
	
	Based on the above definition, the unbiased empirical version of OPAUC is expressed as follows:
		\begin{equation} \label{eq44}
			\hat{\texttt{OPAUC}}(f_\Theta, \beta) = 1 - \sum_{j=1}^{|\hat{\mathcal{D}}_*^+|}\sum_{k=1}^{|\hat{\mathcal{D}}_*^-(\beta)|} \frac{\ell_{0-1}(f_{\Theta}(v_j^+) - f_{\Theta}(v_k^-))}{|\hat{\mathcal{D}}_*^+||\hat{\mathcal{D}}_*^-(\beta)|},
		\end{equation}
		where $N_{\beta}^- := |\hat{\mathcal{D}}_*^-(\beta)| = \lfloor|\hat{\mathcal{D}}_*^-| \cdot \beta \rfloor$ and $\hat{\mathcal{D}}_*^-(\beta)$ is the subset of top-ranked $N_{\beta}^-$ negative samples, i.e., the negative examples with top-$N_{\beta}^-$ largest scores would be leveraged to compute OPAUC within FPR range $[0, \beta]$.
	
	Intuitively, according to (\ref{eq44}), we expect to obtain a well-performed model that induces a large value of OPAUC (preferably equal to $1$). In this sense, to maximize OPAUC, one usually needs to minimize the right term in (\ref{eq44}):
		\begin{equation}
			\begin{aligned}
				\max_{\Theta} \ \ \hat{\texttt{OPAUC}}(f_\Theta, \beta) &= \\ \min_{\Theta} \ \ & \sum_{j=1}^{N^+}\sum_{t=1}^{N_{\beta}^-} \frac{\ell_{0-1}(f_{\Theta}(v_j^+) - f_{\Theta}(v_{[t]}^-))}{N^+N_{\beta}^-},
			\end{aligned}
		\end{equation}
		where for simplicity we set $N^+:= |\hat{\mathcal{D}}_*^+|$, and $v_{[t]}^-$ is the sample induced the top-$t$-th score among all negative data.
	
	\begin{figure*}[!t]
		\centering
		\subfigure[MovieLens-1M]{
			\includegraphics[width=0.488\columnwidth]{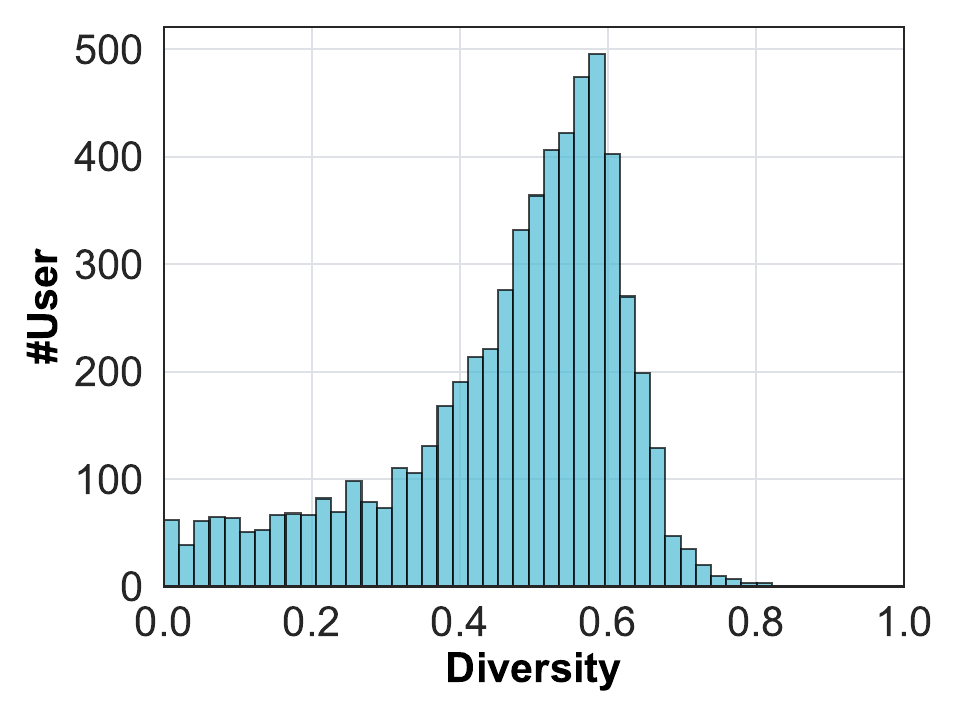}
			\label{ml-1m_div}
		}
		\subfigure[MovieLens-10M]{
			\includegraphics[width=0.488\columnwidth]{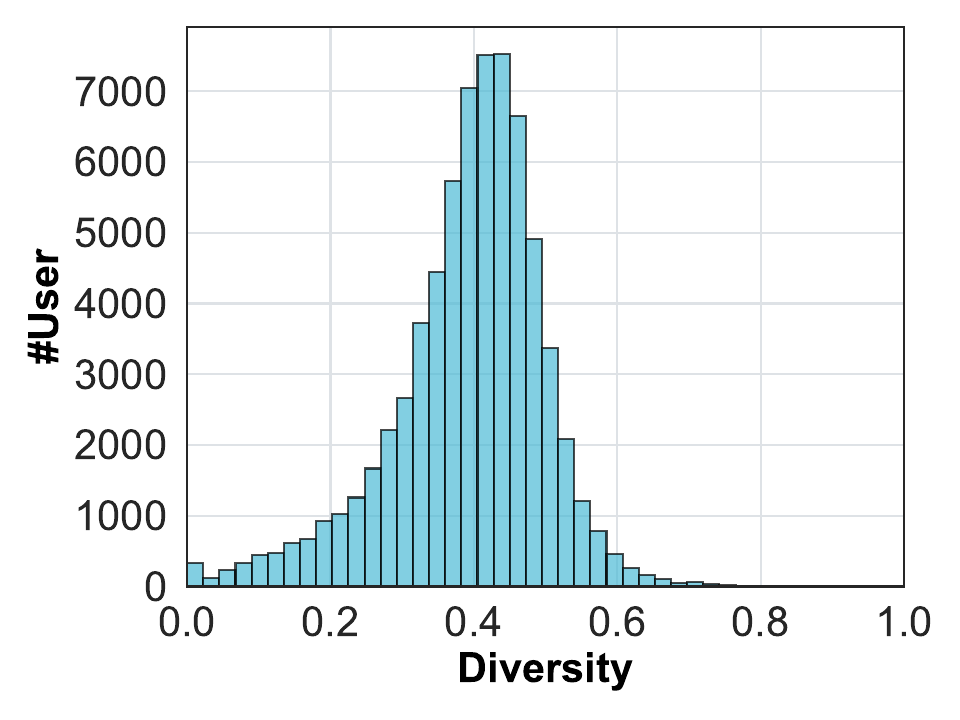}
			\label{ml-10m_div}
		}
		\subfigure[MovieLens-1M]{
			\includegraphics[width=0.478\columnwidth]{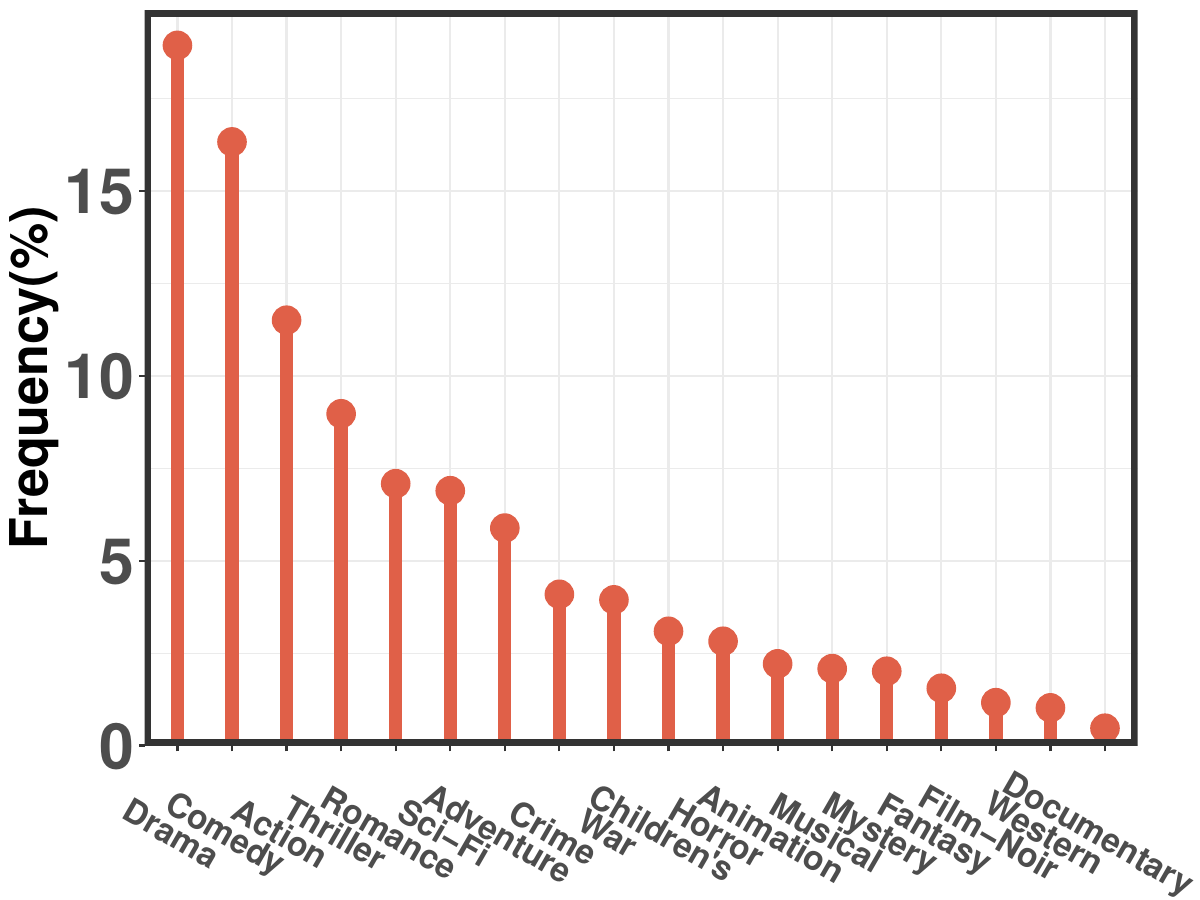}
			\label{ml-1m_freq}
		}
		\subfigure[MovieLens-10M]{
			\includegraphics[width=0.478\columnwidth]{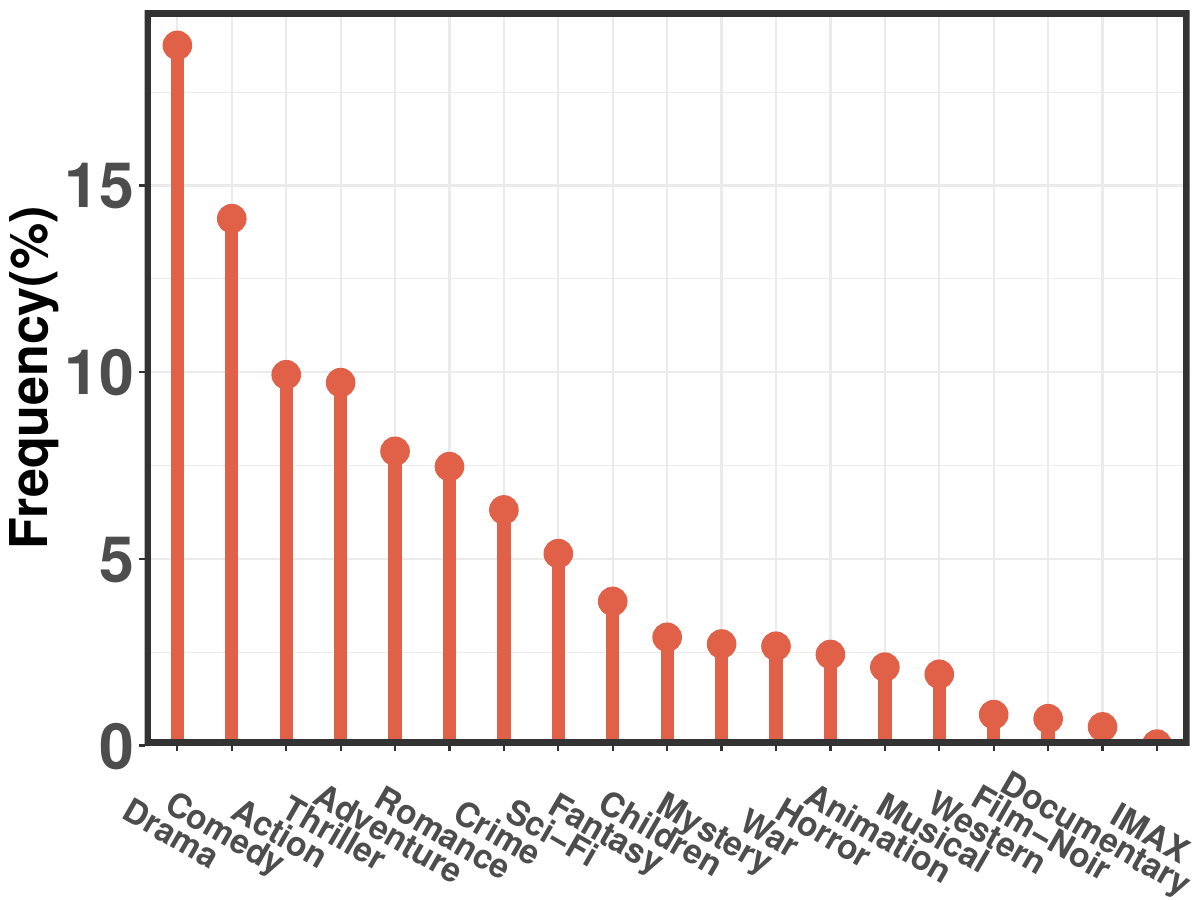}
			\label{ml-10m_freq}
		}
		\caption{Motivating visualizations on MovieLens-1M and MovieLens-10M datasets, where \textbf{(a)}, \textbf{(b)} are the statistics of users' preference diversity and \textbf{(c)}, \textbf{(d)} are the item category distribution, respectively.}
		\label{fig:all}
	\end{figure*}
	
	Furthermore, because $\ell_{0-1}$ is non-differentiable, the existing literature often adopts some \textbf{convex surrogate loss} replacing $\ell_{0-1}$ and thus optimize the following term:
		\begin{equation} \label{opacu_eq6}
			\min_{\Theta} \ \sum_{j=1}^{N^+}\sum_{t=1}^{N_{\beta}^-} \frac{\ell_{surr}(f_{\Theta}(v_j^+) - f_{\Theta}(v_{[t]}^-))}{N^+N_{\beta}^-},
		\end{equation}
		where $\ell_{surr}$ is the surrogate loss, typically using hinge loss and square loss \cite{DBLP:conf/ijcai/GaoZ15,DBLP:conf/nips/YingWL16}. Due to the space limitation, we refer the interested readers to the studies \cite{DBLP:journals/corr/abs-2203-15046,DBLP:journals/corr/abs-2203-14177} for more presentations. 
	
	\section{Methodology}\label{pami:sec4}
	In this section, we first present a motivating example to show the problem of existing CML-based studies. Then, we elaborate on our proposed Diversity-Promoting Collaborative Metric Learning (DPCML) algorithm, including the multi-vector user representation strategy and the Diversity Control Regularization Scheme (DCRS). Finally, we demonstrate that our DPCML could be regarded as a general framework for the joint accessibility problem.
	
	\subsection{Motivating Example} \label{Sec3.2}
	We start with a definition of the preference diversity of users.
	
	\begin{defi} [Preference Diversity] \label{div_defi} Assume that there exists an attribute set $\mathcal{T} = \{\mathcal{T}(v_1), \mathcal{T}(v_2), \dots, \mathcal{T}(v_{|\mci|})\}$ in a typical RS, where $\mathcal{T}(v_j) = \{t_1, t_2, \dots, t_{T_j}\}$ contains the attribute information of item $v_j$ (e.g., the genres of a movie) and $T_j$ is the number of attributes. Given a user $u_i$ and interaction records $\mathcal{D}_{u_i}^+$, the preference diversity is defined as follows:
		\begin{equation} \nonumber
			\begin{aligned}
				\mathsf{Div}(u_i) = \frac{\sum\limits_{v_j, v_k \in\mathcal{D}_{u_i}^+, v_j \neq v_k} \mathbb{I}\left[\mathcal{T}(v_j) \cap \mathcal{T}(v_k) = \varnothing \right]}{|\mathcal{D}_{u_i}^+| (|\mathcal{D}_{u_i}^+| - 1)},
			\end{aligned}
		\end{equation}
		where $\mathbb{I}(x)$ is an indicator function, i.e., returns $1$ if the condition $x$ holds, otherwise $0$ is returned.
	\end{defi}
	
	\begin{rem}
		Intuitively, the range of $\mathsf{Div}(u_i)$ is among $[0, 1]$, and its value measures the diversity of $u_i$'s preference to a certain extent. That is to say, if items among the historical interaction records of users are irrelevant, there should induce a large value (e.g., $\mathsf{Div}(u_i) = 1$), implying the diversity of their preferences. If the opposite is the case, the value is small. This means users may have narrow interests where only some unique attributes appeal to them.
	\end{rem}
	
	Based on Def.\ref{div_defi}, we visualize the user preferences on two real-world benchmark datasets, including \textbf{MovieLens-1M} and \textbf{MovieLens-10M}. The detailed information of datasets is listed in Tab.\ref{table1}. Here we adopt the movie genres as the attribute set $\mathcal{T}$ because such information is easy to obtain. The results are shown in Fig.\ref{fig:all}. From the results, we can make the following observations. First, only a few users have limited interest. Moreover, most of the users have a diversity value spaning $(0, 0.8]$, suggesting that they have multiple categories of interests. Finally, there are very few users with high preference diversity (at the lower-right corner) in both figures. This is a convincing case in the real-world recommendation since most users usually have interests in a couple of movie genres but not all.
	
	\textbf{Motivation and Discussion}. Through the above example, the key information is that users usually have multiple categories of preference in real-world recommendations. This poses a critical challenge to the current CML framework. Specifically, following the convention of RS, the existing CML-based methods leverage unique representations of users to model their preferences. Facing the multiplicity of user intentions, such a paradigm may induce preference bias due to the limited expressiveness, especially when the item category distribution is imbalanced. Fig.\ref{fig:all}-(c) and Fig.\ref{fig:all}-(d) visualize the item distribution on MovieLens-1M and MovieLens-10M datasets. We see that both of them are imbalanced. In this case, as shown in Fig.\ref{motivation}-(b), CML would pay more attention to the \textbf{majority} interest of users, making the unique user embedding close to the items with the science fiction (Sci-Fi) category. In this way, the \textbf{minority}  interest of the user (i.e., Horror movies) would be ignored by the method, inducing performance degradation. This motivates us to explore diversity-promoting strategies on top of CML.
	
	\subsection{Diversity-Promoting Collaborative Metric Learning \label{sec.3.3}}
	
	\subsubsection{Multi-vector Collaborative Metric Learning} \label{major:sec.4.2.1}
	To address the preference bias of CML, we advocate learning a set of multiple representations for each user $u_i$ instead of only unique embeddings, as depicted in Fig.\ref{motivation}-(c). Meanwhile, each item is still represented as one vector in the joint user-item Euclidean metric space. 
	
	Let $C_{u_i}$ ($C_{u_i} \ge 1$) denote the number of vectors for each user $u_i, u_i \in \mcu$. To obtain multiple representations, each user $u_i$ will be projected into the metric space via the following lookup transformations \cite{DBLP:conf/wsdm/WangF0NC21, DBLP:conf/sigir/Wang0WFC19, DBLP:conf/icml/WuZGYZ21}:
	\begin{equation} \label{pami:eq7}
		\boldsymbol{g}_{u_i}^c = \boldsymbol{P}_{c}^\top\boldsymbol{u}_i, \ \ \forall c, u_i, \ c \in [C_{u_i}], \ \  u_i \in \mcu,
	\end{equation}
	where $\boldsymbol{g}_{u_i}^c \in \mathbb{R}^d$ is a representation vector of user $u_i$; $[C_{u_i}]$ is the set $\{1, 2, \dots, C_{u_i}\}$; $\boldsymbol{P}_{c} \in \mathbb{R}^{|\mathcal{U}| \times d}$ is a learned transformation weight; $d$ is the dimension of space and $\boldsymbol{u}_i \in \mathbb{R}^{|\mcu|}$ is a one-hot encoding that the nonzero elements correspond to its index of a particular user ${u}_i$.
	
	Similarly, we apply the following transformation to each item $v_j$:
	\begin{equation} \label{pami:eq8}
		\boldsymbol{g}_{v_j} = \boldsymbol{Q}^\top \boldsymbol{v}_j, \ \ \forall v_j \in \mci,
	\end{equation}
	where $\boldsymbol{g}_{v_j} \in \mathbb{R}^d$ is the embedding of item $v_j$; $\boldsymbol{Q} \in \mathbb{R}^{|\mathcal{I}| \times d}$ is the learned transformation weight and $\boldsymbol{v}_j \in \mathbb{R}^{|\mathcal{I}|}$ is a one-hot embedding of item $v_j$.
	
	After unifying all users and items into a joint metric space, we need to seek a score function to express the target user $u_i$'s preference toward an item under the context of the multiple representation strategy. To do this, we define the score function by taking the minimum item-user Euclidean distance among the user embedding set:
	\begin{equation}\label{eq3}
		s(u_i, v_j) = \min\limits_{c \in [C_{u_i}]} \|\boldsymbol{g}_{u_i}^c - \boldsymbol{g}_{v_j}\|^2, \forall v_j \in \mci.
	\end{equation}
	
	Equipped with this formulation, the model can now pay attention to the potential items that fit one of the user preferences. If user $u_i$ has interacted with item $v_j$, there should be a small value with respect to $s(u_i, v_j)$. If the opposite is the case, we then expect to see a large $s(u_i, v_j)$. Mathematically, the following inequality should be satisfied to reflect the relative preference of $u_i$ in the learned Euclidean space:
	\begin{equation}
		\begin{aligned}
			\ \  s(u_i, v_j^+) < s(u_i, v_k^-), & \ \forall v_j^+ \in \mathcal{D}_{u_i}^+, \ \forall v_k^- \in \mathcal{D}_{u_i}^-. \\
		\end{aligned}\label{Eq3ss}
	\end{equation}  
	
	Therefore, given the whole sample set $\mathcal{D} = \mathop{\cup}\limits_{u_i \in \mathcal{U}} \ \mathcal{D}_{u_i}$, we adopt the following pairwise learning problems \cite{hsieh2017collaborative, tran2019improving,DBLP:conf/nips/LeiLK20,DBLP:journals/corr/abs-2203-15046} to achieve such goal:
	\begin{equation} \label{eqn4}
		\min\limits_{\boldsymbol{g}} \ \ \hat{\mathcal{R}}_{\mcd, \bmg},
	\end{equation}
	where, $\forall v_j^+ \in \mathcal{D}_{u_i}^+, \ \forall v_k^- \in \mathcal{D}_{u_i}^-$, we have
	\begin{equation}\label{cml_eq}
		\hat{\mathcal{R}}_{\mcd, \bmg} = \frac{1}{|\mcu|} \sum_{u_i \in \mcu} \frac{1}{n_i^+n_i^-} \sum_{j=1}^{n_i^+} \sum_{k=1}^{n_i^-} \ell^{(i)}_g(v_j^+, v_k^-),
	\end{equation}
	\begin{equation}\label{margin_eq}
		\ell^{(i)}_g(v_j^+, v_k^-) = [\lambda + s(u_i, v_j^+) - s(u_i, v_k^-)]_+,
	\end{equation}
	$[a]_+ = \max(0, a)$ represents the hinge function, and $\lambda > 0$ is a safe margin.
	
	According to (\ref{eqn4}), we have the following explanations. At first, optimizing the above problem could pull the observed items close to the users and push the unobserved items away from the observed items. This achieves our goal of preserving user preferences in the Euclidean space. Then, as shown in Fig.\ref{motivation}-(c), equipped with multiple representations for each user, DPCML would exploit different user vectors to focus on diverse interest groups. In this sense, the minority interest groups can also be modeled well, alleviating the preference bias issue caused by the traditional CML. Last but not least, one appealing property is that DPCML also preserves the triangle inequality for the items falling into the same interest group.

	\noindent\textbf{Discussions.} We realize that our idea of introducing multiple vectors for each user to capture their diverse preferences is somewhat similar to learning multiple semantic notions studied in the conditional similarity learning (CSL) paradigm \cite{DBLP:conf/cvpr/VeitBK17,DBLP:conf/cvpr/YeSZ22,DBLP:conf/iccv/TanVSP19,DBLP:conf/iccv/NigamTR19}. However, there are a few technical differences between ours and CSL: 1) The primary goal of CSL is to determine the relevances between images (i.e., \textbf{visual-only} similarity) in which all objects will be equipped with multiple representations according to some known/unknown conditional masks. By contrast, we merely \textbf{consider multiple vectors for users} while items are still expressed as \textbf{a single embedding}. 2) Most importantly, under the paradigm of CSL, each notion of similarity will be separately determined in a semantically distinct subspace. Yet, in this work, the user preference toward an item is still \textbf{measured in the unified metric space}, where different embeddings of users are adaptively activated to accommodate different interest clusters.

	\subsubsection{User Representation Assignment Strategies \label{pami:sec6.1}}
	So far, the central challenge is determining a proper $C_{u_i}$ for each user. To this end, we develop two feasible assignment schemes for DPCML, including Basic Preference Assignment (BPA) and Adaptive Preference Assignment (APA). 

	\noindent \textbf{\textit{Basic Preference Assignment (BPA) Strategy.}} A rough way is to assume that all users have the same number of interest clusters in the RS. That is to say, $\forall u_i \in \mcu$, in Sec.\ref{major:sec.4.2.1}, we employ $C_{u_i} = C$ to capture the diverse preferences of users, where $C > 1$.

	Although DPCML with the above BPA strategy has already shown significant improvements in \cite{DPCML}, it is apparent that different users generally demonstrate different preferences (both in quantity and category) in a practical RS with high probability, as shown in Fig.\ref{fig:all}. In this sense, \textit{BPA will fail to capture all preferences accurately, degrading the final recommendation performance.}

\begin{figure}
		\begin{center}
		\includegraphics[width=0.85\columnwidth]{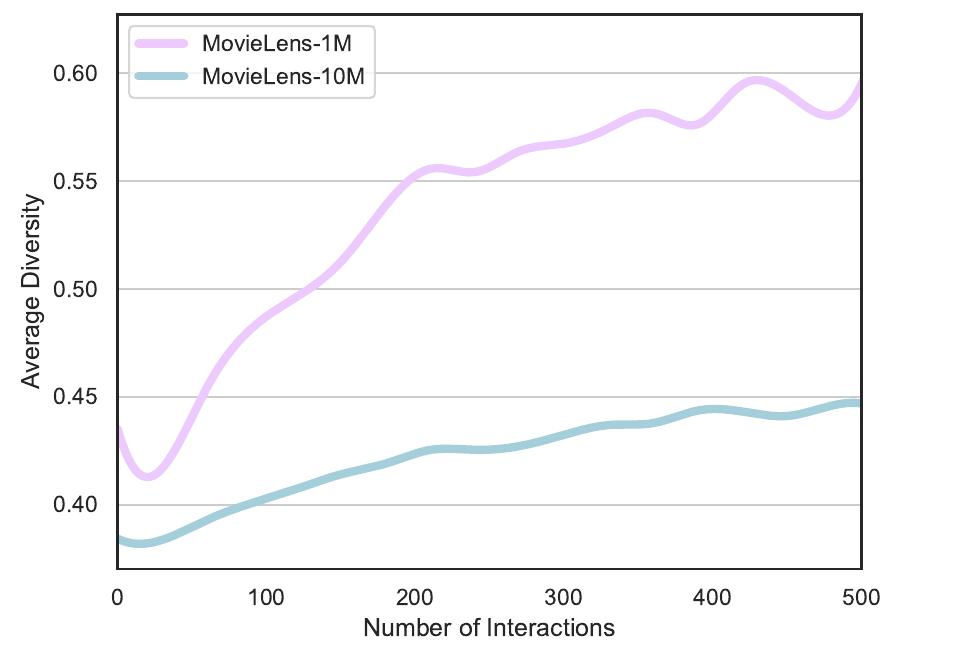}
		\end{center}
		\caption{The relationship between the users' interest diversity and their interaction lengths.}
		\label{APA_demo}
	\end{figure}

	\noindent \textbf{\textit{Adaptive Preference Assignment (APA) Strategy.}} To tackle this challenge, we further explore a more proper assignment rule. Note that, we first realize that it is almost impossible to obtain precisely the ground-truth quantity of each user preference group. This is because their preferences toward those massive unobserved commodities are usually not measurable. Motivated by this fact, we turn to develop a heuristic but effective strategy to determine the value of $C_{u_i}$ for each user adaptively. Our basic intuition here is that the preference patterns of \textbf{users with more observed interaction records are generally more diverse than those fewer ones with a high probability}. We empirically visualize the ``Average Diversity'' of users vs. different ``Interaction Lengths'' in Fig.\ref{APA_demo}, where the results on MovieLens-1M and MovieLens-10M datasets show that the users' preference diversity is almost positively correlated with the lengths of their interactions. As a result, we regard that the number of diverse vectors $C_{u_i}$ should be related to the size of users' historical records to accommodate the users' preferences better. To do this, an \textbf{A}daptive \textbf{P}reference \textbf{A}ssignment (short for APA) strategy is proposed. In practice, such an assignment strategy should satisfy the following properties:
		\begin{enumerate}
			\item[\textbf{(P1)}] The assigned number of clusters ($C_{u_i}$) should be positively correlated with the size of historical observations ($|\mathcal{D}_{u_i}^+|$), i.e., $\frac{dC_{u_i}}{d|\mathcal{D}_{u_i}^+|} > 0$.
			\item[\textbf{(P2)}] The magnitude of $C_{u_i}$ should become saturated gradually as $|\mathcal{D}_{u_i}^+|$ grows. In other words, the marginal benefit of $|\mathcal{D}_{u_i}^+|$ with respect to $C_{u_i}$ should be decreasing. In this sense, we have $\frac{d^2C_{u_i}}{d(|\mathcal{D}_{u_i}^+|)^2} < 0$.
			\item[\textbf{(P3)}] Since there are a finite number of item categories, the number of clusters should also be finite. In this sense, $\exists 0 < C < \infty$, such that $C_{u_i} < C$ holds for all $(u_i, v_j)$. 
		\end{enumerate}
		According to these three properties, we propose the following APA scheme:
		\begin{equation}
			C_{u_i} = \max(C_1,\lfloor\log_{a}(|\mathcal{D}_{u_i}^+|)\rfloor), \ \ \forall u_i \in \mathcal{U},
		\end{equation}
		where $C_1 >0$ is an integer parameter. Here we introduce a new hyperparameter $a$ to adjust the sparsity of the clusters. Specifically, the larger the $a$ is, the more sparse the number of clusters is. 

	\subsubsection{Diversity Control Regularization Scheme} \label{Sec.3.4}
	In practice, we note that a proper regularization scheme is crucial to accommodate the multi-vector representation strategy. Here we focus on the diversity within the embedding sets of a given user. Such diversity is defined as the average pairwise distance among the $C_{u_i}$ user embeddings for user $u_i$, i.e.,
	\begin{equation*}
		\delta_{\bmg, u_i} = \frac{1}{2C_{u_i}(C_{u_i}-1)} \sum_{c_{1}, c_2 \in [C_{u_i}]} \|\boldsymbol{g}_{u_i}^{c_1} - \boldsymbol{g}_{u_i}^{c_2}\|^2. \\
	\end{equation*}
	Based on the definition, we argue that one should attain a proper $\delta_{\bmg, u_i}$ to get a good performance since extremely large/small values of $\delta_{\boldsymbol{g}, u_i}$ might be harmful to the generalization error. It is easy to see that if $\delta_{\boldsymbol{g}, u_i}$ is extremely small, the embeddings for a given user are very close to each other such that the multi-vector representation strategy degenerates to the original single-vector representation. This increases the model complexity with few performance gains and obviously will induce overfitting. On the other hand, a too large diversity might also induce overfitting. It might be a bit confusing at first glance. But, imagine that when some noise observations or extremely rare interests far away from the normal patterns exist in the data, having a large diversity will make it easier to overfit such data. Moreover, it is also a natural assumption that a user's interests should not be too different, as validated in Fig.\ref{fig:all}. In this sense, the distance across different user embeddings should remain at a moderate magnitude. 
	
	Therefore, controlling a proper diversity is essential for the multi-vector representation. To do this, we put forward the following diversity control regularization scheme (DCRS):
	\begin{equation} \label{eqn5}
		\hat{\Omega}_{\mcd, \bmg} = \frac{1}{|\mcu|} \sum_{u_i \in \mcu} \psi_{\bmg}(u_{i}),
	\end{equation}
	where, we have
	\begin{equation}\nonumber
		\begin{aligned}
			\psi_{\bmg}(u_{i}) &= [\delta_1 - \delta_{\bmg, u_i}]_+ + [\delta_{\bmg, u_i} - \delta_2]_+, \\
		\end{aligned}
	\end{equation}
	and $\delta_1$, $\delta_2$ are two threshold parameters with $\delta_1 \le \delta_2$. Intuitively, optimizing (\ref{eqn5}) ensures that the diversity of user's vectors lies between $\delta_1$ and $\delta_2$.

	\subsubsection{Final Optimization Goal of DPCML}\label{major:sec4.2.4}
	Finally, we arrive at the following optimization problem for our proposed DPCML:
	\begin{equation}
		\begin{aligned}
			\min\limits_{\bmg} \hat{\mcl}_{\mcd}(\bmg),
		\end{aligned}
	\end{equation}
	where
	
	\begin{equation}\label{pami:eq9}
		\hat{\mcl}_{\mcd}(\bmg) = \hat{\mathcal{R}}_{\mcd, \bmg} + \eta \cdot 	\hat{\Omega}_{\mcd, \bmg},
	\end{equation}
	and $\eta$ is a trade-off hyper-parameter.
	
	When the training is completed,  one can easily carry out recommendations by choosing the items with the smallest $s(u_i, v_j), \forall v_j, v_j \in \mci$. 
	
	\subsection{General Framework of Joint Accessibility} \label{sec.3.6}
	Now, we expect to provide another intriguing perspective of our proposed method. As we discussed in Sec.\ref{Sec.2.2}, equipped with a multiple set of representations for each user, our proposed algorithm could be treated as a Generalized Framework against the Joint Accessibility (GFJA) issue. To see this, if we restrict the user and item embeddings within a unit sphere, then the score function (\ref{eq3}) degenerates to :
	\begin{equation}\label{eq3123}
		\begin{aligned}
			s(u_i, v_j) &= \min\limits_{c \in [C_{u_i}]} \left(1 - \hat{\bmg}^c_{u_i} \boldsymbol{g}_{v_j}\right), \\
			s.t. \ \ \|\boldsymbol{g}^c_{u_i}\| &= 1,  \forall u_i \in \mcu,\\
			\ \ \|\bmg_{v_j}\| &= 1,  \forall v_j \in \mci,
		\end{aligned}
	\end{equation}
	where $\hat{\bmg}^c_{u_i} \in \mathbb{R}^{1 \times d}$ represents the transpose vector of $\boldsymbol{g}^c_{u_i} \in \mathbb{R}^d$. Therefore, to minimize (\ref{eq3123}), one only needs to maximize the following equivalent problem:
	
	\begin{equation}\label{mf_3123}
		\begin{aligned}
			\hat{s}(u_i, v_j) &= \max\limits_{c \in [C_{u_i}]} \  \hat{\bmg}^c_{u_i} \boldsymbol{g}_{v_j}, \\
			s.t. \ \ \|\hat{\bmg}^c_{u_i}\| &= 1,  \forall u_i \in \mcu,\\
			\ \ \|\bmg_{v_j}\| &= 1,  \forall v_j \in \mci,
		\end{aligned}
	\end{equation}
	which is exactly the original form of the joint accessibility model \cite{DBLP:conf/eaamo/GuoKJG21, DBLP:conf/icml/CurmeiDR21, DBLP:conf/fat/DeanRR20}.

	\section{Generalization Analysis} \label{sec:bound}
	In this section, we present a systematic theoretical analysis of the generalization ability of our proposed algorithm. Following the standard learning theory, deriving a uniform upper bound of the generalization error relies on the proper measure of its complexity over the given hypothesis space $\mathcal{H}$. The most common complexity to achieve this is the Rademacher complexity \cite{DBLP:conf/colt/BartlettM01, DBLP:books/daglib/0034861, DBLP:journals/ijon/LeiDB16}, which is derived from the symmetrization technique as an upper bound for the largest deviation over a given hypothesis space $\mathcal{H}$:
	\[
	\mathbb{E}_{\mathcal{D}}\left[\sup_{f \in \mathcal{H}} \mathbb{E}_{\mathcal{D}}(\hat{\mathcal{R}}_{\mathcal{D}}) - \hat{\mathcal{R}}_{{\mathcal{D}}} \right].
	\]
	However, the standard symmetrization technique requires the empirical risk $\hat{\mathcal{R}}_{\mathcal{D}}$ to be a sum of independent terms, which is not applicable to the CML-based methods. Specifically, we notice that \textbf{each positive (negative) item will be paired with all negative (positive) samples} in (\ref{eqn4}). In this sense, as long as one of them is the same (i.e., $v_j^+=\tilde{v}_j^+$ or $v_k^- = \tilde{v}_k^-$), the terms $\ell^{(i)}_g(v_j^+, v_k^-)$ and $\ell^{(i)}_g(\tilde{v}_j^+, \tilde{v}_k^-)$ would be interdependent. To overcome this challenge, we turn to leverage another complexity measure, i.e., covering number. The necessary notations are summarized as follows. 
	\begin{defi} [$\epsilon$-Covering] \label{def2} \cite{ledoux1991probability} Let $(\mathcal{F}, \rho)$ be a (pseudo) metric space, and $\mathcal{G} \subseteq \mathcal{F}$. $\{f_1, \dots, f_K\}$ is said to be an $\epsilon$-covering of $\mathcal{G}$ if $\mathcal{G} \subseteq \mathop{\cup}\limits_{i=1}^K \mathcal{B}(f_i, \epsilon)$, i.e., $\forall g \in \mathcal{G}$, $\exists i$ such that $\rho(g, f_i) \le \epsilon$.
	\end{defi}
	
	\begin{defi} [Covering Number] \label{def3} \cite{ledoux1991probability} According to the notations in Def.\ref{def2}, the covering number of $\mathcal{G}$ with radius $\epsilon$ is defined as:
		\begin{equation} \nonumber
			\mathcal{N}(\epsilon;\mathcal{G}, \rho) = \min\{n: \exists \epsilon-covering \ over \ \mathcal{G} \ with \ size \ n\}
		\end{equation}
	\end{defi}
	
	With the above definitions, we further have the following assumption and lemma to help us derive the generalization bound.
	\begin{assu} [Basic Assumptions] \label{assu1} We assume that all the embeddings of users and items are chosen from the following embedding hypothesis space:
		\begin{equation}
			\mathcal{H}_R = \left\{\bmg: \bmg \in \mathbb{R}^d, \|\bmg\| \le r\right\},
		\end{equation}
		where $\boldsymbol{g}^c_{u_i} \in \mathcal{H}_R, u_i \in \mcu, c \in [C]$ and $\boldsymbol{g}_{v_j} \in \mathcal{H}_R, v_j \in \mci$.
	\end{assu}
	
	\begin{lem} \label{covering_lem} \cite{DBLP:conf/iclr/LongS20,DBLP:conf/icml/LiL21, DBLP:journals/jc/Zhou02}
		The covering number of the hypothesis class $\mathcal{H}_R$ has the following upper bound: 
		\begin{equation}
			\log \mathcal{N}(\epsilon;\mathcal{H}_R, \rho) \le d \log \left(\frac{3r}{\epsilon}\right),
		\end{equation}
		where $d$ is the dimension of embedding space. 
	\end{lem}
	
	Based on the above introductions, we have the following results. \textbf{\textit{Due to space limitations, please refer to Appendix.\ref{supp.sec.b} for all proofs in detail.}}
	
	\begin{theorem} [Generalization Upper Bound of DPCML] \label{them1} Let $\expe [\hat{\mathcal{L}}_{\mcd}(\bmg)]$ be the population risk of $\hat{\mathcal{L}}_{\mcd}(\bmg)$. Then, $\forall \ \bmg \in \mathcal{H}_R$, with high probability, the following inequation holds:
		\begin{equation}
			\begin{aligned} \label{eq:them1}
				\left| \hat{\mathcal{L}}_{\mcd}(\bmg) - \expe [\hat{\mathcal{L}}_{\mcd}(\bmg)]\right| \le \sqrt{\frac{2d\log \left(3r \tilde{N}\right)}{\tilde{N}}},
			\end{aligned}
		\end{equation}
		where we have
		\begin{small}
			\begin{equation} \nonumber
				\tilde{N} = \left(4r^2\sqrt{\left(\frac{(4 + \eta)^2}{|\mcu|} + \frac{2}{|\mcu|^2} \sum_{u_i \in \mcu} \left(\frac{1}{n_i^+} + \frac{1}{n_i^-}\right)\right)}\right)^{-2}
			\end{equation}
		\end{small}
	\end{theorem}
	Intriguingly, we see that our derived bound does not depend on $C$. This is consistent with the over-parameterization phenomenon \cite{DBLP:journals/corr/abs-2109-02355, DBLP:conf/iclr/NakkiranKBYBS20}. On top of Thm.\ref{them1}, we have the following corollary.
	\begin{coro} \label{cor1} DPCML could enjoy a smaller generalization error than CML.
	\end{coro}
	Therefore, we can conclude that DPCML generalizes to unseen data better than single-vector CML and thus improves the recommendation performance. This supports the superiority of our proposed DPCML from a theoretical perspective. In addition, we also empirically demonstrate this in the experiment Sec.\ref{pami:sec.7.6.4}.

	\section{OPAUC-Oriented Efficient Optimizations \label{pami:sec6.2}}
	Despite the strengths of DPCML in handling multiple user preferences, it would inevitably suffer from the heavy burden of computations due to the pairwise learning paradigm. To be specific, \textit{with respect to each user $u_i$, each positive item $v_j^+$ would be paired with all of the rest negative items $v_k^-$, which brings about an $\mathcal{O}(\sum_{u_i \in \mathcal{U}}n_i^+n_i^-C_{u_i})$ complexity for the full-batch calculation of (\ref{pami:eq9})}. Note that, here we ignore the complexity term of $\hat{\Omega}_{\mcd, \bmg}$ in (\ref{eqn5}) because $C_{u_i}$ is usually far less than the number of observed (unobserved) interactions, i.e., $C_{u_i} < n_i^+$ and $C_{u_i} \llless n_i^-$. Considering that there are tens of thousands of items in real-world recommendation systems, \textbf{directly optimizing (\ref{pami:eq9}) is not affordable.} Practically, this is a common challenge that almost all CML-based approaches have to confront \cite{DBLP:journals/pami/BaoX0CH23,DBLP:conf/www/WangX000C20} as discussed in Sec.\ref{pami:sec2.2}.

	\subsection{Training Acceleration with Negative Sampling}\label{major:sec6.1}
	Over the past decades, the RS community (not only limited to CML-based algorithms) has always been committed to dealing with this efficiency issue. Currently, the mainstream methods usually adopt the so-called \textit{negative sampling strategy} \cite{DBLP:conf/www/HeLZNHC17,DBLP:conf/sigir/Wang0WFC19,DBLP:conf/kdd/Wang00LC19,DBLP:journals/corr/abs-2302-03472,DBLP:conf/iccv/QianSSHTLJ19,DBLP:journals/pami/ZhengLZ21,DBLP:conf/iccv/HarwoodGCRD17,DBLP:conf/iccv/VasudevaDB0C21} to accelerate the training, where a few unobserved items would be selected from $\mathcal{D}_{u_i}^-$ and regarded as negative samples to the subsequent optimization process. Typically, a series of related studies have been successfully deployed in CML-based approaches, such as uniform sampling (UniS) \cite{DBLP:conf/incdm/MatsuiNYN21,DBLP:conf/recsys/RendleKZA20,DBLP:conf/mm/BaoXMYCH19,DBLP:conf/recsys/TranSHM21,DBLP:conf/icdm/PanZCLLSY08}, hard negative sampling (HarS) \cite{DBLP:conf/iccv/HenriquesCCB13,DBLP:conf/sigir/ZhangCWY13,hsieh2017collaborative}, popularity-aware \cite{DBLP:conf/kdd/ChenSSH17,DBLP:conf/sigir/WuVSSR19} and two-stage negative sampling strategies \cite{tran2019improving} (PopS and 2stS, respectively). 
	
	Without loss of generality, in this paper, we first consider two widely used negative sampling strategies for our proposed DPCML framework (i.e., UniS and HarS): 
	
	\noindent\textbf{(1) DPCML with Uniform Negative Sampling.} Assume that we expect to select $U$ unobserved samples as negatives for each user-item $(u_i, v_j^+)$ pair to learn DPCML. Under this circumstance, each positive $(u_i, v_j^+)$ interaction would be paired with $U$ negative items uniformly sampled from the following distribution:
		\begin{equation}\label{pami:eq23}
			\mathbb{P}^{\text{UniS}}(u_i, j) = [\mathbb{P}^{u_i}_{j1}, \mathbb{P}^{u_i}_{j2}, \dots \mathbb{P}^{u_i}_{jn_i^-}],
		\end{equation}
		where $\mathbb{P}^{u_i}_{jk} = \frac{1}{n_i^-}, \forall k\in \{1, \dots, n_i^-\}$ is the probability of item $v_k^-$ sampled as negative sample for user-item $(u_i, v_j^+)$.  
	
	Denote the sampled set for $(u_i, v_j^+)$ as $\mathcal{N}_{ij}^U$, which is obtained via sampling $U$ times without a replacement strategy. Then, we reach a random estimation of (\ref{cml_eq}) with uniform negative sampling:
		\begin{equation}\label{cml_sampling_unis}
			\hat{\mathcal{R}}_{\mcd, \bmg}^{\text{UniS}} = \frac{1}{|\mcu|} \sum_{u_i \in \mcu} \frac{1}{n_i^+U} \sum_{j=1}^{n_i^+} \sum_{v_k^- \in \mathcal{N}_{ij}^U} \ell^{(i)}_g(v_j^+, v_k^-).
		\end{equation}
		
		Performing this estimation in each mini-batch followed with an SGD update, we then reach the standard algorithm for UniS. By doing this, the heavy complexity of the original goal (\ref{pami:eq9}) is now reduced to $\mathcal{O}(\sum_{u_i \in \mathcal{U}}n_i^+UC_{u_i})$, which significantly improves the efficiency since $U \lll n_i^-$.
	
	\noindent\textbf{(2) DPCML with Generic Hardness-aware Negative Sampling.} Different from UniS, the \textbf{generic framework} of HarS aims to employ the top-$U$ informative negative samples for training to pursue better optimization performance. Like (\ref{cml_sampling_unis}), we can obtain a HarS-driven empirical estimation for the original CML objective:
		\begin{equation}\label{cml_sampling_hars}
			\hat{\mathcal{R}}_{\mcd, \bmg}^{\text{HarS}} = \frac{1}{|\mcu|} \sum_{u_i \in \mcu} \frac{1}{n_i^+U} \sum_{j=1}^{n_i^+} \sum_{v_k^- \in \mathcal{S}_{ij}^{\uparrow}(U)} \ell^{(i)}_g(v_j^+, v_k^-), 
		\end{equation}
		where $\mathcal{S}_{ij}^{\uparrow}(U)$ represents the subset of top-$U$ negative items for pair $(u_i, v_j^+)$ sorted by the ascent order of distances (\ref{eq3}). 
		
		\textbf{Note that, we have $U \equiv 1$ in the standard HarS.} In other words, regarding each user-item pair $(u_i, v_j^+)$, merely the closest negative pair $(u_i, v_k^-)$ among all unobserved items is used to compute loss and update the gradient, while the others are discarded \cite{DBLP:conf/sigir/ZhangCWY13,DBLP:conf/iccv/HarwoodGCRD17,DBLP:conf/iccv/VasudevaDB0C21}. However, finding the most useful samples from a tremendous unobserved items pool is challenging. To find the ``hardest'' sample as precisely as possible, current CML studies \cite{hsieh2017collaborative, DBLP:journals/pami/BaoX0CH23,DBLP:conf/mm/BaoXMYCH19} usually split HarS into two stages: For each user-item pair $(u_i, v_j^+)$ \textbf{(1)} uniformly sample $S$ candidates from all unobserved items (like UniS (\ref{pami:eq23})); \textbf{(2)} the item that causes the minimum Euclidean distance towards the target user $u_i$ among the sampled $S$ items is adopted (i.e., $U \equiv 1$) for (\ref{cml_sampling_hars}).
	\begin{rem} In practice, the uniform negative sampling (UniS) might be insufficient to pursue good performance, because UniS cannot constantly yield high-informative examples during optimization. Specifically, as the training progresses, most samples would be satisfied with the preference constraint (\ref{Eq3ss}). In this sense, the contribution from negative examples keeps on vanishing, which eventually leads to sub-optimal solutions\cite{tran2019improving,DBLP:conf/sigir/HeZKC16,DBLP:conf/uai/YuanX0GZCJ18}. By contrast, the conventional HarS, with the "hardest" negative samples (i.e., $U \equiv 1$), could enjoy high-quality model training and induce a competitive performance \cite{DBLP:journals/pami/SongHCLN22,DBLP:conf/iccv/ManmathaWSK17}. This comparison has been validated in the experiment part Sec.\ref{exp}, where CML-based algorithms learning with HarS outperform the UniS counterparts significantly. 
	\end{rem}
	
	\subsection{OPAUC-based Equivalence HarS Reformulation}
	Although HarS-induced CML approaches are promising to obtain a satisfactory recommendation performance, the reason behind their success is still mysterious. In this section, our primary concern is to explore the theoretical foundation of negative sampling from the OPAUC point of view. First, we show that HarS-based CML algorithms are equivalent to OPAUC maximization problems. Then, we derive the performance gap between the Top-$N$ recommendation and OPAUC, indicating that, with a proper FPR range, maximizing OPAUC would directly induce a better Top-$N$ recommendation result. Meanwhile, we intriguingly reveal that the default HarS, only considered the ``hardest" one for training, might degrade the Top-$N$ recommendation performance. Inspired by our findings, we advance a novel Differentiable Hardness-aware negative Sampling (DiHarS) strategy to address this issue.

	Specifically, according to the definition of the OPAUC in Sec.\ref{rw:opauc}, we can realize that HarS-based (DP)CML methods are a particular case of the OPAUC optimization problem. Namely, we have the following proposition:
		\begin{proposition}[Equivalent Reformulation of Generic HarS-based CML Framework] \label{pro1}
			Denote $\mathcal{S}_{ij}^{\uparrow}(U) = \{v_{[t]}^-\}_{t=1}^U$ as the subset of top-$U$ negative items for any $(u_i, v_j^+)$ pair. If we regard the user preference toward a positive/negative item (i.e., $s(u_i, v_j^+)$/$s(u_i, v_{[t]}^-)$) as a positive/negative prediction in (\ref{pami:eq1}) and select hinge loss as the surrogate loss in (\ref{opacu_eq6}), CML-based algorithms (i.e., $C_{u_i} \ge 1$) optimized by Hardness-aware negative sampling (\ref{cml_sampling_hars}) could be reformulated as the following per-user average OPAUC optimization problem: 
			\begin{equation}\label{prop1_eq27}
				\begin{aligned}
					\min\limits_{\bmg} \ \	\hat{\mathcal{R}}_{\mcd, \bmg}^{\text{HarS}} & \Leftrightarrow \max_{\boldsymbol{g}} \ \ \frac{1}{|\mcu|} \sum_{u_i \in \mcu} \hat{\texttt{OPAUC}}^{u_i}(s_{\boldsymbol{g}}, \beta_i)\\
					&\Leftrightarrow\min\limits_{\bmg} \ \ \frac{1}{|\mcu|} \sum_{u_i \in \mcu} \sum_{j=1}^{n_i^+} \sum_{t=1}^{U} \frac{\ell^{(i)}_g(v_j^+, v_{[t]}^-)}{n_i^+U}, 
				\end{aligned}	
			\end{equation}
			where, we define
			\[
			\hat{\texttt{OPAUC}}^{u_i}(s_{\boldsymbol{g}}, \beta_i) := \sum_{j=1}^{n_i^+} \sum_{t=1}^{U} \frac{\ell^{(i)}_g(v_j^+, v_{[t]}^-)}{n_i^+U}, \ \ \forall u_i \in \mcu,
			\]
			$\beta_i=\frac{U}{n_i^-}$ is the specific FPR value, $n_i^- = |\mathcal{D}_{u_i}^-|$ is the number of unobserved items for $u_i$ and $U$ is the sampling number.
	\end{proposition}
	
		\begin{rem}\label{pami:rem3} According to (\ref{prop1_eq27}), we can observe that pursuing preference consistency for each user $u_i$ could be separably regarded as an $\hat{\texttt{OPAUC}}^{u_i}(s_{\boldsymbol{g}}, \beta_i)$ maximization problem with $\beta_i = \frac{U}{n_i^-}$. In this sense, \textbf{the principle of traditional HarS ($U \equiv 1$) is only to consider the OPAUC metric within an FPR range $[0, \frac{1}{n_i^-}]$}. Taking a step further, we derive the performance relationship between the top-$N$ recommendation and OPAUC optimization, presented in Thm.\ref{pami:thm2} (Please refer to Sec.\ref{pami:used_metrics} for the details of these metrics). The result suggests that simply leveraging the single "hardest" sample (i.e., (\ref{cml_sampling_hars})) is insufficient to pursue a reasonable top-$N$ performance when $N > 1$. 
		\end{rem}
	
	\begin{theorem} \label{pami:thm2} Consider a top-$N$ recommendation task evaluated by Precision@N (P@N) and Recall@N (R@N) metrics and assume $n_i^+=|\mathcal{D}_{u_i}^+| \ge N$, $n_i^-=|\mathcal{D}_{u_i}^-| \ge N$, $\forall u_i \in \mathcal{U}$. Then, for any user $u_i$, the following conditions hold:
			\begin{equation}\label{pami:eq28}
				\begin{aligned}
					\text{P@}&N \ge \\ 
					&  \frac{1}{N}\left\lfloor \frac{(n_i^+ + N) - \sqrt{\mathcal{F}(n_i^+, N, -\hat{\texttt{OPAUC}}^{u_i}(s_{\boldsymbol{g}}, \beta_i))}}{2} \right\rfloor,
				\end{aligned}
			\end{equation}
			\begin{equation}\label{pami:eq29}
				\begin{aligned}
					\text{R@}&N  \ge \\ 
					&  \frac{1}{n_i^+}\left\lfloor \frac{(n_i^+ + N) - \sqrt{\mathcal{F}(n_i^+, N, -\hat{\texttt{OPAUC}}^{u_i}(s_{\boldsymbol{g}}, \beta_i))}}{2} \right\rfloor,
				\end{aligned}
			\end{equation}
			where the FPR range $\frac{N}{n_i^-} \le \beta_i \le 1$ and $\mathcal{F}(n_i^+, N, -\hat{\texttt{OPAUC}}^{u_i}(s_{\boldsymbol{g}}, \beta_i))$ represents an essential function that is \textbf{negatively} proportional to the value of $\hat{\texttt{OPAUC}}^{u_i}(s_{\boldsymbol{g}}, \beta_i)$:
			\begin{equation}\nonumber
				\begin{aligned}
					\mathcal{F}(n_i^+, N, -\hat{\texttt{OPAUC}}^{u_i}(s_{\boldsymbol{g}},\beta_i)) &= (n_i^+ + N)^2 - 4 n_i^+N \\ 
					+ 4 n_i^+N_i^{\beta_i} & \times (1 - \hat{\texttt{OPAUC}}^{u_i}(s_{\boldsymbol{g}}, \beta_i)),
				\end{aligned}
			\end{equation}
			and we denote $N_i^{\beta_i} = U = \lfloor n_{i}^- \cdot \beta_i \rfloor$ for the clear expressions of FPR range.
		\end{theorem}

		\begin{rem} Please see Appendix.\ref{pami:proof_thm2} for the proof. Note that, we do not derive the relationship of OPAUC between other ranking metrics for Top-$N$ recommendation (see Sec.\ref{pami:used_metrics}) such as \textbf{Normalized Discounted Cumulative Gain} (NDCG@$N$) and \textbf{Mean Reciprocal Rank} (MRR). Because OPAUC is somewhat consistent with those metrics that expect positive items to achieve higher ranks than those unobserved (or negative) ones. From Thm.\ref{pami:thm2}, we can draw the following important inspirations:
			\begin{enumerate}
				\item The value of OPAUC \textbf{positively correlates with} the performance of the Top-$N$ recommendation. This means that maximizing OPAUC is favorable for promoting the Top-$N$ recommendation results. In particular, when $\hat{\texttt{OPAUC}}^{u_i}(s_{\boldsymbol{g}}, \beta_i)$ tends to $1$, both $\text{P@}N$ and $\text{R@}N$ attain the maximum value, i.e, $\text{P@}N=1$ and $\text{R@}N = \frac{N}{n_i^+}$.
				\item Most importantly, Thm.\ref{pami:thm2} reveals that the FPR range $\beta_i$ be correspondingly adjusted toward different recommendation goals $N$, where $\beta_i$ belongs to $[\frac{N}{n_i^-}, 1]$. However, inspired by Prop.\ref{pro1}, the FPR range $\beta_i$ of conventional HarS ($U \equiv 1$) is always equal to $\frac{1}{n_i^-}$, which might lead to sub-optimal performance, especially when $N > 1$. To address this, we propose a novel Differentiable HarS-based algorithm (DiHarS) in the next section, which can explicitly maximize the OPAUC performance with the expected FPR range.
				\item Moreover, there is a trade-off between $N_i^{\beta_i}$ and $\hat{\texttt{OPAUC}}^{u_i}(s_{\boldsymbol{g}}, \beta_i)$ in the bound (\ref{pami:eq28}) and (\ref{pami:eq29}), where a small $\beta_i$ for $N_i^{\beta_i}$ usually induces a small $\mathcal{F}(n_i^+, N, -\hat{\texttt{OPAUC}}^{u_i}(s_{\boldsymbol{g}}, \beta_i))$ but also increases the difficulty for maximizing OPAUC (i.e., a lager magnitude of $\mathcal{F}(n_i^+, N, -\hat{\texttt{OPAUC}}^{u_i}(s_{\boldsymbol{g}}, \beta_i))$). Thus, in order to obtain a promising recommendation result, one should \textbf{adopt a proper FPR range $\beta_i$} for each user $u_i$ to strike a balance between $N_i^{\beta_i}$ and $\hat{\texttt{OPAUC}}^{u_i}(s_{\boldsymbol{g}}, \beta_i)$. 
			\end{enumerate}
		\end{rem}
	\subsection{Differentiable Hardness-aware Negative Sampling}
	As discussed in Thm.\ref{pami:thm2}, the critical recipe for promising performance is to include a proper number of ``hardest'' negative examples (i.e., $U \ge 1$) during training. This is the significant difference between our proposed approach and the standard HarS that always sets $U \equiv 1$. To do this, without loss of generality, we directly consider the following per-user OPAUC maximization problem from (\ref{prop1_eq27}):
	\begin{equation}\label{ref_eq28}
			\min\limits_{\bmg} \ \ \ \frac{1}{|\mcu|} \sum_{u_i \in \mcu} \sum_{j=1}^{n_i^+} \sum_{t=1}^{N_i^{\beta_i}} \frac{\ell^{(i)}_g(v_j^+, v_{[t]}^-)}{n_i^+N_i^{\beta_i}},
		\end{equation}
		where $N_i^{\beta_i} = U = \lfloor n_{i}^- \cdot \beta_i \rfloor$ with $\beta_i \ge \frac{N}{n_i^-}$ from Thm.\ref{pami:thm2}. 
		
		Nonetheless, directly optimizing (\ref{ref_eq28}) is challenging. To be specific, (\ref{ref_eq28}) requires to determine the top-ranked $N_i^{\beta_i}$ negative items among all unobserved items. A naive way is to use the sort operation to achieve such sample selections. Unfortunately, the sort function is not differentiable \cite{DBLP:conf/nips/SwezeyGCE21}, which cannot be optimized end-to-end, leading to sub-optimal performance.
	
	To avoid this problem, we develop a differentiable algorithm for (\ref{ref_eq28}), which is highly inspired by the recent advances in the sum of top-$k$ learning \cite{DBLP:journals/ipl/OgryczakT03,NIPS2017_6c524f9d,DBLP:conf/uai/LyuY18}. To begin with, we have the following lemma:
		\begin{lem} \label{pami:lem2} $\sum_{t=1}^k z_{[t]}$ is a convex function of $(z_1, \dots, z_n)$ and $z_{[t]}$ represents the top-$t$ element among $(z_1, \dots, z_n)$. Then, we can afford the equivalence of the sum-of-top-$k$ elements with an optimization problem as follows:
			\begin{equation}
				\sum_{t=1}^k z_{[t]} = \min\limits_{\gamma \ge 0} \left\{k\gamma + \sum_{t=1}^n [z_t - \gamma]_+\right\},
			\end{equation}
			where $[a]_+ = \max(0, a)$ is the hinge function.
		\end{lem}
	
	Please refer to Appendix.\ref{pami:proof_lem2} for proof of Lem.\ref{pami:lem2}. 
	
	In light of Lem.\ref{pami:lem2}, we know that any top-$t$ sample selection process could be equivalently reformulated as a differentiable minimization problem. In this sense, we can derive an equivalent surrogate goal of (\ref{ref_eq28}) to eliminate the non-differentiable sort function. The proof of the following Thm.\ref{pami:thm3} is attached in Appendix.\ref{pami:proof_thm3}.
		\begin{theorem} [Differentiable Reformulation of (\ref{ref_eq28})] \label{pami:thm3} Let $\forall u_i \in \mathcal{U}$, $N_i^{\beta_i} = \lfloor n_{i}^- \cdot \beta_i \rfloor$ and $\beta_i \ge \frac{N}{n_i^-}$. Then, based on Lem.\ref{pami:lem2}, (\ref{ref_eq28}) could be equivalently reformulated as a differentiable optimization problem:
			\begin{equation}\nonumber
				\begin{aligned}
					\min\limits_{\bmg} \ \ \frac{1}{|\mcu|} \sum_{u_i \in \mcu} \sum_{j=1}^{n_i^+} \sum_{t=1}^{N_i^{\beta_i}} & \frac{\ell^{(i)}_g(v_j^+, v_{[t]}^-)}{n_i^+N_i^{\beta_i}} \Leftrightarrow \\
					\ \ \min\limits_{\bmg, \bmga \ge \boldsymbol{0}} \frac{1}{|\mcu|} \sum_{u_i \in \mcu} \sum_{j=1}^{n_i^+} &  \left\{\frac{\gamma_{ij}}{n_i^+} + \frac{1}{n_i^+N_i^{\beta_i}}\sum_{k=1}^{n_i^-} d^{(i)}_g(v_j^+, v_{k}^-)\right\}, 
				\end{aligned}
			\end{equation}
			where we denote all learnable $\gamma_{ij}$ parameters as a $\sum\limits_{u_i \in \mcu} n_i^+$ dimensional vector $\bmga$ for ease of expression, and we define
			\begin{equation}\nonumber
				d^{(i)}_g(v_j^+, v_{k}^-) = [\lambda + s(u_i, v_j^+) - s(u_i, v_k^-) - \gamma_{ij}]_+,
			\end{equation}
			$\lambda > 0$ is still the safe margin.
		\end{theorem}
	\noindent \textbf{Optimization Goal.} Based on Thm.\ref{pami:thm3}, the differentiable Hardness-aware Sampling (DiHarS) based DPCML framework could be expressed as the following optimization objective:
	\begin{equation} \label{pami:eq34}
			\begin{aligned}
				\ \ \min\limits_{\bmg, \bmga \ge \boldsymbol{0}} \tilde{{\mcl}}_{\mcd}(\bmg) := \tilde{\mathcal{R}}_{\bmg, \bmga} + \frac{\eta}{|\mcu|} \sum_{u_i \in \mcu} \psi_{\bmg}(u_{i}),
			\end{aligned}
		\end{equation}
		where we define
		\[
		\tilde{\mathcal{R}}_{\bmg, \bmga} := \frac{1}{|\mcu|} \sum_{u_i \in \mcu} \sum_{j=1}^{n_i^+} \frac{1}{n_i^+}\left\{\gamma_{ij} + \frac{1}{N_i^{\beta_i}}\sum_{k=1}^{n_i^-} d^{(i)}_g(v_j^+, v_{k}^-)\right\},
		\]
		and the second part in (\ref{pami:eq34}) is our proposed DCRS regularization in Sec.\ref{Sec.3.4}. The stochastic optimization algorithm for solving (\ref{pami:eq34}) is summarized in Alg.\ref{algorithm1} in Appendix.\ref{supp:alg} due to space limitations.


	\begin{table*}[htbp]
		\centering
		\caption{Performance comparisons on CiteULike. The best and runner-up are highlighted in bold and underlined.}
		\scalebox{0.8}{
			\begin{tabular}{c|c|c|cccccccc}
				\toprule
				& Type & \multicolumn{1}{c}{Method} & P@3 & R@3 & NDCG@3 & P@5 & R@5 & NDCG@5 & MAP & MRR \\
				\midrule
				\multirow{23}[11]{*}{CiteULike} & Item-based & itemKNN & \cellcolor[rgb]{1.0, 1.0, 1.0} 1.20  & \cellcolor[rgb]{1.0, 1.0, 1.0} 0.83  & \cellcolor[rgb]{1.0, 1.0, 1.0} 1.23  & \cellcolor[rgb]{1.0, 1.0, 1.0} 1.15  & \cellcolor[rgb]{1.0, 1.0, 1.0} 0.77  & \cellcolor[rgb]{1.0, 1.0, 1.0} 1.16  & \cellcolor[rgb]{ .98,  .988,  .996} 1.44  & \cellcolor[rgb]{1.0, 1.0, 1.0} 3.78  \\
			\cmidrule{2-11}      & \multirow{6}[2]{*}{MF-based} & BPR & \cellcolor[rgb]{ .812,  .886,  .957} 6.47  & \cellcolor[rgb]{ .808,  .886,  .957} 3.50  & \cellcolor[rgb]{ .812,  .886,  .957} 6.84  & \cellcolor[rgb]{ .816,  .886,  .957} 7.89  & \cellcolor[rgb]{ .827,  .894,  .961} 4.05  & \cellcolor[rgb]{ .808,  .886,  .949} 8.49  & \cellcolor[rgb]{ .812,  .886,  .957} 5.14  & \cellcolor[rgb]{ .804,  .882,  .953} 16.20  \\
			& & GMF & \cellcolor[rgb]{ .98,  .988,  .996} 1.86  & \cellcolor[rgb]{ .992,  .996,  1} 0.96  & \cellcolor[rgb]{ .976,  .988,  .996} 2.05  & \cellcolor[rgb]{ .976,  .988,  .996} 2.15  & \cellcolor[rgb]{ .992,  .996,  1} 0.97  & \cellcolor[rgb]{ .973,  .984,  .996} 2.40  & \cellcolor[rgb]{ .984,  .992,  .996} 1.34  & \cellcolor[rgb]{ .976,  .984,  .996} 5.53  \\
			&   & MLP & \cellcolor[rgb]{ .973,  .984,  .996} 2.06  & \cellcolor[rgb]{ .984,  .992,  1} 1.08  & \cellcolor[rgb]{ .973,  .984,  .996} 2.22  & \cellcolor[rgb]{ .969,  .984,  .996} 2.40  & \cellcolor[rgb]{ .984,  .992,  .996} 1.16  & \cellcolor[rgb]{ .969,  .98,  .996} 2.61  & \cellcolor[rgb]{ .976,  .984,  .996} 1.52  & \cellcolor[rgb]{ .871,  .922,  .973} 12.37  \\
			&   & NeuMF & \cellcolor[rgb]{ .973,  .984,  .996} 2.06  & \cellcolor[rgb]{ .984,  .992,  1} 1.08  & \cellcolor[rgb]{ .973,  .984,  .996} 2.21  & \cellcolor[rgb]{ .973,  .984,  .996} 2.36  & \cellcolor[rgb]{ .984,  .992,  .996} 1.16  & \cellcolor[rgb]{ .969,  .98,  .996} 2.57  & \cellcolor[rgb]{ .976,  .984,  .996} 1.54  & \cellcolor[rgb]{ .875,  .925,  .973} 12.22  \\
			&   & M2F & \cellcolor[rgb]{ .984,  .992,  .996} 1.76  & \cellcolor[rgb]{ .996,  1,  1} 0.90  & \cellcolor[rgb]{ .98,  .988,  .996} 1.97  & \cellcolor[rgb]{ .984,  .992,  .996} 1.87  & \cellcolor[rgb]{ .992,  .996,  1} 0.93  & \cellcolor[rgb]{ .976,  .988,  .996} 2.18  & \cellcolor[rgb]{1.0, 1.0, 1.0} 0.93  & \cellcolor[rgb]{ .992,  .996,  1} 4.53  \\
			&   & MGMF & \cellcolor[rgb]{ .965,  .98,  .992} 2.31  & \cellcolor[rgb]{ .976,  .988,  .996} 1.23  & \cellcolor[rgb]{ .965,  .98,  .992} 2.48  & \cellcolor[rgb]{ .969,  .98,  .996} 2.42  & \cellcolor[rgb]{ .984,  .992,  .996} 1.12  & \cellcolor[rgb]{ .965,  .98,  .992} 2.71  & \cellcolor[rgb]{ .976,  .984,  .996} 1.51  & \cellcolor[rgb]{ .965,  .98,  .992} 6.18  \\
			\cmidrule{2-11}      &VAE-based & Mult-VAE & \cellcolor[rgb]{ .812,  .886,  .957} 6.56  & \cellcolor[rgb]{ .808,  .886,  .957} 3.68  & \cellcolor[rgb]{ .808,  .886,  .957} 6.89  & \cellcolor[rgb]{ .824,  .894,  .961} 7.53  & \cellcolor[rgb]{ .824,  .894,  .957} 4.10  & \cellcolor[rgb]{ .82,  .89,  .953} 8.09  & \cellcolor[rgb]{ .808,  .886,  .957} 5.23  & \cellcolor[rgb]{ .804,  .882,  .953} 16.27  \\
			\cmidrule{2-11}      & GNN-based & LightGCN & \cellcolor[rgb]{ .745,  .847,  .941} 8.33  & \cellcolor[rgb]{ .741,  .843,  .937} 4.64  & \cellcolor[rgb]{ .749,  .847,  .941} 8.68  & \cellcolor[rgb]{ .769,  .859,  .945} 9.58  & \cellcolor[rgb]{ .765,  .859,  .945} 5.23  & \cellcolor[rgb]{ .769,  .859,  .945} 10.23  & \cellcolor[rgb]{ .741,  .843,  .937} 6.32  & \cellcolor[rgb]{ .757,  .855,  .941} 19.14  \\
			\cmidrule{2-11}      & \multirow{8}[2]{*}{CML-based} & UniS & \cellcolor[rgb]{ .804,  .882,  .957} 7.34  & \cellcolor[rgb]{ .82,  .89,  .961} 3.71  & \cellcolor[rgb]{ .812,  .886,  .957} 7.48  & \cellcolor[rgb]{ .784,  .871,  .953} 9.54  & \cellcolor[rgb]{ .784,  .871,  .953} 5.13  & \cellcolor[rgb]{ .792,  .875,  .953} 10.02  & \cellcolor[rgb]{ .792,  .875,  .953} 5.59  & \cellcolor[rgb]{ .796,  .878,  .957} 17.27  \\
			&   & PopS & \cellcolor[rgb]{ .867,  .922,  .973} 5.41  & \cellcolor[rgb]{ .871,  .922,  .973} 2.94  & \cellcolor[rgb]{ .863,  .918,  .969} 5.77  & \cellcolor[rgb]{ .859,  .914,  .969} 6.75  & \cellcolor[rgb]{ .859,  .918,  .969} 3.62  & \cellcolor[rgb]{ .859,  .914,  .969} 7.23  & \cellcolor[rgb]{ .839,  .902,  .965} 4.61  & \cellcolor[rgb]{ .843,  .906,  .965} 14.39  \\
			&   & 2st & \cellcolor[rgb]{ .835,  .902,  .965} 6.40  & \cellcolor[rgb]{ .843,  .906,  .965} 3.35  & \cellcolor[rgb]{ .831,  .898,  .965} 6.77  & \cellcolor[rgb]{ .82,  .89,  .961} 8.27  & \cellcolor[rgb]{ .827,  .898,  .961} 4.29  & \cellcolor[rgb]{ .82,  .894,  .961} 8.81  & \cellcolor[rgb]{ .82,  .89,  .961} 4.99  & \cellcolor[rgb]{ .82,  .89,  .961} 15.87  \\
			&   & HarS & \cellcolor[rgb]{ .769,  .859,  .949} 8.44  & \cellcolor[rgb]{ .776,  .867,  .949} 4.41  & \cellcolor[rgb]{ .769,  .863,  .949} 8.82  & \cellcolor[rgb]{ .765,  .859,  .949} 10.43  & \cellcolor[rgb]{ .761,  .855,  .949} 5.60  & \cellcolor[rgb]{ .765,  .859,  .949} 11.25  & \cellcolor[rgb]{ .745,  .847,  .945} 6.67  & \cellcolor[rgb]{ .757,  .851,  .945} 20.08  \\
			&   & LRML & \cellcolor[rgb]{ .961,  .976,  .992} 2.52  & \cellcolor[rgb]{ .969,  .984,  .996} 1.33  & \cellcolor[rgb]{ .961,  .976,  .992} 2.58  & \cellcolor[rgb]{ .953,  .973,  .992} 3.06  & \cellcolor[rgb]{ .961,  .976,  .992} 1.64  & \cellcolor[rgb]{ .953,  .973,  .992} 3.19  & \cellcolor[rgb]{ .957,  .976,  .992} 1.91  & \cellcolor[rgb]{ .961,  .976,  .992} 6.45  \\
			&   & TransCF & \cellcolor[rgb]{ .855,  .914,  .969} 5.79  & \cellcolor[rgb]{ .863,  .918,  .969} 3.03  & \cellcolor[rgb]{ .855,  .914,  .969} 6.09  & \cellcolor[rgb]{ .839,  .902,  .965} 7.45  & \cellcolor[rgb]{ .847,  .906,  .965} 3.93  & \cellcolor[rgb]{ .843,  .906,  .965} 7.84  & \cellcolor[rgb]{ .839,  .906,  .965} 4.54  & \cellcolor[rgb]{ .839,  .902,  .965} 14.50  \\
			&   & AdaCML & \cellcolor[rgb]{ .816,  .886,  .961} 7.04  & \cellcolor[rgb]{ .82,  .89,  .961} 3.75  & \cellcolor[rgb]{ .816,  .89,  .961} 7.31  & \cellcolor[rgb]{ .808,  .882,  .957} 8.70  & \cellcolor[rgb]{ .816,  .89,  .961} 4.52  & \cellcolor[rgb]{ .812,  .886,  .957} 9.18  & \cellcolor[rgb]{ .796,  .875,  .953} 5.57  & \cellcolor[rgb]{ .796,  .878,  .957} 17.31  \\
			&   & HLR & \cellcolor[rgb]{ .976,  .984,  .996} 2.03  & \cellcolor[rgb]{ .984,  .992,  1} 1.08  & \cellcolor[rgb]{ .973,  .984,  .996} 2.20  & \cellcolor[rgb]{ .973,  .984,  .996} 2.25  & \cellcolor[rgb]{ .984,  .992,  .996} 1.13  & \cellcolor[rgb]{ .969,  .984,  .996} 2.52  & \cellcolor[rgb]{ .98,  .988,  .996} 1.45  & \cellcolor[rgb]{ .969,  .984,  .996} 5.86  \\
			\cmidrule{2-11}      & \multirow{6}[6]{*}{DPCML-based} & BPA+UniS & \cellcolor[rgb]{ .788,  .875,  .953} 7.78  & \cellcolor[rgb]{ .8,  .878,  .957} 4.04  & \cellcolor[rgb]{ .792,  .875,  .953} 8.14  & \cellcolor[rgb]{ .773,  .863,  .949} 10.03  & \cellcolor[rgb]{ .776,  .867,  .949} 5.33  & \cellcolor[rgb]{ .776,  .867,  .949} 10.64  & \cellcolor[rgb]{ .773,  .863,  .949} 6.08  & \cellcolor[rgb]{ .776,  .863,  .949} 18.75  \\
			&   & APA+UniS & \cellcolor[rgb]{ .784,  .871,  .953} 7.99  & \cellcolor[rgb]{ .792,  .875,  .953} 4.17  & \cellcolor[rgb]{ .784,  .871,  .953} 8.36  & \cellcolor[rgb]{ .773,  .863,  .949} 10.00  & \cellcolor[rgb]{ .78,  .867,  .953} 5.23  & \cellcolor[rgb]{ .776,  .867,  .949} 10.69  & \cellcolor[rgb]{ .773,  .863,  .949} 6.08  & \cellcolor[rgb]{ .773,  .863,  .949} 19.03  \\
			\cmidrule{3-11}      &   & BPA+HarS & \cellcolor[rgb]{ .761,  .855,  .949} 8.70  & \cellcolor[rgb]{ .765,  .859,  .949} 4.59  & \cellcolor[rgb]{ .765,  .859,  .949} 9.06  & \cellcolor[rgb]{ .749,  .847,  .945} 10.96  & \cellcolor[rgb]{ .749,  .851,  .945} 5.85  & \cellcolor[rgb]{ .757,  .855,  .945} 11.47  & \cellcolor[rgb]{ .757,  .851,  .945} 6.44  & \cellcolor[rgb]{ .757,  .855,  .945} 19.96  \\
			&   & APA+HarS & \cellcolor[rgb]{ .757,  .855,  .945} 8.82  & \cellcolor[rgb]{ .757,  .855,  .945} 4.73  & \cellcolor[rgb]{ .761,  .855,  .945} 9.18  & \cellcolor[rgb]{ .749,  .847,  .945} \underline{11.02}  & \cellcolor[rgb]{ .749,  .847,  .945} \underline{5.87}  & \cellcolor[rgb]{ .757,  .851,  .945} 11.56  & \cellcolor[rgb]{ .745,  .847,  .945} \underline{6.68}  & \cellcolor[rgb]{ .753,  .851,  .945} 20.30  \\
			\cmidrule{3-11}      &   & BPA+DiHarS & \cellcolor[rgb]{ .749,  .847,  .945} \underline{9.05} & \cellcolor[rgb]{ .753,  .851,  .945} \underline{4.76} & \cellcolor[rgb]{ .753,  .851,  .945} \underline{9.45} & \cellcolor[rgb]{ .757,  .851,  .945} 10.73 & \cellcolor[rgb]{ .761,  .855,  .945} 5.66 & \cellcolor[rgb]{ .757,  .851,  .945} \underline{11.58} & \cellcolor[rgb]{ .753,  .851,  .945} 6.53 & \cellcolor[rgb]{ .753,  .851,  .945} \underline{20.32} \\
			&   & APA+DiHarS & \cellcolor[rgb]{ .741,  .843,  .941} \textbf{9.24} & \cellcolor[rgb]{ .741,  .843,  .941} \textbf{4.94} & \cellcolor[rgb]{ .741,  .843,  .941} \textbf{9.72} & \cellcolor[rgb]{ .741,  .843,  .941} \textbf{11.20} & \cellcolor[rgb]{ .741,  .843,  .941} \textbf{5.99} & \cellcolor[rgb]{ .741,  .843,  .941} \textbf{12.09} & \cellcolor[rgb]{ .741,  .843,  .941} \textbf{6.72} & \cellcolor[rgb]{ .741,  .843,  .941} \textbf{20.88} \\
				\bottomrule
			\end{tabular}
		}
		\label{tab:addlabel}%
	\end{table*}%

	\section{Experiments} \label{exp}
	Due to space limitations, \textbf{\textit{please refer to Appendix.\ref{exp_supp} for a longer version.}} 
	\vspace{-0.2cm}
	\subsection{Overall Performance}
	The experimental results are shown in Tab.\ref{tab:addlabel}, Tab.\ref{results1}, Tab.\ref{results2}, and Tab.\ref{result3} (in Appendix.\ref{suppa.4}). We can draw the following conclusions: a) Our proposed DPCML methods can consistently outperform all competitors significantly on all datasets, in particular with our newly developed APA and DiHarS sampling strategies. This demonstrates the superiority of our proposed algorithms. b) Regarding different preference assignment strategies, as a whole, DPCML+APA optimized by any of the three negative sampling manners (i.e., UniS, HarS, and DiHarS) could achieve better recommendation results than its corresponding counterpart DPCML+BPA. The empirical performance validates the diversity of users' interests and ascertains the effectiveness of the improved adaptive assignment approach. c) Compared with studies targeting joint accessibility (i.e., M2F and MGMF), our proposed methods can perform better on all metrics than M2F and MGMF on all benchmark datasets. This supports the potential advantage of the CML-based paradigm in this direction, which deserves more research attention in future work. d) Concerning CML methods learning with different negative sampling strategies, the HarS-driven CML algorithms demonstrate better than others (say UniS, PopS, and 2stS) in most cases. Most importantly, with respect to the DPCML framework, adopting our proposed DiHarS strategy could further outperform HarS-based DPCML approaches, and the performance gain is sharp. For example, the MRR gaps between BPA+DiHarS and BPA+HarS are $2.4\%$, $7.41\%$ and $2.02\%$ on Steam-200k, MovieLens-10M and RecSys-2 (newly added dataset in this version), respectively. In terms of APA strategy, the enhancements are $2.22\%$, $7.70\%$ and $1.32\%$. This consistently suggests the superiority of DiHarS (Thm.\ref{pami:thm2} and Thm.\ref{pami:thm3}) that can explicitly improve the Top-$N$ recommendation performance from the OPAUC  perspective. e) Finally, we notice that some deep-learning-based methods (such as Mult-VAE and LightGCN) could achieve competitive or even better performance than a few vanilla CML-based methods (such as PopS, TransCF, LRML) to some extent but fail to outperform ours, especially compared to DiHarS-guided DPCML. This shows that our proposed framework could unleash the power of the CML paradigm, contributing to promising recommendation performances.
		\begin{table*}[]
			\centering
			\caption{Performance comparisons on Steam-200k dataset against other diversity-promoting algorithms.}
			\scalebox{0.8}{
			  \begin{tabular}{c|c|c|cccccccc}
			  \toprule
				& Type & Method & P@3 & R@3 & NDCG@3 & P@5 & R@5 & NDCG@5 & MAP & MRR \\
			  \midrule
			  \multirow{10}[6]{*}{Steam-200k} & \multirow{4}[2]{*}{Two-Stage} & UniS+DD & \cellcolor[rgb]{ .902,  .945,  .98} 21.03  & \cellcolor[rgb]{ .839,  .91,  .969} 12.04  & \cellcolor[rgb]{ .902,  .945,  .98} 21.66  & \cellcolor[rgb]{ .882,  .933,  .976} 20.80  & \cellcolor[rgb]{ .859,  .922,  .973} 10.27  & \cellcolor[rgb]{ .89,  .937,  .98} 21.61  & \cellcolor[rgb]{ .827,  .906,  .969} 18.92  & \cellcolor[rgb]{ .89,  .941,  .98} 40.13  \\
			  &   & UniS+PD & \cellcolor[rgb]{ .902,  .945,  .98} 20.89  & \cellcolor[rgb]{ .839,  .91,  .969} 12.04  & \cellcolor[rgb]{ .902,  .945,  .98} 21.56  & \cellcolor[rgb]{ .878,  .933,  .976} 20.89  & \cellcolor[rgb]{ .859,  .922,  .973} 10.34  & \cellcolor[rgb]{ .886,  .937,  .98} 21.62  & \cellcolor[rgb]{ .827,  .906,  .969} 18.92  & \cellcolor[rgb]{ .89,  .941,  .98} 40.19  \\
			  &   & HarS+DD & \cellcolor[rgb]{ .796,  .886,  .961} 27.25  & \cellcolor[rgb]{ .765,  .871,  .953} 15.99  & \cellcolor[rgb]{ .796,  .886,  .961} 28.48  & \cellcolor[rgb]{ .788,  .882,  .961} 26.10  & \cellcolor[rgb]{ .773,  .875,  .957} 13.50  & \cellcolor[rgb]{ .788,  .882,  .961} 27.65  & \cellcolor[rgb]{ .745,  .859,  .949} 23.61  & \cellcolor[rgb]{ .776,  .875,  .957} 49.45  \\
			  &   & HarS+PD & \cellcolor[rgb]{ .808,  .894,  .965} 26.70  & \cellcolor[rgb]{ .769,  .871,  .957} 15.76  & \cellcolor[rgb]{ .804,  .89,  .961} 27.96  & \cellcolor[rgb]{ .808,  .894,  .965} 24.97  & \cellcolor[rgb]{ .792,  .886,  .961} 12.80  & \cellcolor[rgb]{ .808,  .894,  .965} 26.66  & \cellcolor[rgb]{ .753,  .863,  .953} 23.26  & \cellcolor[rgb]{ .784,  .878,  .957} 48.85  \\
		\cmidrule{2-11}      & \multirow{5}[2]{*}{One-Stage} & PRD & \cellcolor[rgb]{ .933,  .965,  .988} 19.01  & \cellcolor[rgb]{ .875,  .929,  .976} 10.27  & \cellcolor[rgb]{ .933,  .965,  .988} 19.56  & \cellcolor[rgb]{ .886,  .937,  .98} 20.57  & \cellcolor[rgb]{ .867,  .925,  .976} 10.02  & \cellcolor[rgb]{ .89,  .941,  .98} 21.49  & \cellcolor[rgb]{ .871,  .929,  .976} 16.52  & \cellcolor[rgb]{ .918,  .957,  .984} 38.02  \\
			  &   & RecNet & \cellcolor[rgb]{ .965,  .98,  .996} 17.20  & \cellcolor[rgb]{ .882,  .937,  .976} 9.75  & \cellcolor[rgb]{ .961,  .98,  .992} 17.93  & \cellcolor[rgb]{ .949,  .973,  .992} 16.91  & \cellcolor[rgb]{ .914,  .953,  .984} 8.31  & \cellcolor[rgb]{ .949,  .973,  .992} 17.83  & \cellcolor[rgb]{ .898,  .945,  .98} 14.83  & \cellcolor[rgb]{ .957,  .976,  .992} 34.80  \\
			  &   & DP-RecNet & \cellcolor[rgb]{ .992,  .996,  1} 15.59  & \cellcolor[rgb]{ .886,  .937,  .98} 9.57  & \cellcolor[rgb]{ .988,  .996,  1} 16.12  & \cellcolor[rgb]{1.0, 1.0,  1.0} 13.88  & \cellcolor[rgb]{ .941,  .969,  .988} 7.31  & \cellcolor[rgb]{1.0, 1.0,  1.0} 14.64  & \cellcolor[rgb]{ .898,  .941,  .98} 14.94  & \cellcolor[rgb]{ .996,  1,  1} 31.85  \\
			  &   & IDCF & \cellcolor[rgb]{ .843,  .914,  .969} 24.45  & \cellcolor[rgb]{ .804,  .89,  .961} 13.92  & \cellcolor[rgb]{ .843,  .914,  .969} 25.41  & \cellcolor[rgb]{ .824,  .902,  .965} 24.11  & \cellcolor[rgb]{ .816,  .898,  .965} 11.94  & \cellcolor[rgb]{ .827,  .902,  .969} 25.38  & \cellcolor[rgb]{ .788,  .882,  .961} 21.12  & \cellcolor[rgb]{ .827,  .906,  .969} 45.29  \\
			  &   & GraphDiv & \cellcolor[rgb]{1.0, 1.0,  1.0} 15.01  & \cellcolor[rgb]{ .918,  .953,  .984} 7.89  & \cellcolor[rgb]{1.0, 1.0,  1.0} 15.29  & \cellcolor[rgb]{ .965,  .98,  .996} 15.92  & \cellcolor[rgb]{ .922,  .957,  .984} 7.98  & \cellcolor[rgb]{ .965,  .98,  .996} 16.88  & \cellcolor[rgb]{ .969,  .984,  .996} 10.84  & \cellcolor[rgb]{1.0, 1.0,  1.0} 31.31  \\
		\cmidrule{2-11}      & Ours & APA+DiHarS & \cellcolor[rgb]{ .706,  .835,  .941} \textbf{32.58} & \cellcolor[rgb]{ .706,  .835,  .941} \textbf{19.09} & \cellcolor[rgb]{ .706,  .835,  .941} \textbf{33.98} & \cellcolor[rgb]{ .706,  .835,  .941} \textbf{30.81} & \cellcolor[rgb]{ .706,  .835,  .941} \textbf{15.99} & \cellcolor[rgb]{ .706,  .835,  .941} \textbf{32.68} & \cellcolor[rgb]{ .706,  .835,  .941} \textbf{25.78} & \cellcolor[rgb]{ .706,  .835,  .941} \textbf{54.90} \\
			  \bottomrule
			  \end{tabular}%
			}
			\label{tab:diversity_comp}%
		  \end{table*}%
	
	\vspace{-0.3cm}
	\subsection{Diversity-promoting Performance Comparison}\label{major:sec7.3}
	\subsubsection{Compared to other Diversity-promoting Methods}
	Since this paper aims to develop a diversity-promoting algorithm only accessing the collaborative data, we evaluate its performance with other $9$ diversity-promoting baselines that can perform well without requiring external information. \textbf{Please see Appendix.\ref{marjor:supp_C.6.1} for the detailed introductions.} 

	\noindent \textbf{Performance Comparison.} Partial results are summarized in Tab.\ref{tab:diversity_comp}, and the remains are attached in Appendix.\ref{marjor:supp_C.6.1}. Firstly, although re-ranking techniques improve the recommendation performance to some degree, DPCML could still significantly outperform all of them. Secondly, compared to one-stage methods, DPCML still achieves the best towards all metrics. Besides, neural-network-based algorithms (such as RecNet and GCN-AccDiv) show relatively low performance due to the data sparsity. To sum up, the above results consistently demonstrate the potential of the CML-based paradigm in diversity-promoting aspects.


	\begin{figure}[]
		\centering
		\subfigure{
			\includegraphics[width=0.23\textwidth]{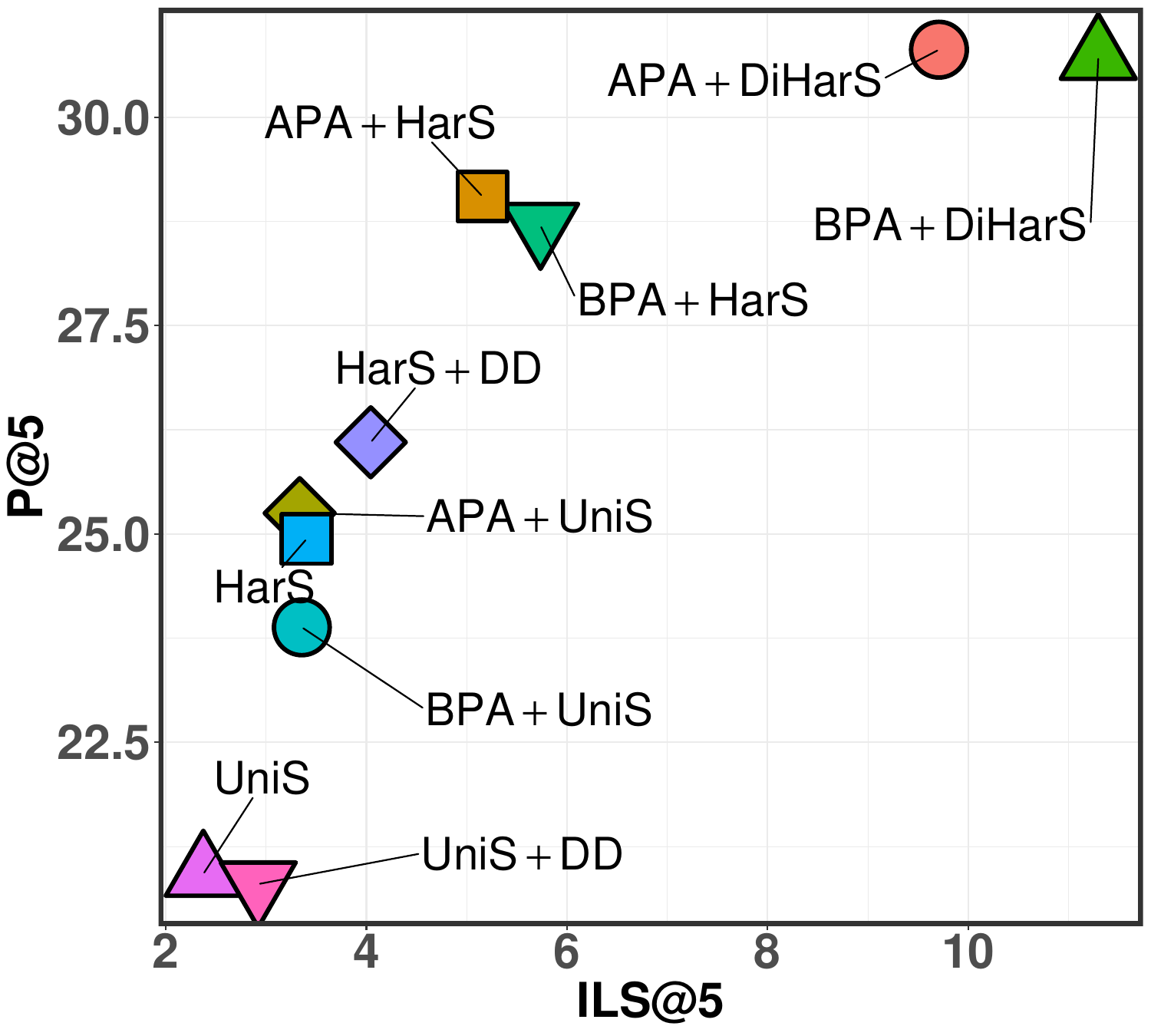}}
		\subfigure{
			\includegraphics[width=0.23\textwidth]{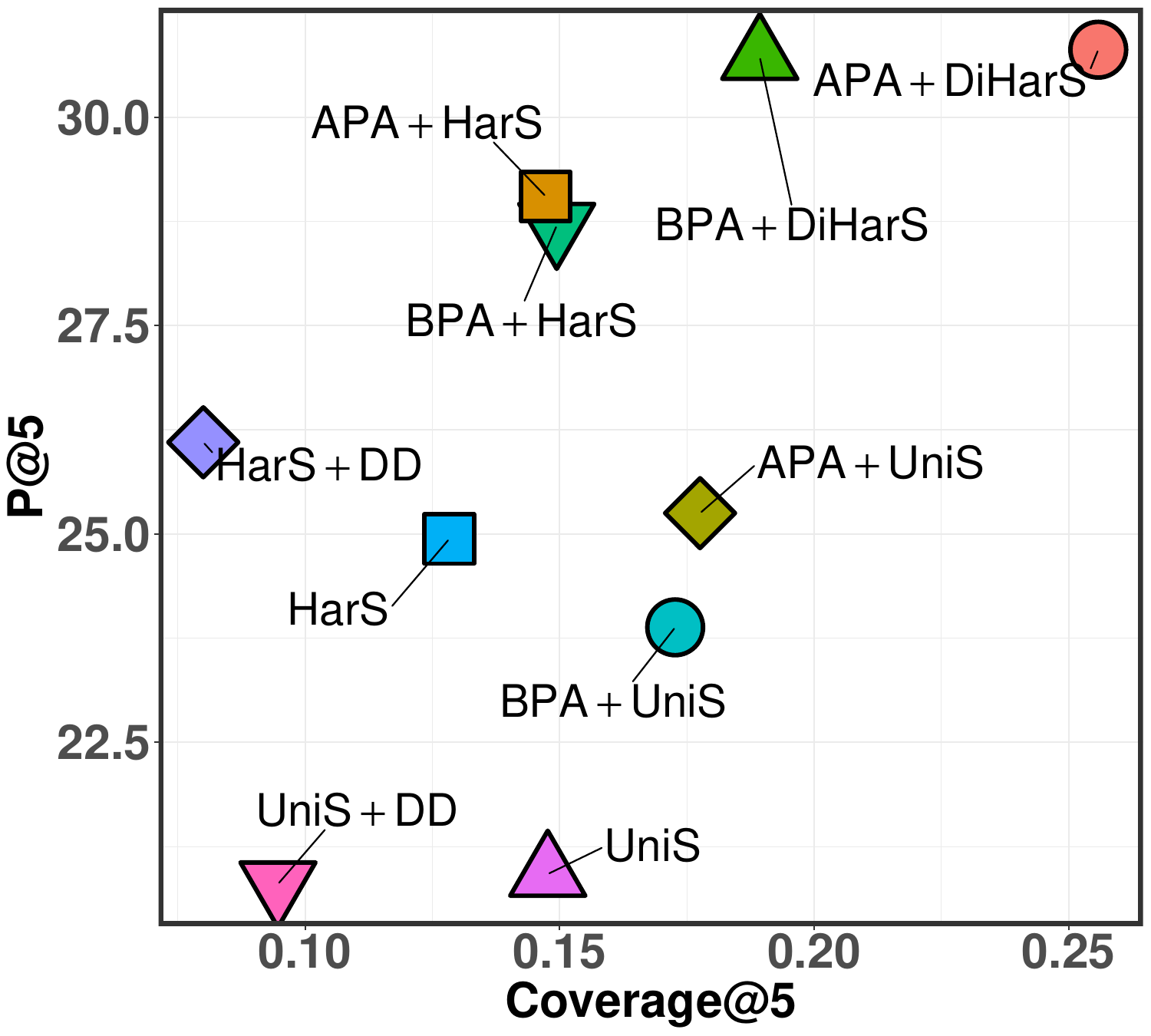}}
		\caption{Diversity vs. performance on Steam-200k.}
		\label{fig:div}
	\end{figure}
	\vspace{-0.22cm}
	\subsubsection{Recommendation Diversity Performance} 
	Besides performance evaluations, recommendation diversity \cite{DBLP:journals/tbd/XieLLZ00L22,DBLP:conf/coling/RazaBN22} is another significant concern. In this sense, we test the diversity performance with a series of widely adopted diversity metrics, including \textit{Max-sum Diversification (MaxDiv)} \cite{10.1145/2213556.2213580}, \textit{Intra-List Similarity (ILS)} \cite{DBLP:conf/www/ZieglerMKL05,DBLP:journals/csur/ZangerleB23} and \textit{Coverage} \cite{DBLP:journals/tiis/KaminskasB17}. Please refer to Appendix.\ref{pami:supp_4.2} for the detailed introductions. The experiments are conducted on Steam-200k and MovieLens-1M datasets. The empirical results are provided in Fig.\ref{fig:div} and Fig.\ref{fig:supp_div}. The elaborate diversity results are attached in Appendix.\ref{pami:supp_4.2}. From these results, we can conclude: a) Within the same negative sampling strategy, DPCML could achieve better diversity in most cases, even CML using the reranking trick \textbf{DD}. b) More significantly, our proposed DiHarS strategy could further boost recommendation diversity. This suggests the effectiveness of promoting recommendation diversity. c) Even without the regularization term, DPCML still outperforms CML. Most importantly, equipped with DCRS, DPCML could achieve better diversification results against w/o DCRS in most cases. Overall, DPCML could perform better than traditional CML in recommendation accuracy and diversity.


	\subsection{Quantitative Analysis} \label{qqaa}
	\subsubsection{Ablation Study for DiHarS Framework}
	We investigate the performance of different DiHarS variants. At first, we consider the usage of DiHarS for the CML framework (i.e., \textbf{CML+DiHarS}) and regard the HarS approach (\textbf{CML+HarS}) as the benchmark. Furthermore, we also consider the non-differentiable version of DiHarS (short for \textbf{NDiHarS}), i.e., directly using the sort operation to achieve the sparse sample selections in (\ref{ref_eq28}). Compared with the traditional HarS fixing $U\equiv1$ in (\ref{cml_sampling_hars}), the major difference of NDiHarS is its parameter $U=\lfloor n_{i}^- \cdot \beta_i \rfloor \ge 1$ determined by the FPR range $\beta_i$ in Thm.2. The hyper-parameter setups stay the same as DiHarS. The empirical results are presented in Fig.\ref{fig:supp_ab_for_DiHarS} in Appendix.\ref{pami:supp_ab_for_DiHarS}. Please refer to Appendix.\ref{pami:supp_ab_for_DiHarS} for more evidence. Our proposed DiHarS could outperform its sort-based counterpart (i.e., NDiHarS-driven methods) significantly because the non-differentiable loss function might be challenging to optimize. Besides, we can observe that applying DiHarS to the standard CML could also perform better than the conventional HarS trick in most cases. These results consistently provide evidence for the superiority of our proposed DiHarS. 
	
	\begin{figure*}[]
		\centering
		\subfigure[Different $U$ for UniS]{
		  \includegraphics[width=0.30\textwidth]{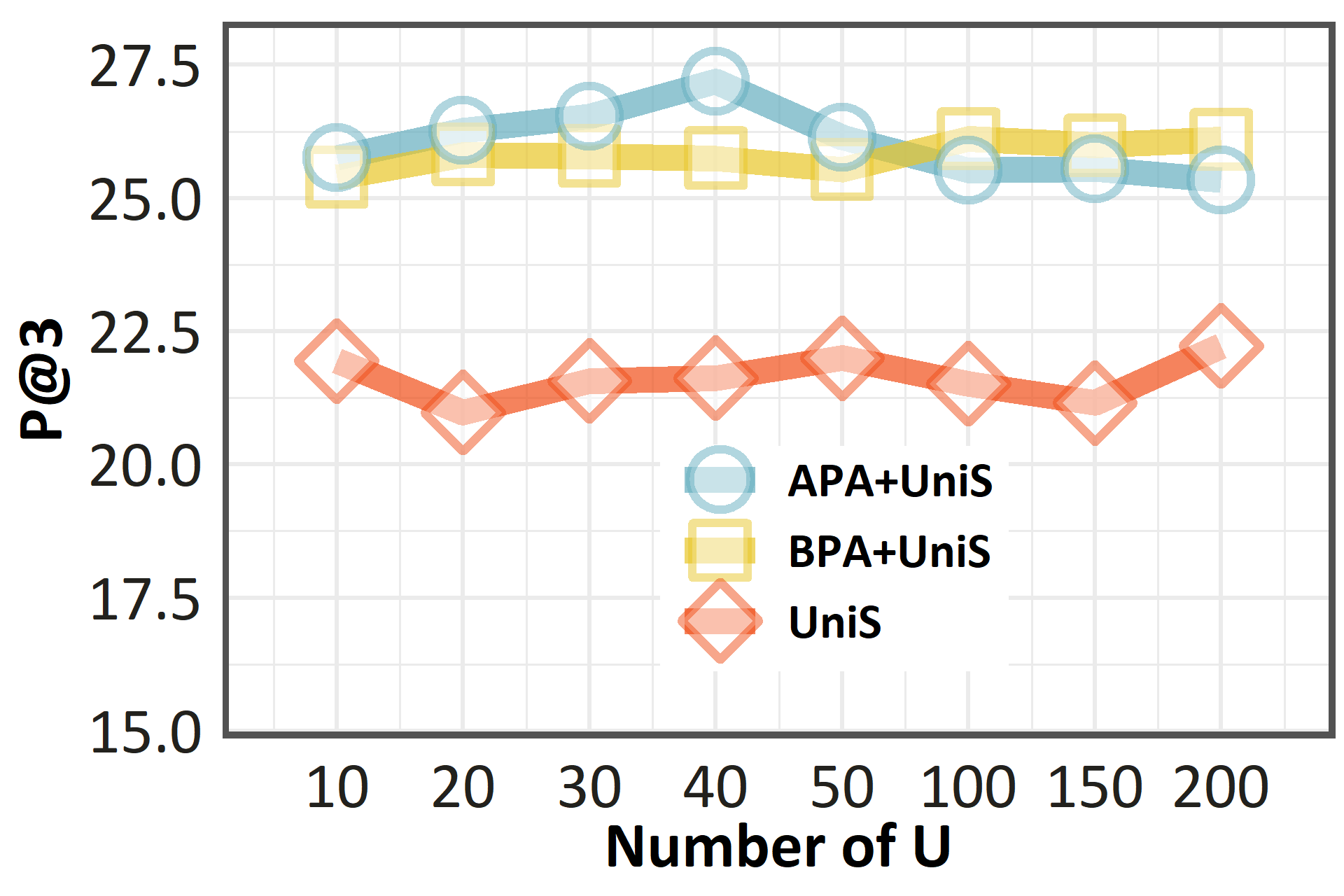}
		}
		\subfigure[Different $S$ for HarS]{
		  \includegraphics[width=0.30\textwidth]{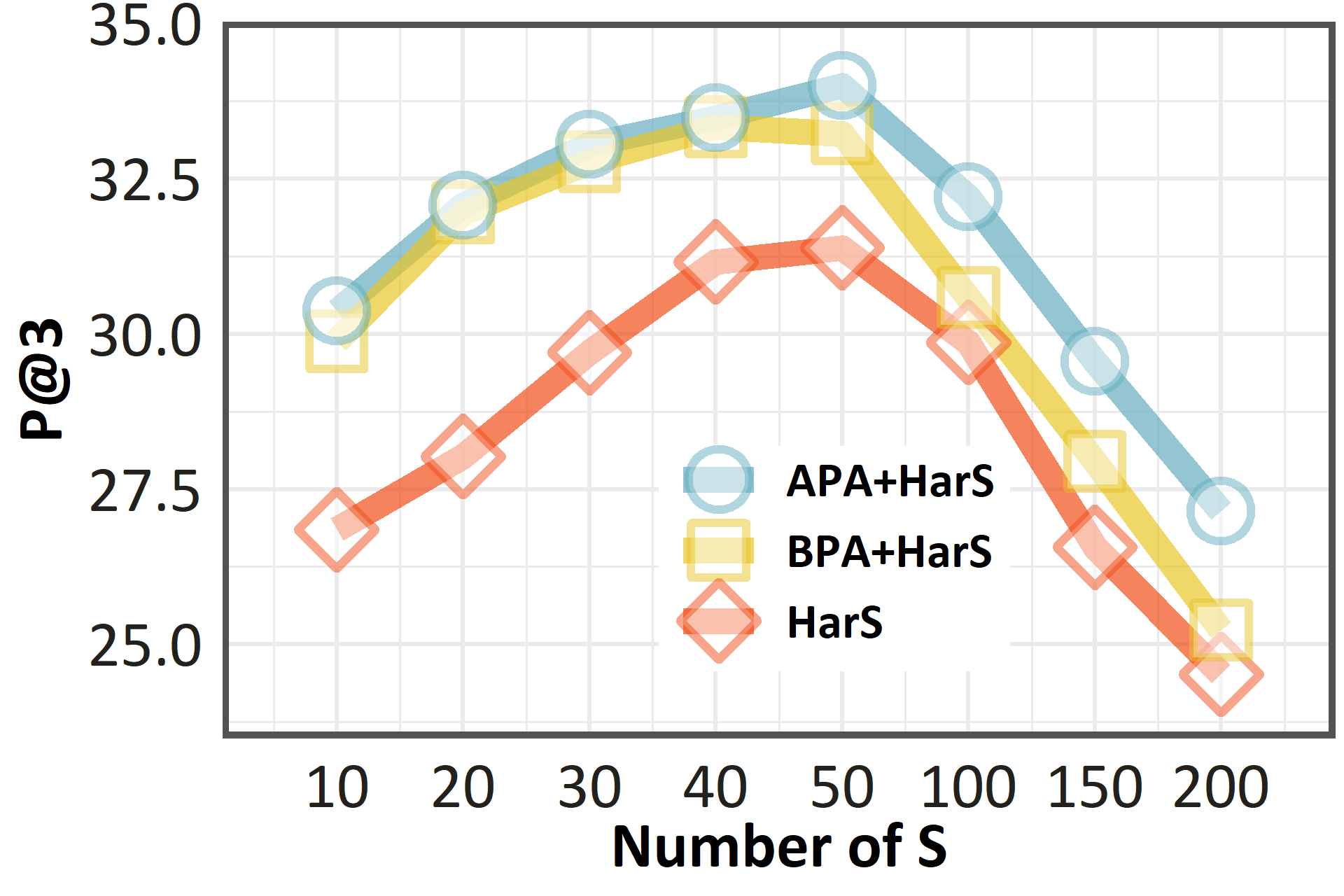}
		}
		\subfigure[Different $U$ for HarS]{
		  \includegraphics[width=0.30\textwidth]{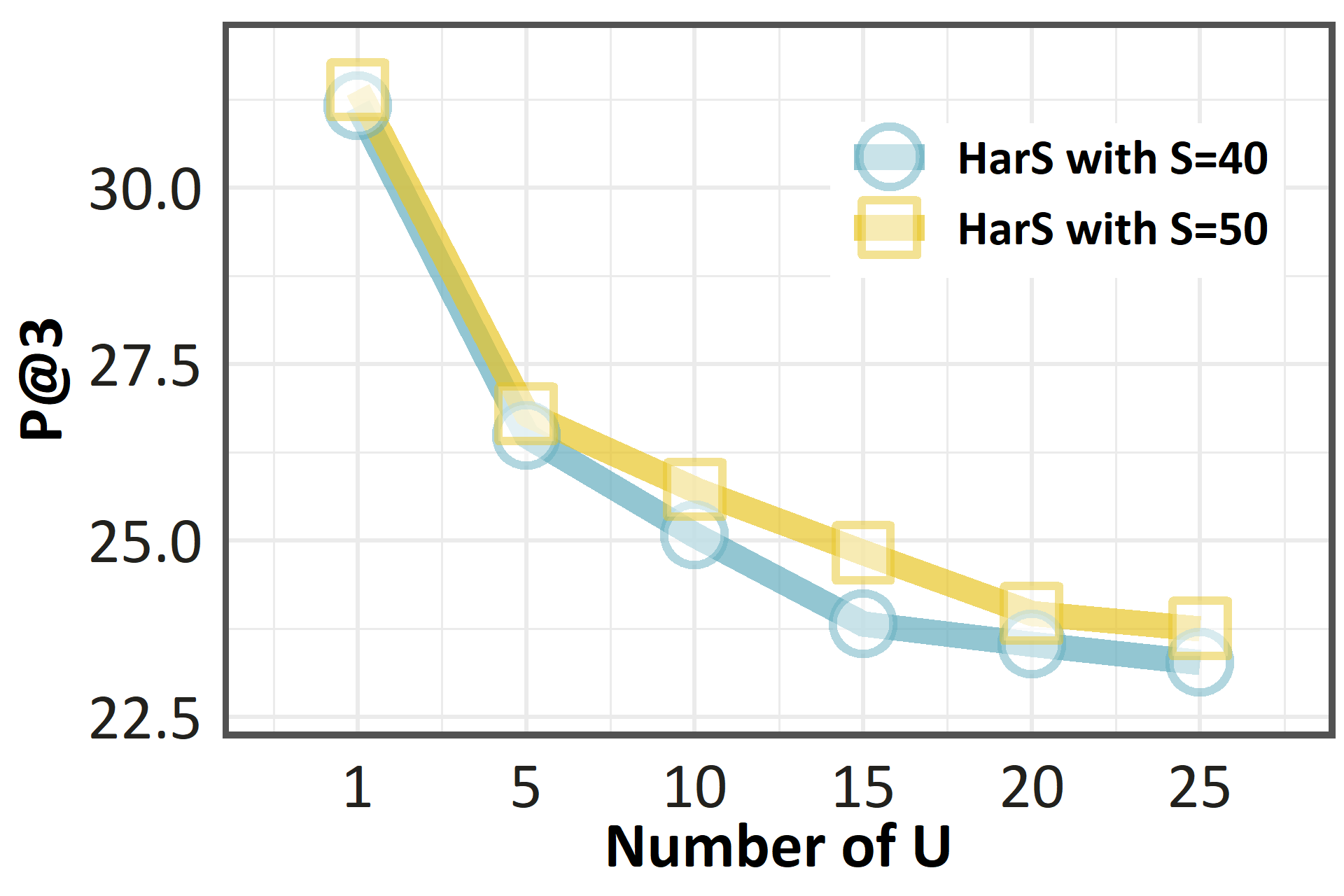}
		}
		\caption{Ablation studies for sampling parameters on Steam-200k dataset.}
		\label{fig:ab_for_HarS}
	  \end{figure*}

		\subsubsection{Ablation Study for Sampling Parameters}
		We compare \textbf{CML} and our proposed \textbf{BPA} and \textbf{APA-based DPCML} approaches under various sampling numbers. The experiments are performed on Steam-200k, where all methods are optimized by \textbf{UniS} and \textbf{HarS}, respectively. Specifically, the parameter $U$ for UniS and $S$ for HarS are conducted among $\{10, 20, 30, 40, 50, 100, 150, 200\}$, respectively. The results are summarized in Fig.\ref{fig:ab_for_HarS}-(a) and (b). Although determining a proper sampling parameter is non-trivial \cite{DBLP:journals/pami/BaoX0CH23,DBLP:conf/icml/XuRKKA21}, we see that DPCML could always outperform CML-based counterparts at all different sampling parameters. In addition, we also conduct the sensitive analysis of another parameter $U \in \{1, 5, 10, 15, 20, 25\}$ included in HarS, where another parameter $S$ is fixed as $S \in \{40, 50\}$ suggested by Fig.\ref{fig:ab_for_HarS}-(b). However, simply adopting a larger number of $U$ would not improve the performance of HarS as depicted in Fig.\ref{fig:ab_for_HarS}-(c). Its performance will gradually worsen because it will degrade to UniS when $U$ approaches $S$. Let alone surpass DPCML. 
	\subsubsection{Fine-grained Performance Comparison} 
	Fig.\ref{supp:per_arrtribute_performance} in the Appendix reports the fine-grained MAP performance over each interest group (i.e., movie genre) on MovieLens-10M. We can observe that our proposed framework could not only significantly outperform their single-vector counterparts in the majority interests but also improve the performance of minority groups in most cases. Especially compared with HarS, the performance improvement of DPCML on minority interests is sharp. This shows that DPCML could reasonably focus on potentially interesting items even with the imbalanced item distribution.
	

	\subsubsection{Empirical Justification of Corol.\ref{cor1}} \label{pami:sec.7.6.4}
	We conduct empirical studies on Steam-200k to show the correctness of Corol.\ref{cor1}. The results are summarized in Fig.\ref{just_thm1} in Appendix.\ref{pami:supp_sec.7.6.4}. With the increase of $C$, the empirical risk (i.e., training loss) of DPCML ($C>1$) with any of three sampling strategies could be significantly smaller than the corresponding CML ($C=1$) counterpart. Meanwhile, the performance on the validation/test set is also improved. This suggests that DPCML could induce a smaller generalization error. 
	
	\subsubsection{Effect of the DCRS}
	Appendix.\ref{ab_stu} studies the influence of two main hyper-parameters in DCRS, i.e., $\delta_1$ and $\delta_2$ and sensitive analysis of different DCRS variants. The experimental results show that DCRS could significantly boost the final performance. 

	\vspace{-0.5pt}
	\subsubsection{Sensitivity analysis of $\eta$} 
	Appendix.\ref{major:ab_eta} presents the sensitivity analysis of $\eta \in \{0, 1, 3, 5, 10, 20, 30\}$ on Steam-200k. The results shown in Tab.\ref{tab:sen_dpcml1} and Tab.\ref{tab:sen_dpcml2} consistently prove that controlling a proper $\eta$ is essential for promising performances.
	
	
	
	\subsubsection{Training \& Inference Efficiency}
	Appendix.\ref{major:supp_efficiency} investigates the training/inference overheads among CML-based approaches. According to Fig.\ref{app:runtime} and Tab.\ref{app:inference} in the Appendix, we can observe that DPCML could achieve promising performance with acceptable efficiency in general.

	\subsubsection{Effectiveness of DCRS for Joint Accessibility Model}
	Appendix.\ref{supp:C.7.5} explores the effectiveness of DCRS for GFJA (\ref{eq3123}) and M2F \cite{DBLP:conf/recsys/WestonWY13,DBLP:conf/eaamo/GuoKJG21}. The experimental results presented in Tab.\ref{tab:reg_for_MF} show the potential of DCRS, which deserves more research attention in the future.
	
		\begin{figure*}[!t]
			\centering
			\subfigure[WarmStart (JT)]{
			  \includegraphics[width=0.15\textwidth]{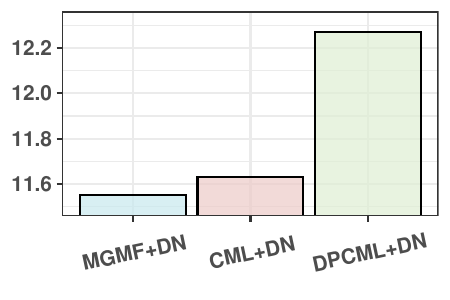}
			}
			\subfigure[Cold Users (JT)]{
			  \includegraphics[width=0.15\textwidth]{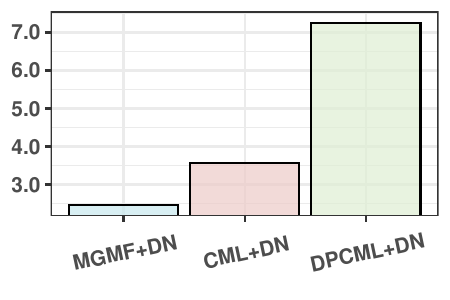}
			}
			\subfigure[Cold Items (JT)]{
			  \includegraphics[width=0.15\textwidth]{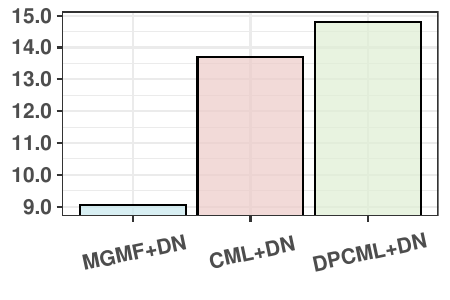}
			}
			\subfigure[WarmStart (PT)]{
			  \includegraphics[width=0.15\textwidth]{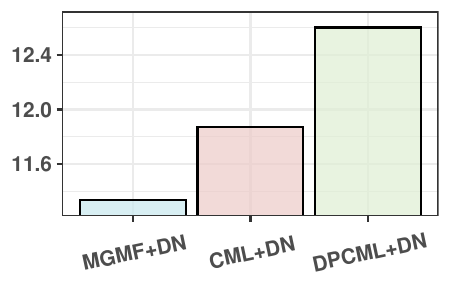}
			}
			\subfigure[Cold Users (PT)]{
			  \includegraphics[width=0.15\textwidth]{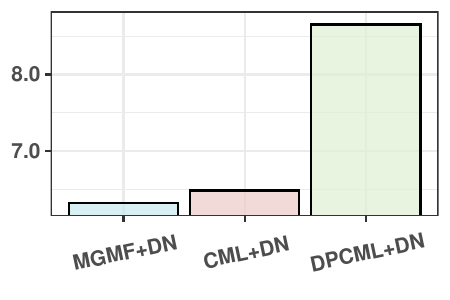}
			}
			\subfigure[Cold Items (PT)]{
			  \includegraphics[width=0.15\textwidth]{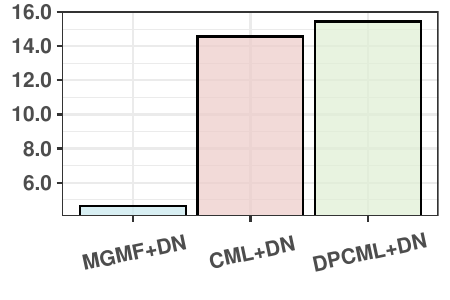}
			}
			\subfigure[WarmStart (JT)]{
			  \includegraphics[width=0.15\textwidth]{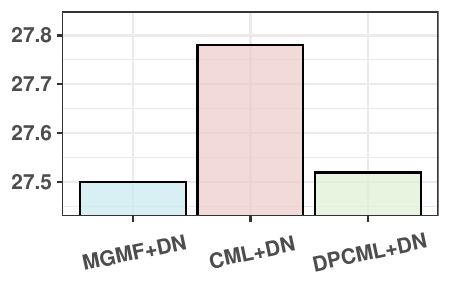}
			}
			\subfigure[Cold Users (JT)]{
			  \includegraphics[width=0.15\textwidth]{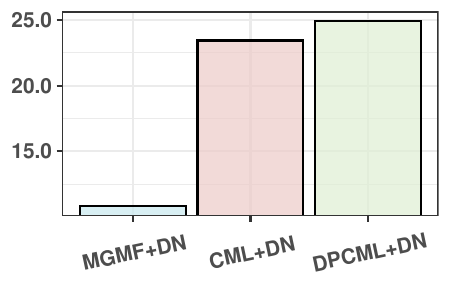}
			}
			\subfigure[Cold Items (JT)]{
			  \includegraphics[width=0.15\textwidth]{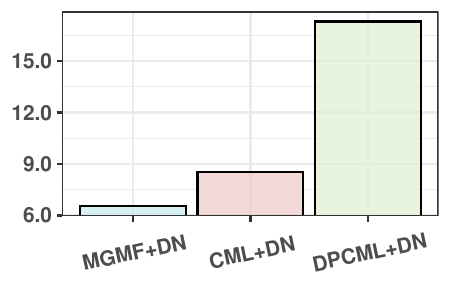}
			}
			\subfigure[WarmStart (PT)]{
			  \includegraphics[width=0.15\textwidth]{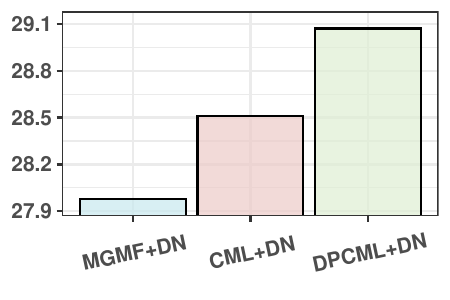}
			}
			\subfigure[Cold Users (PT)]{
			  \includegraphics[width=0.15\textwidth]{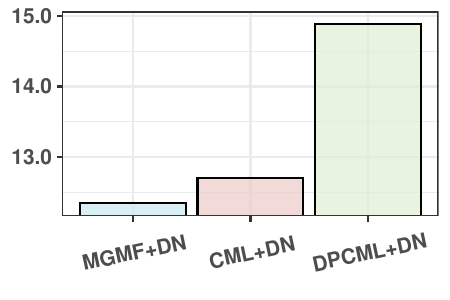}
			  }
			  \subfigure[Cold Items (PT)]{
			  \includegraphics[width=0.148\textwidth]{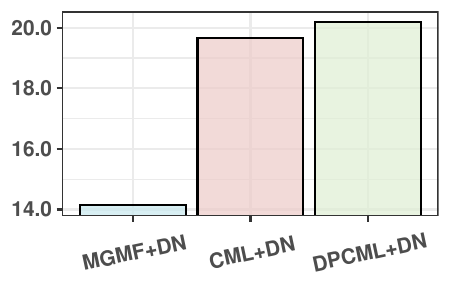}
			}
			\caption{Performance comparisons (P@$3$) on RecSys, where results for subsets 1 and 2 are shown in the first and second rows.}
			\label{fig:cold}
		  \end{figure*}

	\subsection{Potential Challenges and Solutions of DPCML}\label{major:Sec7.5}
	Despite the superiority of DPCML, two limitations might hinder its deployments: \textcolor{blue}{\textbf{(L1)}} DPCML cannot include other content features (i.e., side information) to learn users' and items' representations. \textcolor{blue}{\textbf{(L2)}} DPCML will lose efficacy when no interest records are available for some users (i.e., cold start users). Note that \textcolor{blue}{\textbf{(L1)}} and \textcolor{blue}{\textbf{(L2)}} widely exist for most latent collaborative filtering models \cite{DBLP:conf/kdd/LiuBXG022,DBLP:journals/corr/abs-1907-08674}. We explore combining DPCML with a simple but effective framework called DropoutNet (DN) \cite{DBLP:journals/jmlr/SrivastavaHKSS14} to solve \textcolor{blue}{\textbf{(L1)}} and \textcolor{blue}{\textbf{(L2)}} simultaneously. Given the preference and content inputs, the fundamental idea of DN is to randomly sample a fraction of users and items through Dropout \cite{DBLP:journals/jmlr/SrivastavaHKSS14} and then mask their corresponding preference inputs as $\bm{0}$ during training. After that, during the test phase, the model could generate a reasonable representation of the object even if its latent input is not supplied (i.e., the cold start case). \textbf{Please refer to Appendix.\ref{major:coldstartC.8} for detailed introductions and discussions.}

	\noindent\textbf{Performance Comparisons.} We evaluate the effectiveness of our proposed DPCML+DN on two RecSys subsets and compare its performance with \textbf{MGMF+DN}, and \textbf{CML+DN} with the UniS technique. Partial results are shown in Fig.\ref{fig:cold}. We also report the detailed performance in Tab.\ref{tab:cold} in Appendix.\ref{major:C.8.2}. Our proposed DPCML+DN could significantly outperform the competitors in cold start cases while achieving competitive or even better performance toward most warm start cases. This shows the potential of DPCML and deserves more research attention in the future. 
   
	\section{Conclusion}
		This paper proposes a novel DPCML method to capture users' multiple categories of interests. The success secret is introducing multiple representations for each user in the model design. To do this, two practical multi-vector assignment strategies, i.e., BPA and APA, are proposed. Meanwhile, a novel DCRS is specifically tailored to serve our purpose better. Theoretically, we present a high probability upper bound, showing that DPCML could generalize well to unseen data. Furthermore, we equivalently reformulate HarS-based (DP)CML to a per-user averaged OPAUC maximization problem. By doing so, we show that the standard HarS is insufficient to pursue promising top-$N$ recommendation performance. To alleviate this, we develop a novel OPAUC-guided hardness-aware negative sampling technique (DiHarS) from the OPAUC maximization point of view, which can enjoy better performance than HarS with acceptable efficiency. Finally, comprehensive experiments over a range of benchmark datasets demonstrate the effectiveness of DPCML.
\section{Acknowledgments}
This work was supported in part by the National Key R\&D Program of China under Grant 2018AAA0102000, in part by National Natural Science Foundation of China: 62236008, U21B2038, U23B2051, 61931008, 62122075, 61976202,  62025604, 62206264 and 92370102, in part by Youth Innovation Promotion Association CAS, in part by the Strategic Priority Research Program of the Chinese Academy of Sciences, Grant No. XDB0680000, in part by the Innovation Funding of ICT, CAS under Grant No.E000000.

	\bibliographystyle{IEEEtran}
	\bibliography{ref}

\begin{thebibliography}{100}
\providecommand{\url}[1]{#1}
\csname url@samestyle\endcsname
\providecommand{\newblock}{\relax}
\providecommand{\bibinfo}[2]{#2}
\providecommand{\BIBentrySTDinterwordspacing}{\spaceskip=0pt\relax}
\providecommand{\BIBentryALTinterwordstretchfactor}{4}
\providecommand{\BIBentryALTinterwordspacing}{\spaceskip=\fontdimen2\font plus
\BIBentryALTinterwordstretchfactor\fontdimen3\font minus \fontdimen4\font\relax}
\providecommand{\BIBforeignlanguage}[2]{{%
\expandafter\ifx\csname l@#1\endcsname\relax
\typeout{** WARNING: IEEEtran.bst: No hyphenation pattern has been}%
\typeout{** loaded for the language `#1'. Using the pattern for}%
\typeout{** the default language instead.}%
\else
\language=\csname l@#1\endcsname
\fi
#2}}
\providecommand{\BIBdecl}{\relax}
\BIBdecl

\bibitem{DBLP:conf/pakdd/WangZ0HC20}
C.~Wang, T.~Zhou, C.~Chen, T.~Hu, and G.~Chen, ``Off-policy recommendation system without exploration,'' in \emph{PAKDD}, vol. 12084.\hskip 1em plus 0.5em minus 0.4em\relax Springer, 2020, pp. 16--27.

\bibitem{DBLP:conf/nips/MaZ0Y019}
J.~Ma, C.~Zhou, P.~Cui, H.~Yang, and W.~Zhu, ``Learning disentangled representations for recommendation,'' in \emph{NeurIPS}, 2019, pp. 5712--5723.

\bibitem{DBLP:conf/kdd/MaZYCW020}
J.~Ma, C.~Zhou, H.~Yang, P.~Cui, X.~Wang, and W.~Zhu, ``Disentangled self-supervision in sequential recommenders,'' in \emph{{KDD}}, 2020, pp. 483--491.

\bibitem{DBLP:journals/aei/LvZWWW20}
Y.~Lv, Y.~Zheng, F.~Wei, C.~Wang, and C.~Wang, ``{AICF:} attention-based item collaborative filtering,'' \emph{Adv. Eng. Informatics}, vol.~44, pp. 101\,090:1--11, 2020.

\bibitem{DBLP:journals/tkde/JiangCCW0Y15}
M.~Jiang, P.~Cui, X.~Chen, F.~Wang, W.~Zhu, and S.~Yang, ``Social recommendation with cross-domain transferable knowledge,'' \emph{{IEEE} TKDE}, vol.~27, no.~11, pp. 3084--3097, 2015.

\bibitem{DBLP:conf/nips/WangGZZ18}
M.~Wang, M.~Gong, X.~Zheng, and K.~Zhang, ``Modeling dynamic missingness of implicit feedback for recommendation,'' in \emph{NeurIPS}, 2018, pp. 6670--6679.

\bibitem{DBLP:conf/sigir/AskariSS21}
B.~Askari, J.~Szlichta, and A.~Salehi{-}Abari, ``Variational autoencoders for top-k recommendation with implicit feedback,'' in \emph{{SIGIR}}, 2021, pp. 2061--2065.

\bibitem{DBLP:conf/www/TogashiKOS21}
R.~Togashi, M.~Kato, M.~Otani, and S.~Satoh, ``Density-ratio based personalised ranking from implicit feedback,'' in \emph{{WWW}}, 2021, pp. 3221--3233.

\bibitem{DBLP:conf/icml/XuRKKA21}
D.~Xu, C.~Ruan, E.~K{\"{o}}rpeoglu, S.~Kumar, and K.~Achan, ``Rethinking neural vs. matrix-factorization collaborative filtering: the theoretical perspectives,'' in \emph{{ICML}}, 2021, pp. 11\,514--11\,524.

\bibitem{DBLP:conf/icml/ZhengTDZ16}
Y.~Zheng, B.~Tang, W.~Ding, and H.~Zhou, ``A neural autoregressive approach to collaborative filtering,'' in \emph{{ICML}}, 2016, pp. 764--773.

\bibitem{DBLP:conf/aaai/WangWSSL20}
X.~Wang, R.~Wang, C.~Shi, G.~Song, and Q.~Li, ``Multi-component graph convolutional collaborative filtering,'' in \emph{{AAAI}}, 2020, pp. 6267--6274.

\bibitem{DBLP:conf/icdm/PanZCLLSY08}
R.~Pan, Y.~Zhou, B.~Cao, N.~N. Liu, R.~M. Lukose, M.~Scholz, and Q.~Yang, ``One-class collaborative filtering,'' in \emph{{ICDM}}, 2008, pp. 502--511.

\bibitem{DBLP:journals/ijon/ZhangR21}
Q.~Zhang and F.~Ren, ``Prior-based bayesian pairwise ranking for one-class collaborative filtering,'' \emph{Neurocomputing}, vol. 440, pp. 365--374, 2021.

\bibitem{DBLP:conf/aaai/ChenL019}
J.~Chen, D.~Lian, and K.~Zheng, ``Improving one-class collaborative filtering via ranking-based implicit regularizer,'' in \emph{AAAI}, 2019, pp. 37--44.

\bibitem{hsieh2017collaborative}
C.-K. Hsieh, L.~Yang, Y.~Cui, T.-Y. Lin, S.~Belongie, and D.~Estrin, ``Collaborative metric learning,'' in \emph{WWW}, 2017, pp. 193--201.

\bibitem{DBLP:conf/iccv/KumarTZ07}
M.~P. Kumar, P.~H.~S. Torr, and A.~Zisserman, ``An invariant large margin nearest neighbour classifier,'' in \emph{{ICCV}}, 2007, pp. 1--8.

\bibitem{tran2019improving}
V.-A. Tran, R.~Hennequin, J.~Royo-Letelier, and M.~Moussallam, ``Improving collaborative metric learning with efficient negative sampling,'' in \emph{{SIGIR}}, 2019, pp. 1201--1204.

\bibitem{DBLP:conf/www/TayTH18}
Y.~Tay, L.~A. Tuan, and S.~C. Hui, ``Latent relational metric learning via memory-based attention for collaborative ranking,'' in \emph{{WWW}}, 2018, pp. 729--739.

\bibitem{DBLP:conf/icdm/ParkKXY18}
C.~Park, D.~Kim, X.~Xie, and H.~Yu, ``Collaborative translational metric learning,'' in \emph{{ICDM}}, 2018, pp. 367--376.

\bibitem{DBLP:conf/mm/BaoXMYCH19}
S.~Bao, Q.~Xu, K.~Ma, Z.~Yang, X.~Cao, and Q.~Huang, ``Collaborative preference embedding against sparse labels,'' in \emph{ACM {MM}}, 2019, pp. 2079--2087.

\bibitem{DBLP:journals/nn/WuZNC20}
H.~Wu, Q.~Zhou, R.~Nie, and J.~Cao, ``Effective metric learning with co-occurrence embedding for collaborative recommendations,'' \emph{Neural Networks}, vol. 124, pp. 308--318, 2020.

\bibitem{wang2019group}
H.~Wang, Y.~Li, and F.~Frimpong, ``Group recommendation via self-attention and collaborative metric learning model,'' \emph{IEEE Access}, vol.~7, pp. 164\,844--164\,855, 2019.

\bibitem{DBLP:conf/ijcai/ZhouLL019}
X.~Zhou, D.~Liu, J.~Lian, and X.~Xie, ``Collaborative metric learning with memory network for multi-relational recommender systems,'' in \emph{{IJCAI}}, 2019, pp. 4454--4460.

\bibitem{DBLP:conf/dasfaa/ZhangZLXF0SC19}
T.~Zhang, P.~Zhao, Y.~Liu, J.~Xu, J.~Fang, L.~Zhao, V.~S. Sheng, and Z.~Cui, ``Adacml: Adaptive collaborative metric learning for recommendation,'' in \emph{{DASFAA}}, vol. 11447, 2019, pp. 301--316.

\bibitem{DBLP:conf/recsys/TranSHM21}
V.~Tran, G.~Salha{-}Galvan, R.~Hennequin, and M.~Moussallam, ``Hierarchical latent relation modeling for collaborative metric learning,'' in \emph{RecSys}, 2021, pp. 302--309.

\bibitem{DBLP:conf/colt/BartlettM01}
P.~L. Bartlett and S.~Mendelson, ``Rademacher and gaussian complexities: Risk bounds and structural results,'' in \emph{{COLT}}, vol. 2111, 2001, pp. 224--240.

\bibitem{DBLP:books/daglib/0034861}
M.~Mohri, A.~Rostamizadeh, and A.~Talwalkar, \emph{Foundations of Machine Learning}.\hskip 1em plus 0.5em minus 0.4em\relax {MIT} Press, 2012.

\bibitem{DBLP:conf/iccv/VasudevaDB0C21}
B.~Vasudeva, P.~Deora, S.~Bhattacharya, U.~Pal, and S.~Chanda, ``Loop: Looking for optimal hard negative embeddings for deep metric learning,'' in \emph{{ICCV}}, 2021, pp. 10\,614--10\,623.

\bibitem{DBLP:journals/corr/abs-2210-03967}
H.~Shao, Q.~Xu, Z.~Yang, S.~Bao, and Q.~Huang, ``Asymptotically unbiased instance-wise regularized partial {AUC} optimization: Theory and algorithm,'' 2022.

\bibitem{DBLP:conf/icml/0001XBHCH21}
Z.~Yang, Q.~Xu, S.~Bao, Y.~He, X.~Cao, and Q.~Huang, ``When all we need is a piece of the pie: {A} generic framework for optimizing two-way partial {AUC},'' in \emph{{ICML}}, 2021, pp. 11\,820--11\,829.

\bibitem{DPCML}
S.~Bao, Q.~Xu, Z.~Yang, Y.~He, X.~Cao, and Q.~Huang, ``The minority matters: A diversity-promoting collaborative metric learning algorithm,'' in \emph{NeurIPS}, 2022.

\bibitem{DBLP:conf/kdd/YangDZYZT20}
Z.~Yang, M.~Ding, C.~Zhou, H.~Yang, J.~Zhou, and J.~Tang, ``Understanding negative sampling in graph representation learning,'' in \emph{{KDD}}, 2020, pp. 1666--1676.

\bibitem{DBLP:conf/wsdm/RendleF14}
S.~Rendle and C.~Freudenthaler, ``Improving pairwise learning for item recommendation from implicit feedback,'' in \emph{{WSDM}}, 2014, pp. 273--282.

\bibitem{DBLP:conf/nips/VolkovsYP17}
M.~Volkovs, G.~W. Yu, and T.~Poutanen, ``Dropoutnet: Addressing cold start in recommender systems,'' in \emph{NeurIPS}, 2017, pp. 4957--4966.

\bibitem{DBLP:journals/www/YaoTYXZSL19}
Y.~Yao, H.~Tong, G.~Yan, F.~Xu, X.~Zhang, B.~K. Szymanski, and J.~Lu, ``Dual-regularized one-class collaborative filtering with implicit feedback,'' \emph{WWW}, pp. 1099--1129, 2019.

\bibitem{DBLP:conf/icml/HeckelR17}
R.~Heckel and K.~Ramchandran, ``The sample complexity of online one-class collaborative filtering,'' in \emph{{ICML}}, 2017, pp. 1452--1460.

\bibitem{DBLP:journals/kbs/ZhangR21}
Q.~Zhang and F.~Ren, ``Double bayesian pairwise learning for one-class collaborative filtering,'' \emph{Knowl. Based Syst.}, vol. 229, p. 107339, 2021.

\bibitem{DBLP:conf/sigir/LeeKJPY21}
D.~Lee, S.~Kang, H.~Ju, C.~Park, and H.~Yu, ``Bootstrapping user and item representations for one-class collaborative filtering,'' in \emph{{SIGIR}}, 2021, pp. 1513--1522.

\bibitem{DBLP:conf/ijcai/0001DWTTC18}
X.~He, X.~Du, X.~Wang, F.~Tian, J.~Tang, and T.~Chua, ``Outer product-based neural collaborative filtering,'' in \emph{{IJCAI}}, 2018, pp. 2227--2233.

\bibitem{DBLP:conf/uai/RendleFGS09}
S.~Rendle, C.~Freudenthaler, Z.~Gantner, and L.~Schmidt{-}Thieme, ``{BPR:} bayesian personalized ranking from implicit feedback,'' in \emph{{UAI}}, 2009, pp. 452--461.

\bibitem{DBLP:conf/sigir/HeZKC16}
X.~He, H.~Zhang, M.~Kan, and T.~Chua, ``Fast matrix factorization for online recommendation with implicit feedback,'' in \emph{{SIGIR}}, 2016, pp. 549--558.

\bibitem{DBLP:conf/www/HeLZNHC17}
X.~He, L.~Liao, H.~Zhang, L.~Nie, X.~Hu, and T.~Chua, ``Neural collaborative filtering,'' in \emph{{WWW}}, 2017, pp. 173--182.

\bibitem{DBLP:journals/symmetry/KayaB19}
M.~Kaya and H.~S. Bilge, ``Deep metric learning: {A} survey,'' \emph{Symmetry}, vol.~11, no.~9, p. 1066, 2019.

\bibitem{DBLP:journals/pami/MilbichRBO22}
T.~Milbich, K.~Roth, B.~Brattoli, and B.~Ommer, ``Sharing matters for generalization in deep metric learning,'' \emph{{IEEE} Trans. Pattern Anal. Mach. Intell.}, vol.~44, no.~1, pp. 416--427, 2022.

\bibitem{DBLP:journals/pami/EleziSWVTPL23}
I.~Elezi, J.~Seidenschwarz, L.~Wagner, S.~Vascon, A.~Torcinovich, M.~Pelillo, and L.~Leal{-}Taix{\'{e}}, ``The group loss++: {A} deeper look into group loss for deep metric learning,'' \emph{{IEEE} Trans. Pattern Anal. Mach. Intell.}, vol.~45, no.~2, pp. 2505--2518, 2023.

\bibitem{DBLP:conf/sigir/Wang0WFC19}
X.~Wang, X.~He, M.~Wang, F.~Feng, and T.~Chua, ``Neural graph collaborative filtering,'' in \emph{{ACM} {SIGIR}}, 2019, pp. 165--174.

\bibitem{DBLP:conf/kdd/Wang00LC19}
X.~Wang, X.~He, Y.~Cao, M.~Liu, and T.~Chua, ``{KGAT:} knowledge graph attention network for recommendation,'' in \emph{{SIGKDD}}, 2019, pp. 950--958.

\bibitem{DBLP:journals/corr/abs-2302-03472}
W.~Shi, J.~Chen, F.~Feng, J.~Zhang, J.~Wu, C.~Gao, and X.~He, ``On the theories behind hard negative sampling for recommendation,'' 2023.

\bibitem{DBLP:conf/iccv/QianSSHTLJ19}
Q.~Qian, L.~Shang, B.~Sun, J.~Hu, T.~Tacoma, H.~Li, and R.~Jin, ``Softtriple loss: Deep metric learning without triplet sampling,'' in \emph{{ICCV}}, 2019, pp. 6449--6457.

\bibitem{DBLP:journals/pami/ZhengLZ21}
W.~Zheng, J.~Lu, and J.~Zhou, ``Hardness-aware deep metric learning,'' \emph{{IEEE} Trans. Pattern Anal. Mach. Intell.}, vol.~43, no.~9, pp. 3214--3228, 2021.

\bibitem{DBLP:conf/iccv/HarwoodGCRD17}
B.~Harwood, V.~K.~B. G, G.~Carneiro, I.~D. Reid, and T.~Drummond, ``Smart mining for deep metric learning,'' in \emph{{ICCV}}, 2017, pp. 2840--2848.

\bibitem{DBLP:conf/iclr/RobinsonCSJ21}
J.~D. Robinson, C.~Chuang, S.~Sra, and S.~Jegelka, ``Contrastive learning with hard negative samples,'' in \emph{{ICLR}}, 2021.

\bibitem{DBLP:journals/corr/abs-2206-01197}
A.~Tabassum, M.~Wahed, H.~Eldardiry, and I.~Lourentzou, ``Hard negative sampling strategies for contrastive representation learning,'' in \emph{{ICLR}}, 2023.

\bibitem{DBLP:conf/sigir/WuVSSR19}
G.~Wu, M.~Volkovs, C.~L. Soon, S.~Sanner, and H.~Rai, ``Noise contrastive estimation for one-class collaborative filtering,'' in \emph{{SIGIR}}, 2019, pp. 135--144.

\bibitem{DBLP:journals/pr/GajicAG21}
B.~Gajic, A.~Amato, and C.~Gatta, ``Fast hard negative mining for deep metric learning,'' \emph{Pattern Recognition}, vol. 112, p. 107795, 2021.

\bibitem{DBLP:journals/kbs/KunaverP17}
M.~Kunaver and T.~Pozrl, ``Diversity in recommender systems - {A} survey,'' \emph{Knowl. Based Syst.}, vol. 123, pp. 154--162, 2017.

\bibitem{DBLP:journals/csur/ZangerleB23}
E.~Zangerle and C.~Bauer, ``Evaluating recommender systems: Survey and framework,'' \emph{{ACM} Comput. Surv.}, vol.~55, no.~8, pp. 170:1--170:38, 2023.

\bibitem{DBLP:reference/sp/2022rsh}
F.~Ricci, L.~Rokach, and B.~Shapira, Eds., \emph{Recommender Systems Handbook}.\hskip 1em plus 0.5em minus 0.4em\relax Springer {US}, 2022.

\bibitem{DBLP:conf/coling/RazaBN22}
S.~Raza, S.~R. Bashir, and U.~Naseem, ``Accuracy meets diversity in a news recommender system,'' in \emph{{COLING}}, 2022, pp. 3778--3787.

\bibitem{DBLP:journals/tbd/XieLLZ00L22}
R.~Xie, Q.~Liu, S.~Liu, Z.~Zhang, P.~Cui, B.~Zhang, and L.~Lin, ``Improving accuracy and diversity in matching of recommendation with diversified preference network,'' \emph{{IEEE} Trans. Big Data}, vol.~8, no.~4, pp. 955--967, 2022.

\bibitem{DBLP:journals/tkde/AdomaviciusK12}
G.~Adomavicius and Y.~Kwon, ``Improving aggregate recommendation diversity using ranking-based techniques,'' \emph{{IEEE} Trans. Knowl. Data Eng.}, vol.~24, no.~5, pp. 896--911, 2012.

\bibitem{vaishnavi2013ranking}
S.~Vaishnavi, A.~Jayanthi, and S.~Karthik, ``Ranking technique to improve diversity in recommender systems,'' \emph{IJCA}, vol.~68, no.~2, 2013.

\bibitem{DBLP:conf/ijcai/ShaWN16}
C.~Sha, X.~Wu, and J.~Niu, ``A framework for recommending relevant and diverse items,'' in \emph{Proceedings of the Twenty-Fifth International Joint Conference on Artificial Intelligence, {IJCAI} 2016, New York, NY, USA, 9-15 July 2016}, 2016, pp. 3868--3874.

\bibitem{DBLP:conf/ijcai/AshkanKBW15}
A.~Ashkan, B.~Kveton, S.~Berkovsky, and Z.~Wen, ``Optimal greedy diversity for recommendation,'' in \emph{Proceedings of the Twenty-Fourth International Joint Conference on Artificial Intelligence, {IJCAI} 2015, Buenos Aires, Argentina, July 25-31, 2015}, 2015, pp. 1742--1748.

\bibitem{DBLP:conf/nips/ChenZZ18}
L.~Chen, G.~Zhang, and E.~Zhou, ``Fast greedy {MAP} inference for determinantal point process to improve recommendation diversity,'' in \emph{NeurIPS}, 2018, pp. 5627--5638.

\bibitem{DBLP:conf/icml/GillenwaterKMV19}
J.~Gillenwater, A.~Kulesza, Z.~Mariet, and S.~Vassilvitskii, ``A tree-based method for fast repeated sampling of determinantal point processes,'' in \emph{{ICML}}, 2019, pp. 2260--2268.

\bibitem{DBLP:conf/ah/BridgeK06}
D.~G. Bridge and J.~P. Kelly, ``Ways of computing diverse collaborative recommendations,'' in \emph{{AH}}, 2006, pp. 41--50.

\bibitem{DBLP:conf/compsac/PremchaiswadiPJP13}
W.~Premchaiswadi, P.~Poompuang, N.~Jongsawat, and N.~Premchaiswadi, ``Enhancing diversity-accuracy technique on user-based top-n recommendation algorithms,'' in \emph{{IEEE} {COMPSAC}}, 2013, pp. 403--408.

\bibitem{DBLP:conf/recsys/SuYCY13}
R.~Su, L.~Yin, K.~Chen, and Y.~Yu, ``Set-oriented personalized ranking for diversified top-n recommendation,'' in \emph{RecSys}, 2013, pp. 415--418.

\bibitem{DBLP:conf/www/ChengWMSX17}
P.~Cheng, S.~Wang, J.~Ma, J.~Sun, and H.~Xiong, ``Learning to recommend accurate and diverse items,'' in \emph{{WWW}}, 2017, pp. 183--192.

\bibitem{DBLP:conf/dasc/LiZZZL17}
S.~Li, Y.~Zhou, D.~Zhang, Y.~Zhang, and X.~Lan, ``Learning to diversify recommendations based on matrix factorization,'' in \emph{DataCom}, 2017, pp. 68--74.

\bibitem{DBLP:conf/aaai/0020X0MZZT20}
Y.~Liu, Y.~Xiao, Q.~Wu, C.~Miao, J.~Zhang, B.~Zhao, and H.~Tang, ``Diversified interactive recommendation with implicit feedback,'' in \emph{{AAAI}}, 2020, pp. 4932--4939.

\bibitem{DBLP:conf/sigir/LiangQLY21}
Y.~Liang, T.~Qian, Q.~Li, and H.~Yin, ``Enhancing domain-level and user-level adaptivity in diversified recommendation,'' in \emph{{SIGIR}}, 2021, pp. 747--756.

\bibitem{DBLP:conf/recsys/Hurley13}
N.~J. Hurley, ``Personalised ranking with diversity,'' in \emph{{ACM}}, 2013, pp. 379--382.

\bibitem{DBLP:conf/uic/YangFW18}
W.~Yang, S.~Fan, and H.~Wang, ``An item-diversity-based collaborative filtering algorithm to improve the accuracy of recommender system,'' in \emph{SCALCOM}, 2018, pp. 106--110.

\bibitem{DBLP:conf/www/ZhengGCJL21}
Y.~Zheng, C.~Gao, L.~Chen, D.~Jin, and Y.~Li, ``{DGCN:} diversified recommendation with graph convolutional networks,'' in \emph{{WWW}}, 2021, pp. 401--412.

\bibitem{DBLP:journals/ipm/IsufiPH21}
E.~Isufi, M.~Pocchiari, and A.~Hanjalic, ``Accuracy-diversity trade-off in recommender systems via graph convolutions,'' \emph{Inf. Process. Manag.}, vol.~58, no.~2, p. 102459, 2021.

\bibitem{DBLP:conf/eaamo/GuoKJG21}
W.~Guo, K.~Krauth, M.~I. Jordan, and N.~Garg, ``The stereotyping problem in collaboratively filtered recommender systems,'' in \emph{{EAAMO}}, 2021, pp. 6:1--6:10.

\bibitem{DBLP:conf/icml/CurmeiDR21}
M.~Curmei, S.~Dean, and B.~Recht, ``Quantifying availability and discovery in recommender systems via stochastic reachability,'' in \emph{{ICML}}, 2021, pp. 2265--2275.

\bibitem{DBLP:conf/fat/DeanRR20}
S.~Dean, S.~Rich, and B.~Recht, ``Recommendations and user agency: the reachability of collaboratively-filtered information,'' in \emph{FAT}, 2020, pp. 436--445.

\bibitem{DBLP:conf/recsys/TakacsT12}
G.~Tak{\'{a}}cs and D.~Tikk, ``Alternating least squares for personalized ranking,'' in \emph{RecSys}, 2012, pp. 83--90.

\bibitem{DBLP:conf/eccv/MensinkVPC12}
T.~Mensink, J.~Verbeek, F.~Perronnin, and G.~Csurka, ``Metric learning for large scale image classification: Generalizing to new classes at near-zero cost,'' in \emph{{ECCV}}, 2012, pp. 488--501.

\bibitem{DBLP:journals/tmm/YaoSZYCW21}
X.~Yao, D.~She, H.~Zhang, J.~Yang, M.~Cheng, and L.~Wang, ``Adaptive deep metric learning for affective image retrieval and classification,'' \emph{{IEEE} Trans. Multim.}, vol.~23, pp. 1640--1653, 2021.

\bibitem{DBLP:conf/cvpr/WangSLRWPCW14}
J.~Wang, Y.~Song, T.~Leung, C.~Rosenberg, J.~Wang, J.~Philbin, B.~Chen, and Y.~Wu, ``Learning fine-grained image similarity with deep ranking,'' in \emph{{CVPR}}, 2014, pp. 1386--1393.

\bibitem{DBLP:conf/iccv/WangZWLL17}
J.~Wang, F.~Zhou, S.~Wen, X.~Liu, and Y.~Lin, ``Deep metric learning with angular loss,'' in \emph{{ICCV}}, 2017, pp. 2612--2620.

\bibitem{DBLP:conf/cvpr/KarlinskySHSAFG19}
L.~Karlinsky, J.~Shtok, S.~Harary, E.~Schwartz, A.~Aides, R.~S. Feris, R.~Giryes, and A.~M. Bronstein, ``Repmet: Representative-based metric learning for classification and few-shot object detection,'' in \emph{{CVPR}}, 2019, pp. 5197--5206.

\bibitem{DBLP:conf/ijcai/FengLZCCY15}
S.~Feng, X.~Li, Y.~Zeng, G.~Cong, Y.~M. Chee, and Q.~Yuan, ``Personalized ranking metric embedding for next new {POI} recommendation,'' in \emph{{IJCAI}}, 2015, pp. 2069--2075.

\bibitem{DBLP:journals/pami/BaoX0CH23}
S.~Bao, Q.~Xu, Z.~Yang, X.~Cao, and Q.~Huang, ``Rethinking collaborative metric learning: Toward an efficient alternative without negative sampling,'' \emph{{IEEE} Trans. Pattern Anal. Mach. Intell.}, vol.~45, no.~1, pp. 1017--1035, 2023.

\bibitem{DBLP:journals/jmlr/WeinbergerS09}
K.~Q. Weinberger and L.~K. Saul, ``Distance metric learning for large margin nearest neighbor classification,'' \emph{J. Mach. Learn. Res.}, vol.~10, pp. 207--244, 2009.

\bibitem{DBLP:conf/nips/CortesM03}
C.~Cortes and M.~Mohri, ``{AUC} optimization vs. error rate minimization,'' in \emph{NeurIPS}, 2003, pp. 313--320.

\bibitem{DBLP:journals/tnn/GultekinSRP20}
S.~Gultekin, A.~Saha, A.~Ratnaparkhi, and J.~W. Paisley, ``{MBA:} mini-batch {AUC} optimization,'' \emph{{IEEE} Trans. Neural Networks Learn. Syst.}, vol.~31, no.~12, pp. 5561--5574, 2020.

\bibitem{DBLP:journals/corr/abs-2206-00260}
Q.~Hu, Y.~Zhong, and T.~Yang, ``Multi-block min-max bilevel optimization with applications in multi-task deep {AUC} maximization,'' 2022.

\bibitem{DBLP:conf/uai/LyuY18}
S.~Lyu and Y.~Ying, ``A univariate bound of area under {ROC},'' in \emph{{UAI}}, 2018, pp. 43--52.

\bibitem{DBLP:journals/pami/YangXBCH22}
Z.~Yang, Q.~Xu, S.~Bao, X.~Cao, and Q.~Huang, ``Learning with multiclass {AUC:} theory and algorithms,'' \emph{{IEEE} Trans. Pattern Anal. Mach. Intell.}, vol.~44, no.~11, pp. 7747--7763, 2022.

\bibitem{DBLP:conf/nips/WenXYHH21}
P.~Wen, Q.~Xu, Z.~Yang, Y.~He, and Q.~Huang, ``When false positive is intolerant: End-to-end optimization with low {FPR} for multipartite ranking,'' in \emph{NeurIPS}, 2021, pp. 5025--5037.

\bibitem{DBLP:conf/iccvw/LiuHQGLJ19}
X.~Liu, X.~Han, Y.~Qiao, Y.~Ge, S.~Li, and L.~Jun, ``Unimodal-uniform constrained wasserstein training for medical diagnosis,'' in \emph{{ICCV}}, 2019, pp. 332--341.

\bibitem{DBLP:conf/icml/ZhuLWWY22}
D.~Zhu, G.~Li, B.~Wang, X.~Wu, and T.~Yang, ``When {AUC} meets {DRO:} optimizing partial {AUC} for deep learning with non-convex convergence guarantee,'' in \emph{{ICML}}, 2022, pp. 27\,548--27\,573.

\bibitem{DBLP:conf/ijcai/GaoZ15}
W.~Gao and Z.~Zhou, ``On the consistency of {AUC} pairwise optimization,'' in \emph{{IJCAI}}, 2015, pp. 939--945.

\bibitem{DBLP:conf/nips/YingWL16}
Y.~Ying, L.~Wen, and S.~Lyu, ``Stochastic online {AUC} maximization,'' in \emph{NeurIPS}, 2016, pp. 451--459.

\bibitem{DBLP:journals/corr/abs-2203-15046}
T.~Yang and Y.~Ying, ``{AUC} maximization in the era of big data and {AI:} {A} survey,'' \emph{ACM Computing Surveys}, 2022.

\bibitem{DBLP:journals/corr/abs-2203-14177}
D.~Zhu, X.~Wu, and T.~Yang, ``Benchmarking deep {AUROC} optimization: Loss functions and algorithmic choices,'' \emph{Arxiv}, 2022.

\bibitem{DBLP:conf/wsdm/WangF0NC21}
W.~Wang, F.~Feng, X.~He, L.~Nie, and T.~Chua, ``Denoising implicit feedback for recommendation,'' in \emph{{WSDM}}, 2021, pp. 373--381.

\bibitem{DBLP:conf/icml/WuZGYZ21}
Q.~Wu, H.~Zhang, X.~Gao, J.~Yan, and H.~Zha, ``Towards open-world recommendation: An inductive model-based collaborative filtering approach,'' in \emph{{ICML}}, 2021, pp. 11\,329--11\,339.

\bibitem{DBLP:conf/nips/LeiLK20}
Y.~Lei, A.~Ledent, and M.~Kloft, ``Sharper generalization bounds for pairwise learning,'' in \emph{NeurIPs}, 2020.

\bibitem{DBLP:conf/cvpr/VeitBK17}
A.~Veit, S.~J. Belongie, and T.~Karaletsos, ``Conditional similarity networks,'' in \emph{{CVPR}}, 2017, pp. 1781--1789.

\bibitem{DBLP:conf/cvpr/YeSZ22}
H.~Ye, Y.~Shi, and D.~Zhan, ``Identifying ambiguous similarity conditions via semantic matching,'' in \emph{{CVPR}}, 2022, pp. 16\,589--16\,598.

\bibitem{DBLP:conf/iccv/TanVSP19}
R.~Tan, M.~I. Vasileva, K.~Saenko, and B.~A. Plummer, ``Learning similarity conditions without explicit supervision,'' in \emph{{ICCV}}, 2019, pp. 10\,372--10\,381.

\bibitem{DBLP:conf/iccv/NigamTR19}
I.~Nigam, P.~Tokmakov, and D.~Ramanan, ``Towards latent attribute discovery from triplet similarities,'' in \emph{{ICCV}}, 2019, pp. 402--410.

\bibitem{DBLP:journals/ijon/LeiDB16}
Y.~Lei, L.~Ding, and Y.~Bi, ``Local rademacher complexity bounds based on covering numbers,'' \emph{Neurocomputing}, vol. 218, pp. 320--330, 2016.

\bibitem{ledoux1991probability}
M.~Ledoux and M.~Talagrand, \emph{Probability in Banach Spaces: isoperimetry and processes}, 1991.

\bibitem{DBLP:conf/iclr/LongS20}
P.~M. Long and H.~Sedghi, ``Generalization bounds for deep convolutional neural networks,'' in \emph{{ICLR} 2020}, 2020.

\bibitem{DBLP:conf/icml/LiL21}
S.~Li and Y.~Liu, ``Sharper generalization bounds for clustering,'' in \emph{{ICML}}, 2021, pp. 6392--6402.

\bibitem{DBLP:journals/jc/Zhou02}
D.~Zhou, ``The covering number in learning theory,'' \emph{J. Complex.}, vol.~18, no.~3, pp. 739--767, 2002.

\bibitem{DBLP:journals/corr/abs-2109-02355}
Y.~Dar, V.~Muthukumar, and R.~G. Baraniuk, ``A farewell to the bias-variance tradeoff? an overview of the theory of overparameterized machine learning,'' 2021.

\bibitem{DBLP:conf/iclr/NakkiranKBYBS20}
P.~Nakkiran, G.~Kaplun, Y.~Bansal, T.~Yang, B.~Barak, and I.~Sutskever, ``Deep double descent: Where bigger models and more data hurt,'' in \emph{{ICLR}}, 2020.

\bibitem{DBLP:conf/www/WangX000C20}
X.~Wang, Y.~Xu, X.~He, Y.~Cao, M.~Wang, and T.~Chua, ``Reinforced negative sampling over knowledge graph for recommendation,'' in \emph{{WWW}}, 2020, pp. 99--109.

\bibitem{DBLP:conf/incdm/MatsuiNYN21}
R.~Matsui, T.~Naito, S.~Yaginuma, and K.~Nakata, ``Confident collaborative metric learning,'' in \emph{{ICDM}}, 2021, pp. 246--253.

\bibitem{DBLP:conf/recsys/RendleKZA20}
S.~Rendle, W.~Krichene, L.~Zhang, and J.~R. Anderson, ``Neural collaborative filtering vs. matrix factorization revisited,'' in \emph{RecSys}, 2020, pp. 240--248.

\bibitem{DBLP:conf/iccv/HenriquesCCB13}
J.~F. Henriques, J.~Carreira, R.~Caseiro, and J.~Batista, ``Beyond hard negative mining: Efficient detector learning via block-circulant decomposition,'' in \emph{{ICCV}}, 2013, pp. 2760--2767.

\bibitem{DBLP:conf/sigir/ZhangCWY13}
W.~Zhang, T.~Chen, J.~Wang, and Y.~Yu, ``Optimizing top-n collaborative filtering via dynamic negative item sampling,'' in \emph{{SIGIR}}, 2013, pp. 785--788.

\bibitem{DBLP:conf/kdd/ChenSSH17}
T.~Chen, Y.~Sun, Y.~Shi, and L.~Hong, ``On sampling strategies for neural network-based collaborative filtering,'' in \emph{{SIGKDD}}, 2017, pp. 767--776.

\bibitem{DBLP:conf/uai/YuanX0GZCJ18}
F.~Yuan, X.~Xin, X.~He, G.~Guo, W.~Zhang, T.~Chua, and J.~M. Joemon, ``f\({}_{\mbox{bgd}}\): Learning embeddings from positive unlabeled data with {BGD},'' in \emph{{UAI}}, 2018, pp. 198--207.

\bibitem{DBLP:journals/pami/SongHCLN22}
K.~Song, J.~Han, G.~Cheng, J.~Lu, and F.~Nie, ``Adaptive neighborhood metric learning,'' \emph{T-PAMI}, vol.~44, no.~9, pp. 4591--4604, 2022.

\bibitem{DBLP:conf/iccv/ManmathaWSK17}
R.~Manmatha, C.~Wu, A.~J. Smola, and P.~Kr{\"{a}}henb{\"{u}}hl, ``Sampling matters in deep embedding learning,'' in \emph{ICCV}, 2017, pp. 2859--2867.

\bibitem{DBLP:conf/nips/SwezeyGCE21}
R.~M.~E. Swezey, A.~Grover, B.~Charron, and S.~Ermon, ``Pirank: Scalable learning to rank via differentiable sorting,'' in \emph{NeurIPS}, 2021, pp. 21\,644--21\,654.

\bibitem{DBLP:journals/ipl/OgryczakT03}
W.~Ogryczak and A.~Tamir, ``Minimizing the sum of the k largest functions in linear time,'' \emph{Inf. Process. Lett.}, vol.~85, no.~3, pp. 117--122, 2003.

\bibitem{NIPS2017_6c524f9d}
Y.~Fan, S.~Lyu, Y.~Ying, and B.~Hu, ``Learning with average top-k loss,'' in \emph{NeurIPS}, 2017.

\bibitem{10.1145/2213556.2213580}
A.~Borodin, H.~C. Lee, and Y.~Ye, ``Max-sum diversification, monotone submodular functions and dynamic updates,'' in \emph{SIGMOD/PODS}, 2012, p. 155–166.

\bibitem{DBLP:conf/www/ZieglerMKL05}
C.~Ziegler, S.~M. McNee, J.~A. Konstan, and G.~Lausen, ``Improving recommendation lists through topic diversification,'' in \emph{{WWW}}, 2005, pp. 22--32.

\bibitem{DBLP:journals/tiis/KaminskasB17}
M.~Kaminskas and D.~Bridge, ``Diversity, serendipity, novelty, and coverage: {A} survey and empirical analysis of beyond-accuracy objectives in recommender systems,'' \emph{{ACM} Trans. Interact. Intell. Syst.}, vol.~7, no.~1, pp. 2:1--2:42, 2017.

\bibitem{DBLP:conf/recsys/WestonWY13}
J.~Weston, R.~J. Weiss, and H.~Yee, ``Nonlinear latent factorization by embedding multiple user interests,'' in \emph{RecSys}, 2013, pp. 65--68.

\bibitem{DBLP:conf/kdd/LiuBXG022}
B.~Liu, B.~Bai, W.~Xie, Y.~Guo, and H.~Chen, ``Task-optimized user clustering based on mobile app usage for cold-start recommendations,'' in \emph{{KDD}}, 2022, pp. 3347--3356.

\bibitem{DBLP:journals/corr/abs-1907-08674}
K.~Rama, P.~Kumar, and B.~Bhasker, ``Deep learning to address candidate generation and cold start challenges in recommender systems: {A} research survey,'' \emph{CoRR}, vol. abs/1907.08674, 2019.

\bibitem{DBLP:journals/jmlr/SrivastavaHKSS14}
N.~Srivastava, G.~E. Hinton, A.~Krizhevsky, I.~Sutskever, and R.~Salakhutdinov, ``Dropout: a simple way to prevent neural networks from overfitting,'' \emph{J. Mach. Learn. Res.}, vol.~15, no.~1, pp. 1929--1958, 2014.

\bibitem{mc}
C.~McDiarmid, ``Concentration,'' in \emph{Probabilistic methods for algorithmic discrete mathematics}, 1998, pp. 195--248.

\bibitem{DBLP:journals/corr/abs-2203-01505}
Y.~Yao, Q.~Lin, and T.~Yang, ``Large-scale optimization of partial {AUC} in a range of false positive rates,'' 2022.

\bibitem{DBLP:conf/ijcai/WangCL13}
H.~Wang, B.~Chen, and W.~Li, ``Collaborative topic regression with social regularization for tag recommendation,'' in \emph{{IJCAI}}, 2013, pp. 2719--2725.

\bibitem{DBLP:conf/recsys/AbelDEK17}
F.~Abel, Y.~Deldjoo, M.~Elahi, and D.~Kohlsdorf, ``Recsys challenge 2017: Offline and online evaluation,'' in \emph{{ACM} RecSys}, 2017, pp. 372--373.

\bibitem{linden2003amazon}
G.~Linden, B.~Smith, and J.~York, ``Amazon. com recommendations: Item-to-item collaborative filtering,'' \emph{IEEE Internet computing}, no.~1, pp. 76--80, 2003.

\bibitem{DBLP:conf/www/LiangKHJ18}
D.~Liang, R.~G. Krishnan, M.~D. Hoffman, and T.~Jebara, ``Variational autoencoders for collaborative filtering,'' in \emph{{WWW}}, 2018, pp. 689--698.

\bibitem{DBLP:conf/sigir/0001DWLZ020}
X.~He, K.~Deng, X.~Wang, Y.~Li, Y.~Zhang, and M.~Wang, ``Lightgcn: Simplifying and powering graph convolution network for recommendation,'' in \emph{{SIGIR}}, 2020, pp. 639--648.

\bibitem{paszke2017automatic}
A.~Paszke, S.~Gross, S.~Chintala, G.~Chanan, E.~Yang, Z.~DeVito, Z.~Lin, A.~Desmaison, L.~Antiga, and A.~Lerer, ``Automatic differentiation in pytorch,'' 2017.

\bibitem{DBLP:journals/corr/KingmaB14}
D.~P. Kingma and J.~Ba, ``Adam: {A} method for stochastic optimization,'' in \emph{{ICLR}}, 2015.

\bibitem{DBLP:conf/iccbr/SmythM01}
B.~Smyth and P.~McClave, ``Similarity vs. diversity,'' in \emph{{ICCBR}}, 2001, pp. 347--361.

\bibitem{DBLP:conf/recsys/VargasC11}
S.~Vargas and P.~Castells, ``Rank and relevance in novelty and diversity metrics for recommender systems,'' in \emph{{ACM} RecSys}, 2011, pp. 109--116.

\bibitem{DBLP:journals/jmlr/JahrerT12a}
M.~Jahrer and A.~T{\"{o}}scher, ``Collaborative filtering ensemble for ranking,'' in \emph{{KDD}}, 2012, pp. 153--167.

\bibitem{DBLP:journals/corr/TrofimovSHLMA17}
M.~Trofimov, S.~Sidana, O.~Horodnitskii, C.~Laclau, Y.~Maximov, and M.~Amini, ``Representation learning and pairwise ranking for implicit and explicit feedback in recommendation systems,'' in \emph{Arxiv}, 2017.

\bibitem{DBLP:conf/recsys/SidanaLA18}
S.~Sidana, C.~Laclau, and M.~Amini, ``Learning to recommend diverse items over implicit feedback on {PANDOR},'' in \emph{{ACM} RecSys}, 2018, pp. 427--431.

\bibitem{DBLP:conf/recsys/Shi13}
L.~Shi, ``Trading-off among accuracy, similarity, diversity, and long-tail: a graph-based recommendation approach,'' in \emph{{ACM} RecSys}, 2013, pp. 57--64.

\bibitem{DBLP:conf/recsys/CovingtonAS16}
P.~Covington, J.~Adams, and E.~Sargin, ``Deep neural networks for youtube recommendations,'' in \emph{{ACM} RecSys}, 2016, pp. 191--198.

\end{thebibliography}
	\begin{IEEEbiography}[{\includegraphics[width=1in,height=1.25in,clip,keepaspectratio]{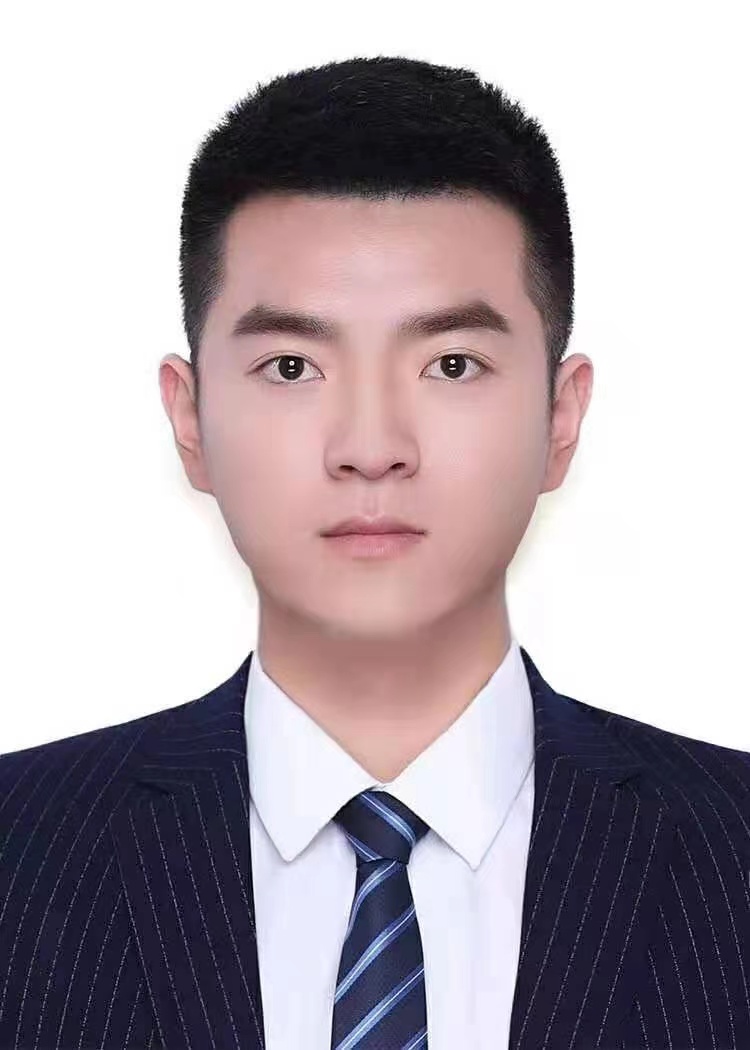}}]{\textbf{Shilong Bao}} 
		received the B.S. degree in College of Computer Science and Technology from Qingdao University in 2019. He is currently pursuing the Ph.D. degree with University of Chinese Academy of Sciences. His research interest is machine learning and data mining. He has authored or coauthored several academic papers in top-tier international conferences and journals including T-PAMI, NeurIPS, ICML, and ACM Multimedia. He also served as a reviewer for several top-tier conferences and journals, such as ICML, NeurIPS, ICLR and IEEE Transactions on Multimedia.
	\end{IEEEbiography}
	
	\begin{IEEEbiography}
		[{\includegraphics[width=1in,height=1.25in,clip,keepaspectratio]{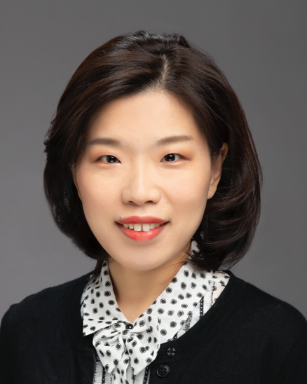}}]{Qianqian Xu} received the B.S. degree in computer science from China University of Mining and Technology in 2007 and the Ph.D. degree in computer science from University of Chinese Academy of Sciences in 2013. She is currently a Professor with the Institute of Computing Technology, Chinese Academy of Sciences, Beijing, China. Her research interests include statistical machine learning, with applications in multimedia and computer vision. She has authored or coauthored 70+ academic papers in prestigious international journals and conferences (including T-PAMI, IJCV, T-IP, NeurIPS, ICML, CVPR, AAAI, etc). Moreover, she serves as an associate editor of IEEE Transactions on Circuits and Systems for Video Technology, IEEE Transactions on Multimedia, and ACM Transactions on Multimedia Computing, Communications, and Applications.
	\end{IEEEbiography}
	
	\begin{IEEEbiography}[{\includegraphics[width=1in,height=1.25in,clip,keepaspectratio]{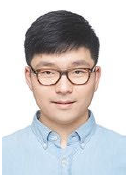}}]{Zhiyong Yang} received the M.Sc. degree in computer science and technology from University of Science and Technology Beijing (USTB) in 2017, and the Ph.D. degree from University of Chinese Academy of Sciences (UCAS) in 2021. He is currently a Tenure-track Assistant Professor with the UCAS. His research interests lie in machine learning and learning theory, with special focus on AUC optimization, meta-learning/multi-task learning, and learning theory for recommender systems. He has authored or coauthored several academic papers in top-tier international conferences and journals including T-PAMI/ICML/NeurIPS/CVPR. He served as a reviewer for several top-tier journals and conferences such as T-PAMI, ICML, NeurIPS and ICLR.
	\end{IEEEbiography}
	
	\begin{IEEEbiography}[{\includegraphics[width=1in,height=1.25in,clip,keepaspectratio]{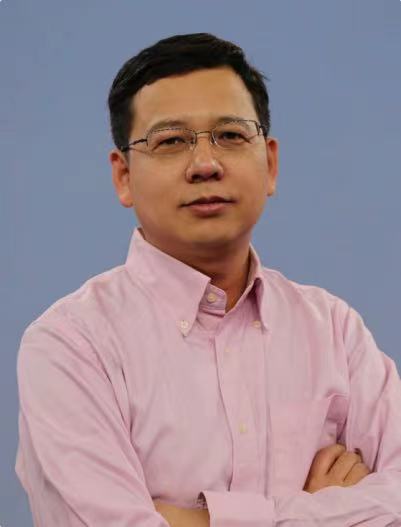}}]{\textbf{Yuan He}} received his B.S. degree and Ph.D. degree from Tsinghua University, P.R. China. He is a Senior Staff Engineer in the Security Department of Alibaba Group, and working on artificial intelligence-based content moderation and intellectual property protection systems. Before joining Alibaba, he was a research manager at Fujitsu working on document analysis system. He has published more than 30 papers in computer vision and machine learning related conferences and journals including CVPR, ICCV, ICML, NeurIPS, AAAI and ACM MM. His research interests include computer vision, machine learning, and AI security.
	\end{IEEEbiography}
	
	\begin{IEEEbiography}
		[{\includegraphics[width=1in,height=1.25in,clip,keepaspectratio]{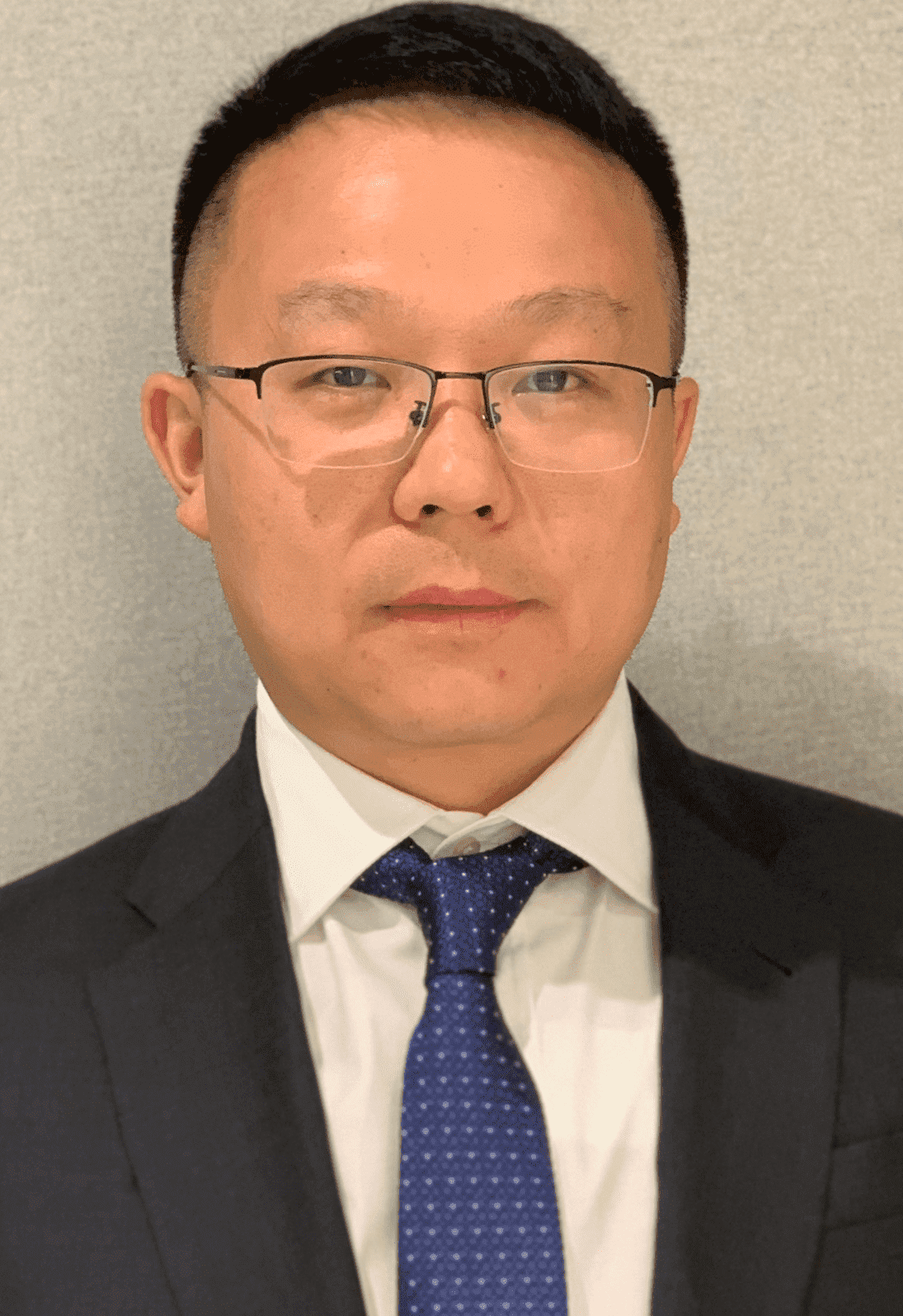}}]{Xiaochun Cao}, is a Professor of  School of Cyber Science and Technology, Shenzhen Campus, Sun Yat-sen University. He received the B.E. and M.E. degrees both in computer science from Beihang University (BUAA), China, and the Ph.D. degree in computer science from the University of Central Florida, USA, with his dissertation nominated for the university level Outstanding Dissertation Award. After graduation, he spent about three years at ObjectVideo Inc. as a Research Scientist. From 2008 to 2012, he was a professor at Tianjin University. From 2012 to 2022, he was a professor at Institute of Information Engineering, Chinese Academy of Sciences.  He has authored and coauthored over 200 journal and conference papers.  In 2004 and 2010, he was the recipients of the Piero Zamperoni best student paper award at the International Conference on Pattern Recognition. He is a fellow of IET and a Senior Member of IEEE. He is an associate editor of IEEE Transactions on Image Processing, IEEE Transactions on Circuits and Systems for Video Technology and IEEE Transactions on Multimedia.
	\end{IEEEbiography}
	
	\begin{IEEEbiography}
		[{\includegraphics[width=1in,height=1.25in,clip,keepaspectratio]{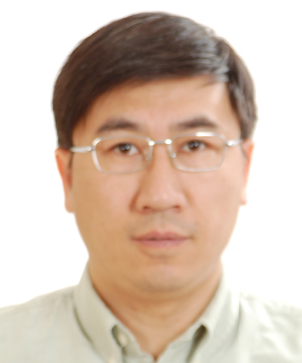}}] {Qingming Huang} is a chair professor in the University of Chinese Academy of Sciences and an adjunct research professor in the Institute of Computing Technology, Chinese Academy of Sciences. He graduated with a Bachelor degree in Computer Science in 1988 and Ph.D. degree in Computer Engineering in 1994, both from Harbin Institute of Technology, China. His research areas include multimedia computing, image processing, computer vision and pattern recognition. He has authored or coauthored more than 400 academic papers in prestigious international journals and top-level international conferences. He was the associate editor of IEEE Trans. on CSVT and Acta Automatica Sinica, and the reviewer of various international journals including IEEE Trans. on PAMI, IEEE Trans. on Image Processing, IEEE Trans. on Multimedia, etc. He is a Fellow of IEEE and has served as general chair, program chair, area chair and TPC member for various conferences, including ACM Multimedia, CVPR, ICCV, ICME, ICMR, PCM, BigMM, PSIVT, etc.
	\end{IEEEbiography}

	\clearpage
	\newpage
	\onecolumn
	\appendices
	
	\section*{\textcolor{blue}{\Large{Contents}}}
	\startcontents[sections]
	\printcontents[sections]{l}{1}{\setcounter{tocdepth}{2}}
	
	\clearpage
	\section{Generalization Bounds and its proofs} 
	This section will show the detailed results and proofs for the generalization. Without loss of generality, the following presentations merely consider the DPCML with the Basic Preference Assignment (BPA) strategy. Note that similar conclusions are still satisfied with the Adaptive Preference Assignment (APA) scheme because our bound (Thm.\ref{them1}) is independent of the number of preference embeddings assigned to each user.
	\label{supp.sec.b}
	\subsection{Preliminary Lemmas}
	In this section, we first briefly review some preparatory knowledge for the proof.
	\begin{defi} [Bounded Difference Property]\label{def:bdp}
		Given a group of independent random variables $X_1, X_2, \cdots, X_n$ where $X_t \in \mathbb{X}, \forall t$, $f(X_1,X_2,\cdots, X_n)$ is satisfied with the bounded difference property, if there exists some non-negative constants $c_1, c_2,\cdots, c_n$, such that: 
		\begin{equation}
			\sup_{x_1,x_2,\cdots,x_n, x'_t} \left|f(x_1,\cdots,x_n) - f(x_1,\cdots, x_{t-1},x'_t,\cdots,x_n)\right| \le c_t, ~ \forall t,  1 \le t \le n.
		\end{equation}
	\end{defi}
	Hereafter, if any function $f$ holds the Bounded Difference Property, the following Mcdiarmid's inequality is always satisfied.
	\begin{lem}[Mcdiarmid's Inequality \cite{mc}] \label{lem:mc} Assume we have $n$ independent random variables $X_1, X_2, \dots, X_n$ that all of them are chosen from the set $\mathcal{X}$. For a function $f: \mathcal{X} \rightarrow \mathbb{R}$, $\forall t, 1 \le t \le n$, if the following inequality holds:
		\[
		\sup_{x_1,x_2,\cdots,x_n, x'_t} \left|f(x_1,\cdots,x_n) - f(x_1,\cdots, x_{t-1},x'_t,\cdots,x_n)\right| \le c_t, ~ \forall t,  1 \le t \le n.
		\]with $\boldsymbol{x} \neq \boldsymbol{x}'$, then for all $\epsilon >0$, we have
		\[\mathbb{P}[ \mathbb{E}(f) - f \ge \epsilon ] \le \exp\left(\dfrac{-2\epsilon^2}{\sum_{t=1}^nc_t^2} \right), \]
		\[
		\mathbb{P}[ f - \mathbb{E}(f) \ge \epsilon ] \le \exp\left(\dfrac{-2\epsilon^2}{\sum_{t=1}^nc_t^2} \right).
		\]
	\end{lem}
	\begin{lem}[Union bound/Boole’s inequality] \label{lem46} Given the countable or finite set of events $E_i$, the probability that at least one event happens is less than or equal to the sum of all probabilities of the events happened individually, i.e.,
		\begin{equation}
			\begin{aligned}
				\mathbb{P}\left[\mathop{\cup}\limits_{i} E_i\right] &\le \sum_{i} \mathbb{P}\left[E_i\right]
			\end{aligned}
		\end{equation}
		
	\end{lem}
	\begin{lem} [$\phi$-Lipschitz Continuous] \label{lem2} Given a set $\mathcal{X}$ and a function $f: \mathcal{X} \rightarrow \mathbb{R}$, if $f$ is continuously differentiable on $\mathcal{X}$ such that, $\forall x, y \in \mathcal{X}$, the following condition holds with a real constant $\phi$:
		\[
		\left\|f(x) - f(y)\right\| \le \phi \left\|x - y\right\|.
		\]
		Thereafter, $f$ is said to be a $\phi$-Lipschitz continuous function.
	\end{lem}
	
	\subsection{Key Lemmas}
	\begin{rdef2}
		[$\epsilon$-Covering] \label{rdef2} \cite{ledoux1991probability} Let $(\mathcal{F}, \rho)$ be a (pesudo) metric space, and $\mathcal{G} \subseteq \mathcal{F}$. $\{f_1, \dots, f_n\}$ is said to be an $\epsilon$-covering of $\mathcal{G}$ if $\mathcal{G} \subseteq \mathop{\cup}\limits_{i=1}^n \mathcal{B}(f_i, \epsilon)$, i.e., $\forall g \in \mathcal{G}$, $\exists i$ such that $\rho(g, f_i) \le \epsilon$.
	\end{rdef2}
	\begin{rdef3} [Covering Number] \label{rdef3} \cite{ledoux1991probability} According to the notations in Def.\ref{rdef2}, the covering number of $\mathcal{G}$ with radius $\epsilon$ is defined as:
		\begin{equation} \nonumber
			\mathcal{N}(\epsilon;\mathcal{G}, \rho) = \min\{n: \exists \epsilon-covering \ over \ \mathcal{G} \ with \ size \ n\}
		\end{equation}
	\end{rdef3}
	\begin{rassu1}
		[Basic Assumptions] \label{rassu1} We assume that all the embeddings of users and items are chosen from the following embedding hypothesis space:
		\begin{equation}
			\mathcal{H}_R = \left\{\bmg: \bmg \in \mathbb{R}^d, \|\bmg\| \le r\right\},
		\end{equation}
		where $\boldsymbol{g}^c_{u_i} \in \mathcal{H}_R, u_i \in \mcu, c \in [C]$ and $\boldsymbol{g}_{v_j} \in \mathcal{H}_R, v_j \in \mci$.
	\end{rassu1}
	\begin{rlem1} \label{covering_rlem} \cite{DBLP:conf/iclr/LongS20,DBLP:conf/icml/LiL21, DBLP:journals/jc/Zhou02}
		The covering number of the hypothesis class $\mathcal{H}_R$ has the following upper bound: 
		\begin{equation}
			\log \mathcal{N}(\epsilon;\mathcal{H}_R, \rho) \le d \log \left(\frac{3r}{\epsilon}\right),
		\end{equation}
		where $d$ is the dimension of embedding space. 
	\end{rlem1}
	
	In what follows, we will present the key lemmas to derive the upper bounds.
	
	\begin{lem} \label{lem8} Let $\varepsilon$ be the generalization error between $\hat{\mathcal{L}}_{\mcd}(\bmg)$ and $\expe[\hat{\mathcal{L}}_{\mcd}(\bmg)]$. Then by constructing an $\sigma$-covering $\{\bmg_1, \bmg_2, \dots, \bmg_n\}$ of $\mathcal{H}_R$ with $\sigma = \frac{\varepsilon}{16r(4 + \eta)}$, the following inequality holds
		\begin{equation}
			\begin{aligned} \label{eqlem8}
				\mathbb{P} \left[ \sup\limits_{\bmg \in \mathcal{B}(\bmg_l, \sigma)}\left|\hat{\mathcal{L}}_{\mcd}(\bmg) - \expe [\hat{\mathcal{L}}_{\mcd}(\bmg)]\right|\le \varepsilon \right] &\ge \mathbb{P}\left[\left|\hat{\mathcal{L}}_{\mcd}(\bmg_l) - \expe [\hat{\mathcal{L}}_{\mcd}(\bmg_l)]\right| \le \frac{\varepsilon}{2}\right], \ \ \forall l \in [n],
			\end{aligned}
		\end{equation}

		
	\end{lem}
	\begin{proof} Assume there exists an $\sigma$-covering $\{\bmg_1, \bmg_2, \dots, \bmg_n\}$ of $\mathcal{H}_R$. To prove (\ref{eqlem8}), we turn to prove the following inequality:
		\begin{equation}
			\begin{aligned} \label{eq122}
				\left||\hat{\mathcal{L}}_{\mcd}(\bmg) - \expe [\hat{\mathcal{L}}_{\mcd}(\bmg)]| - |\hat{\mathcal{L}}_{\mcd}(\bmg_l) - \expe [\hat{\mathcal{L}}_{\mcd}(\bmg_l)]|\right| &\le \frac{\varepsilon}{2}, \ \forall l \in [n].
			\end{aligned}
		\end{equation}
		
		Note that, we have
		\begin{equation}
			\begin{aligned}
				\left||\hat{\mathcal{L}}_{\mcd}(\bmg) - \expe [\hat{\mathcal{L}}_{\mcd}(\bmg)]| - |\hat{\mathcal{L}}_{\mcd}(\bmg_l) - \expe [\hat{\mathcal{L}}_{\mcd}(\bmg_l)]|\right|
				&\overset{\textcolor{orange}{(**)}}{\le} \left|\hat{\mathcal{L}}_{\mcd}(\bmg) - \expe [\hat{\mathcal{L}}_{\mcd}(\bmg)] - \left(\hat{\mathcal{L}}_{\mcd}(\bmg_l) - \expe [\hat{\mathcal{L}}_{\mcd}(\bmg_l)]\right)\right| \\
				&\overset{\textcolor{orange}{(*)}}{\le} \left|\hat{\mathcal{L}}_{\mcd}(\bmg) - \hat{\mathcal{L}}_{\mcd}(\bmg_l)\right| + \left|\expe [\hat{\mathcal{L}}_{\mcd}(\bmg_l)] - \expe [\hat{\mathcal{L}}_{\mcd}(\bmg)]\right|,
			\end{aligned}
		\end{equation}
		where $\textcolor{orange}{(*)}$ and $\textcolor{orange}{(**)}$ follows the facts $|x + y| \le |x| + |y|$ and $||x| - |y|| \le |x - y|$, respectively.
		
		Then, to achieve (\ref{eq122}), we only need to show that the following inequation holds: 
		\begin{equation}
			\begin{aligned}\label{eq177}
				\left|\hat{\mathcal{L}}_{\mcd}(\bmg) - \hat{\mathcal{L}}_{\mcd}(\bmg_l)\right| &\le \frac{\varepsilon}{4}, \ \ \forall l \in [n].\\
			\end{aligned}
		\end{equation}
		
		Recall that 
		\begin{equation} \label{eq15}
			\hat{\mcl}_{\mcd}(\bmg) = \hat{\mathcal{R}}_{\mcd, \bmg} + \eta \cdot \hat{\Omega}_{\mcd, \bmg},
		\end{equation}
		where 
		\begin{equation}
			\begin{aligned}
				\hat{\mathcal{R}}_{\mcd, \bmg} &= \frac{1}{|\mcu|} \sum_{u_i \in \mcu} \frac{1}{n_i^+n_i^-} \sum_{j=1}^{n_i^+} \sum_{k=1}^{n_i^-} \ell^{(i)}_g(v_j^+, v_k^-), \\
				\ell^{(i)}_g(v_j^+, v_k^-) &= \max (0, \lambda + s(u_i, v_j^+) - s(u_i, v_k^-)), \\
				s(u_i, v_j) &= \min\limits_{c \in [C]} \|\boldsymbol{g}_{u_i}^c - \boldsymbol{g}_{v_j}\|^2, \forall \ v_j, v_j \in \mci
			\end{aligned}
		\end{equation}
		and 
		\begin{equation}
			\begin{aligned}
				\hat{\Omega}_{\mcd, \bmg} &= \frac{1}{|\mcu|} \sum_{u_i \in \mcu} \left(\max\left(0, \delta_1 - \delta_{\bmg, u_i}) + \max(0, \delta_{\bmg, u_i} - \delta_2\right)\right), \\
				\delta_{\bmg, u_i} &= \frac{1}{2C(C-1)} \sum_{c_{1}, c_2 \in C} \|\boldsymbol{g}_{u_i}^{c_1} - \boldsymbol{g}_{u_i}^{c_2}\|^2.
			\end{aligned}
		\end{equation}
		
		Let us define some intermediate variables:
		\begin{equation}
			\begin{aligned}
				\hat{\mathcal{R}}_{\mathcal{D}_{u_i}, \bmg} &= \frac{1}{n_i^+n_i^-} \sum_{j=1}^{n_i^+}\sum_{k=1}^{n_i^-}\ell^{(i)}_g(v_j^+, v_k^-), \\
				\hat{\Omega}_{\mcd_{u_i}, \bmg} &= \max\left(0, \delta_1 - \delta_{\bmg, u_i}\right) + \max\left(0, \delta_{\bmg, u_i} - \delta_2\right), \\
				\Delta_{\bmg, \bmg_l}(c_1,c_2) &= \left(\|\boldsymbol{g}_{u_i}^{c_1} - \boldsymbol{g}_{v_j^+}\|^2 - \|\tilde{\boldsymbol{g}}_{u_i}^{c_2} - \tilde{\boldsymbol{g}}_{v_j^+}\|^2\right).
			\end{aligned}
		\end{equation}
		
		In this sense, we have 
		\begin{equation}
			\begin{aligned}
				\left|\hat{\mathcal{L}}_{\mcd}(\bmg) - \hat{\mathcal{L}}_{\mcd}(\bmg_l)\right| &= \left|\hat{\mathcal{R}}_{\mcd, \bmg} + \eta \cdot \hat{\Omega}_{\mcd, \bmg} -\hat{\mathcal{R}}_{\mcd, \bmg_l} - \eta \cdot \hat{\Omega}_{\mcd, \bmg_l} \right| \\
				& \le \underbrace{\left|\hat{\mathcal{R}}_{\mcd, \bmg} - \hat{\mathcal{R}}_{\mcd, \bmg_l}\right|}_{\textcolor{orange}{(1)}} + \underbrace{\eta\left|\Omega_{\mcd, \bmg} - \cdot \Omega_{\mcd, \bmg_l}\right|}_{\textcolor{orange}{(2)}}. \\
			\end{aligned}
		\end{equation}
		
		Subsequently, in terms of $\textcolor{orange}{(1)}$, we first consider a specific user $u_i$ with her/his corresponding interaction records $\mathcal{D}_{u_i}$. We have
		\begin{equation}
			\begin{aligned}
				\left|\hat{\mathcal{R}}_{\mcd_{u_i}, \bmg} - \hat{\mathcal{R}}_{\mcd_{u_i}, \bmg_l}\right| &= \left|\frac{1}{n_i^+n_i^-} \sum_{j=1}^{n_i^+}\sum_{k=1}^{n_i^-}\ell^{(i)}_{\bmg}(v_j^+, v_k^-) - \frac{1}{n_i^+n_i^-} \sum_{j=1}^{n_i^+}\sum_{k=1}^{n_i^-}\ell^{(i)}_{\bmg_l}(v_j^+, v_k^-) \right| \\
				& \le \frac{1}{n_i^+n_i^-}  \sum_{j=1}^{n_i^+}\sum_{k=1}^{n_i^-} \left|\ell^{(i)}_{\bmg}(v_j^+, v_k^-) - \ell^{(i)}_{\bmg_l}(v_j^+, v_k^-) \right| \\
				& \overset{\textcolor{orange}{(a)}}{\le} \frac{1}{n_i^+n_i^-} \sum_{j=1}^{n_i^+}\sum_{k=1}^{n_i^-} \left|s(u_i, v_j^+) - s(u_i, v_k^-) - \tilde{s}_l(u_i, v_j^+) + \tilde{s}_l(u_i, v_k^-)\right| \\
				& \overset{\textcolor{orange}{(*)}}{\le} \frac{1}{n_i^+n_i^-} \sum_{j=1}^{n_i^+}\sum_{k=1}^{n_i^-} \left(\left|s(u_i, v_j^+)  - \tilde{s}_l(u_i, v_j^+) \right| +\left|\tilde{s}_l(u_i, v_k^-) - s(u_i, v_k^-) \right|\right) \\
			\end{aligned}
		\end{equation}
		
		where \textcolor{orange}{(a)} follows the Lem.\ref{lem2} and $\ell^{(i)}_{\bmg}$ is apparently a $1$-Lipschitz continuous function. 
		
		In terms of $\left|s(u_i, v_j^+)  - \tilde{s}_l(u_i, v_j^+) \right|$, the following equation holds:
		\begin{equation}
			\begin{aligned}
				\left|s(u_i, v_j^+)  - \tilde{s}_l(u_i, v_j^+) \right| &= \left|\min\limits_{c_1 \in [C]} \|\boldsymbol{g}_{u_i}^{c_1} - \boldsymbol{g}_{v_j^+}\|^2 - \min\limits_{c_2 \in [C]} \|\tilde{\boldsymbol{g}}_{u_i}^{c_2} - \tilde{\boldsymbol{g}}_{v_j^+}\|^2 \right| \\
				&= \left|\min\limits_{c_1 \in [C]} \max\limits_{c_2 \in [C]} \Delta_{\bmg, \bmg_l}(c_1,c_2)\right| \\
				&= \max \left\{\min\limits_{c_1 \in [C]} \max\limits_{c_2 \in [C]} \Delta_{\bmg, \bmg_l}(c_1,c_2), \max\limits_{c_1 \in [C]} \min\limits_{c_2 \in [C]} \Delta_{\bmg_l, \bmg}(c_2,c_1)\right\}.
			\end{aligned}
		\end{equation}
		
		Moreover, we have 
		\begin{equation}
			\begin{aligned}
				\ \ \ \ & \min\limits_{c_1 \in [C]} \max\limits_{c_2 \in [C]} \Delta_{\bmg, \bmg_l}(c_1,c_2) \\
				&\le \max\limits_{c_1 = c2, c_1,c_2 \in [C]} \Delta_{\bmg, \bmg_l}(c_1,c_2) \\
				&\le \max\limits_{c_1 = c2, c_1,c_2 \in [C]} \left|\Delta_{\bmg, \bmg_l}(c_1,c_2) \right|\\
				&= \max\limits_{c \in [C]} \left|\left(\|\boldsymbol{g}_{u_i}^{c} - \boldsymbol{g}_{v_j^+}\| +\|\tilde{\boldsymbol{g}}_{u_i}^{c} - \tilde{\boldsymbol{g}}_{v_j^+}\| \right)\left(\|\boldsymbol{g}_{u_i}^{c} - \boldsymbol{g}_{v_j^+}\| -\|\tilde{\boldsymbol{g}}_{u_i}^{c} - \tilde{\boldsymbol{g}}_{v_j^+}\|\right)\right| \\
				& \overset{\textcolor{orange}{(**)}}{\le} \max\limits_{c \in [C]}  \left(\|\boldsymbol{g}_{u_i}^{c} - \boldsymbol{g}_{v_j^+}\| +\|\tilde{\boldsymbol{g}}_{u_i}^{c} - \tilde{\boldsymbol{g}}_{v_j^+}\| \right)\left(\|\boldsymbol{g}_{u_i}^{c} - \boldsymbol{g}_{v_j^+} -\tilde{\boldsymbol{g}}_{u_i}^{c} + \tilde{\boldsymbol{g}}_{v_j^+}\|\right) \\
				& \le 4r\left(\max\limits_{c\in [C]} \|\boldsymbol{g}_{u_i}^{c} - \tilde{\boldsymbol{g}}_{u_i}^{c}\| +  \|\tilde{\boldsymbol{g}}_{v_j^+} - \boldsymbol{g}_{v_j^+}\|\right) \\
				& \le 8r \sigma
			\end{aligned}
		\end{equation}
		
		where $\textcolor{orange}{(**)}$ follows the fact $||x| - |y|| \le |x - y|$.
		
		Similarly, we have
		\begin{equation}
			\begin{aligned}
				\left|\tilde{s}_l(u_i, v_k^-) - s(u_i, v_k^-) \right| &\le 4r\left(\max\limits_{c\in [C]} \|\boldsymbol{g}_{u_i}^{c} - \tilde{\boldsymbol{g}}_{u_i}^{c}\| +  \|\tilde{\boldsymbol{g}}_{v_k^-} - \boldsymbol{g}_{v_k^-}\|\right)  \\
				& \le 8r \sigma.
			\end{aligned}
		\end{equation}

		Thus, we have
		
		\begin{equation}
			\begin{aligned}
				\left|\hat{\mathcal{R}}_{\mcd, g} - \hat{\mathcal{R}}_{\mcd, \tilde{g}_l}\right| = 
				\left|\hat{\mathcal{R}}_{\mcd_{u_i}, \bmg} - \hat{\mathcal{R}}_{\mcd_{u_i}, \bmg_l}\right| & \le 16r \sigma.
			\end{aligned}
		\end{equation}
		
		Therefore, for all users, we also have
		\begin{equation} \label{proof1}
			\left|\hat{\mathcal{R}}_{\mcd, \bmg} - \hat{\mathcal{R}}_{\mcd, \bmg_l}\right| \le 16r \sigma.
		\end{equation}
		{\noindent} \rule[-5pt]{17.5cm}{0.05em}
		
		With respect to $\textcolor{orange}{(2)}$, we also first consider a specific user $u_i$, i.e.,
		\begin{equation}
			\begin{aligned}
				& \eta\left|\hat{\Omega}_{\mcd_{u_i}, \bmg} - \hat{\Omega}_{\mcd_{u_i}, \bmg_l}\right| \overset{\textcolor{orange}{(a)}}{\le} 2\eta \left|\delta_{\bmg, u_i} - \delta_{\bmg_l, u_i}\right| \\
				&= \frac{\eta}{C(C-1)} \left|\sum_{c_{1}, c_2 \in C} \left\|\boldsymbol{g}_{u_i}^{c_1} - \boldsymbol{g}_{u_i}^{c_2}\right\|^2 - \sum_{c_{1}, c_2 \in C} \left\|\tilde{\boldsymbol{g}}_{u_i}^{c_1} - \tilde{\boldsymbol{g}}_{u_i}^{c_2}\right\|^2\right| \\
				& \le \frac{\eta}{C(C-1)}\left|\sum_{c_{1}, c_2 \in C} \left(\left\|\boldsymbol{g}_{u_i}^{c_1} - \boldsymbol{g}_{u_i}^{c_2}\right\| + \left\|\tilde{\boldsymbol{g}}_{u_i}^{c_1} - \tilde{\boldsymbol{g}}_{u_i}^{c_2}\right\|\right)\left(\left\|\boldsymbol{g}_{u_i}^{c_1} - \boldsymbol{g}_{u_i}^{c_2}\right\| - \left\|\tilde{\boldsymbol{g}}_{u_i}^{c_1} - \tilde{\boldsymbol{g}}_{u_i}^{c_2}\right\|\right) \right| \\
				& \le \frac{\eta}{C(C-1)}\sum_{c_{1}, c_2 \in C}\left|  \left(\left\|\boldsymbol{g}_{u_i}^{c_1} - \boldsymbol{g}_{u_i}^{c_2}\right\| + \left\|\tilde{\boldsymbol{g}}_{u_i}^{c_1} - \tilde{\boldsymbol{g}}_{u_i}^{c_2}\right\|\right)\left(\left\|\boldsymbol{g}_{u_i}^{c_1} - \boldsymbol{g}_{u_i}^{c_2}\right\| - \left\|\tilde{\boldsymbol{g}}_{u_i}^{c_1} - \tilde{\boldsymbol{g}}_{u_i}^{c_2}\right\|\right)\right| \\
				& \overset{\textcolor{orange}{(**)}}{\le} \frac{4\eta r }{C(C-1)}\sum_{c_{1}, c_2 \in C}\left(\|\boldsymbol{g}_{u_i}^{c_1} -\tilde{\boldsymbol{g}}_{u_i}^{c_1} \| + \|\tilde{\boldsymbol{g}}_{u_i}^{c_2} - \boldsymbol{g}_{u_i}^{c_2}\|\right) \\
				& \le 4\eta r \left(\max\limits_{c \in [C]} \|\boldsymbol{g}_{u_i}^{c} -\tilde{\boldsymbol{g}}_{u_i}^{c}\|\right) \\
				& \le 4\eta r \sigma
			\end{aligned}
		\end{equation}
		where \textcolor{orange}{(a)} follows the Lem.\ref{lem2} and $\textcolor{orange}{(**)}$ follows $||x| - |y|| \le |x - y|$.
		
		
		In like wise, we have 
		\begin{equation}
			\begin{aligned} \label{proof2}
				\eta\left|\hat{\Omega}_{\mcd, \bmg} -  \hat{\Omega}_{\mcd, \bmg_l}\right|\le 4\eta r \sigma.
			\end{aligned}
		\end{equation}
		
		Finally, based on (\ref{proof1}) and (\ref{proof2}), we have
		\begin{equation}
			\left|\hat{\mathcal{L}}_{\mcd}(\bmg) - \hat{\mathcal{L}}_{\mcd}(\bmg_l)\right| \le 4r\sigma(4 + \eta).
		\end{equation}
		Based on this, by further choosing $\sigma = \frac{\varepsilon}{16r(4 + \eta)}$, we could construct the covering number $\mathcal{N}_1$ and $\mathcal{N}_2$ with respect to users and items, respectively, i.e., 
		\begin{equation}
			\begin{aligned}
				\mathcal{N}_1\left(\frac{\varepsilon}{16r(4 + \eta)}, \mathcal{H}_R,\rho_1\right), \ \ \rho_1 &= \max\limits_{c \in [C]} \|\boldsymbol{g}_{u_i}^{c} -\tilde{\boldsymbol{g}}_{u_i}^{c}\|, \ \ \forall u_i \in \mcu, \\
				\mathcal{N}_2\left(\frac{\varepsilon}{16r(4 + \eta)},\mathcal{H}_R, \rho_2\right), \ \ \rho_2 &=  \|\tilde{\boldsymbol{g}}_{v_j} - \boldsymbol{g}_{v_j}\|, \ \  \forall v_j \in \mci,
			\end{aligned}
		\end{equation}
		
		such that the following inequality holds:
		\[
		\left|\hat{\mathcal{L}}_{\mcd}(\bmg) - \hat{\mathcal{L}}_{\mcd}(\bmg_l)\right| \le \frac{\varepsilon}{4}.
		\]
		This completed the proof.
	\end{proof}
	
	{\noindent} \rule[-5pt]{17.5cm}{0.05em} \\
	
	\begin{lem} [Bounded Difference Property of DPCML] \label{lem9} Let $\mcd$ and $\mcd'$ be two independent datasets where exactly one instance is different instead of a term. We conclude that $\hat{\mathcal{L}}_{\mcd}(\bmg)$ satisfies the bounded difference property (Lem.\ref{def:bdp}).
	\end{lem}
	
	\begin{proof} We need to seek the upper bound of
		\[
		\sup\limits_{\bmg \in \mathcal{H}_R}\left|\hat{\mcl}_{\mathcal{D}'}(\bmg) - \hat{\mcl}_{\mcd}(\bmg)\right|.
		\]
		
		To achieve this, notice that, such difference between $\mathcal{D}$ and $\mathcal{D}'$  could be caused by either the user side or the item side. Therefore, we have the following three possible cases:
		\begin{itemize}
			\item \textbf{Case 1:} Only one user is different, i.e., 
			\begin{equation}
				\mcd = \mathop{\cup}\limits_{u_i \in \mathcal{U}} \ \mathcal{D}_{u_i}, \ \ \ \, \mathcal{D}' = \left(\mcd \backslash \mathcal{D}_{u_t}\right) \cup \mathcal{D}_{u'_t}, \ \ \forall t, t = 1, 2, \dots, |\mcu|.
			\end{equation}
			Under this circumstance, we have
			\begin{equation}
				\begin{aligned}\label{eq17}
					\sup\limits_{\bmg \in \mathcal{H}_R}\left|\hat{\mcl}_{\mathcal{D}'}(\bmg) - \hat{\mcl}_{\mcd}(\bmg)\right| &\overset{\textcolor{orange}{(b)}}{\le} \underbrace{\sup\limits_{\bmg \in \mathcal{H}_R} \left|\hat{\mcr}_{\mcd, \bmg} - \hat{\mcr}_{\mcd', \bmg} \right|}_{\textcolor{orange}{(3)}} + \underbrace{\sup\limits_{\bmg \in \mathcal{H}_R} \left|\eta \hat{\Omega}_{\mcd, \bmg} - \eta \hat{\Omega}_{\mathcal{D}', \bmg}\right|}_{\textcolor{orange}{(4)}} \\
				\end{aligned}  
			\end{equation}
			where $\textcolor{orange}{(b)}$ is achieved by the inequality: $\sup (x + y) \le \sup (x) + \sup (y)$.
			
			Based on (\ref{eq17}), in what follows, we will show the upper bound of term $\textcolor{orange}{(3)}$ and $\textcolor{orange}{(4)}$, respectively.
			
			At first, we define some intermediate variables:
			\begin{equation}\nonumber
				\begin{aligned}
					\hat{\mathcal{R}}_{\mathcal{D}_{u_i}, \bmg} &= \frac{1}{n_i^+n_i^-} \sum_{j=1}^{n_i^+}\sum_{k=1}^{n_i^-}\ell^{(i)}_{\bmg}(v_j^+, v_k^-), \\
					\phi_{\bmg}(c_1, c_2) &= \|\boldsymbol{g}^{c_1}_{u_t} - \boldsymbol{g}_{v_j^+}\|^2 - \|\boldsymbol{g}^{c_2}_{u'_t} - \boldsymbol{g}_{v_j^+}\|^2, \forall c_1, c_2, c_1, c_2 \in\ [C]\\
				\end{aligned}
			\end{equation}
			Then, with respect to term $\textcolor{orange}{(3)}$, we have 
			\begin{equation}
				\begin{aligned} \label{eq333}
					&\sup\limits_{\bmg \in \mathcal{H}_R} \left|\hat{\mcr}_{\mcd, \bmg} - \hat{\mcr}_{\mcd', \bmg} \right| = \frac{1}{|\mcu|}\sup\limits_{\bmg \in \mathcal{H}_R} \left|\hat{\mcr}_{\mcd_{u_t}, \bmg} - \hat{\mcr}_{\mcd_{u'_t}, \bmg} \right| \\
					=& \frac{1}{|\mcu|} \sup\limits_{\bmg \in \mathcal{H}_R} \left|\frac{1}{n_i^+n_i^-} \sum_{j=1}^{n_i^+}\sum_{k=1}^{n_i^-}\ell^{(t)}_{\bmg}(v_j^+, v_k^-) - \frac{1}{n_i^+n_i^-} \sum_{j=1}^{n_i^+}\sum_{k=1}^{n_i^-}\ell^{(t')}_{\bmg}(v_j^+, v_k^-)\right| \\
					\le& \frac{1}{|\mcu|} \frac{1}{n_i^+n_i^-} \sup\limits_{\bmg \in \mathcal{H}_R} \sum_{j=1}^{n_i^+}\sum_{k=1}^{n_i^-} \left|\ell^{(t)}_{\bmg}(v_j^+, v_k^-) - \ell^{(t')}_{\bmg}(v_j^+, v_k^-)\right| \\
					\overset{\textcolor{orange}{(b)}}{\le}& \frac{1}{|\mcu|} \frac{1}{n_i^+n_i^-} \sum_{j=1}^{n_i^+}\sum_{k=1}^{n_i^-} \sup\limits_{\bmg \in \mathcal{H}_R} \left|\ell^{(t)}_{\bmg}(v_j^+, v_k^-) - \ell^{(t')}_{\bmg}(v_j^+, v_k^-)\right| \\
					\overset{\textcolor{orange}{(a)}}{\le}& \frac{1}{|\mcu|} \frac{1}{n_i^+n_i^-}  \sum_{j=1}^{n_i^+}\sum_{k=1}^{n_i^-} \sup\limits_{\bmg \in \mathcal{H}_R} \left|s(u_t, v_j^+) - s(u_t, v_k^-) - \left(s(u_t', v_j^+) - s(u_t', v_k^-)\right) \right| \\
					\overset{\textcolor{orange}{(*)}}{\le}& \frac{1}{|\mcu|} \frac{1}{n_i^+n_i^-}  \sum_{j=1}^{n_i^+}\sum_{k=1}^{n_i^-} \sup\limits_{g \in \mathcal{H}_R} \left(
					\left|s(u_t, v_j^+) - s(u_t', v_j^+)\right| + \left|s(u_t', v_k^-) - s(u_t, v_k^-)\right|\right) \\
					\overset{\textcolor{orange}{(b)}}{\le}& \frac{1}{|\mcu|} \frac{1}{n_i^+n_i^-}  \sum_{j=1}^{n_i^+}\sum_{k=1}^{n_i^-} \left(\sup\limits_{\bmg \in \mathcal{H}_R} 
					\left|s(u_t, v_j^+) - s(u_t', v_j^+)\right| + \sup\limits_{\bmg \in \mathcal{H}_R} 
					\left|s(u_t', v_k^-) - s(u_t, v_k^-)\right| \right)
				\end{aligned}
			\end{equation}
			
			For $\sup\limits_{\bmg \in \mathcal{H}_R} 
			\left|s(u_t, v_j^+) - s(u_t', v_j^+)\right|$, the following results hold:
			\begin{equation}
				\begin{aligned} \label{eq322}
					\sup\limits_{\bmg \in \mathcal{H}_R} \left|s(u_t, v_j^+) - s(u_t', v_j^+)\right| &=
					\sup\limits_{\bmg \in \mathcal{H}_R}  \left|\min\limits_{c_1 \in [C]} \|\boldsymbol{g}^{c_1}_{u_t} - \boldsymbol{g}_{v_j^+}\|^2 - \min\limits_{c_2 \in [C]} \|\boldsymbol{g}^{c_2}_{u'_t} - \boldsymbol{g}_{v_j^+}\|^2\right| \\
					& \le \sup\limits_{\bmg \in \mathcal{H}_R}  \left|\min\limits_{c_1 \in [C]} \max\limits_{c_2 \in [C]} \left(\|\boldsymbol{g}^{c_1}_{u_t} - \boldsymbol{g}_{v_j^+}\|^2 - \|\boldsymbol{g}^{c_2}_{u'_t} - \boldsymbol{g}_{v_j^+}\|^2 \right)\right| \\
					& \le \max \left\{\min\limits_{c_1 \in [C]} \max\limits_{c_2 \in [C]} \phi_{\bmg}(c_1, c_2), \max\limits_{c_1 \in [C]} \min\limits_{c_2 \in [C]} \phi_{\bmg}(c_2, c_1)\right\}\
				\end{aligned}
			\end{equation}
			
			According to (\ref{eq322}), we can go a step further:
			\begin{equation}
				\begin{aligned} \label{eq355}
					\min\limits_{c_1 \in [C]} \max\limits_{c_2 \in [C]} \phi_{\bmg}(c_1, c_2) & \le \max\limits_{c_1 = c_2, c1,c2 \in [C]} \phi_{\bmg}(c_1, c_1) \\
					& \le \max\limits_{c \in [C]} \left|\phi_{\bmg}(c, c)\right| \\
					&= \max\limits_{c \in [C]} \left|\|\boldsymbol{g}^{c}_{u_t} - \boldsymbol{g}_{v_j^+}\|^2 - \|\boldsymbol{g}^{c}_{u'_t} - \boldsymbol{g}_{v_j^+}\|^2\right| \\
					&= \max\limits_{c \in [C]} \left|\left(\|\boldsymbol{g}^{c}_{u_t} - \boldsymbol{g}_{v_j^+}\| + \|\boldsymbol{g}^{c}_{u'_t} - \boldsymbol{g}_{v_j^+}\| \right)\left(\|\boldsymbol{g}^{c}_{u_t} - \boldsymbol{g}_{v_j^+}\| - \|\boldsymbol{g}^{c}_{u'_t} - \boldsymbol{g}_{v_j^+}\| \right) \right| \\
					& \overset{\textcolor{orange}{(**)}}{\le} 4 r \max\limits_{c \in [C]} \|\boldsymbol{g}^{c}_{u_t} -\boldsymbol{g}^{c}_{u'_t} \| \\
					& \le 8r^2
				\end{aligned}
			\end{equation}
			
			Based on the result of (\ref{eq355}), we have the following result for (\ref{eq333})
			
			\begin{equation}
				\begin{aligned}
					\frac{1}{|\mcu|}\sup\limits_{\bmg \in \mathcal{H}_R} \left|\hat{\mcr}_{\mcd_{u_t}, \bmg} - \hat{\mcr}_{\mcd_{u'_t}, \bmg} \right| & \le \frac{16r^2}{|\mcu|}
				\end{aligned}
			\end{equation}
			
			
			With respect to $\textcolor{orange}{(4)}$, recall that, we have
			\[
			\hat{\Omega}_{\mcd, \bmg} = \frac{1}{|\mcu|} \sum_{u_i \in \mcu} \psi_{\bmg}(u_{i}),
			\]
			where 
			\begin{equation}\nonumber
				\begin{aligned}
					\psi_{\bmg}(u_{i}) &= \max\left(0, \delta_1 - \delta_{\bmg, u_i}\right) + \max\left(0, \delta_{\bmg, u_i} - \delta_2\right), \\
					\delta_{\bmg, u_i} &= \frac{1}{2C(C-1)} \sum_{c_{1}, c_2 \in C} \|\boldsymbol{g}_{u_i}^{c_1} - \boldsymbol{g}_{u_i}^{c_2}\|^2, \\
				\end{aligned}
			\end{equation}
			Moreover, let us define some intermediate variables:
			\begin{equation}\nonumber
				\begin{aligned}
					\psi_{\bmg, \delta_1}(u_{i}, u_j) &= \max\left(0, \delta_1 - \delta_{\bmg, u_i}\right) - \max\left(0, \delta_1 - \delta_{\bmg, u_j}\right), \\
					\psi_{\bmg, \delta_2}(u_{i}, u_j) &= \max\left(0, \delta_{\bmg, u_i} - \delta_2\right) - \max\left(0, \delta_{\bmg, u_j} - \delta_2\right).
				\end{aligned}
			\end{equation}
			In this sense, in terms of (\ref{eq17}), the following result holds:
			\begin{equation}
				\begin{aligned}\label{eq19}
					\sup\limits_{\bmg \in \mathcal{H}_R}\left|\hat{\mcl}_{\mathcal{D}'}(\bmg) - \hat{\mcl}_{\mcd}(\bmg)\right| & = \eta \cdot \left|\frac{1}{|\mcu|}\psi_{\bmg}(u_{t}) - \frac{1}{|\mcu|}\psi_{\bmg}(u'_{t})\right| \\
					&= \frac{\eta}{|\mcu|} \cdot \left|\psi_{\bmg}(u_{t}) - \psi_{\bmg}(u'_{t})\right| \\
					& \overset{\textcolor{orange}{(*)}}{\le} \frac{\eta}{|\mcu|} \cdot \left(\left|\psi_{\bmg, \delta_1}(u_t, u'_t)\right| + \left|\psi_{\bmg, \delta_2}(u_t, u'_t)\right|\right) \\
					& \overset{\textcolor{orange}{(a)}}{\le} \frac{2\eta}{|\mcu|} \cdot \left|\delta_{\bmg, u_t} - \delta_{\bmg, u'_t}\right| \\
					& = \frac{\eta}{C(C-1)|\mcu|} \left|\sum_{c_{1}, c_2 \in C} \left( \|\boldsymbol{g}_{u_t}^{c_1} - \boldsymbol{g}_{u_t}^{c_2}\|^2 - \|\boldsymbol{g}_{u'_t}^{c_1} - \boldsymbol{g}_{u'_t}^{c_2}\|^2\right)\right| \\
					& \overset{\textcolor{orange}{(*)}}{\le} \frac{\eta}{C(C-1)|\mcu|}\sum_{c_{1}, c_2 \in C} \left|\|\boldsymbol{g}_{u_t}^{c_1} - \boldsymbol{g}_{u_t}^{c_2}\|^2 - \|\boldsymbol{g}_{u'_t}^{c_1} - \boldsymbol{g}_{u'_t}^{c_2}\|^2\right| \\
					& \le \frac{4r^2\eta}{|\mcu|}
				\end{aligned}
			\end{equation}
			where $\textcolor{orange}{(*)}$ achieves via the inequality $|x + y| \le |x| + |y|$ and $\textcolor{orange}{(a)}$ follows the Lem.\ref{lem2}.
			
			Finally, in this case, we have
			\begin{equation}
				\begin{aligned}
					\sup\limits_{\bmg \in \mathcal{H}_R}\left|\hat{\mcl}_{\mathcal{D}'}(\bmg) - \hat{\mcl}_{\mcd}(\bmg)\right| \le \frac{16r^2 + 4r^2\eta}{|\mcu|}.
				\end{aligned}
			\end{equation}
			
			\item \textbf{Case 2:} \label{SFCML:case1} Only one positive item is different. In this case, we consider such a difference occurs in the positive item $v_{t_1}^+$ with respect to a specific user $u_i$ and there are $|\mcu|$ cases for all users. Mathematically, we have
			\begin{equation}
				\begin{aligned}
					\mathcal{D}_{u_i} = \{v_j^+\}_{j=1}^{n_i^+} \cup \{v_k^-\}_{k=1}^{n_i^-}, \ \ \ \ \mathcal{D}_{u_i}' = (\mathcal{D}_{u_i} \backslash	\{v_{t_1}^+\}) \cup \{\tilde{v}_{t_1}^{+}\}, \label{SFCML:eq26}
				\end{aligned}
			\end{equation}
			where $\forall t_1, t_1 = 1, 2, \dots, n_i^+$ and $n_i^+ + n_i^- = |\mci|$.
			Then, it is obvious that in this case only the first term in (\ref{eq15}) contributes to the upper bound. According to this observation, the upper bound could be simplified as follows:
			\begin{equation}
				\begin{aligned}
					\sup\limits_{\bmg \in \mathcal{H}_R}\left|\hat{\mcl}_{\mathcal{D}'}(\bmg) - \hat{\mcl}_{\mcd}(\bmg)\right| 
					&= \sup\limits_{\bmg \in \mathcal{H}_R} \left|\hat{\mathcal{R}}_{\mcd}(\bmg) - \hat{\mathcal{R}}_{\mathcal{D}'}(\bmg)\right|	\\
					&= \sup\limits_{\bmg \in \mathcal{H}_R} \left|\hat{\mathcal{R}}_{\mathcal{D}_{u_i}}(\bmg) - \hat{\mathcal{R}}_{\mathcal{D}'_{u_i}}(\bmg)\right|,\\
					\label{SFCML:eq28}
				\end{aligned}
			\end{equation}
			where again we denote
			\[
			\hat{\mathcal{R}}_{\mathcal{D}_{u_i}}(\bmg) = \frac{1}{|\mcu|} \cdot \sum_{j=1}^{n_i^+} \sum_{k=1}^{n_i^-} \ell^{(i)}_{\bmg}(v_j^+, v_k^-),
			\]
			and 
			\[
			\ell^{(i)}_{\bmg}(v_j^+, v_k^-) = \max (0, \lambda + s(u_i, v_j^+) - s(u_i, v_k^-)).
			\]
			Let
			\begin{equation} \label{eqqq}
				\Delta_{\bmg}(c_1, c_2) = \|\boldsymbol{g}_{u_i}^{c_1} - \boldsymbol{g}_{v_{t_1}^+}\|^2 -\|\boldsymbol{g}_{u_i}^{c_2} - \boldsymbol{g}_{\tilde{v}_{t_1}^+}\|^2.
			\end{equation}
			
			Then, since $v_j^+$ and $\tilde{v}_j^{+}$ are different in this case, we have
			\begin{equation}
				\begin{aligned}
					\sup\limits_{\bmg \in \mathcal{H}_R}\left|\hat{\mcl}_{\mathcal{D}'}(\bmg) - \hat{\mcl}_{\mcd}(\bmg)\right| 
					& = \frac{1}{|\mcu|}\sup\limits_{\bmg \in \mathcal{H}_R} \left| \frac{1}{n_i^+n_i^-} \sum_{k=1}^{n_i^-} \ell^{(i)}_{\bmg}(v_{t_1}^+, v_k^-) -  \frac{1}{n_i^+n_i^-} \sum_{k=1}^{n_i^-} \ell^{(i)}_{\bmg}(\tilde{v}_{t_1}^{+}, v_k^-)\right|\\ 
					&  \overset{\textcolor{orange}{(*)}}{\le} \frac{1}{|\mcu|n_i^+n_i^-}  \sup\limits_{\bmg \in \mathcal{H}_R} \sum_{k=1}^{n_i^-} \left|\ell^{(i)}_{\bmg}(v_{t_1}^+, v_k^-) - \ell^{(i)}_{\bmg}(\tilde{v}_{t_1}^{+}, v_k^-) \right| \\
					& \overset{\textcolor{orange}{(b)}}{\le} \frac{1}{|\mcu|n_i^+n_i^-}  \sum_{k=1}^{n_i^-} \left(\sup\limits_{\bmg \in \mathcal{H}_R} \left|\ell^{(i)}_{\bmg}(v_{t_1}^+, v_k^-) - \ell^{(i)}_{\bmg}(\tilde{v}_{t_1}^{+}, v_k^-) \right|\right) \\
					& \overset{\textcolor{orange}{(a)
					}}{\le} \frac{1}{|\mcu|n_i^+n_i^-}   \sum_{k=1}^{n_i^-}\sup\limits_{\bmg \in \mathcal{H}_R}\left|s(u_i, v_{t_1}^+)-s(u_i, \tilde{v}_{t_1}^{+}) \right| \\
					&= \frac{1}{|\mcu|n_i^+n_i^-}  \sum_{k=1}^{n_i^-} \sup\limits_{\bmg \in \mathcal{H}_R} \left|\min\limits_{c_1 \in [C]} \|\boldsymbol{g}_{u_i}^{c_1} - \boldsymbol{g}_{v_{t_1}^+}\|^2 - \min\limits_{c_2 \in [C]} \|\boldsymbol{g}_{u_i}^{c_2} - \boldsymbol{g}_{\tilde{v}_{t_1}^+}\|^2\right| \\
					&= \frac{1}{|\mcu|n_i^+n_i^-}  \sum_{k=1}^{n_i^-} \sup\limits_{\bmg \in \mathcal{H}_R} \left| \min\limits_{c_1 \in [C]} \max\limits_{c_2 \in [C]} \left(\|\boldsymbol{g}_{u_i}^{c_1} - \boldsymbol{g}_{v_{t_1}^+}\|^2 -\|\boldsymbol{g}_{u_i}^{c_2} - \boldsymbol{g}_{\tilde{v}_{t_1}^+}\|^2 \right)\right|. \\
					\label{SFCML:thoe3:1}
				\end{aligned}
			\end{equation}
			
			According to (\ref{eqqq}), we have
			\begin{equation}
				\begin{aligned}\label{eq11}
					\sup\limits_{\bmg \in \mathcal{H}_R}\left|\hat{\mcl}_{\mathcal{D}'}(\bmg) - \hat{\mcl}_{\mcd}(\bmg)\right| 
					& \le \frac{1}{|\mcu|n_i^+n_i^-}  \sum_{k=1}^{n_i^-} \sup\limits_{\bmg \in \mathcal{H}_R} \left|\min\limits_{c_1 \in [C]} \max\limits_{c_2 \in [C]} \Delta_{\bmg}(c_1, c_2)\right| \\
					&= \frac{1}{|\mcu|n_i^+n_i^-}  \sum_{k=1}^{n_i^-}\left|  \max\left\{\min\limits_{c_1 \in [C]} \max\limits_{c_2 \in [C]} \Delta_{\bmg}(c_1, c_2), \max\limits_{c_1 \in [C]} \min\limits_{c_2 \in [C]} \Delta_{\bmg}(c_2, c_1)\right\} \right|. \\
				\end{aligned}
			\end{equation}
			
			It is easy to show that, 
			\begin{equation}
				\begin{aligned}\label{eq12}
					\min\limits_{c_1 \in [C]} \max\limits_{c_2 \in [C]} \Delta_{\bmg}(c_1, c_2) & \le \max\limits_{c_1 = c_2, c_1, c_2 \in [C]} \Delta_{\bmg}(c_1, c_2) \\
					& \le \max\limits_{c_1 = c_2, c_1, c_2 \in [C]} \left| \Delta_{\bmg}(c_1, c_2)\right| \\
					&= \max\limits_{c \in [C]}\left|\|\boldsymbol{g}_{u_i}^{c} - \boldsymbol{g}_{v_{t_1}^+}\|^2 -\|\boldsymbol{g}_{u_i}^{c} - \boldsymbol{g}_{\tilde{v}_{t_1}^+}\|^2\right|\\
					&= \max\limits_{c\in [C]}\left| \left(\|\boldsymbol{g}_{u_i}^{c} - \boldsymbol{g}_{v_{t_1}^+}\| + \|\boldsymbol{g}_{u_i}^{c} - \boldsymbol{g}_{\tilde{v}_{t_1}^+}\|\right) \left(\|\boldsymbol{g}_{u_i}^{c} - \boldsymbol{g}_{v_{t_1}^+}\| - \|\boldsymbol{g}_{u_i}^{c} - \boldsymbol{g}_{\tilde{v}_{t_1}^+}\|\right) \right|\\
					& \overset{\textcolor{orange}{(**)}}{\le} \left(\|\boldsymbol{g}_{u_i}^{c} - \boldsymbol{g}_{v_{t_1}^+}\| + \|\boldsymbol{g}_{u_i}^{c} - \boldsymbol{g}_{\tilde{v}_{t_1}^+}\|\right) \left(\|\boldsymbol{g}_{u_i}^{c} - \boldsymbol{g}_{v_{t_1}^+} -\boldsymbol{g}_{u_i}^{c} +  \boldsymbol{g}_{\tilde{v}_{t_1}^+}\|\right) \\
					& \le 8r^2.
				\end{aligned}
			\end{equation}
			In the same way, we also have 
			\begin{equation}\label{eq13}
				\max\limits_{c_1 \in [C]} \min\limits_{c_2 \in [C]} \Delta_{\bmg}(c_2, c_1) \le 8r^2
			\end{equation}
			
			Therefore, applying (\ref{eq12}) and (\ref{eq13}) to (\ref{eq11}), we have
			\begin{equation}
				\begin{aligned}\label{eq14}
					\sup\limits_{\bmg \in \mathcal{H}_R}\left|\hat{\mcl}_{\mathcal{D}'}(\bmg) - \hat{\mcl}_{\mcd}(\bmg)\right|  \le \frac{8r^2}{|\mcu|n_i^+}.
				\end{aligned}
			\end{equation}
			\item \textbf{Case 3:} \label{SFCML:case2} Only one negative item is different. In this case, we assume such a difference occurs in the negative item $v_{t_2}^-$ with respect to a specific user $u_i$, and there are also $|\mcu|$ cases for all users. Mathematically, we have
			\begin{equation}
				\mathcal{D}_i = \{v_j^+\}_{j=1}^{n_i^+} \cup \{v_k^-\}_{k=1}^{n_i^-}, \ \ \ \ \mathcal{D}_i' = (\mathcal{D}_i \backslash \{v_{t_2}^-\}) \cup \{\tilde{v}_{t_2}^{-}\}. \label{SFCML:eq27}
			\end{equation}
			where $\forall t_2, t_2 = 1, 2, \dots, n_i^-$.
			
			Similarly, if $v_k^-$ and $\tilde{v}_k^{-}$ are different, we can also hold
			\begin{equation}
				\sup\limits_{\bmg \in \mathcal{H}_R}\left|\hat{\mcl}_{\mathcal{D}'}(\bmg) - \hat{\mcl}_{\mcd}(\bmg)\right|  \le \frac{8r^2}{|\mcu|n_i^-}. \label{SFCML:thoe3:2}
			\end{equation}
		\end{itemize}
		
		Finally, taking all above three cases into account, one can conclude that $\hat{\mathcal{L}}_{\mcd}(\bmg)$ is satisfied with the bounded difference property (Lem.\ref{def:bdp}). 
		
		This completed the proof.
	\end{proof}
	
	\begin{lem} \label{lem6} Equipped with Lem.\ref{lem8} and Lem.\ref{lem9}, the following inequality holds:
		\begin{equation}\nonumber
			\begin{aligned}
				\mathbb{P}\left[\left|\hat{\mathcal{L}}_{\mcd}(\bmg_l) - \expe [\hat{\mathcal{L}}_{\mcd}(\bmg_l)]\right| \ge \frac{\varepsilon}{2}\right] & \le 2 \exp \left(\frac{- \varepsilon^2 \tilde{N}}{2}\right),
			\end{aligned}
		\end{equation}
		where 
		\[
		\tilde{N} = \left(4r^2\sqrt{\left(\frac{(4 + \eta)^2}{|\mcu|} + \frac{2}{|\mcu|^2} \sum_{u_i \in \mcu} \left(\frac{1}{n_i^+} + \frac{1}{n_i^-}\right)\right)}\right)^{-2}.
		\]
	\end{lem}
	\begin{proof}
		The proof could be easily achieved by applying Lem.\ref{lem:mc} on top of Lem.\ref{lem8} and Lem.\ref{lem9}. 
	\end{proof}
	
	{\noindent} \rule[-10pt]{17.5cm}{0.05em}	
	
	\subsection{Proof of the Main Result}
	\subsubsection{Proof of Thm.\ref{them1}}
	\begin{rthm1} [Generalization Upper Bound of DPCML] \label{rthem1} Let $\expe [\hat{\mathcal{L}}_{\mcd}(\bmg)]$ be the population risk of $\hat{\mathcal{L}}_{\mcd}(\bmg)$. Then, $\forall \ \bmg, \bmg \in \mathcal{H}_R$, with high probability, the following inequation holds:
		\begin{equation}
			\begin{aligned}
				\left| \hat{\mathcal{L}}_{\mcd}(\bmg) - \expe [\hat{\mathcal{L}}_{\mcd}(\bmg)]\right| \le \sqrt{\frac{2d\log \left(3r \tilde{N}\right)}{\tilde{N}}},
			\end{aligned}
		\end{equation}
		where we have
		\begin{equation} \nonumber
			\tilde{N} = \left(4r^2\sqrt{\left(\frac{(4 + \eta)^2}{|\mcu|} + \frac{2}{|\mcu|^2} \sum_{u_i \in \mcu} \left(\frac{1}{n_i^+} + \frac{1}{n_i^-}\right)\right)}\right)^{-2}.
		\end{equation}
	\end{rthm1}
	
	\begin{proof} \textcolor{blue}{Step 1}. In order to obtain the generalization bound, we need to first figure out the following probability:
		
		\[
		\mathbb{P}\left[\sup_{\bmg \in \mathcal{H}_R} \left| \hat{\mathcal{L}}_{\mcd}(\bmg) - \expe [\hat{\mathcal{L}}_{\mcd}(\bmg)]\right|\ge \varepsilon\right],
		\]
		where $\varepsilon$ is the generalization error and usually a very small value.
		
		%
		Denote the covering number of $\sigma$-covering in Lem.\ref{lem8} as $\mathcal{N}_3(\sigma; \mathcal{H}_R, \rho_3)$. Then, according to Def.\ref{rdef2}, Def.\ref{rdef3}, Lem.\ref{lem46} and Lem.\ref{lem8}, we have
		\begin{equation}
			\begin{aligned}\label{eq10}
				\mathbb{P}\left[\sup_{\bmg \in \mathcal{H}_R} \left| \hat{\mathcal{L}}_{\mcd}(\bmg) - \expe [\hat{\mathcal{L}}_{\mcd}(\bmg)]\right| \ge \varepsilon \right] 
				&\le \mathbb{P}\left[\sup\limits_{\bmg \in \mathop{\cup}\limits_{l=1}^{\mathcal{N}_3} \mathcal{B}(\bmg_l, \sigma)}  \left|\hat{\mathcal{L}}_{\mcd}(\bmg) - \expe [\hat{\mathcal{L}}_{\mcd}(\bmg)]\right| \ge \varepsilon\right]\\
				& \overset{\textcolor{orange}{Lem.\ref{lem46}}}{\le} \sum_{l=1}^{\mathcal{N}_3} \mathbb{P} \left[ \sup\limits_{\bmg \in \mathcal{B}(\bmg_l, \sigma)}\left|\hat{\mathcal{L}}_{\mcd}(\bmg) - \expe [\hat{\mathcal{L}}_{\mcd}(\bmg)]\right|\ge \varepsilon \right] \\
				& \overset{\textcolor{orange}{Lem.\ref{lem8}}}{\le} \sum_{l=1}^{\mathcal{N}_3} \mathbb{P}\left[\left|\hat{\mathcal{L}}_{\mcd}(\bmg_l) - \expe [\hat{\mathcal{L}}_{\mcd}(\bmg_l)]\right| \ge \frac{\varepsilon}{2}\right] \\				
			\end{aligned}
		\end{equation}
		where, without the loss of generality, we denote the covering number as $\mathcal{N}_3$ for short.
		
		Note that, from Lem.\ref{lem8} we have $\sigma = \frac{\varepsilon}{16r(4 + \eta)}$, and 
		\[
		\mathcal{N}_3(\sigma; \mathcal{H}_R, \rho_3) \le \mathcal{N}_1\left(\frac{\varepsilon}{16r(4 + \eta)}, \mathcal{H}_R,\rho_1\right) \cdot \mathcal{N}_2\left(\frac{\varepsilon}{16r(4 + \eta)}, \mathcal{H}_R,\rho_2\right).
		\]
		Therefore, we further have
		
		\begin{equation}
			\begin{aligned}
				\mathbb{P}\left[\sup_{\bmg \in \mathcal{H}_R} \left| \hat{\mathcal{L}}_{\mcd}(\bmg) - \expe [\hat{\mathcal{L}}_{\mcd}(\bmg)]\right| \ge \varepsilon \right] & \le \mathcal{N}_1 \cdot \mathcal{N}_2 \cdot \mathbb{P}\left[\left|\hat{\mathcal{L}}_{\mcd}(\bmg_l) - \expe [\hat{\mathcal{L}}_{\mcd}(\bmg_l)]\right| \ge \frac{\varepsilon}{2}\right]
			\end{aligned}
		\end{equation}
		
		\textcolor{blue}{\textbf{Step 2}}. Now, according to Lem.\ref{lem6}, we have 
		
		\begin{equation}
			\begin{aligned}
				\mathbb{P}\left[\sup_{\bmg \in \mathcal{H}_R} \left| \hat{\mathcal{L}}_{\mcd}(\bmg) - \expe [\hat{\mathcal{L}}_{\mcd}(\bmg)]\right| \ge \varepsilon \right] & \le 2 \mathcal{N}_1 \cdot \mathcal{N}_2 \cdot \exp \left(\frac{- \varepsilon^2 \tilde{N}}{2}\right).
			\end{aligned}
		\end{equation}
		
		Then with Lem.\ref{covering_lem} and by further choosing 
		\begin{equation} \nonumber
			\begin{aligned}
				\varepsilon = \sqrt{\frac{2d}{\tilde{N}} \log \left(3r \tilde{N}\right)},
			\end{aligned}
		\end{equation}
		we have:
		\begin{equation}
			\begin{aligned}
				\mathbb{P}\left[\sup_{\bmg \in \mathcal{H}_R} \left| \hat{\mathcal{L}}_{\mcd}(\bmg) - \expe [\hat{\mathcal{L}}_{\mcd}(\bmg)]\right|\ge \sqrt{\frac{2d\log \left(3r \tilde{N}\right)}{\tilde{N}}}\right] &\le 2\left(\frac{3rB^2}{2d\log\left(3r \tilde{N}\right)}\right)^d,
			\end{aligned}
		\end{equation}
		where again 
		\[
		\tilde{N} = \left(4r^2\sqrt{\left(\frac{(4 + \eta)^2}{|\mcu|} + \frac{2}{|\mcu|^2} \sum_{u_i \in \mcu} \left(\frac{1}{n_i^+} + \frac{1}{n_i^-}\right)\right)}\right)^{-2},
		\] 
		and 
		\begin{equation}
			\begin{aligned}
				B &= 16r(4 + \eta).
			\end{aligned}
		\end{equation}
		
		Therefore, we can conclude that, with high probability, 
		\begin{equation}
			\begin{aligned}
				\left| \hat{\mathcal{L}}_{\mcd}(\bmg) - \expe [\hat{\mathcal{L}}_{\mcd}(\bmg)]\right| \le \sqrt{\frac{2d\log \left(3r \tilde{N}\right)}{\tilde{N}}}, \ \ \forall \ \bmg, \bmg \in \mathcal{H}_R.
			\end{aligned}
		\end{equation}
		
		This completed the proof.
	\end{proof}
	
	\subsubsection{Proof of Corol.\ref{cor1}}
	\begin{rcor1} \label{rcor1} On the top of Thm.\ref{them1}, DPCML could enjoy a smaller generalization error than CML.
	\end{rcor1}
	\begin{proof} For simplification of notations, let $\mathcal{X}_{=1}$ and $\mathcal{X}_{>1}$ be the feasible regions of CML ($C=1$) and DPCML ($C > 1$), and $\hat{\mathcal{L}}_{=1}(\boldsymbol{g})$ and $\hat{\mathcal{L}}_{>1}(\boldsymbol{g})$ be the empirical risks of CML ($C=1$) and DPCML ($C > 1$), respectively. Then, since DPCML leverages $\min\limits_{c \in C} \|\boldsymbol{g}_{u_i}^c - \boldsymbol{g}_{v_j}\|^2$ as the distance, which can be regarded as a minimum of multiple single version CML, it is easy to know that the feasible solution of CML is also included in DPCML, i.e., $\mathcal{X}_{=1}\subseteq \mathcal{X}_{>1}$. Therefore, we can conclude that $\hat{\mathcal{L}}_{>1}(\boldsymbol{g}) \le \hat{\mathcal{L}}_{=1}(\boldsymbol{g})$. Denote $\Delta = \sqrt{\frac{2d\log \left(3r \tilde{N}\right)}{\tilde{N}}}$ as the residuals between $\mathbb{E}[\hat{\mathcal{L}}(\boldsymbol{g})]$ and $\hat{\mathcal{L}}(\boldsymbol{g})$. Moreover, we have $\Delta_{\text{DPCML}} = \Theta(\Delta_{\text{CML}})$ since $\Delta$ in our bound does not depend on $C$. This is consistent with the over-parameterization phenomenon \cite{DBLP:journals/corr/abs-2109-02355, DBLP:conf/iclr/NakkiranKBYBS20}. According to Thm.\ref{them1}, we see that $\mathbb{E}[\hat{\mathcal{L}_*}(\boldsymbol{g})] \le \hat{\mathcal{L}_*}(\boldsymbol{g}) + \Delta$, where $*$ represents $=1$ or $>1$. Therefore, we can conclude that DPCML could enjoy a smaller generalization error than the traditional CML. We also empirically demonstrate this in the experiment Sec.\ref{qqaa}.
	\end{proof}
	
	\section{Proofs and algorithms for DiHarS Framework}
	
	\subsection{Proof of the Lem.\ref{pami:lem2} \label{pami:proof_lem2}}
	\begin{rlem2} \label{pami:rlem2} $\sum_{t=1}^k z_{[t]}$ is a convex function of $(z_1, \dots, z_n)$ and $z_{[t]}$ represents the top-$t$ element among $(z_1, \dots, z_n)$. Then, we can afford the equivalence of the sum-of-top-$k$ elements with an optimization problem as follows:
		\begin{equation}
			\sum_{t=1}^k z_{[t]} = \min\limits_{\gamma \ge 0} \left\{k\gamma + \sum_{t=1}^n [z_t - \gamma]_+\right\},
		\end{equation}
		where $[a]_+ = \max(0, a)$ is the hinge function.
	\end{rlem2}
	
	\noindent\rule[0.05\baselineskip]{\columnwidth}{1pt}
	\begin{proof} Note that the proof is directly followed from \cite{NIPS2017_6c524f9d}. To begin with, we define the following linear programming problem: 
		\begin{equation}\label{pami:eq87}
			\max\limits_{\boldsymbol{\rho}} \ \ \boldsymbol{\rho}^\top \boldsymbol{z}, \ \ s.t. \ \ \boldsymbol{\rho}^\top \boldsymbol{1} = k, \ \ \boldsymbol{0} \le \boldsymbol{\rho} \le \boldsymbol{1},
		\end{equation}
		where both $\boldsymbol{0}$ and $\boldsymbol{1}$ are $n$-dimension vectors.
		
		Obviously, we can see that $\sum_{t=1}^k z_{[t]}$ is exactly the solution of (\ref{pami:eq87}). To solve this, we can adopt the Lagrangian multiplier method and thus have
		\begin{equation}\label{pami:eq88}
			L(\boldsymbol{\rho}, \boldsymbol{r}, \boldsymbol{t}, \gamma) = - \boldsymbol{\rho}^\top \boldsymbol{z} - \boldsymbol{t}^\top \boldsymbol{\rho} + \boldsymbol{r}^\top(\boldsymbol{\rho} - \boldsymbol{1}) + \gamma(\boldsymbol{\rho}^\top\boldsymbol{1} - k),
		\end{equation}
		where $\boldsymbol{r} \ge 0, \boldsymbol{t} \ge 0$ and $\gamma$ are our introduced Lagrangian multipliers. 
		
		Subsequently, to solve (\ref{pami:eq88}), the following condition holds by taking the derivative concerning $\boldsymbol{\rho}$ and forcing it to $\boldsymbol{0}$:
		\begin{equation}\label{pami:eq89}
			\boldsymbol{t} = \boldsymbol{r} - \boldsymbol{z} + \gamma\boldsymbol{1}.
		\end{equation}
		
		According to (\ref{pami:eq89}), we can derive the dual problem of (\ref{pami:eq87}):
		\begin{equation}
			\min\limits_{\boldsymbol{r}, \gamma} \ \ \boldsymbol{r}^\top\boldsymbol{1} + \gamma k, \ \ s.t. \ \ \boldsymbol{r} \ge \boldsymbol{0}, \ \ \boldsymbol{r} + \gamma \boldsymbol{1} - \boldsymbol{z} \ge \boldsymbol{0}.
		\end{equation}
		
		In this sense, we have 
		\begin{equation}\label{pami:eq91}
			\sum_{t=1}^k z_{[t]} = \min\limits_{\gamma} \left\{k\gamma + \sum_{t=1}^n [z_t - \gamma]_+\right\}.
		\end{equation}
		
		Finally, the following result directly holds because $\gamma = z_{[k]}$ is always one optimal solution for (\ref{pami:eq91})
		\begin{equation}\label{pami:eq92}
			\sum_{t=1}^k z_{[t]} = \min\limits_{\gamma \ge 0} \left\{k\gamma + \sum_{t=1}^n [z_t - \gamma]_+\right\}.
		\end{equation}
		This completed the proof.
	\end{proof}
	
	\subsection{Proof of the Lem.\ref{pami:lem3} \label{pami:proof_lem3}}
	\begin{rlem3} \label{pami:rlem3} In terms of $c_1 \ge 0, c_2 \ge 0$, we have $[[c_1 - s]_+ - c_2]_+ = [c_1 - c_2 - s]_+$. 
	\end{rlem3}
	\noindent\rule[0.05\baselineskip]{\columnwidth}{1pt}
	\begin{proof} To prove this, we will separately consider the following two cases for any $c_1 \ge 0, c_2 \ge 0$:
		\begin{itemize}
			\item \textcolor{blue}{\textbf{Case 1:}} $c_1 - s \ge 0$. In this case, we can directly hold $[[c_1 - s]_+ - c_2]_+ = [c_1 - c_2 - s]_+$. 
			
			\item\textcolor{blue}{\textbf{Case 2:}} $c_1 - s < 0$. Since $c_1, c_2 \ge 0$, now we have $[[c_1 - s]_+ - c_2]_+ = [0 - c_2]_+ = 0$. Meanwhile, we notice that $[c_1 - c_2 - s]_+$ is also equal to $0$, implying $[[c_1 - s]_+ - c_2]_+ = [c_1 - c_2 - s]_+$. 
		\end{itemize}
		This completed the proof.
	\end{proof}
	\subsection{Proof of the Thm.\ref{pami:thm2} \label{pami:proof_thm2}}
	\begin{rthm2} \label{pami:4thm2} Consider a top-$N$ recommendation task evaluated by Precision@N (P@N) and Recall@N (R@N) metrics and assume $n_i^+=|\mathcal{D}_{u_i}^+| \ge N$, $n_i^-=|\mathcal{D}_{u_i}^-| \ge N$, $\forall u_i \in \mathcal{U}$. Then, for any user $u_i$, the following conditions hold:
		\begin{equation}\label{pami:req28}
			\begin{aligned}
				\text{P@}N \ge \frac{1}{N}\left\lfloor \frac{(n_i^+ + N) - \sqrt{\mathcal{F}(n_i^+, N, -\hat{\texttt{OPAUC}}^{u_i}(s_{\boldsymbol{g}}, \beta_i))}}{2} \right\rfloor,
			\end{aligned}
		\end{equation}
		\begin{equation}\label{pami:req29}
			\begin{aligned}
				\text{R@}N  \ge \frac{1}{n_i^+}\left\lfloor \frac{(n_i^+ + N) - \sqrt{\mathcal{F}(n_i^+, N, -\hat{\texttt{OPAUC}}^{u_i}(s_{\boldsymbol{g}}, \beta_i))}}{2} \right\rfloor,
			\end{aligned}
		\end{equation}
		where the FPR range $\frac{N}{n_i^-} \le \beta_i \le 1$ and $\mathcal{F}(n_i^+, N, -\hat{\texttt{OPAUC}}^{u_i}(s_{\boldsymbol{g}}, \beta_i))$ represents an essential function that is \textbf{negatively} proportional to the value of $\hat{\texttt{OPAUC}}^{u_i}(s_{\boldsymbol{g}}, \beta_i)$, namely,
		\begin{equation}\nonumber
			\begin{aligned}
				\mathcal{F}(n_i^+, N, -\hat{\texttt{OPAUC}}^{u_i}(s_{\boldsymbol{g}},\beta_i)) = (n_i^+ + N)^2 - 4 n_i^+N + 4 n_i^+N_i^{\beta_i} \times (1 - \hat{\texttt{OPAUC}}^{u_i}(s_{\boldsymbol{g}}, \beta_i)),
			\end{aligned}
		\end{equation}
		and $N_i^{\beta_i} = U = \lfloor n_{i}^- \cdot \beta_i \rfloor$.
	\end{rthm2}
	
	\noindent\rule[0.05\baselineskip]{\columnwidth}{1pt}
	\begin{proof} The proofs of the lower bound for P@$N$ and R@$N$ are similar because we can see that P@$N = \frac{n_i^+}{N}$R@$N$. Thus, here we merely present the proof for P@$N$. 
		
		Suppose that there are $n$ ($n \le N$) positive items among the Top-$N$ recommendation list. Then, with respect to any permutation of $n$ positive items, we now have P@$N = \frac{n}{N}$. 
		
		Meanwhile, under this circumstance, if $n_i^+ \ge N$, $n_i^- \ge N$ and, for any $\beta_i$, $N \le N_i^{\beta_i} \le n_i^- \rightarrow \frac{N}{n_i^-} \le \beta_i \le 1$, we can definitely determine that the maximum value of $\hat{\texttt{OPAUC}}(s_{\boldsymbol{g}}, \beta_i)$ is $\frac{nN_i^{\beta_i} + (n_i^+ - n)(N_i^{\beta_i} - N + n)}{n_i^+N_i^{\beta_i}}$, expressed as follows:
		\begin{equation}
			\underbrace{\textcolor{blue}{\oplus \dots \oplus}}_{\textcolor{blue}{n}} \big\vert \underbrace{\textcolor{orange}{\ominus \dots \ominus}}_{\textcolor{orange}{N-n}}\big\vert \underbrace{\textcolor{blue}{\oplus \dots \oplus}}_{\textcolor{blue}{n_i^+ -n}} \big\vert \underbrace{\textcolor{orange}{\ominus \dots \ominus}}_{\textcolor{orange}{N_i^{\beta_i} - N + n}} \big\vert \underbrace{\textcolor{orange}{\ominus \dots \ominus}}_{\textcolor{orange}{n_i^--N_i^{\beta_i}}} .
		\end{equation} 
		
		After that, if one proceeds to maximize the OPAUC value, the corresponding performance of P@$N$ would also be improved, i.e., \textbf{the number of positive items $n$ must increase among the Top-$N$ recommendation list.} Based on this, we can derive the following performance condition to make sure that $n$ must be an integer:
		\begin{equation}
			\begin{aligned}
				\text{P@}N \ge \frac{1}{N}\left\lfloor \frac{(n_i^+ + N) - \sqrt{(n_i^+ + N)^2 - 4 n_i^+N + 4 n_i^+N_i^{\beta_i} \times (1 - \hat{\texttt{OPAUC}}(s_{\boldsymbol{g}}, \beta_i))}}{2} \right\rfloor.
			\end{aligned}
		\end{equation}
		Finally, defining \[\mathcal{F}(n_i^+, N, -\hat{\texttt{OPAUC}}(s_{\boldsymbol{g}}, \beta_i)) = (n_i^+ + N)^2 - 4 n_i^+N + 4 n_i^+N_i^{\beta_i} \times (1 - \hat{\texttt{OPAUC}}(s_{\boldsymbol{g}}, \beta_i))\] completed the proof.
	\end{proof}
	
	\subsection{Proof of the Thm.\ref{pami:thm3} \label{pami:proof_thm3}}
	
	\begin{rthm3} [Differentiable Reformulation of (\ref{ref_eq28})] \label{pami:rthm3} Let $\forall u_i \in \mathcal{U}$, $N_i^{\beta_i} = \lfloor n_{i}^- \cdot \beta_i \rfloor$ and $\beta_i \ge \frac{1}{n_i^-}$. Then, based on Lem.\ref{pami:lem2}, (\ref{ref_eq28}) could be equivalently reformulated as a differentiable optimization problem:
		\begin{equation}
			\begin{aligned}
				\min\limits_{\bmg} \ \ \frac{1}{|\mcu|} \sum_{u_i \in \mcu} \sum_{j=1}^{n_i^+} \sum_{t=1}^{N_i^{\beta_i}}  \frac{\ell^{(i)}_g(v_j^+, v_{[t]}^-)}{n_i^+N_i^{\beta_i}} \Leftrightarrow
				\ \ \min\limits_{\bmg, \bmga \ge \boldsymbol{0}} \frac{1}{|\mcu|} \sum_{u_i \in \mcu} \sum_{j=1}^{n_i^+} \left\{\gamma_{ij} + \frac{1}{N_i^{\beta_i}}\sum_{k=1}^{n_i^-} d^{(i)}_g(v_j^+, v_{k}^-)\right\}, 
			\end{aligned}
		\end{equation}
		where we denote all learnable $\gamma_{ij}$ parameters as a $\sum\limits_{u_i \in \mcu} n_i^+$ dimensional vector $\bmga$ for ease of expression, and we define
		\begin{equation}\nonumber
			d^{(i)}_g(v_j^+, v_{k}^-) = [\lambda + s(u_i, v_j^+) - s(u_i, v_k^-) - \gamma_{ij}]_+,
		\end{equation}
		$\lambda > 0$ is still the safe margin.
	\end{rthm3}
	
	\noindent\rule[0.05\baselineskip]{\columnwidth}{1pt}
	\begin{proof} To prove Thm.\ref{pami:thm3}, we can first realize that the following property holds: 
		\begin{property} \label{pami:property1} For each $(u_i, v_j^+)$ pair, $\ell_{g}^{(i)}$ is a non-increasing function with respect to $s(u_i, v_k^-), \forall v_k^- \in \mathcal{D}_{u_i}^-$. Hence, selecting the negative item with $t$-th minimum $s(u_i, v_k^-)$ score is equivalent to find the $t$-th maximum loss, i.e.,
			\begin{equation} \nonumber
				\ell^{(i)}_g(v_j^+, v_{[t]}^-) \ \  \Leftrightarrow \ \ \ell^{(i)}_g(v_j^+, v_{k}^-)[t], \ \ \exists v_{k}^- \in \mathcal{D}_{u_i}^-,  
			\end{equation}
			where $\ell^{(i)}_g(v_j^+, v_{k}^-)[t]$ represents the $t$-th largest loss induced by the unobserved items $v_{k}^-$.  
			
		\end{property}
		Then, based on Proty.\ref{pami:property1}, we can see that (\ref{ref_eq28}) is equivalent to the following minimization problem, i.e.,
		\begin{equation}\label{ref_eqq28}
			\min\limits_{\bmg} \ \ \frac{1}{|\mcu|} \sum_{u_i \in \mcu} \sum_{j=1}^{n_i^+} \sum_{t=1}^{N_i^{\beta_i}}  \frac{\ell^{(i)}_g(v_j^+, v_{[t]}^-)}{n_i^+N_i^{\beta_i}} \Leftrightarrow \min\limits_{\bmg} \ \ \frac{1}{|\mcu|} \sum_{u_i \in \mcu} \sum_{j=1}^{n_i^+} \sum_{t=1}^{N_i^{\beta_i}}  \frac{\ell^{(i)}_g(v_j^+, v_{k}^-)[t]}{n_i^+N_i^{\beta_i}}.  
		\end{equation}
		
		Subsequently, given that $\ell^{(i)}_g$ is a convex function, the sparse negative sampling selection process of (\ref{ref_eqq28}) could be further rewritten as a differentiable variant by applying Lem.\ref{pami:lem2}:
		\begin{equation} \label{pami:eq31}
			\min\limits_{\bmg, \bmga \ge \boldsymbol{0}} \frac{1}{|\mcu|} \sum_{u_i \in \mcu} \sum_{j=1}^{n_i^+} \frac{1}{n_i^+} \left\{\gamma_{ij} + \frac{1}{N_i^{\beta_i}}\sum_{k=1}^{n_i^-} [\ell^{(i)}_g(v_j^+, v_{k}^-) - \gamma_{ij}]_{+}\right\}.
		\end{equation}
		
		In addition, we proceed to leverage the following result to eliminate the inner hinge function in (\ref{pami:eq31}). Please see Appendix.\ref{pami:proof_lem3} for the proof.
		\begin{lem} \label{pami:lem3} In terms of $c_1 \ge 0, c_2 \ge 0$, we have $[[c_1 - s]_+ - c_2]_+ = [c_1 - c_2 - s]_+$. 
		\end{lem}
		
		Recall that, we have 
		\[
		\ell^{(i)}_g(v_j^+, v_{k}^-) = [\lambda + s(u_i, v_j^+) - s(u_i, v_k^-)]_+,
		\]
		$\lambda > 0$ and $s(u_i, v_j^+) \ge 0$.
		
		Thus, according to Lem.\ref{pami:lem3}, the following condition holds:
		\begin{equation}\label{pami:eq32}
			\begin{aligned}
				[\ell^{(i)}_g(v_j^+, v_{k}^-) - \gamma_{ij}]_{+} = [\lambda + s(u_i, v_j^+) - s(u_i, v_k^-) - \gamma_{ij}]_+.
			\end{aligned}	
		\end{equation}
		
		Finally, applying (\ref{pami:eq32}) to (\ref{pami:eq31}) completes the proof.
	\end{proof}
	
	\subsection{Optimization Algorithm} \label{supp:alg}

	\begin{algorithm}[!t]
		\caption{Differentiable Hardness-aware Negative Sampling (DiHarS) Framework}
		\label{algorithm1}
		\LinesNumbered
		\KwIn{User set $\mathcal{U}=\{u_1, u_2, \dots, u_{|\mathcal{U}|}\}$}
		\KwIn{Item set $\mathcal{I} = \{v_1, v_2\, \dots, v_{|\mathcal{I}|}\}$}
		\KwIn{Observed item sets $\{\mathcal{D}_{u_i}^+\}_{n=1}^{|\mathcal{U}|}$}
		\KwIn{Unobserved item sets $\{\mathcal{D}_{u_i}^-\}_{n=1}^{|\mathcal{U}|}$}
		\KwIn{Safe margin $\lambda$, diversity number $C_{u_i}$, FPR \textbf{$\beta_i$}} 
		\KwIn{Threshold parameters $\delta_1, \delta_2, \delta_1 \le \delta_2$} \KwIn{Regularization parameter $\eta$}
		\KwIn{Sample size $J_1$ and $J_2$}
		\KwOut{User transformation matrix: $\boldsymbol{P}_{c}, c\in [C_{u_i}]$}
		\KwOut{Item transformation matrix: $\boldsymbol{Q}$}
		Initialize $\boldsymbol{P}_{c}, c\in [C_{u_i}]$\;
		Initialize $\boldsymbol{Q}$\;
		Construct $\mathcal{S} = \{(u_i, v_j^+)| \forall \ \ u_i \in \mathcal{U}, v_j^+ \in \mathcal{D}_{u_i}^+\}_{s=1}^{N_s}$ \;
		Initialize $\gamma_{ij}^{0}$ for all pair $(u_i, v_j^+) \in \mathcal{S}$ \;
		Compute $N_i^{\beta_i} = \lfloor n_{i}^- \cdot \beta_i \rfloor$ for $u_i \in \mathcal{U}$\;
		\While{Not Converged}{
			Sample subset $\tilde{\mathcal{S}} \subset \mathcal{S}$ with $|\tilde{\mathcal{S}}| = J_1$ \;
			Sample subset $\tilde{\mathcal{D}_{u_i}^-} \subset \mathcal{D}_{u_i}^-$ with $|\tilde{\mathcal{D}_{u_i}^-}| = J_2$ for all $(u_i, v_j^+) \in \tilde{\mathcal{S}}$ \;
			Compute $\boldsymbol{g}_{u_i}^{c}, \boldsymbol{g}_{v_j^+}$ by (\ref{pami:eq7}) and (\ref{pami:eq8}), respectively, for all $(u_i, v_j^+) \in \tilde{\mathcal{S}}$\;
			Compute $\boldsymbol{g}_{v_k^-}$ by (\ref{pami:eq8}) for all $v_k^- \in \tilde{\mathcal{D}_{u_i}^-}$\;
			Update $\boldsymbol{g}_{u_i}^{c}, \boldsymbol{g}_{v_j^+}$ by $\nabla \tilde{{\mcl}}_{\mcd}(\bmg)$ for all $(u_i, v_j^+) \in \tilde{\mathcal{S}}$\;
			Update $\boldsymbol{g}_{v_k^-}$ by $\nabla \tilde{{\mcl}}_{\mcd}(\bmg)$ for all $v_k^- \in \tilde{\mathcal{D}_{u_i}^-}$\;
			Update $\gamma_{ij}$ by $\nabla \tilde{\mathcal{R}}_{\bmg, \bmga}$ for all $(u_i, v_j^+) \in \tilde{\mathcal{S}}$\;
			Project $\gamma_{ij}$ to ensure $\gamma_{ij} \ge 0$\;
		}
		
		\Return $\boldsymbol{P}_{c}^{T}, c \in [C_{u_i}], \forall u_i \in \mcu$ and $\boldsymbol{Q}$
	\end{algorithm}

	Following the top-$k$ learning paradigms \cite{DBLP:journals/corr/abs-2210-03967,DBLP:journals/corr/abs-2203-01505,NIPS2017_6c524f9d}, the stochastic optimization algorithm for solving (\ref{pami:eq34}) is summarized in Alg.\ref{algorithm1}. At first, the transformation matrices $\boldsymbol{P}_{c}$ and $\boldsymbol{Q}$ are randomly initialized (row $1$-$2$). We randomly initialize $\gamma_{ij}$ for each user-item positive pair (row $3$-$4$) and determine the number of hard negative samples $N_i^{\beta_i}$ according to the given FPR range $\beta_i$ (row $5$). Subsequently, the stochastic (projected) gradients of $\bmg$ and $\bmga$ (row $7$-$10$) are constructed by sampling positive user-item pairs $(u_i, v_j^+)$ and unknown item $v_k^-$. Row $11$-$13$ compute the stochastic gradient based on (\ref{pami:eq34}) and then update these learnable parameters. Meanwhile, row $14$ is further conducted to guarantee the limitation $\boldsymbol{\gamma} \ge \boldsymbol{0}$. Our algorithm can significantly reduce the computation burden caused by the vast number of unobserved item sets (i.e., $n_i^-$ is usually large) such that (\ref{pami:eq34}) can be efficiently optimized.

	\section{Experiments} 
	\label{exp_supp}
	

	\subsection{Dataset}\label{major:supp_C.1}
	We perform empirical studies on $6$ public and real-world benchmark datasets, including \textbf{MovieLens-1M}\footnote{\url{https://grouplens.org/datasets/movielens/1m/}}, \textbf{Steam-200k\footnote{\url{https://www.kaggle.com/tamber/steam-video-games}}}, \textbf{CiteULike}\footnote{\url{http://www.citeulike.org/faq/data.adp}} \cite{DBLP:conf/ijcai/WangCL13}, \textbf{MovieLens-10M}\footnote{\url{https://grouplens.org/datasets/movielens/10m/}} and two subsets of \textbf{RecSys} \cite{DBLP:conf/recsys/AbelDEK17}\footnote{\url{https://www.recsyschallenge.com/2017/}}:
	\begin{itemize}[leftmargin=*]
		\item \textbf{MovieLens}\footnote{\url{https://grouplens.org/datasets/movielens/}} - One of the most popular benchmark datasets with many versions. Specifically, it includes explicit user-item ratings ranging from 1 to 5 and movie types in terms of various movies. We adopt \textbf{MovieLens-1M}\footnote{\url{https://grouplens.org/datasets/movielens/1m/}} and \textbf{MovieLens-10M}\footnote{\url{https://grouplens.org/datasets/movielens/10m/}} here to evaluate the performance. To obtain the implicit preference feedback, if the score of item $v_j$ rated by user $u_i$ is no less than 4, we regard item $v_j$ as a positive item for user $u_i$ following the
		previous and successful research \cite{DBLP:conf/www/HeLZNHC17}. 
		
		\item \textbf{CiteULike}\footnote{\url{http://www.citeulike.org/faq/data.adp}} \cite{DBLP:conf/ijcai/WangCL13} - An implicit feedback dataset that includes the preferences of users toward different articles. There are two configurations of CiteULike collected from CiteULike and Google Scholar. Following \cite{hsieh2017collaborative}, we adopt \textbf{CiteULike-T} here to evaluate the performance.
		\item \textbf{Steam-200k\footnote{\url{https://www.kaggle.com/tamber/steam-video-games}}} - This dataset is collected from Steam which is the world's most popular PC gaming hub. The observed behaviors of users include 'purchase' and 'play' signals. In order to obtain the implicit feedback, if a user has purchased a game as well as the playing hours $play > 0$, we regard this game as a positive item. 
		\item \textbf{RecSys} - We employ two different scales of implicit feedback datasets generated by the released data from the ACM RecSys 2017 Challenge \cite{DBLP:conf/recsys/AbelDEK17}. Specifically, we remove duplicate actions by reserving the latest user-item interactions and also delete users with interaction lengths less than $25$ to ensure a reasonable dataset sparsity. For the sake of expressions, we denote these two subsets as \textbf{RecSys-1} and \textbf{RecSys-2}, respectively. 
	\end{itemize}
	The detailed statistics in terms of these datasets are summarized in Tab.\ref{table1}.

\begin{table*}[!t]
	\centering
	\setlength{\abovecaptionskip}{10pt}    
	\setlength{\belowcaptionskip}{20pt}    
	\setlength{\tabcolsep}{12pt}
	\caption{Basic Information of the Datasets. \%Density is defined as $\frac{\#Ratings}{\#Users \times \#Items} \times 100\% $.}	
	\label{table1}
	\scalebox{1.1}{\begin{tabular}{c|ccccccc}
			\toprule
			Datasets & MovieLens-1M & Steam-200k & CiteULike-T & MovieLens-10M & RecSys-1 & RecSys-2\\
			\midrule
			Domain   & Movie & Game & Paper & Movie & Job & Job\\
			\#Users   & 6,034  & 3,757 & 5,219 & 69,167 & 2,799 & 20,134\\
			\#Items   & 3,953 & 5,113 & 25,975 & 10,019 & 12,612 & 42,214\\
			\#Ratings   & 575,271  & 115,139 & 125,580 & 5,003,437 & 94,016 & 639,742\\
			\%Density  & 2.4118\% & 0.5994\% & 0.0926\%  & 0.7220\% & 0.2633\% & 0.0753\%\\
			\bottomrule
		\end{tabular}
	}
\end{table*}

	\subsection{Competitors \label{pami:supp_competitors}}
	The involved competitors roughly fall into five groups here, including:
	
	\noindent\textbf{1) Item-based collaborative filtering algorithm.} 
	\begin{itemize}[leftmargin=*]
		\item \textbf{itemKNN} \cite{linden2003amazon} is designed on the criterion of the k-nearest neighborhood (KNN), which directly considers the similarity (such as cosine similarity) between the candidate and the previously interacted items to make the recommendations.
	\end{itemize}
	%
	
	\noindent\textbf{2) MF-based algorithms including the combination of MF and deep learning network and multi-vector MF-based methods.}
	
	\begin{itemize}[leftmargin=*]
		\item \textbf{Bayesian Personalized Ranking} (BPR) \cite{DBLP:conf/uai/RendleFGS09} is a classical MF-based approach, which leverages a pairwise log-sigmoid loss to directly optimize the AUC ranking.
		\item \textbf{Generalized Matrix Factorization} (GMF) adopts a linear kernel to capture the preference of users such that it is more expressive than the traditional MF algorithms.
		\item \textbf{Multi-Layer Perceptron} (MLP) leverages a multi-layer perceptron endowed with reasonable flexibility and non-linearity to model the users' preference toward items.
		\item \textbf{Neural network-based Collaborative Filtering} (NeuMF)  \footnote{\url{https://github.com/guoyang9/NCF}} \cite{DBLP:conf/www/HeLZNHC17} is a seminal and competitive deep learning based recommendation framework. Specifically, NCF integrates the GMF and MLP algorithms and makes recommendation via regarding the recommendation task as a regression problem.
		\item \textbf{Multi-vector MF} (M2F) \cite{DBLP:conf/eaamo/GuoKJG21} is a state-of-the-art MF-based recommendation algorithm, which models the diversity preference of users by assigning them multiple embeddings in the dot-product space. This could be regarded as a competitive baseline to figure out the superiority of our proposed algorithm.
		\item \textbf{Multi-vector GMF} (MGMF). Considering that the original algorithm \cite{DBLP:conf/eaamo/GuoKJG21} might be specifically tailored for the explicit feedback rather than the implicit signals, we further apply a multiple set of users' representations to GMF \cite{DBLP:conf/www/HeLZNHC17}.
	\end{itemize}
	
	\noindent\textbf{3) VAE-based representative algorithm.}
	\begin{itemize}[leftmargin=*]
		\item \textbf{Mult-VAE} \cite{DBLP:conf/www/LiangKHJ18} is a successful attempt to apply non-linear probabilistic model (variational autoencoders, VAE) to collaborative filtering for implicit feedback. 
	\end{itemize}
	
	\noindent\textbf{4) GNN-based collaborative filtering framework.}
	\begin{itemize}[leftmargin=*]
		\item \textbf{LightGCN} \cite{DBLP:conf/sigir/0001DWLZ020} recently proposes a simple but competitive Graph Convolution Network (GCN) for recommendations. It captures users' preferences toward items by linearly propagating them on a constructed user-item interaction graph.
	\end{itemize}
	
	\noindent\textbf{5) CML-based recommendation competitors.}
	\begin{itemize}[leftmargin=*]
		\item \textbf{Uniform Negative Sampling} (UniS) \cite{DBLP:conf/icdm/PanZCLLSY08} in terms of each user, uniformly samples $S$ items from unobserved interactions as negative instances to optimize the pairwise ranking loss.
		\item \textbf{Popularity-based Negative Sampling} (PopS) \cite{DBLP:conf/sigir/WuVSSR19} samples $S$ negative candidates from unobserved interactions based on their popularity/frequencies. 
		\item  \textbf{Two-Stage Negative Sampling} (2stS) \footnotemark \cite{tran2019improving} adopts a two-stage sampling strategy. 1) A candidate set of items is sampled based on their popularity; 2) according to their inner product values with anchors (positive items), the most informative samples are selected from this candidate.
		\item \textbf{Hard Negative Sampling} (HarS) \footnote{\url{https://github.com/changun/CollMetric}} is similar to the negative sample mining process broadly used in metric learning \cite{DBLP:journals/pr/GajicAG21,DBLP:journals/pami/ZhengLZ21}. To achieve (\ref{cml_sampling_hars}), it can be divided into two steps: a) uniformly sample $S$ candidates from unobserved items; b) select the hardest item (i.e., $U \equiv 1$) from the candidates as negative according to the distance between the targeted user and each item. 
		\item \textbf{Collaborative Translational Metric Learning} (TransCF) \cite{DBLP:conf/icdm/ParkKXY18} is a translation-based method. Specifically, such translation-based algorithms employ $\bm{d}(i, j) = ||\boldsymbol{g}_{u_i} + \boldsymbol{g}_{r_{ij}} - \boldsymbol{g}_{v_j}||^2$ as the distance/score between user $u_i$ and item $v_j$ instead of $||\boldsymbol{g}_{u_i} - \boldsymbol{g}_{v_j}||^2$, where $\boldsymbol{g}_{r_{ij}}$ is a specific translation vector for $u_i$ and $v_j$. In light of this, TransCF discovers such user–item translation vectors via the users' relationships with their neighbor items.
		\item \textbf{Latent Relational Metric Learning} (LRML) \cite{DBLP:conf/www/TayTH18} is also a translation-based CML method. As a whole, the key idea of LRML is similar to TransCF. The main difference is how to access the translation vectors effectively. Concretely, TransCF leverages the neighborhood information of users and items to acquire the translation vectors while LRML introduces an attention-based memory-augmented neural architecture to learn the exclusive and optimal translation vectors. 
		\item \textbf{Adaptive Collaborative Metric Learning} (AdaCML) \cite{DBLP:conf/dasfaa/ZhangZLXF0SC19} learns an adaptive user representation via a memory component and an attention mechanism to accurately model the implicit relationships of user-item pairs and users’ interests.
		\item \textbf{Hierarchical Latent Relation modeling} (HLR) \cite{DBLP:conf/recsys/TranSHM21} is a state-of-the-art CML-based approach that employs memory-based attention networks to hierarchically capture users' preferences from both latent user-item and item-item relations. 
	\end{itemize}
	Finally, we consider DPCML with both BPA and APA discussed in Sec.\ref{pami:sec6.1} for clear demonstrations of our proposed methods. As introduced in Sec.\ref{pami:sec6.2}, we apply all three sampling techniques to DPCML to avoid the heavy learning burden, where the UniS-driven DPCMLs are abbreviated by \textbf{BPA+UniS} and \textbf{APA+UniS}, respectively; the HarS-driven DPCMLs are named as \textbf{BPA+HarS} and \textbf{APA+HarS}, respectively; DPCMLs optimized by our proposed DiHarS algorithm are abbreviated by \textbf{BPA+DiHarS} and \textbf{APA+DiHarS}, respectively.

	\subsection{Evaluation Metrics \label{pami:used_metrics}}
	In some typical recommendation systems, users often care about the top-$N$ items in recommendation lists, so the most relevant items should be ranked first as much as possible. Motivated by this, we evaluate the performance of competitors and our algorithm with the following extensively adopted metrics, including:
	\begin{itemize}
		\item \textbf{Precision} (P@$N$) counts the proportion that the ground-truth items are among the Top-$N$ recommended list.
		\[
		\text{P@}N = \frac{1}{|\mcu|} \sum_{u_i \in \mathcal{U}}\frac{|\mathcal{D}_{u_i}^+ \cap I^{u_i}_N|}{N}
		\]
		where again $\mathcal{D}_{u_i}^+$ is the set of ground-truth items of user $u_i$; $I^{u_i}_N$ is the top-$N$ recommendation list for user $u_i$; and $|\cdot|$ means the size of the set.
		\item \textbf{Recall} (R@$N$) is defined as the number of the ground-truth items in top-$N$ recommendation list divided by the amount of totally ground-truth items. This reflects the ability of the model to find the relevant items.
		\[
		\text{R@}N = \frac{1}{|\mcu|} \sum_{u_i \in \mathcal{U}} \frac{|\mathcal{D}_{u_i}^+ \cap I^{u_i}_N|}{|\mathcal{D}_{u_i}^+|}
		\]
		\item \textbf{Normalized Discounted Cumulative Gain} (NDCG@$N$) counts the ground-truth items in the top-$N$ recommendation list with a position weighting strategy, i.e., assigning a larger value on top items than bottom ones.
		\[
		\text{NDCG@}N = \frac{1}{|\mcu|}\sum_{u_i \in \mathcal{U}} \frac{\text{DCG}_{u_i}\text{@}N}{\text{IDCG}_{u_i}\text{@}N}
		\]
		Specifically, the $\text{DCG}_{u_i}\text{@}N$ and $\text{IDCG}_{u_i}\text{@}N$ are defined as:
		\[
		\text{DCG}_{u_i}\text{@}N = \sum_{j = 1} ^ {N}\frac{1 \cdot \mathbb{I}(I^{u_i}_{N, j} \in \mathcal{D}_{u_i}^+)}{\log_2(j + 1)},
		\]
		\[
		\text{IDCG}_{u_i}\text{@}N = \sum_{k = 1}^{\min(N,|\mathcal{D}_{u_i}^+|)} \frac{1}{\log_2(k + 1)},
		\]
		where $I^{u_i}_{N, j}$ respresents the $j$-th item in the top-$N$ recommendation list; $\mathbb{I}(\cdot)$ is an indicator function that returns $1$ if the statement is true and returns $0$, otherwise.
		\item \textbf{Mean Average Precision} (MAP) is an extension of Average Precision(AP). AP is the average of precision values at all positions where ground-truth items are found.
		\[
		\text{AP}_{u_i} = \frac{1}{|\mathcal{D}_{u_i}^+|}\sum_{j = 1}^{|\hat{I}_{u_i}|}\frac{|\mathcal{D}_{u_i}^+ \cap \hat{I}_{u_i, 1: j}| \cdot \mathbb{I}(j \in \mathcal{D}_{u_i}^+)}{rank_j^{u_i}}
		\] 
		\[
		\text{MAP} = \frac{1}{|\mcu|} \sum_{u_i \in \mathcal{U}} \text{AP}_{u_i}
		\]
		where different from $I^{u_i}_N$, $\hat{I}_{u_i}$ is the recommendation rankings in terms of all items for user $u_i$; $\hat{I}_{u_i, 1:j}$ represents the top-$j$ recommendation list for user $u_i$; and $rank_j^{u_i}$ means the ranking of item $j$ in $\hat{I}_{u_i}$.
		\item \textbf{Mean Reciprocal Rank} (MRR) takes the rank of each recommended item into account. It is the average of reciprocal ranks of the desired item:
		\[
		\text{MRR} = \frac{1}{|\mcu|} \sum_{u_i\in \mathcal{U}} \sum_{j = 1}^{|\hat{I}_{u_i}|} \frac{1}{rank_j^{u_i}} \cdot \mathbb{I}(\hat{I}_{u_i, j} \in \mathcal{D}_{u_i}^+)
		\]
	\end{itemize}
	
	Note that, for all the above metrics, the higher the metric is, the better the algorithm achieves.
		\subsection{Implementation Details} \label{app:train}
		We implement our model with PyTorch\footnote{\url{https://pytorch.org/}} \cite{paszke2017automatic} and employ \textit{Adam} \cite{DBLP:journals/corr/KingmaB14} as the optimizer. In terms of all benchmark datasets, user interactions are divided into training/validation/test sets with a $0.6:0.2:0.2$ split ratio. According to this, to ensure that each user has at least one positive interaction in training/validation/test, users who have less than five interactions are filtered out from these datasets. We adopt grid search for all methods to select the best parameters based on the validation set and report the corresponding performance on the test set. To be specific, the batch size is set to $256$ and the learning rate is searched within $\{3 \times 10^{-4}, 5 \times 10^{-4}, 1 \times 10^{-3}, 3 \times 10^{-3}, 1 \times 10^{-2}\}$. The number of epochs is set as $100$. The dimension of embedding $d$ is fixed as $100$, and the margin $\lambda$ is searched within $\{1.0, 1.5, 2.0\}$. Besides, for our proposed \textbf{DPCML with BPA scheme}, the number of user representations $C$ is tuned among $\{2, 3, 4, 5\}$. For the regularization term, $\eta$ is searched within $\{10, 20, 30\}$, $\delta_1 \in 
		\{0.,0.05, 0.1, 0.2, 0.5\}$ and $\delta_2 \in \{0.1, 0.25, 0.35, 0.5, 0.8\}$. With respect to the \textbf{APA strategy}, $C_1$ is searched within $\{1, 2, 3, 4, 5\}$ and $a$ is tuned among $\{2, 3, 5, 10\}$. To ensure a reasonable comparison, we set the sampling constant $U = 10$ for all UniS-based methods and $S=10$ for HarS-based approaches. For the other parameters of baseline models, we follow their tuning strategies in the original papers. Moreover, our proposed DiHarS strategy is also applied to both versions of the DPCML framework to show the effectiveness compared with the traditional HarS sampling technique. Concretely, we fix $J_1$ as $256$, search $J_2$ within $\{10, 20, 30, 50, 100, 120, 200\}$ and tune the FPR range $\beta$ among $\{1 \times 10^{-5}, 5 \times 10^{-5}, 1 \times 10^{-4}, 3 \times 10^{-4}, 5 \times 10^{-4}, 1 \times 10^{-3}, 1.5 \times 10^{-3}\}$. Finally, in terms of the top-$N$ recommendation, we evaluate the performance at $N \in \{3, 5\}$, respectively.
	
	\subsection{Overall Performance} \label{suppa.4}
	The experimental results of all the involved competitors are shown in Tab.\ref{tab:addlabel} and Tab.\ref{results2}. Consequently, we can draw the following conclusions: 
		\begin{enumerate}[leftmargin=*]
			\item Our proposed DPCML methods can consistently outperform all competitors significantly on all datasets, in particular with our newly developed APA and DiHarS sampling strategies. This demonstrates the superiority of our proposed algorithms. \item Regarding different preference assignment strategies, as a whole, DPCML+APA optimized by any of the three negative sampling manners (i.e., UniS, HarS, and DiHarS) could achieve better recommendation results than its corresponding counterpart DPCML+BPA. The empirical performance validates the diversity of users' interests and ascertains the effectiveness of the improved adaptive assignment approach. 
			\item Compared with studies targeting joint accessibility (i.e., M2F and MGMF), our proposed methods can perform better on all metrics than M2F and MGMF on all benchmark datasets. This supports the potential advantage of the CML-based paradigm in this direction, which deserves more research attention in future work. 
			\item Concerning CML methods learning with different negative sampling strategies, the HarS-driven CML algorithms demonstrate better than others (say UniS, PopS, and 2stS) in most cases. Most importantly, with respect to the DPCML framework, adopting our proposed DiHarS strategy could further outperform HarS-based DPCML approaches, and the performance gain is sharp. This consistently suggests the superiority of DiHarS (Thm.\ref{pami:thm2} and Thm.\ref{pami:thm3}) that can explicitly improve the Top-$N$ recommendation performance from the OPAUC  perspective. 
			\item We notice that some deep-learning-based methods (such as Mult-VAE and LightGCN) could achieve competitive or even better performance than a few vanilla CML-based methods (such as PopS, TransCF, LRML) to some extent but fail to outperform ours, especially compared to DiHarS-guided DPCML. This shows that our proposed framework could unleash the power of the CML paradigm, contributing to promising recommendation performances.
		\end{enumerate}
	
	
	\begin{table*}[htbp]
		\centering
		\caption{Performance comparisons on MovieLens-1M and Steam-200k. The best and second-best are highlighted in bold and underlined, respectively.}
		\begin{tabular}{c|c|c|cccccccc}
			\toprule
			& Type & Method & P@3 & R@3 & NDCG@3 & P@5 & R@5 & NDCG@5 & MAP & MRR \\
			\midrule
			\multirow{23}[12]{*}{MovieLens-1M} & Item-based & itemKNN & \cellcolor[rgb]{ .988,  .957,  .957} 12.24  & \cellcolor[rgb]{ .984,  .949,  .949} 2.90  & \cellcolor[rgb]{ .988,  .965,  .965} 12.41  & \cellcolor[rgb]{ .984,  .945,  .945} 12.43  & \cellcolor[rgb]{ .98,  .937,  .937} 4.29  & \cellcolor[rgb]{ .984,  .953,  .953} 12.79  & \cellcolor[rgb]{ .973,  .918,  .918} 8.34  & \cellcolor[rgb]{ .988,  .957,  .957} 26.16  \\
				\cmidrule{2-11}      & \multirow{6}[2]{*}{MF-based} 
				& BPR & \cellcolor[rgb]{ .945,  .827,  .831} 22.06  & \cellcolor[rgb]{ .949,  .843,  .847} 4.87  & \cellcolor[rgb]{ .945,  .827,  .831} 22.60  & \cellcolor[rgb]{ .941,  .824,  .827} 22.26  & \cellcolor[rgb]{ .945,  .835,  .835} 6.96  & \cellcolor[rgb]{ .929,  .788,  .792} 23.09  & \cellcolor[rgb]{ .941,  .82,  .824} 13.88  & \cellcolor[rgb]{ .941,  .824,  .824} 41.45  \\
				& & GMF & \cellcolor[rgb]{ .98,  .933,  .933} 14.10  & \cellcolor[rgb]{ .984,  .953,  .953} 2.81  & \cellcolor[rgb]{ .98,  .941,  .941} 14.33  & \cellcolor[rgb]{ .976,  .925,  .925} 14.28  & \cellcolor[rgb]{ .98,  .941,  .941} 4.08  & \cellcolor[rgb]{ .98,  .933,  .933} 14.73  & \cellcolor[rgb]{ .973,  .918,  .918} 8.29  & \cellcolor[rgb]{ .976,  .929,  .929} 29.51  \\
				&   & MLP & \cellcolor[rgb]{ .98,  .937,  .937} 13.95  & \cellcolor[rgb]{ .988,  .957,  .957} 2.78  & \cellcolor[rgb]{ .98,  .941,  .941} 14.22  & \cellcolor[rgb]{ .976,  .925,  .925} 14.06  & \cellcolor[rgb]{ .984,  .945,  .945} 3.98  & \cellcolor[rgb]{ .98,  .933,  .933} 14.56  & \cellcolor[rgb]{ .973,  .918,  .918} 8.30  & \cellcolor[rgb]{ .976,  .929,  .929} 29.39  \\
				&   & NeuMF & \cellcolor[rgb]{ .969,  .906,  .906} 16.43  & \cellcolor[rgb]{ .98,  .933,  .933} 3.20  & \cellcolor[rgb]{ .973,  .91,  .91} 16.87  & \cellcolor[rgb]{ .965,  .894,  .894} 16.73  & \cellcolor[rgb]{ .976,  .922,  .922} 4.68  & \cellcolor[rgb]{ .969,  .902,  .902} 17.40  & \cellcolor[rgb]{ .965,  .894,  .894} 9.69  & \cellcolor[rgb]{ .969,  .898,  .898} 33.23  \\
				&   & M2F & \cellcolor[rgb]{1.0, 1.0, 1.0} 8.61  &   \cellcolor[rgb]{1.0, 1.0, 1.0} 1.84  &  \cellcolor[rgb]{1.0, 1.0, 1.0} 9.36  &  \cellcolor[rgb]{1.0, 1.0, 1.0} 7.60  &  \cellcolor[rgb]{1.0, 1.0, 1.0} 2.30  &  \cellcolor[rgb]{1.0, 1.0, 1.0} 8.67  &  \cellcolor[rgb]{1.0, 1.0, 1.0} 2.95  &  \cellcolor[rgb]{1.0, 1.0, 1.0} 20.40  \\
				&   & MGMF & \cellcolor[rgb]{ .965,  .894,  .894} 17.38  & \cellcolor[rgb]{ .976,  .922,  .922} 3.51  & \cellcolor[rgb]{ .965,  .894,  .894} 18.08  & \cellcolor[rgb]{ .965,  .886,  .886} 17.63  & \cellcolor[rgb]{ .973,  .91,  .91} 5.05  & \cellcolor[rgb]{ .965,  .886,  .886} 18.52  & \cellcolor[rgb]{ .965,  .886,  .886} 10.12  & \cellcolor[rgb]{ .961,  .882,  .882} 35.15  \\
				\cmidrule{2-11} & VAE-based & Mult-VAE & \cellcolor[rgb]{ .945,  .831,  .831} 21.82  & \cellcolor[rgb]{ .937,  .804,  .808} 5.59  & \cellcolor[rgb]{ .945,  .831,  .835} 22.23  & \cellcolor[rgb]{ .945,  .827,  .831} 21.70  & \cellcolor[rgb]{ .937,  .812,  .816} 7.60  & \cellcolor[rgb]{ .941,  .831,  .831} 22.39  & \cellcolor[rgb]{ .929,  .788,  .792} 15.42  & \cellcolor[rgb]{ .929,  .788,  .792} 42.07  \\
				\cmidrule{2-11} & GNN-based & LightGCN & \cellcolor[rgb]{ .937,  .804,  .808} 23.81  & \cellcolor[rgb]{ .929,  .788,  .792} 5.67  & \cellcolor[rgb]{ .937,  .804,  .808} 24.39  & \cellcolor[rgb]{ .933,  .796,  .8} 24.28  & \cellcolor[rgb]{ .933,  .796,  .8} 8.08  & \cellcolor[rgb]{ .933,  .8,  .804} 25.03  & \cellcolor[rgb]{ .929,  .788,  .792} 15.82  & \cellcolor[rgb]{ .933,  .796,  .8} 44.37  \\
				\cmidrule{2-11}      & \multirow{8}[2]{*}{CML-based} & UniS & \cellcolor[rgb]{ .965,  .894,  .894} 17.56  & \cellcolor[rgb]{ .973,  .91,  .91} 3.71  & \cellcolor[rgb]{ .969,  .898,  .898} 17.89  & \cellcolor[rgb]{ .961,  .878,  .878} 18.34  & \cellcolor[rgb]{ .965,  .89,  .89} 5.60  & \cellcolor[rgb]{ .965,  .886,  .886} 18.79  & \cellcolor[rgb]{ .953,  .851,  .851} 12.40  & \cellcolor[rgb]{ .961,  .878,  .878} 35.77  \\
				&   & PopS & \cellcolor[rgb]{ .984,  .949,  .949} 12.96  & \cellcolor[rgb]{ .98,  .941,  .941} 3.11  & \cellcolor[rgb]{ .984,  .953,  .953} 13.30  & \cellcolor[rgb]{ .98,  .941,  .941} 12.82  & \cellcolor[rgb]{ .98,  .933,  .933} 4.41  & \cellcolor[rgb]{ .984,  .945,  .945} 13.40  & \cellcolor[rgb]{ .976,  .929,  .929} 7.59  & \cellcolor[rgb]{ .98,  .937,  .937} 28.61  \\
				&   & 2st & \cellcolor[rgb]{ .953,  .851,  .851} 21.07  & \cellcolor[rgb]{ .953,  .855,  .855} 4.84  & \cellcolor[rgb]{ .953,  .855,  .855} 21.35  & \cellcolor[rgb]{ .945,  .835,  .835} 21.81  & \cellcolor[rgb]{ .949,  .843,  .843} 7.07  & \cellcolor[rgb]{ .949,  .843,  .843} 22.29  & \cellcolor[rgb]{ .941,  .82,  .82} 14.42  & \cellcolor[rgb]{ .949,  .843,  .843} 40.36  \\
				&   & HarS & \cellcolor[rgb]{ .937,  .804,  .804} 24.88  & \cellcolor[rgb]{ .937,  .804,  .804} 5.86  & \cellcolor[rgb]{ .937,  .808,  .808} 25.38  & \cellcolor[rgb]{ .933,  .8,  .8} 24.89  & \cellcolor[rgb]{ .937,  .804,  .804} 8.25  & \cellcolor[rgb]{ .937,  .804,  .804} 25.77  & \cellcolor[rgb]{ .933,  .796,  .796} 15.74  & \cellcolor[rgb]{ .937,  .804,  .804} 45.15  \\
				&   & LRML & \cellcolor[rgb]{ .969,  .898,  .898} 17.15  & \cellcolor[rgb]{ .976,  .922,  .922} 3.52  & \cellcolor[rgb]{ .969,  .902,  .902} 17.56  & \cellcolor[rgb]{ .965,  .886,  .886} 17.45  & \cellcolor[rgb]{ .973,  .91,  .91} 5.12  & \cellcolor[rgb]{ .965,  .894,  .894} 18.08  & \cellcolor[rgb]{ .961,  .882,  .882} 10.42  & \cellcolor[rgb]{ .965,  .89,  .89} 34.36  \\
				&   & TransCF & \cellcolor[rgb]{ .996,  .984,  .984} 10.03  & \cellcolor[rgb]{ 1,  .992,  .992} 1.84  & \cellcolor[rgb]{ 1,  .992,  .992} 10.31  & \cellcolor[rgb]{ .988,  .965,  .965} 10.90  & \cellcolor[rgb]{ .992,  .976,  .976} 3.09  & \cellcolor[rgb]{ .992,  .973,  .973} 11.20  & \cellcolor[rgb]{ .98,  .937,  .937} 7.07  & \cellcolor[rgb]{ .992,  .976,  .976} 23.66  \\
				&   & AdaCML & \cellcolor[rgb]{ .961,  .875,  .875} 19.06  & \cellcolor[rgb]{ .965,  .89,  .89} 4.12  & \cellcolor[rgb]{ .961,  .882,  .882} 19.31  & \cellcolor[rgb]{ .953,  .859,  .859} 19.74  & \cellcolor[rgb]{ .957,  .871,  .871} 6.23  & \cellcolor[rgb]{ .957,  .867,  .867} 20.20  & \cellcolor[rgb]{ .945,  .835,  .835} 13.30  & \cellcolor[rgb]{ .957,  .867,  .867} 37.36  \\
				&   & HLR & \cellcolor[rgb]{ .953,  .851,  .851} 21.10  & \cellcolor[rgb]{ .953,  .855,  .855} 4.80  & \cellcolor[rgb]{ .953,  .855,  .855} 21.53  & \cellcolor[rgb]{ .949,  .839,  .839} 21.61  & \cellcolor[rgb]{ .949,  .843,  .843} 7.06  & \cellcolor[rgb]{ .949,  .843,  .843} 22.28  & \cellcolor[rgb]{ .945,  .827,  .827} 13.95  & \cellcolor[rgb]{ .949,  .839,  .839} 40.71  \\
				\cmidrule{2-11}      & \multirow{6}[6]{*}{DPCML (Ours)} & BPA+UniS & \cellcolor[rgb]{ .961,  .875,  .875} 19.12  & \cellcolor[rgb]{ .965,  .89,  .89} 4.14  & \cellcolor[rgb]{ .961,  .878,  .878} 19.34  & \cellcolor[rgb]{ .953,  .859,  .859} 19.90  & \cellcolor[rgb]{ .957,  .871,  .871} 6.27  & \cellcolor[rgb]{ .957,  .867,  .867} 20.29  & \cellcolor[rgb]{ .949,  .839,  .839} 13.24  & \cellcolor[rgb]{ .957,  .867,  .867} 37.55  \\
				&   & APA+UniS & \cellcolor[rgb]{ .957,  .867,  .867} 19.56  & \cellcolor[rgb]{ .961,  .882,  .882} 4.26  & \cellcolor[rgb]{ .961,  .875,  .875} 19.72  & \cellcolor[rgb]{ .953,  .855,  .855} 20.13  & \cellcolor[rgb]{ .957,  .867,  .867} 6.30  & \cellcolor[rgb]{ .957,  .863,  .863} 20.55  & \cellcolor[rgb]{ .945,  .835,  .835} 13.29  & \cellcolor[rgb]{ .957,  .863,  .863} 37.98  \\
				\cmidrule{3-11}      &   & BPA+HarS & \cellcolor[rgb]{ .933,  .8,  .8} 25.18  & \cellcolor[rgb]{ .933,  .796,  .796} 6.06  & \cellcolor[rgb]{ .937,  .804,  .804} 25.64  & \cellcolor[rgb]{ .933,  .796,  .796} 25.35  & \cellcolor[rgb]{ .933,  .796,  .796} 8.51  & \cellcolor[rgb]{ .933,  .8,  .8} 26.16  & \cellcolor[rgb]{ .933,  .792,  .792} \underline{16.09}  & \cellcolor[rgb]{ .937,  .804,  .804} 45.32  \\
				&   & APA+HarS & \cellcolor[rgb]{ .933,  .796,  .796} 25.49  & \cellcolor[rgb]{ .933,  .792,  .792} 6.08  & \cellcolor[rgb]{ .933,  .796,  .796} 26.08  & \cellcolor[rgb]{ .933,  .792,  .792} \underline{25.53}  & \cellcolor[rgb]{ .933,  .792,  .792} 8.56  & \cellcolor[rgb]{ .933,  .796,  .796} \underline{26.48}  & \cellcolor[rgb]{ .929,  .788,  .788} \textbf{16.19}  & \cellcolor[rgb]{ .933,  .796,  .796} 46.07  \\
				\cmidrule{3-11}      &   & BPA+DiHarS & \cellcolor[rgb]{ .933,  .796,  .796} \underline{25.60}  & \cellcolor[rgb]{ .933,  .792,  .792} \underline{6.11}  & \cellcolor[rgb]{ .933,  .796,  .796} \underline{26.17}  & \cellcolor[rgb]{ .933,  .792,  .792} 25.44  & \cellcolor[rgb]{ .933,  .792,  .792} \underline{8.57}  & \cellcolor[rgb]{ .933,  .796,  .796} 26.45  & \cellcolor[rgb]{ .933,  .796,  .796} 15.83  & \cellcolor[rgb]{ .933,  .796,  .796} \underline{46.21}  \\
				&   & APA+DiHarS & \cellcolor[rgb]{ .929,  .788,  .788} \textbf{25.98} & \cellcolor[rgb]{ .929,  .788,  .788} \textbf{6.16} & \cellcolor[rgb]{ .929,  .788,  .788} \textbf{26.71} & \cellcolor[rgb]{ .929,  .788,  .788} \textbf{25.74} & \cellcolor[rgb]{ .929,  .788,  .788} \textbf{8.65} & \cellcolor[rgb]{ .929,  .788,  .788} \textbf{26.90} & \cellcolor[rgb]{ .933,  .796,  .796} 15.82 & \cellcolor[rgb]{ .929,  .788,  .788} \textbf{46.92} \\
			\midrule
				\multirow{23}[12]{*}{Steam-200k} & Item-based & itemKNN & \cellcolor[rgb]{ 1,  .992,  .984} 12.58  & \cellcolor[rgb]{ .996,  .945,  .91} 9.47  & \cellcolor[rgb]{ 1,  .992,  .984} 13.23  & \cellcolor[rgb]{ 1,  .992,  .984} 6.47  & \cellcolor[rgb]{ 1,  .992,  .984} 3.90  & \cellcolor[rgb]{ 1,  .992,  .984} 7.23  & \cellcolor[rgb]{ .996,  .945,  .91} 11.74  & \cellcolor[rgb]{ 1,  .992,  .984} 23.33  \\
				\cmidrule{2-11}      & \multirow{6}[2]{*}{MF-based} & BPR & \cellcolor[rgb]{ .98,  .855,  .765} 22.88  & \cellcolor[rgb]{ .984,  .855,  .765} 13.11  & \cellcolor[rgb]{ .984,  .855,  .769} 23.92  & \cellcolor[rgb]{ .984,  .835,  .733} 22.32  & \cellcolor[rgb]{ .98,  .831,  .725} 11.46  & \cellcolor[rgb]{ .98,  .835,  .729} 23.63  & \cellcolor[rgb]{ .984,  .835,  .737} 20.33  & \cellcolor[rgb]{ .984,  .839,  .741} 43.94  \\
   & & GMF & \cellcolor[rgb]{ 1,  .992,  .984} 12.57  & \cellcolor[rgb]{ 1,  .996,  .992} 6.17  & \cellcolor[rgb]{ 1,  .988,  .984} 13.29  & \cellcolor[rgb]{ .996,  .941,  .898} 14.22  & \cellcolor[rgb]{ .996,  .957,  .925} 6.86  & \cellcolor[rgb]{ .996,  .941,  .898} 15.39  & \cellcolor[rgb]{ 1,  .969,  .945} 9.72  & \cellcolor[rgb]{ 1,  .969,  .949} 28.38  \\
				&   & MLP & \cellcolor[rgb]{ .996,  .949,  .918} 17.07  & \cellcolor[rgb]{ .996,  .945,  .906} 9.63  & \cellcolor[rgb]{ .996,  .953,  .922} 17.49  & \cellcolor[rgb]{ .992,  .918,  .867} 16.89  & \cellcolor[rgb]{ .996,  .929,  .882} 8.49  & \cellcolor[rgb]{ .992,  .922,  .871} 17.67  & \cellcolor[rgb]{ .992,  .91,  .851} 15.15  & \cellcolor[rgb]{ .996,  .933,  .886} 34.54  \\
				&   & NeuMF & \cellcolor[rgb]{ .996,  .945,  .91} 17.36  & \cellcolor[rgb]{ .996,  .945,  .906} 9.65  & \cellcolor[rgb]{ .996,  .949,  .914} 17.95  & \cellcolor[rgb]{ .992,  .914,  .859} 17.41  & \cellcolor[rgb]{ .996,  .925,  .875} 8.79  & \cellcolor[rgb]{ .992,  .918,  .863} 18.45  & \cellcolor[rgb]{ .992,  .91,  .855} 15.11  & \cellcolor[rgb]{ .996,  .925,  .878} 35.55  \\
				&   & M2F & \cellcolor[rgb]{ 1,  1, 1} 11.33  &\cellcolor[rgb]{ 1,  1, 1} 5.69  & \cellcolor[rgb]{ 1,  1, 1} 11.95  & \cellcolor[rgb]{ .996,  .961,  .937} 11.44  & \cellcolor[rgb]{ 1,  .973,  .953} 5.73  & \cellcolor[rgb]{ .996,  .957,  .929} 12.98  & \cellcolor[rgb]{ 1,  1, 1} 6.43  & \cellcolor[rgb]{ 1,  .992,  .984} 25.05  \\
				&   & MGMF & \cellcolor[rgb]{ 1,  .992,  .984} 12.51  & \cellcolor[rgb]{ 1,  .996,  .992} 6.14  & \cellcolor[rgb]{ 1,  .992,  .984} 13.25  & \cellcolor[rgb]{ .996,  .937,  .898} 14.45  & \cellcolor[rgb]{ .996,  .953,  .925} 6.88  & \cellcolor[rgb]{ .996,  .937,  .898} 15.55  & \cellcolor[rgb]{ 1,  .969,  .949} 9.63  & \cellcolor[rgb]{ 1,  .969,  .949} 28.40  \\
				\cmidrule{2-11}      & VAE-based & Mult-VAE & \cellcolor[rgb]{ .98,  .831,  .725} 24.95  & \cellcolor[rgb]{ .98,  .804,  .682} 15.62  & \cellcolor[rgb]{ .98,  .827,  .725} 26.11  & \cellcolor[rgb]{ .984,  .847,  .749} 21.33  & \cellcolor[rgb]{ .98,  .835,  .733} 11.28  & \cellcolor[rgb]{ .98,  .843,  .741} 22.75  & \cellcolor[rgb]{ .98,  .816,  .702} 22.05  & \cellcolor[rgb]{ .98,  .824,  .714} 46.21  \\
				\cmidrule{2-11}      & GNN-based & LightGCN & \cellcolor[rgb]{ .976,  .8,  .678} 27.33  & \cellcolor[rgb]{ .976,  .8,  .678} 15.98  & \cellcolor[rgb]{ .976,  .8,  .678} 28.36  & \cellcolor[rgb]{ .976,  .8,  .678} 25.49  & \cellcolor[rgb]{ .976,  .8,  .678} 12.81  & \cellcolor[rgb]{ .976,  .8,  .678} 26.89  & \cellcolor[rgb]{ .98,  .804,  .686} 23.00  & \cellcolor[rgb]{ .98,  .804,  .682} 48.73  \\
				\cmidrule{2-11}      & \multirow{8}[2]{*}{CML-based} & UniS & \cellcolor[rgb]{ .992,  .918,  .863} 20.71  & \cellcolor[rgb]{ .992,  .91,  .851} 11.97  & \cellcolor[rgb]{ .992,  .918,  .863} 21.42  & \cellcolor[rgb]{ .992,  .886,  .812} 20.92  & \cellcolor[rgb]{ .992,  .898,  .835} 10.36  & \cellcolor[rgb]{ .992,  .89,  .82} 21.61  & \cellcolor[rgb]{ .988,  .871,  .788} 18.88  & \cellcolor[rgb]{ .992,  .898,  .831} 40.10  \\
				&   & PopS & \cellcolor[rgb]{ .996,  .941,  .902} 18.05  & \cellcolor[rgb]{ .992,  .918,  .859} 11.58  & \cellcolor[rgb]{ .996,  .941,  .902} 18.76  & \cellcolor[rgb]{ .996,  .933,  .89} 14.94  & \cellcolor[rgb]{ .996,  .937,  .894} 7.98  & \cellcolor[rgb]{ .996,  .937,  .894} 15.78  & \cellcolor[rgb]{ .992,  .91,  .855} 15.13  & \cellcolor[rgb]{ .996,  .933,  .89} 34.04  \\
				&   & 2st & \cellcolor[rgb]{ .988,  .875,  .796} 25.20  & \cellcolor[rgb]{ .988,  .871,  .788} 14.62  & \cellcolor[rgb]{ .988,  .875,  .796} 26.20  & \cellcolor[rgb]{ .988,  .863,  .773} 23.97  & \cellcolor[rgb]{ .988,  .875,  .792} 11.91  & \cellcolor[rgb]{ .988,  .863,  .773} 25.35  & \cellcolor[rgb]{ .984,  .843,  .745} 21.48  & \cellcolor[rgb]{ .988,  .859,  .769} 46.17  \\
				&   & HarS & \cellcolor[rgb]{ .988,  .863,  .773} 26.66  & \cellcolor[rgb]{ .988,  .855,  .761} 15.74  & \cellcolor[rgb]{ .988,  .859,  .769} 27.93  & \cellcolor[rgb]{ .988,  .855,  .761} 24.94  & \cellcolor[rgb]{ .988,  .863,  .773} 12.78  & \cellcolor[rgb]{ .988,  .855,  .757} 26.63  & \cellcolor[rgb]{ .984,  .827,  .714} 23.25  & \cellcolor[rgb]{ .984,  .843,  .741} 48.84  \\
				&   & LRML & \cellcolor[rgb]{ 1,  .969,  .949} 14.91  & \cellcolor[rgb]{ 1,  .976,  .961} 7.48  & \cellcolor[rgb]{ 1,  .973,  .953} 15.43  & \cellcolor[rgb]{ .992,  .922,  .871} 16.49  & \cellcolor[rgb]{ .996,  .937,  .894} 8.06  & \cellcolor[rgb]{ .992,  .922,  .875} 17.51  & \cellcolor[rgb]{ .996,  .941,  .902} 12.24  & \cellcolor[rgb]{ .996,  .949,  .914} 31.89  \\
				&   & TransCF & \cellcolor[rgb]{ 1,  .984,  .973} 13.30  & \cellcolor[rgb]{ 1,  .988,  .98} 6.61  & \cellcolor[rgb]{ 1,  .988,  .976} 13.58  & \cellcolor[rgb]{ .996,  .933,  .886} 15.26  & \cellcolor[rgb]{ .996,  .953,  .918} 7.09  & \cellcolor[rgb]{ .996,  .937,  .894} 15.89  & \cellcolor[rgb]{ .996,  .953,  .922} 11.08  & \cellcolor[rgb]{ 1,  .984,  .973} 26.29  \\
				&   & AdaCML & \cellcolor[rgb]{ .992,  .894,  .827} 23.02  & \cellcolor[rgb]{ .992,  .894,  .824} 13.19  & \cellcolor[rgb]{ .992,  .902,  .835} 23.38  & \cellcolor[rgb]{ .988,  .875,  .792} 22.35  & \cellcolor[rgb]{ .988,  .882,  .808} 11.31  & \cellcolor[rgb]{ .988,  .878,  .8} 23.23  & \cellcolor[rgb]{ .988,  .863,  .773} 19.88  & \cellcolor[rgb]{ .992,  .886,  .812} 42.03  \\
				&   & HLR & \cellcolor[rgb]{ .992,  .922,  .867} 20.30  & \cellcolor[rgb]{ .992,  .914,  .859} 11.65  & \cellcolor[rgb]{ .992,  .922,  .871} 20.96  & \cellcolor[rgb]{ .992,  .894,  .827} 19.79  & \cellcolor[rgb]{ .992,  .906,  .847} 9.88  & \cellcolor[rgb]{ .992,  .898,  .831} 20.94  & \cellcolor[rgb]{ .992,  .89,  .82} 17.06  & \cellcolor[rgb]{ .992,  .902,  .839} 39.26  \\
				\cmidrule{2-11}      & \multirow{6}[6]{*}{DPCML (Ours)} & BPA+UniS & \cellcolor[rgb]{ .988,  .875,  .792} 25.39  & \cellcolor[rgb]{ .988,  .867,  .784} 14.84  & \cellcolor[rgb]{ .988,  .875,  .788} 26.56  & \cellcolor[rgb]{ .988,  .863,  .773} 23.88  & \cellcolor[rgb]{ .988,  .871,  .788} 12.11  & \cellcolor[rgb]{ .988,  .863,  .776} 25.25  & \cellcolor[rgb]{ .984,  .835,  .729} 22.26  & \cellcolor[rgb]{ .988,  .855,  .761} 46.79  \\
				&   & APA+UniS & \cellcolor[rgb]{ .988,  .871,  .784} 25.76  & \cellcolor[rgb]{ .988,  .867,  .776} 15.07  & \cellcolor[rgb]{ .988,  .871,  .784} 26.91  & \cellcolor[rgb]{ .988,  .851,  .753} 25.25  & \cellcolor[rgb]{ .988,  .859,  .769} 12.90  & \cellcolor[rgb]{ .988,  .855,  .761} 26.49  & \cellcolor[rgb]{ .984,  .835,  .725} 22.49  & \cellcolor[rgb]{ .988,  .851,  .757} 47.37  \\
				\cmidrule{3-11}      &   & BPA+HarS & \cellcolor[rgb]{ .984,  .831,  .725} 29.88  & \cellcolor[rgb]{ .984,  .835,  .729} 17.13  & \cellcolor[rgb]{ .984,  .831,  .722} 31.22  & \cellcolor[rgb]{ .984,  .824,  .71} 28.70  & \cellcolor[rgb]{ .984,  .835,  .725} 14.51  & \cellcolor[rgb]{ .984,  .824,  .71} 30.56  & \cellcolor[rgb]{ .984,  .82,  .698} 24.10  & \cellcolor[rgb]{ .984,  .824,  .71} 51.95  \\
				&   & APA+HarS & \cellcolor[rgb]{ .984,  .827,  .718} 30.37  & \cellcolor[rgb]{ .984,  .827,  .714} 17.64  & \cellcolor[rgb]{ .984,  .827,  .714} 31.73  & \cellcolor[rgb]{ .984,  .82,  .706} 29.05  & \cellcolor[rgb]{ .984,  .831,  .722} 14.60  & \cellcolor[rgb]{ .984,  .82,  .706} 30.85  & \cellcolor[rgb]{ .98,  .804,  .678} \underline{25.24}  & \cellcolor[rgb]{ .984,  .82,  .702} 52.68  \\
				\cmidrule{3-11}      &   & BPA+DiHarS & \cellcolor[rgb]{ .984,  .808,  .682} \textbf{32.62}  & \cellcolor[rgb]{ .984,  .808,  .686} \underline{18.91}  & \cellcolor[rgb]{ .984,  .808,  .682} \underline{33.85}  & \cellcolor[rgb]{ .984,  .808,  .682} \underline{30.72}  & \cellcolor[rgb]{ .984,  .812,  .686} \underline{15.98}  & \cellcolor[rgb]{ .984,  .808,  .682} \textbf{32.71}  & \cellcolor[rgb]{ .984,  .812,  .69} 24.63  & \cellcolor[rgb]{ .984,  .808,  .682} \underline{54.35}  \\
				&   & APA+DiHarS & \cellcolor[rgb]{ .98,  .804,  .678} \underline{32.58} & \cellcolor[rgb]{ .98,  .804,  .678} \textbf{19.09} & \cellcolor[rgb]{ .98,  .804,  .678} \textbf{33.98} & \cellcolor[rgb]{ .98,  .804,  .678} \textbf{30.81} & \cellcolor[rgb]{ .98,  .804,  .678} \textbf{15.99} & \cellcolor[rgb]{ .98,  .804,  .678} \underline{32.68} & \cellcolor[rgb]{ .984,  .808,  .682} \textbf{25.78} & \cellcolor[rgb]{ .98,  .804,  .678} \textbf{54.90} \\
			\bottomrule
		\end{tabular}%
		\label{results1}%
	\end{table*}%
	
	\begin{table*}[htbp]
		\centering
		\caption{Performance comparisons on MovieLens-10M. The best and second-best are highlighted in bold and underlined, respectively.}
		\begin{tabular}{c|c|c|cccccccc}
			\toprule
			& Type & Method & P@3 & R@3 & NDCG@3 & P@5 & R@5 & NDCG@5 & MAP & MRR \\
			\midrule
			\multirow{21}[12]{*}{MovieLens-10M} & Item-based & itemKNN & \cellcolor[rgb]{ .941,  .969,  .922} 11.44  & \cellcolor[rgb]{ .918,  .957,  .89} 3.70  & \cellcolor[rgb]{ .941,  .969,  .922} 11.78  & \cellcolor[rgb]{ .937,  .965,  .918} 12.27  & \cellcolor[rgb]{ .922,  .957,  .898} 4.93  & \cellcolor[rgb]{ .941,  .969,  .922} 12.63  & \cellcolor[rgb]{ .91,  .953,  .882} 8.25  & \cellcolor[rgb]{ .922,  .957,  .894} 25.85  \\
			\cmidrule{2-11}      & \multirow{5}[2]{*}{MF-based} & BPR & \cellcolor[rgb]{ .859,  .922,  .812} 14.62  & \cellcolor[rgb]{ .843,  .918,  .796} 4.42  & \cellcolor[rgb]{ .855,  .922,  .808} 14.98  & \cellcolor[rgb]{ .855,  .922,  .808} 15.77  & \cellcolor[rgb]{ .843,  .914,  .792} 6.17  & \cellcolor[rgb]{ .851,  .922,  .804} 16.29  & \cellcolor[rgb]{ .851,  .922,  .804} 10.38  & \cellcolor[rgb]{ .843,  .918,  .796} 30.56  \\
			& & GMF & \cellcolor[rgb]{ .91,  .953,  .882} 13.55  & \cellcolor[rgb]{ .91,  .953,  .882} 3.87  & \cellcolor[rgb]{ .914,  .953,  .886} 13.91  & \cellcolor[rgb]{ .902,  .949,  .875} 14.67  & \cellcolor[rgb]{ .906,  .949,  .878} 5.41  & \cellcolor[rgb]{ .906,  .949,  .878} 15.13  & \cellcolor[rgb]{ .894,  .945,  .863} 9.14  & \cellcolor[rgb]{ .898,  .945,  .863} 28.91  \\
			&   & MLP & \cellcolor[rgb]{ .886,  .941,  .851} 15.27  & \cellcolor[rgb]{ .871,  .929,  .831} 4.93  & \cellcolor[rgb]{ .89,  .941,  .859} 15.46  & \cellcolor[rgb]{ .886,  .937,  .847} 16.08  & \cellcolor[rgb]{ .875,  .933,  .835} 6.53  & \cellcolor[rgb]{ .89,  .941,  .855} 16.38  & \cellcolor[rgb]{ .839,  .914,  .784} 12.77  & \cellcolor[rgb]{ .871,  .929,  .831} 32.21  \\
			&   & NeuMF & \cellcolor[rgb]{ .89,  .941,  .851} 15.19  & \cellcolor[rgb]{ .867,  .929,  .824} 5.02  & \cellcolor[rgb]{ .894,  .945,  .859} 15.27  & \cellcolor[rgb]{ .882,  .937,  .847} 16.09  & \cellcolor[rgb]{ .871,  .929,  .827} 6.65  & \cellcolor[rgb]{ .894,  .941,  .859} 16.24  & \cellcolor[rgb]{ .839,  .914,  .784} 12.76  & \cellcolor[rgb]{ .875,  .933,  .831} 31.87  \\
			&   & M2F & \cellcolor[rgb]{ 1.0, 1.0, 1.0} 7.03  & \cellcolor[rgb]{ 1.0, 1.0, 1.0} 1.41  & \cellcolor[rgb]{ 1.0, 1.0, 1.0} 7.21  & \cellcolor[rgb]{ 1.0, 1.0, 1.0} 7.55  & \cellcolor[rgb]{ 1.0, 1.0, 1.0} 2.23  & \cellcolor[rgb]{ 1.0, 1.0, 1.0} 7.98  & \cellcolor[rgb]{ 1.0, 1.0, 1.0} 2.50  &  \cellcolor[rgb]{ 1.0, 1.0, 1.0} 15.17  \\
			&   & MGMF & \cellcolor[rgb]{ .898,  .945,  .863} 14.62  & \cellcolor[rgb]{ .894,  .945,  .863} 4.26  & \cellcolor[rgb]{ .898,  .945,  .863} 15.15  & \cellcolor[rgb]{ .89,  .941,  .859} 15.53  & \cellcolor[rgb]{ .89,  .941,  .855} 5.96  & \cellcolor[rgb]{ .894,  .941,  .859} 16.26  & \cellcolor[rgb]{ .878,  .933,  .839} 10.30  & \cellcolor[rgb]{ .878,  .937,  .839} 31.07  \\
			\cmidrule{2-11}      & VAE-based & Mult-VAE & \cellcolor[rgb]{ .776,  .878,  .706} 21.95  & \cellcolor[rgb]{ .776,  .878,  .706} 7.09  & \cellcolor[rgb]{ .776,  .878,  .706} 22.60  & \cellcolor[rgb]{ .776,  .878,  .706} 22.60  & \cellcolor[rgb]{ .776,  .878,  .706} 9.44  & \cellcolor[rgb]{ .776,  .878,  .706} 23.56  & \cellcolor[rgb]{ .776,  .878,  .706} 17.10  & \cellcolor[rgb]{ .776,  .878,  .706} 42.31  \\
			\cmidrule{2-11}      & GNN-based & LightGCN & \cellcolor[rgb]{ .776,  .878,  .706} 22.49  & \cellcolor[rgb]{ .776,  .878,  .706} 7.23  & \cellcolor[rgb]{ .776,  .878,  .706} 23.18  & \cellcolor[rgb]{ .776,  .878,  .706} 22.78  & \cellcolor[rgb]{ .776,  .878,  .706} 9.28  & \cellcolor[rgb]{ .776,  .878,  .706} 23.87  & \cellcolor[rgb]{ .776,  .878,  .706} 16.44  & \cellcolor[rgb]{ .776,  .878,  .706} 42.52  \\
			\cmidrule{2-11}      & \multirow{8}[2]{*}{CML-based} & UniS & \cellcolor[rgb]{ .957,  .976,  .945} 10.15  & \cellcolor[rgb]{ .949,  .973,  .933} 2.84  & \cellcolor[rgb]{ .961,  .98,  .949} 10.33  & \cellcolor[rgb]{ .953,  .976,  .937} 11.19  & \cellcolor[rgb]{ .945,  .973,  .929} 4.08  & \cellcolor[rgb]{ .957,  .976,  .941} 11.38  & \cellcolor[rgb]{ .898,  .945,  .867} 8.92  & \cellcolor[rgb]{ .933,  .965,  .91} 24.24  \\
			&   & PopS & \cellcolor[rgb]{ .98,  .988,  .973} 8.61  & \cellcolor[rgb]{ .941,  .969,  .922} 3.06  & \cellcolor[rgb]{ .98,  .988,  .973} 8.96  & \cellcolor[rgb]{ .992,  .996,  .988} 8.34  & \cellcolor[rgb]{ .957,  .976,  .941} 3.76  & \cellcolor[rgb]{ .992,  .996,  .988} 8.84  & \cellcolor[rgb]{ .945,  .973,  .925} 6.08  & \cellcolor[rgb]{ .957,  .976,  .945} 20.97  \\
			&   & 2st & \cellcolor[rgb]{ .871,  .929,  .827} 16.47  & \cellcolor[rgb]{ .871,  .933,  .831} 4.89  & \cellcolor[rgb]{ .875,  .933,  .835} 16.72  & \cellcolor[rgb]{ .863,  .925,  .82} 17.62  & \cellcolor[rgb]{ .863,  .925,  .82} 6.87  & \cellcolor[rgb]{ .871,  .929,  .827} 18.06  & \cellcolor[rgb]{ .835,  .91,  .784} 12.89  & \cellcolor[rgb]{ .859,  .925,  .816} 33.75  \\
			&   & HarS & \cellcolor[rgb]{ .863,  .925,  .82} 17.00  & \cellcolor[rgb]{ .871,  .929,  .827} 4.97  & \cellcolor[rgb]{ .871,  .929,  .827} 17.16  & \cellcolor[rgb]{ .855,  .922,  .808} 18.34  & \cellcolor[rgb]{ .863,  .925,  .816} 6.96  & \cellcolor[rgb]{ .859,  .925,  .816} 18.70  & \cellcolor[rgb]{ .831,  .91,  .776} 13.14  & \cellcolor[rgb]{ .855,  .922,  .808} 34.20  \\
			&   & LRML & \cellcolor[rgb]{ .91,  .953,  .878} 13.72  & \cellcolor[rgb]{ .906,  .949,  .878} 3.96  & \cellcolor[rgb]{ .91,  .953,  .882} 13.98  & \cellcolor[rgb]{ .906,  .949,  .875} 14.53  & \cellcolor[rgb]{ .902,  .949,  .871} 5.58  & \cellcolor[rgb]{ .91,  .949,  .878} 15.08  & \cellcolor[rgb]{ .898,  .945,  .867} 8.99  & \cellcolor[rgb]{ .898,  .945,  .863} 28.77  \\
			&   & TransCF & \cellcolor[rgb]{ .945,  .973,  .929} 11.00  & \cellcolor[rgb]{ .918,  .957,  .89} 3.70  & \cellcolor[rgb]{ .953,  .976,  .937} 10.91  & \cellcolor[rgb]{ .945,  .973,  .929} 11.62  & \cellcolor[rgb]{ .922,  .957,  .894} 4.94  & \cellcolor[rgb]{ .953,  .976,  .937} 11.61  & \cellcolor[rgb]{ .914,  .953,  .886} 7.99  & \cellcolor[rgb]{ .937,  .965,  .918} 23.67  \\
			&   & AdaCML & \cellcolor[rgb]{ .91,  .953,  .882} 13.65  & \cellcolor[rgb]{ .906,  .949,  .875} 4.00  & \cellcolor[rgb]{ .914,  .953,  .886} 13.82  & \cellcolor[rgb]{ .906,  .949,  .875} 14.64  & \cellcolor[rgb]{ .906,  .949,  .875} 5.52  & \cellcolor[rgb]{ .91,  .953,  .878} 14.98  & \cellcolor[rgb]{ .863,  .925,  .82} 11.13  & \cellcolor[rgb]{ .89,  .941,  .855} 29.58  \\
			&   & HLR & \cellcolor[rgb]{ .89,  .941,  .855} 15.13  & \cellcolor[rgb]{ .863,  .925,  .82} 5.12  & \cellcolor[rgb]{ .898,  .945,  .867} 14.94  & \cellcolor[rgb]{ .878,  .937,  .843} 16.40  & \cellcolor[rgb]{ .859,  .925,  .816} 7.00  & \cellcolor[rgb]{ .894,  .941,  .859} 16.23  & \cellcolor[rgb]{ .827,  .906,  .773} 13.40  & \cellcolor[rgb]{ .875,  .933,  .835} 31.66  \\
			\cmidrule{2-11}      & \multirow{6}[6]{*}{DPCML-based} & BPA+UniS & \cellcolor[rgb]{ .922,  .961,  .898} 12.73  & \cellcolor[rgb]{ .914,  .953,  .882} 3.82  & \cellcolor[rgb]{ .925,  .961,  .898} 13.05  & \cellcolor[rgb]{ .925,  .961,  .902} 13.12  & \cellcolor[rgb]{ .918,  .957,  .89} 5.07  & \cellcolor[rgb]{ .925,  .961,  .902} 13.72  & \cellcolor[rgb]{ .878,  .933,  .839} 10.32  & \cellcolor[rgb]{ .898,  .945,  .867} 28.65  \\
			&   & APA+UniS & \cellcolor[rgb]{ .918,  .957,  .89} 13.17  & \cellcolor[rgb]{ .91,  .953,  .878} 3.91  & \cellcolor[rgb]{ .918,  .957,  .894} 13.42  & \cellcolor[rgb]{ .914,  .953,  .886} 13.83  & \cellcolor[rgb]{ .91,  .953,  .878} 5.37  & \cellcolor[rgb]{ .918,  .957,  .89} 14.31  & \cellcolor[rgb]{ .875,  .933,  .835} 10.51  & \cellcolor[rgb]{ .894,  .945,  .863} 29.01  \\
			\cmidrule{3-11}      &   & BPA+HarS & \cellcolor[rgb]{ .851,  .918,  .8} 18.00  & \cellcolor[rgb]{ .851,  .922,  .804} 5.46  & \cellcolor[rgb]{ .855,  .922,  .808} 18.37  & \cellcolor[rgb]{ .843,  .918,  .796} 18.97  & \cellcolor[rgb]{ .851,  .918,  .8} 7.37  & \cellcolor[rgb]{ .847,  .918,  .8} 19.57  & \cellcolor[rgb]{ .82,  .902,  .761} 14.01  & \cellcolor[rgb]{ .839,  .914,  .788} 36.44  \\
			&   & APA+HarS & \cellcolor[rgb]{ .839,  .914,  .788} 18.76  & \cellcolor[rgb]{ .843,  .914,  .792} 5.69  & \cellcolor[rgb]{ .843,  .918,  .796} 19.06  & \cellcolor[rgb]{ .831,  .91,  .776} 19.93  & \cellcolor[rgb]{ .839,  .914,  .784} 7.77  & \cellcolor[rgb]{ .839,  .914,  .784} 20.43  & \cellcolor[rgb]{ .816,  .898,  .753} 14.27  & \cellcolor[rgb]{ .835,  .91,  .78} 37.04  \\
			\cmidrule{3-11}      &   & BPA+DiHarS & \cellcolor[rgb]{ .78,  .882,  .71} \underline{23.47}  & \cellcolor[rgb]{ .78,  .882,  .71} \underline{7.50}  & \cellcolor[rgb]{ .78,  .882,  .71} \underline{24.17}  & \cellcolor[rgb]{ .78,  .882,  .714} \underline{23.71}  & \cellcolor[rgb]{ .784,  .882,  .718} \underline{9.66}  & \cellcolor[rgb]{ .784,  .882,  .714} \underline{24.86}  & \cellcolor[rgb]{ .78,  .882,  .714} \underline{16.34}  & \cellcolor[rgb]{ .784,  .882,  .714} \underline{43.85}  \\
			&   & APA+DiHarS & \cellcolor[rgb]{ .776,  .878,  .706} \textbf{24.02} & \cellcolor[rgb]{ .776,  .878,  .706} \textbf{7.73} & \cellcolor[rgb]{ .776,  .878,  .706} \textbf{24.79} & \cellcolor[rgb]{ .776,  .878,  .706} \textbf{24.17} & \cellcolor[rgb]{ .776,  .878,  .706} \textbf{9.89} & \cellcolor[rgb]{ .776,  .878,  .706} \textbf{25.38} & \cellcolor[rgb]{ .776,  .878,  .706} \textbf{16.72} & \cellcolor[rgb]{ .776,  .878,  .706} \textbf{44.74} \\
			\bottomrule
		\end{tabular}%
		\label{results2}%
	\end{table*}%

\begin{table}[htbp]
	\centering
	\caption{Performance comparisons on two different scaled RecSys datasets. The best and second-best are highlighted in bold and underlined, respectively.}
	  \begin{tabular}{c|c|c|cccccccc}
	  \toprule
		& Type & \multicolumn{1}{c}{Method} & P@3 & R@3 & NDCG@3 & P@5 & R@5 & NDCG@5 & MAP & MRR \\
	  \midrule
	  \multirow{23}[16]{*}{RecSys-1} & Item-based & itemKNN & \cellcolor[rgb]{ .992,  .996,  1} 9.68  & \cellcolor[rgb]{ .996,  1,  1} 3.87  & \cellcolor[rgb]{ 1, 1,  1} 9.19  & \cellcolor[rgb]{ .91,  .949,  .984} 9.96  & \cellcolor[rgb]{ .918,  .953,  .984} 6.60  & \cellcolor[rgb]{ .973,  .988,  .996} 9.53  & \cellcolor[rgb]{ .761,  .867,  .953} 14.22  & \cellcolor[rgb]{ .992,  .996,  1} 22.72  \\
  \cmidrule{2-11}      & \multirow{6}[2]{*}{MF-based} & BPR & \cellcolor[rgb]{ .773,  .875,  .957} 12.10  & \cellcolor[rgb]{ .78,  .878,  .957} 4.78  & \cellcolor[rgb]{ .765,  .871,  .957} 12.10  & \cellcolor[rgb]{ .737,  .855,  .949} 11.86  & \cellcolor[rgb]{ .737,  .855,  .949} 7.85  & \cellcolor[rgb]{ .753,  .863,  .953} 11.94  & \cellcolor[rgb]{ .678,  .824,  .937} 16.45  & \cellcolor[rgb]{ .745,  .859,  .953} 27.63  \\
		&   & GMF & \cellcolor[rgb]{ .91,  .953,  .984} 10.58  & \cellcolor[rgb]{ .965,  .98,  .996} 4.00  & \cellcolor[rgb]{ .878,  .933,  .976} 10.70  & \cellcolor[rgb]{ .855,  .918,  .973} 10.58  & \cellcolor[rgb]{ .902,  .945,  .98} 6.69  & \cellcolor[rgb]{ .867,  .925,  .976} 10.68  & \cellcolor[rgb]{ .761,  .867,  .953} 14.22  & \cellcolor[rgb]{ .91,  .949,  .984} 24.38  \\
		&   & MLP & \cellcolor[rgb]{ .812,  .898,  .965} 11.66  & \cellcolor[rgb]{ .82,  .902,  .965} 4.62  & \cellcolor[rgb]{ .804,  .89,  .961} 11.64  & \cellcolor[rgb]{ .757,  .863,  .953} 11.66  & \cellcolor[rgb]{ .765,  .871,  .957} 7.66  & \cellcolor[rgb]{ .78,  .878,  .957} 11.65  & \cellcolor[rgb]{ .694,  .831,  .941} 16.04  & \cellcolor[rgb]{ .788,  .882,  .961} 26.76  \\
		&   & NeuMF & \cellcolor[rgb]{ .808,  .894,  .965} 11.71  & \cellcolor[rgb]{ .824,  .902,  .969} 4.60  & \cellcolor[rgb]{ .792,  .886,  .961} 11.77  & \cellcolor[rgb]{ .776,  .875,  .957} 11.46  & \cellcolor[rgb]{ .788,  .882,  .961} 7.50  & \cellcolor[rgb]{ .784,  .882,  .961} 11.58  & \cellcolor[rgb]{ .698,  .831,  .941} 15.96  & \cellcolor[rgb]{ .773,  .875,  .957} 27.12  \\
		&   & M2F & \cellcolor[rgb]{ .886,  .937,  .98} 10.85  & \cellcolor[rgb]{ .933,  .965,  .988} 4.13  & \cellcolor[rgb]{ .867,  .925,  .976} 10.87  & \cellcolor[rgb]{ .875,  .929,  .976} 10.37  & \cellcolor[rgb]{ .91,  .953,  .984} 6.63  & \cellcolor[rgb]{ .882,  .937,  .976} 10.52  & \cellcolor[rgb]{ .792,  .886,  .961} 13.37  & \cellcolor[rgb]{ .929,  .961,  .988} 23.99  \\
		&   & MGMF & \cellcolor[rgb]{ .835,  .91,  .969} 11.43  & \cellcolor[rgb]{ .882,  .933,  .976} 4.36  & \cellcolor[rgb]{ .82,  .902,  .965} 11.42  & \cellcolor[rgb]{ .847,  .914,  .973} 10.65  & \cellcolor[rgb]{ .89,  .937,  .98} 6.79  & \cellcolor[rgb]{ .851,  .918,  .973} 10.88  & \cellcolor[rgb]{ .761,  .867,  .953} 14.25  & \cellcolor[rgb]{ .89,  .941,  .98} 24.77  \\
  \cmidrule{2-11}      & VAE-based & Multi-VAE & \cellcolor[rgb]{ .78,  .878,  .957} 12.03  & \cellcolor[rgb]{ .78,  .878,  .957} 4.78  & \cellcolor[rgb]{ .776,  .875,  .957} 11.95  & \cellcolor[rgb]{ .753,  .863,  .953} 11.68  & \cellcolor[rgb]{ .753,  .863,  .953} 7.74  & \cellcolor[rgb]{ .773,  .875,  .957} 11.72  & \cellcolor[rgb]{ .682,  .824,  .937} 16.32  & \cellcolor[rgb]{ .765,  .871,  .957} 27.22  \\
  \cmidrule{2-11}      & GNN-based & LightGCN & \cellcolor[rgb]{ .769,  .871,  .957} 12.14  & \cellcolor[rgb]{ .784,  .878,  .957} 4.77  & \cellcolor[rgb]{ .765,  .871,  .957} 12.11  & \cellcolor[rgb]{ .722,  .843,  .945} 12.06  & \cellcolor[rgb]{ .729,  .851,  .949} 7.92  & \cellcolor[rgb]{ .741,  .855,  .949} 12.07  & \cellcolor[rgb]{ .675,  .82,  .937} 16.56  & \cellcolor[rgb]{ .725,  .847,  .949} 28.04  \\
  \cmidrule{2-11}      & \multirow{8}[2]{*}{CML-based} & UniS & \cellcolor[rgb]{ .808,  .894,  .965} 11.70  & \cellcolor[rgb]{ .816,  .898,  .965} 4.63  & \cellcolor[rgb]{ .792,  .886,  .961} 11.77  & \cellcolor[rgb]{ .729,  .851,  .949} 11.95  & \cellcolor[rgb]{ .737,  .855,  .949} 7.87  & \cellcolor[rgb]{ .753,  .863,  .953} 11.93  & \cellcolor[rgb]{ .678,  .824,  .937} 16.43  & \cellcolor[rgb]{ .757,  .867,  .953} 27.37  \\
		&   & PopS & \cellcolor[rgb]{ 1, 1,  1} 9.57  & \cellcolor[rgb]{ 1, 1,  1} 3.84  & \cellcolor[rgb]{ .961,  .98,  .992} 9.70  & \cellcolor[rgb]{ 1, 1,  1} 8.94  & \cellcolor[rgb]{ 1, 1,  1} 5.99  & \cellcolor[rgb]{ 1, 1,  1} 9.22  & \cellcolor[rgb]{ 1, 1,  1} 7.71  & \cellcolor[rgb]{ 1, 1,  1} 22.57  \\
		&   & 2st & \cellcolor[rgb]{ .8,  .89,  .961} 11.81  & \cellcolor[rgb]{ .804,  .89,  .961} 4.69  & \cellcolor[rgb]{ .792,  .886,  .961} 11.76  & \cellcolor[rgb]{ .769,  .871,  .957} 11.51  & \cellcolor[rgb]{ .776,  .875,  .957} 7.58  & \cellcolor[rgb]{ .784,  .882,  .961} 11.57  & \cellcolor[rgb]{ .78,  .878,  .957} 13.70  & \cellcolor[rgb]{ .776,  .875,  .957} 27.01  \\
		&   & HarS & \cellcolor[rgb]{ .773,  .875,  .957} 12.10  & \cellcolor[rgb]{ .788,  .882,  .961} 4.75  & \cellcolor[rgb]{ .757,  .867,  .953} 12.18  & \cellcolor[rgb]{ .69,  .827,  .941} 12.40  & \cellcolor[rgb]{ .698,  .831,  .941} 8.14  & \cellcolor[rgb]{ .714,  .839,  .945} 12.36  & \cellcolor[rgb]{ .675,  .82,  .937} 16.56  & \cellcolor[rgb]{ .722,  .847,  .945} 28.09  \\
		&   & LRML & \cellcolor[rgb]{ .796,  .886,  .961} 11.84  & \cellcolor[rgb]{ .796,  .886,  .961} 4.72  & \cellcolor[rgb]{ .796,  .886,  .961} 11.72  & \cellcolor[rgb]{ .757,  .863,  .953} 11.66  & \cellcolor[rgb]{ .753,  .863,  .953} 7.76  & \cellcolor[rgb]{ .78,  .878,  .957} 11.64  & \cellcolor[rgb]{ .682,  .824,  .937} 16.31  & \cellcolor[rgb]{ .796,  .886,  .961} 26.66  \\
		&   & TransCF & \cellcolor[rgb]{ .82,  .902,  .965} 11.59  & \cellcolor[rgb]{ .824,  .902,  .969} 4.60  & \cellcolor[rgb]{ .808,  .894,  .965} 11.60  & \cellcolor[rgb]{ .749,  .863,  .953} 11.73  & \cellcolor[rgb]{ .753,  .863,  .953} 7.74  & \cellcolor[rgb]{ .776,  .875,  .957} 11.68  & \cellcolor[rgb]{ .69,  .827,  .941} 16.15  & \cellcolor[rgb]{ .788,  .882,  .961} 26.81  \\
		&   & AdaCML & \cellcolor[rgb]{ .922,  .957,  .988} 10.45  & \cellcolor[rgb]{ .937,  .965,  .988} 4.11  & \cellcolor[rgb]{ .898,  .945,  .98} 10.48  & \cellcolor[rgb]{ .843,  .914,  .969} 10.71  & \cellcolor[rgb]{ .851,  .918,  .973} 7.07  & \cellcolor[rgb]{ .871,  .929,  .976} 10.66  & \cellcolor[rgb]{ .706,  .835,  .945} 15.70  & \cellcolor[rgb]{ .859,  .922,  .973} 25.37  \\
		&   & HLR & \cellcolor[rgb]{ .949,  .973,  .992} 10.14  & \cellcolor[rgb]{ .988,  .996,  1} 3.89  & \cellcolor[rgb]{ .922,  .957,  .984} 10.17  & \cellcolor[rgb]{ .925,  .961,  .988} 9.79  & \cellcolor[rgb]{ .965,  .98,  .996} 6.25  & \cellcolor[rgb]{ .937,  .969,  .988} 9.91  & \cellcolor[rgb]{ .776,  .875,  .957} 13.84  & \cellcolor[rgb]{ .973,  .984,  .996} 23.18  \\
  \cmidrule{2-11}      & \multirow{6}[6]{*}{DPCML-based (Ours)} & BPA+UniS & \cellcolor[rgb]{ .698,  .831,  .941} 12.93  & \cellcolor[rgb]{ .694,  .831,  .941} 5.16  & \cellcolor[rgb]{ .682,  .824,  .937} 13.13  & \cellcolor[rgb]{ .702,  .835,  .941} 12.28  & \cellcolor[rgb]{ .698,  .831,  .941} 8.13  & \cellcolor[rgb]{ .69,  .827,  .941} 12.61  & \cellcolor[rgb]{ .667,  .816,  .937} \underline{16.77}  & \cellcolor[rgb]{ .675,  .82,  .937} 29.02  \\
		&   & APA+UniS & \cellcolor[rgb]{ .675,  .82,  .937} \underline{13.18}  & \cellcolor[rgb]{ .678,  .824,  .937} \underline{5.22}  & \cellcolor[rgb]{ .682,  .824,  .937} 13.10  & \cellcolor[rgb]{ .69,  .827,  .941} 12.40  & \cellcolor[rgb]{ .694,  .831,  .941} 8.17  & \cellcolor[rgb]{ .694,  .827,  .941} 12.59  & \cellcolor[rgb]{ .671,  .816,  .937} 16.70  & \cellcolor[rgb]{ .694,  .831,  .941} 28.63  \\
  \cmidrule{3-11}      &   & BPA+HarS & \cellcolor[rgb]{ .722,  .847,  .945} 12.66  & \cellcolor[rgb]{ .722,  .847,  .945} 5.04  & \cellcolor[rgb]{ .706,  .839,  .945} 12.81  & \cellcolor[rgb]{ .718,  .843,  .945} 12.08  & \cellcolor[rgb]{ .722,  .847,  .945} 7.97  & \cellcolor[rgb]{ .714,  .839,  .945} 12.36  & \cellcolor[rgb]{ .667,  .816,  .937} 16.73  & \cellcolor[rgb]{ .702,  .835,  .941} 28.51  \\
		&   & APA+HarS & \cellcolor[rgb]{ .682,  .824,  .937} 13.10  & \cellcolor[rgb]{ .682,  .824,  .937} 5.20  & \cellcolor[rgb]{ .675,  .82,  .937} \underline{13.19}  & \cellcolor[rgb]{ .69,  .827,  .941} 12.39  & \cellcolor[rgb]{ .694,  .831,  .941} 8.17  & \cellcolor[rgb]{ .686,  .824,  .941} \underline{12.67} & \cellcolor[rgb]{ .663,  .812,  .933} \textbf{16.83} & \cellcolor[rgb]{ .667,  .816,  .937} \underline{29.21}  \\
  \cmidrule{3-11}      &   & BPA+DiHarS & \cellcolor[rgb]{ .686,  .827,  .941} 13.07  & \cellcolor[rgb]{ .698,  .831,  .941} 5.14  & \cellcolor[rgb]{ .698,  .831,  .941} 12.92  & \cellcolor[rgb]{ .678,  .82,  .937} \underline{12.54}  & \cellcolor[rgb]{ .678,  .824,  .937} \underline{8.26}  & \cellcolor[rgb]{ .69,  .827,  .941} 12.59  & \cellcolor[rgb]{ .671,  .82,  .937} 16.62  & \cellcolor[rgb]{ .718,  .843,  .945} 28.19  \\
		&   & APA+DiHarS & \cellcolor[rgb]{ .663,  .812,  .933} \textbf{13.30} & \cellcolor[rgb]{ .663,  .812,  .933} \textbf{5.28} & \cellcolor[rgb]{ .663,  .812,  .933} \textbf{13.33} & \cellcolor[rgb]{ .663,  .812,  .933} \textbf{12.68} & \cellcolor[rgb]{ .663,  .812,  .933} \textbf{8.38} & \cellcolor[rgb]{ .663,  .812,  .933} \textbf{12.89} & \cellcolor[rgb]{ .667,  .816,  .937} 16.73  & \cellcolor[rgb]{ .663,  .812,  .933} \textbf{29.22} \\
	  \midrule
	  \multirow{23}[16]{*}{RecSys-2} & Item-based & itemKNN & \cellcolor[rgb]{ 1,  .976,  .894} 25.85  & \cellcolor[rgb]{ 1,  .98,  .914} 11.93  & \cellcolor[rgb]{ 1,  .98,  .902} 25.64  & \cellcolor[rgb]{ 1,  .973,  .867} 26.17  & \cellcolor[rgb]{ 1,  .976,  .878} 20.13  & \cellcolor[rgb]{ 1,  .976,  .882} 25.92  & \cellcolor[rgb]{ 1,  .965,  .831} 33.05  & \cellcolor[rgb]{ 1,  .976,  .882} 44.61  \\
  \cmidrule{2-11}      & \multirow{6}[2]{*}{MF-based} & BPR & \cellcolor[rgb]{ 1,  .973,  .875} 26.50  & \cellcolor[rgb]{ 1,  .98,  .898} 12.16  & \cellcolor[rgb]{ 1,  .973,  .878} 26.52  & \cellcolor[rgb]{ 1,  .965,  .839} 27.09  & \cellcolor[rgb]{ 1,  .969,  .851} 20.77  & \cellcolor[rgb]{ 1,  .969,  .855} 26.91  & \cellcolor[rgb]{ 1,  .957,  .804} 34.71  & \cellcolor[rgb]{ 1,  .969,  .851} 46.37  \\
		&   & GMF & \cellcolor[rgb]{ 1,  .98,  .906} 25.41  & \cellcolor[rgb]{ 1,  .976,  .882} 12.41  & \cellcolor[rgb]{ 1,  .98,  .91} 25.47  & \cellcolor[rgb]{ 1,  .98,  .91} 24.53  & \cellcolor[rgb]{ 1,  .976,  .886} 19.97  & \cellcolor[rgb]{ 1,  .98,  .91} 24.86  & \cellcolor[rgb]{ 1,  .969,  .851} 31.67  & \cellcolor[rgb]{ 1,  .988,  .941} 41.43  \\
		&   & MLP & \cellcolor[rgb]{ 1,  .965,  .843} 27.70  & \cellcolor[rgb]{ 1,  .969,  .855} 12.77  & \cellcolor[rgb]{ 1,  .965,  .843} 27.87  & \cellcolor[rgb]{ 1,  .965,  .831} 27.33  & \cellcolor[rgb]{ 1,  .965,  .843} 21.02  & \cellcolor[rgb]{ 1,  .965,  .835} 27.57  & \cellcolor[rgb]{ 1,  .957,  .796} 35.18  & \cellcolor[rgb]{ 1,  .965,  .831} 47.45  \\
		&   & NeuMF & \cellcolor[rgb]{ 1,  .965,  .835} 28.05  & \cellcolor[rgb]{ 1,  .969,  .847} 12.89  & \cellcolor[rgb]{ 1,  .965,  .835} 28.11  & \cellcolor[rgb]{ 1,  .961,  .816} 27.95  & \cellcolor[rgb]{ 1,  .961,  .827} 21.40  & \cellcolor[rgb]{ 1,  .961,  .824} 28.03  & \cellcolor[rgb]{ 1,  .957,  .792} 35.34  & \cellcolor[rgb]{ 1,  .961,  .824} 47.91  \\
		&   & M2F & \cellcolor[rgb]{ 1, 1, 1} 21.87  & \cellcolor[rgb]{ 1, 1, 1} 10.66  & \cellcolor[rgb]{ 1, 1, 1} 22.12  & \cellcolor[rgb]{ 1, 1, 1} 21.29  & \cellcolor[rgb]{ 1, 1, 1} 17.29  & \cellcolor[rgb]{ 1, 1, 1} 21.65  & \cellcolor[rgb]{ 1, 1, 1} 22.46  & \cellcolor[rgb]{ 1, 1, 1} 38.01  \\
		&   & MGMF & \cellcolor[rgb]{ 1,  .976,  .886} 26.05  & \cellcolor[rgb]{ 1,  .976,  .894} 12.22  & \cellcolor[rgb]{ 1,  .976,  .886} 26.27  & \cellcolor[rgb]{ 1,  .976,  .886} 25.40  & \cellcolor[rgb]{ 1,  .976,  .894} 19.83  & \cellcolor[rgb]{ 1,  .976,  .886} 25.76  & \cellcolor[rgb]{ 1,  .961,  .827} 33.27  & \cellcolor[rgb]{ 1,  .976,  .882} 44.64  \\
  \cmidrule{2-11}      & VAE-based & Multi-VAE & \cellcolor[rgb]{ 1,  .973,  .863} 26.98  & \cellcolor[rgb]{ 1,  .976,  .882} 12.38  & \cellcolor[rgb]{ 1,  .973,  .867} 26.95  & \cellcolor[rgb]{ 1,  .965,  .835} 27.18  & \cellcolor[rgb]{ 1,  .969,  .851} 20.81  & \cellcolor[rgb]{ 1,  .969,  .847} 27.10  & \cellcolor[rgb]{ 1,  .957,  .796} 35.01  & \cellcolor[rgb]{ 1,  .965,  .843} 46.75  \\
  \cmidrule{2-11}      & GNN-based & LightGCN & \cellcolor[rgb]{ 1,  .961,  .824} 28.49  & \cellcolor[rgb]{ 1,  .965,  .827} 13.17  & \cellcolor[rgb]{ 1,  .961,  .82} 28.64  & \cellcolor[rgb]{ 1,  .961,  .812} 28.14  & \cellcolor[rgb]{ 1,  .961,  .816} 21.66  & \cellcolor[rgb]{ 1,  .961,  .812} 28.36  & \cellcolor[rgb]{ 1,  .953,  .788} 35.50  & \cellcolor[rgb]{ 1,  .961,  .812} 48.45  \\
  \cmidrule{2-11}      & \multirow{8}[2]{*}{CML-based} & UniS & \cellcolor[rgb]{ 1,  .961,  .82} 28.51  & \cellcolor[rgb]{ 1,  .965,  .831} 13.13  & \cellcolor[rgb]{ 1,  .961,  .824} 28.55  & \cellcolor[rgb]{ 1,  .961,  .812} 28.12  & \cellcolor[rgb]{ 1,  .961,  .82} 21.57  & \cellcolor[rgb]{ 1,  .961,  .816} 28.27  & \cellcolor[rgb]{ 1,  .953,  .788} 35.58  & \cellcolor[rgb]{ 1,  .961,  .812} 48.57  \\
		&   & PopS & \cellcolor[rgb]{ 1,  .988,  .933} 24.39  & \cellcolor[rgb]{ 1,  .996,  .969} 11.16  & \cellcolor[rgb]{ 1,  .984,  .929} 24.77  & \cellcolor[rgb]{ 1,  .992,  .953} 23.00  & \cellcolor[rgb]{ 1,  1,  .992} 17.49  & \cellcolor[rgb]{ 1,  .988,  .945} 23.71  & \cellcolor[rgb]{ 1,  1,  .992} 22.97  & \cellcolor[rgb]{ 1,  .98,  .902} 43.53  \\
		&   & 2st & \cellcolor[rgb]{ 1,  .961,  .824} 28.36  & \cellcolor[rgb]{ 1,  .965,  .831} 13.11  & \cellcolor[rgb]{ 1,  .965,  .827} 28.37  & \cellcolor[rgb]{ 1,  .961,  .816} 27.98  & \cellcolor[rgb]{ 1,  .961,  .82} 21.53  & \cellcolor[rgb]{ 1,  .961,  .82} 28.11  & \cellcolor[rgb]{ 1,  .984,  .925} 27.11  & \cellcolor[rgb]{ 1,  .961,  .824} 47.90  \\
		&   & HarS & \cellcolor[rgb]{ 1,  .969,  .851} 27.48  & \cellcolor[rgb]{ 1,  .973,  .867} 12.61  & \cellcolor[rgb]{ 1,  .969,  .851} 27.55  & \cellcolor[rgb]{ 1,  .965,  .831} 27.36  & \cellcolor[rgb]{ 1,  .969,  .847} 20.93  & \cellcolor[rgb]{ 1,  .965,  .839} 27.45  & \cellcolor[rgb]{ 1,  .957,  .796} 34.99  & \cellcolor[rgb]{ 1,  .965,  .831} 47.55  \\
		&   & LRML & \cellcolor[rgb]{ 1,  .984,  .922} 24.88  & \cellcolor[rgb]{ 1,  .98,  .898} 12.17  & \cellcolor[rgb]{ 1,  .984,  .922} 25.00  & \cellcolor[rgb]{ 1,  .984,  .918} 24.26  & \cellcolor[rgb]{ 1,  .976,  .898} 19.75  & \cellcolor[rgb]{ 1,  .984,  .922} 24.54  & \cellcolor[rgb]{ 1,  .969,  .855} 31.53  & \cellcolor[rgb]{ 1,  .988,  .949} 41.01  \\
		&   & TransCF & \cellcolor[rgb]{ 1,  .969,  .855} 27.23  & \cellcolor[rgb]{ 1,  .973,  .875} 12.48  & \cellcolor[rgb]{ 1,  .969,  .859} 27.23  & \cellcolor[rgb]{ 1,  .965,  .839} 27.16  & \cellcolor[rgb]{ 1,  .969,  .855} 20.76  & \cellcolor[rgb]{ 1,  .969,  .847} 27.18  & \cellcolor[rgb]{ 1,  .957,  .8} 34.76  & \cellcolor[rgb]{ 1,  .965,  .843} 46.77  \\
		&   & AdaCML & \cellcolor[rgb]{ 1,  .973,  .871} 26.76  & \cellcolor[rgb]{ 1,  .976,  .89} 12.27  & \cellcolor[rgb]{ 1,  .973,  .871} 26.82  & \cellcolor[rgb]{ 1,  .969,  .851} 26.72  & \cellcolor[rgb]{ 1,  .973,  .867} 20.43  & \cellcolor[rgb]{ 1,  .969,  .859} 26.78  & \cellcolor[rgb]{ 1,  .957,  .804} 34.65  & \cellcolor[rgb]{ 1,  .965,  .843} 46.85  \\
		&   & HLR & \cellcolor[rgb]{ 1,  .965,  .843} 27.74  & \cellcolor[rgb]{ 1,  .969,  .851} 12.83  & \cellcolor[rgb]{ 1,  .965,  .835} 28.09  & \cellcolor[rgb]{ 1,  .965,  .843} 26.99  & \cellcolor[rgb]{ 1,  .969,  .851} 20.78  & \cellcolor[rgb]{ 1,  .965,  .839} 27.47  & \cellcolor[rgb]{ 1,  .965,  .827} 33.07  & \cellcolor[rgb]{ 1,  .961,  .82} 48.08  \\
  \cmidrule{2-11}      & \multirow{6}[6]{*}{DPCML-based (Ours)} & BPA+UniS & \cellcolor[rgb]{ 1,  .961,  .812} 28.81  & \cellcolor[rgb]{ 1,  .961,  .82} 13.28  & \cellcolor[rgb]{ 1,  .961,  .812} 28.98  & \cellcolor[rgb]{ 1,  .957,  .804} 28.33  & \cellcolor[rgb]{ 1,  .961,  .812} 21.73  & \cellcolor[rgb]{ 1,  .957,  .804} 28.60  & \cellcolor[rgb]{ 1,  .953,  .784} 35.76  & \cellcolor[rgb]{ 1,  .957,  .8} 49.26  \\
		&   & APA+UniS & \cellcolor[rgb]{ 1,  .961,  .812} 28.85  & \cellcolor[rgb]{ 1,  .961,  .824} 13.27  & \cellcolor[rgb]{ 1,  .961,  .816} 28.85  & \cellcolor[rgb]{ 1,  .957,  .8} 28.54  & \cellcolor[rgb]{ 1,  .957,  .808} 21.86  & \cellcolor[rgb]{ 1,  .957,  .804} 28.64  & \cellcolor[rgb]{ 1,  .953,  .784} 35.82  & \cellcolor[rgb]{ 1,  .957,  .808} 48.88  \\
  \cmidrule{3-11}      &   & BPA+HarS & \cellcolor[rgb]{ 1,  .961,  .824} 28.46  & \cellcolor[rgb]{ 1,  .965,  .835} 13.07  & \cellcolor[rgb]{ 1,  .961,  .824} 28.56  & \cellcolor[rgb]{ 1,  .961,  .816} 28.02  & \cellcolor[rgb]{ 1,  .961,  .824} 21.47  & \cellcolor[rgb]{ 1,  .961,  .816} 28.22  & \cellcolor[rgb]{ 1,  .953,  .788} 35.58  & \cellcolor[rgb]{ 1,  .961,  .812} 48.57  \\
		&   & APA+HarS & \cellcolor[rgb]{ 1,  .961,  .816} 28.75  & \cellcolor[rgb]{ 1,  .961,  .827} 13.20  & \cellcolor[rgb]{ 1,  .961,  .816} 28.79  & \cellcolor[rgb]{ 1,  .957,  .804} 28.39  & \cellcolor[rgb]{ 1,  .961,  .812} 21.76  & \cellcolor[rgb]{ 1,  .957,  .808} 28.54  & \cellcolor[rgb]{ 1,  .953,  .784} 35.74  & \cellcolor[rgb]{ 1,  .957,  .808} 48.78  \\
  \cmidrule{3-11}      &   & BPA+DiHarS & \cellcolor[rgb]{ 1,  .953,  .784} \underline{29.91}  & \cellcolor[rgb]{ 1,  .953,  .788} \underline{13.77}  & \cellcolor[rgb]{ 1,  .953,  .776} \underline{30.24}  & \cellcolor[rgb]{ 1,  .953,  .78} \underline{29.22}  & \cellcolor[rgb]{ 1,  .953,  .78} \underline{22.42}  & \cellcolor[rgb]{ 1,  .953,  .776} \underline{29.68}  & \cellcolor[rgb]{ 1,  .949,  .773} \textbf{36.42} & \cellcolor[rgb]{ 1,  .949,  .773} \textbf{50.61} \\
		&   & APA+DiHarS & \cellcolor[rgb]{ 1,  .949,  .773} \textbf{30.23} & \cellcolor[rgb]{ 1,  .949,  .773} \textbf{13.95} & \cellcolor[rgb]{ 1,  .949,  .773} \textbf{30.27} & \cellcolor[rgb]{ 1,  .949,  .773} \textbf{29.42} & \cellcolor[rgb]{ 1,  .949,  .773} \textbf{22.59} & \cellcolor[rgb]{ 1,  .949,  .773} \textbf{29.71} & \cellcolor[rgb]{ 1,  .953,  .776} \textbf{36.42} & \cellcolor[rgb]{ 1,  .953,  .784} \underline{50.10}  \\
	  \bottomrule
	  \end{tabular}%
	\label{result3}%
  \end{table}%

	\begin{figure*}[!t]
		\centering
		\subfigure[Drama]{
			\includegraphics[width=0.231\textwidth]{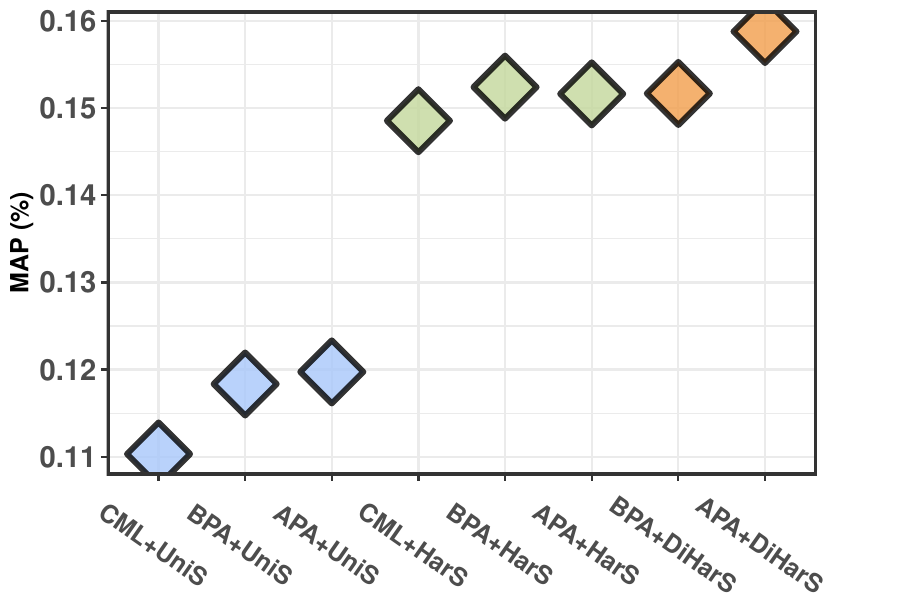}
		}
		\subfigure[Comedy]{
			\includegraphics[width=0.231\textwidth]{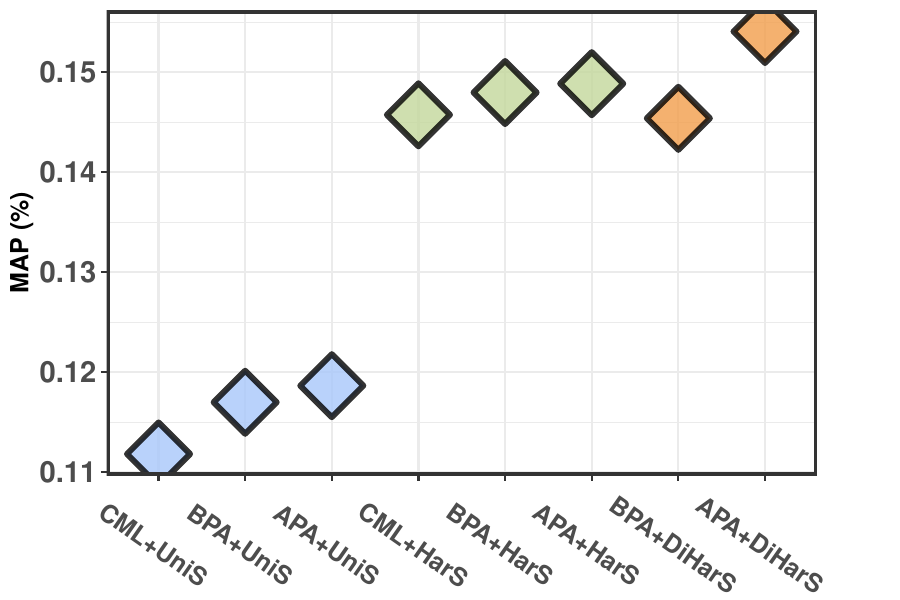}
		}
		\subfigure[Action]{
			\includegraphics[width=0.231\textwidth]{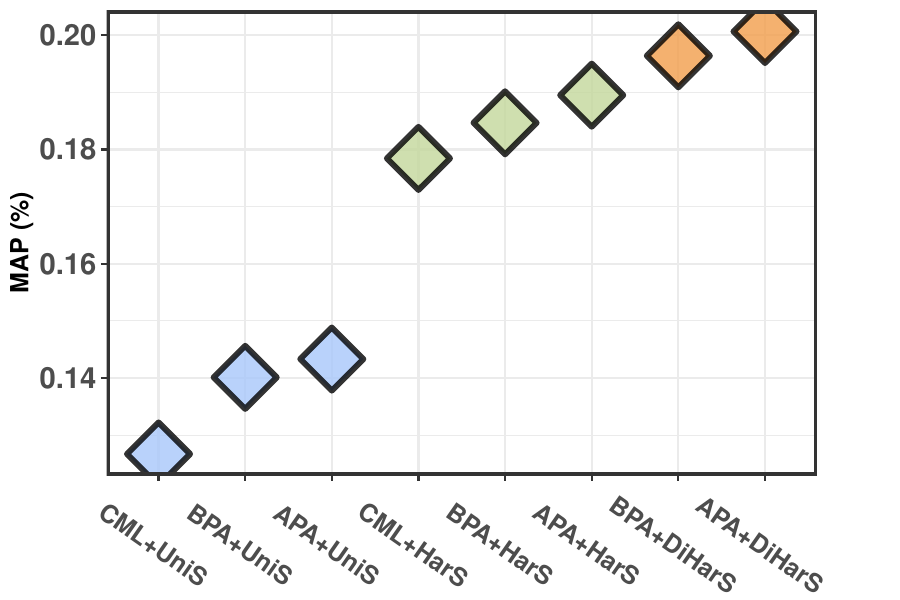}
		}
		\subfigure[Thriller]{
			\includegraphics[width=0.231\textwidth]{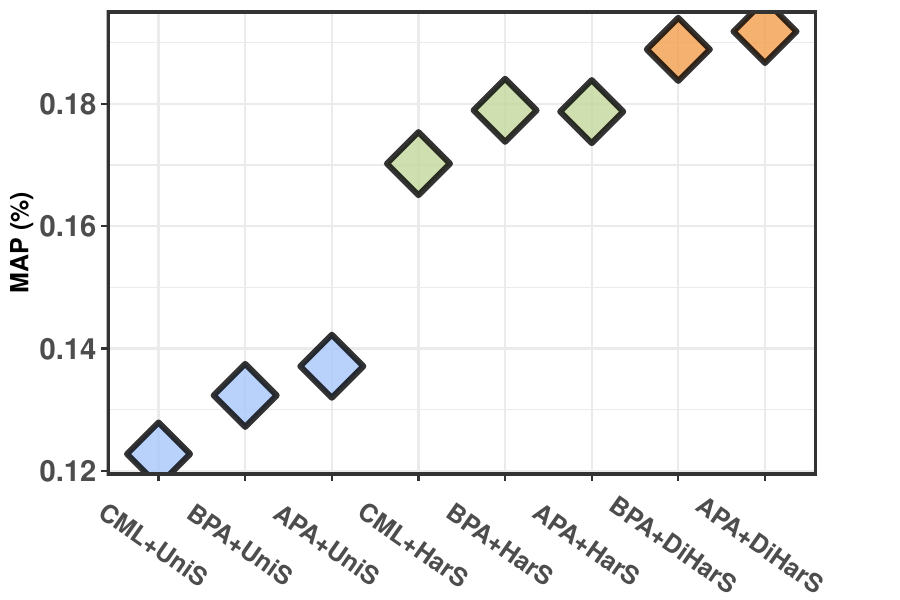}
		}
		\subfigure[Adventure]{
			\includegraphics[width=0.231\textwidth]{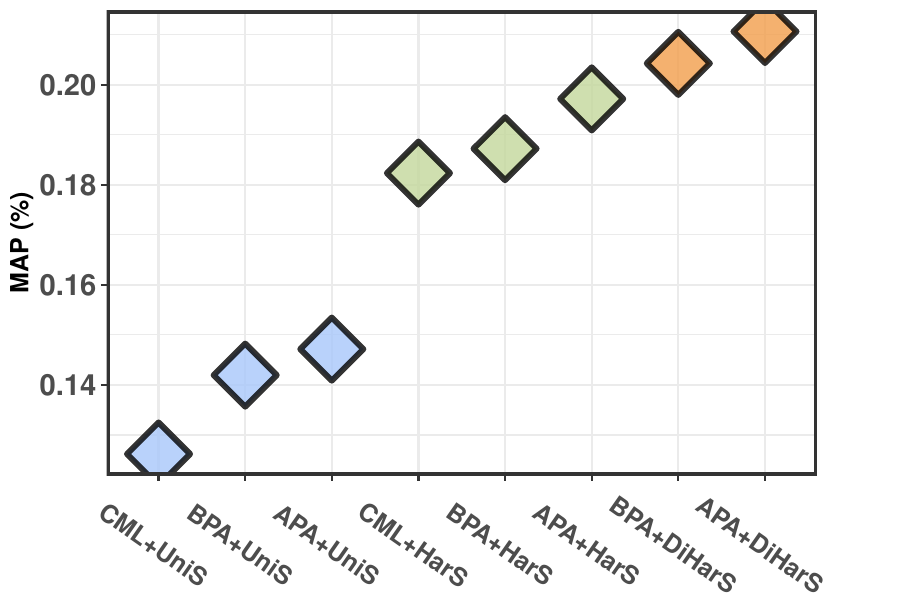}
		}
		\subfigure[Romance]{
			\includegraphics[width=0.231\textwidth]{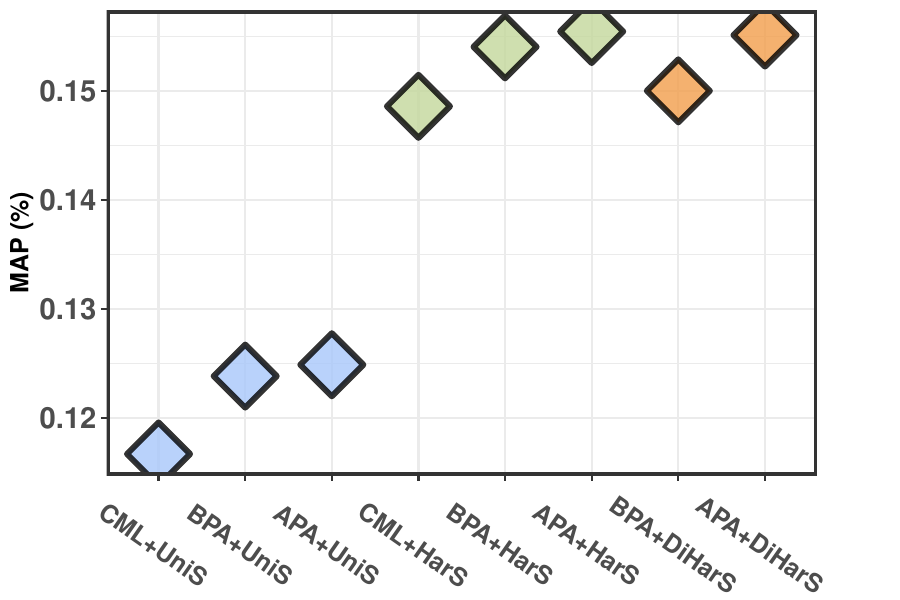}
		}
		\subfigure[Crime]{
			\includegraphics[width=0.231\textwidth]{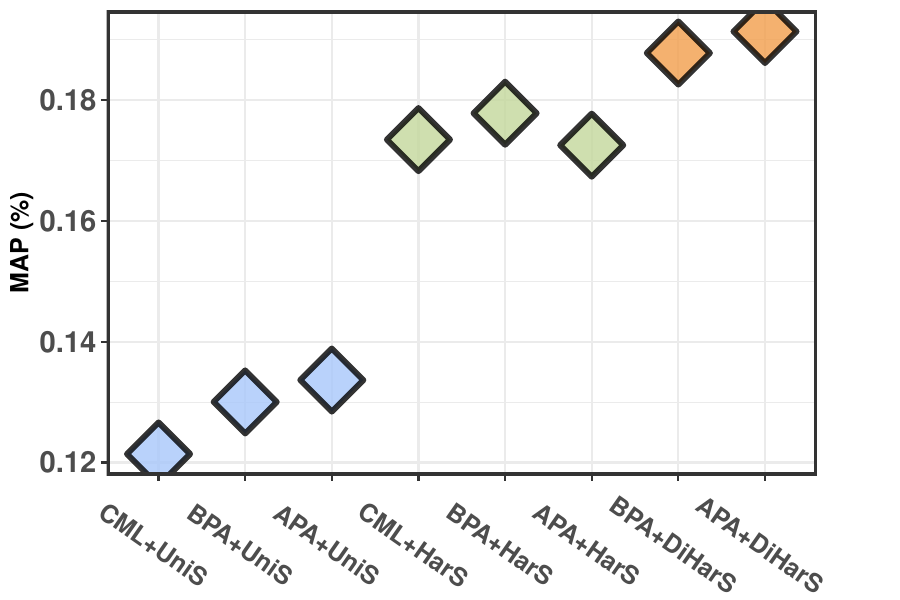}
		}
		\subfigure[Sci-Fi]{
			\includegraphics[width=0.231\textwidth]{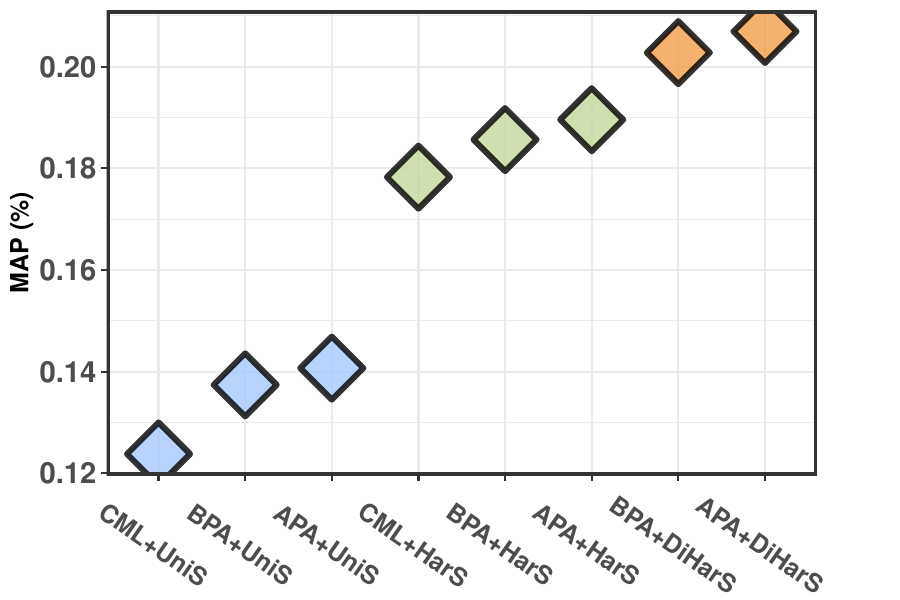}
		}
		\subfigure[Fantasy]{
			\includegraphics[width=0.231\textwidth]{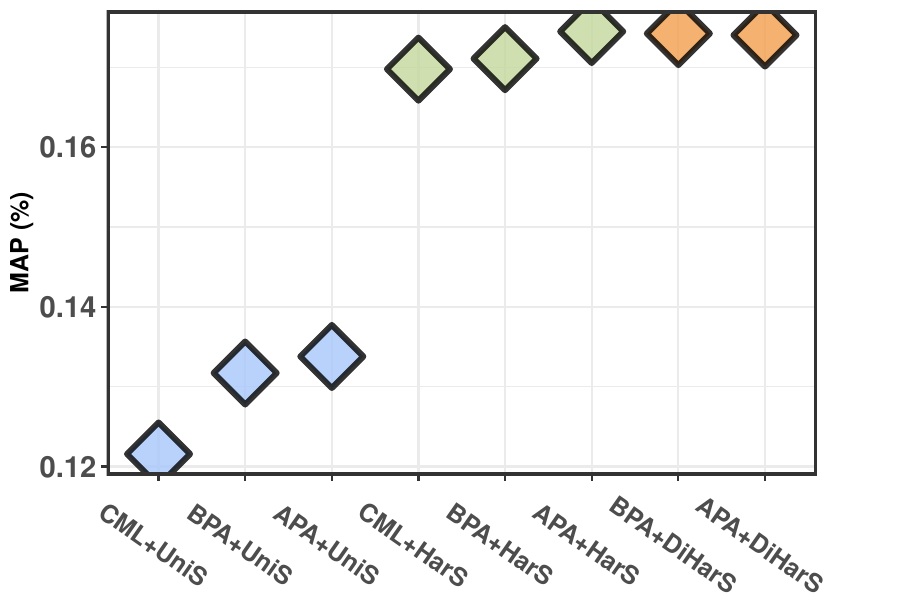}
		}
		\subfigure[Children]{
			\includegraphics[width=0.231\textwidth]{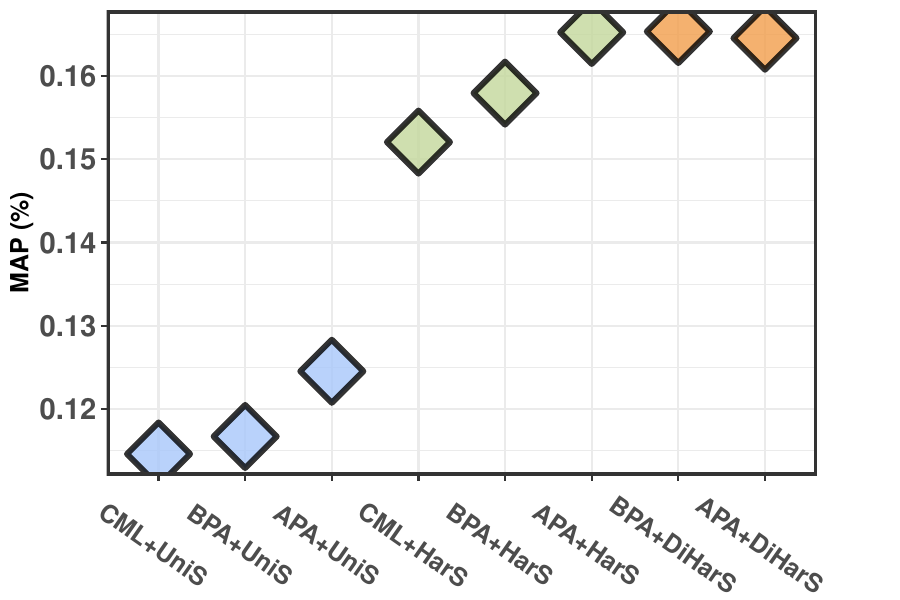}
		}
		\subfigure[Mystery]{
			\includegraphics[width=0.231\textwidth]{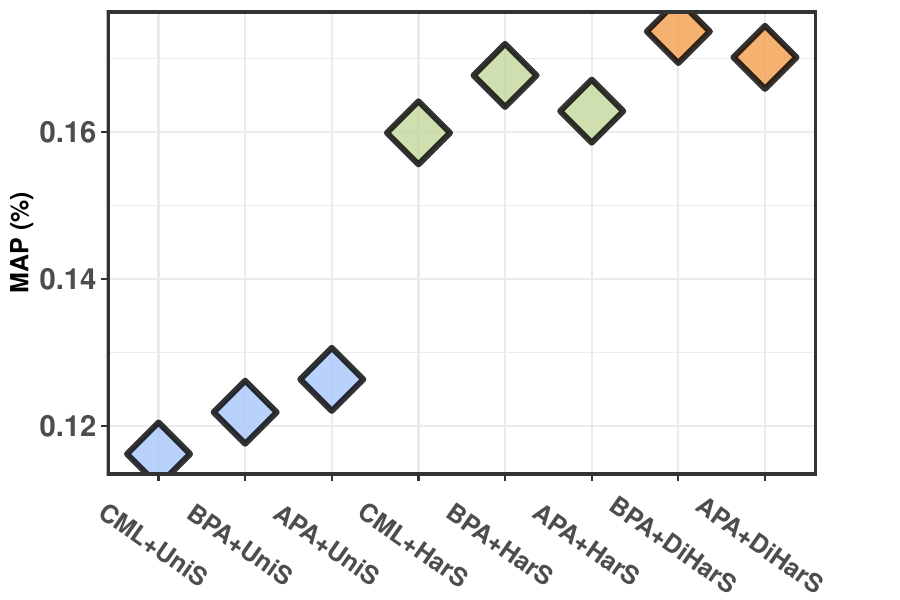}
		}
		\subfigure[War]{
			\includegraphics[width=0.231\textwidth]{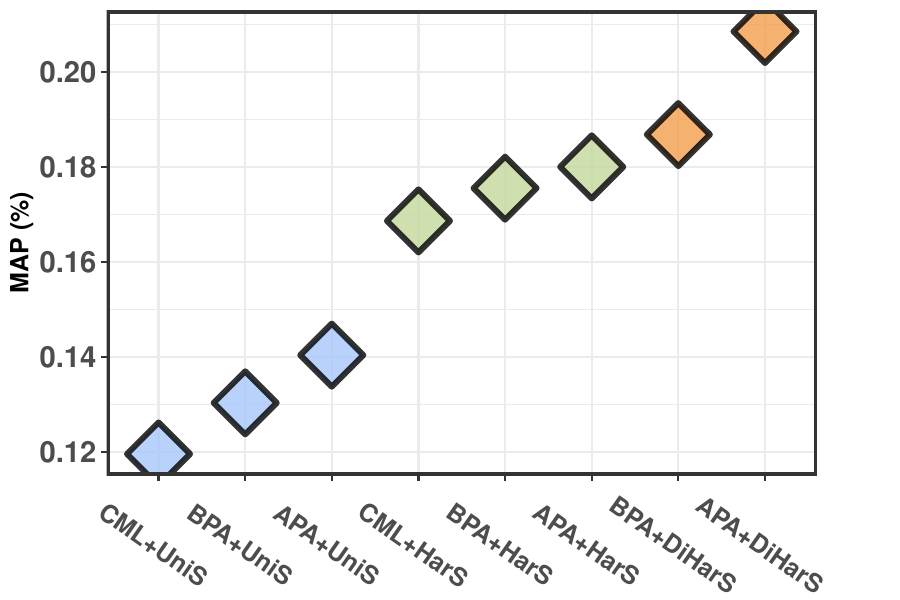}
		}
		\subfigure[Horror]{
			\includegraphics[width=0.231\textwidth]{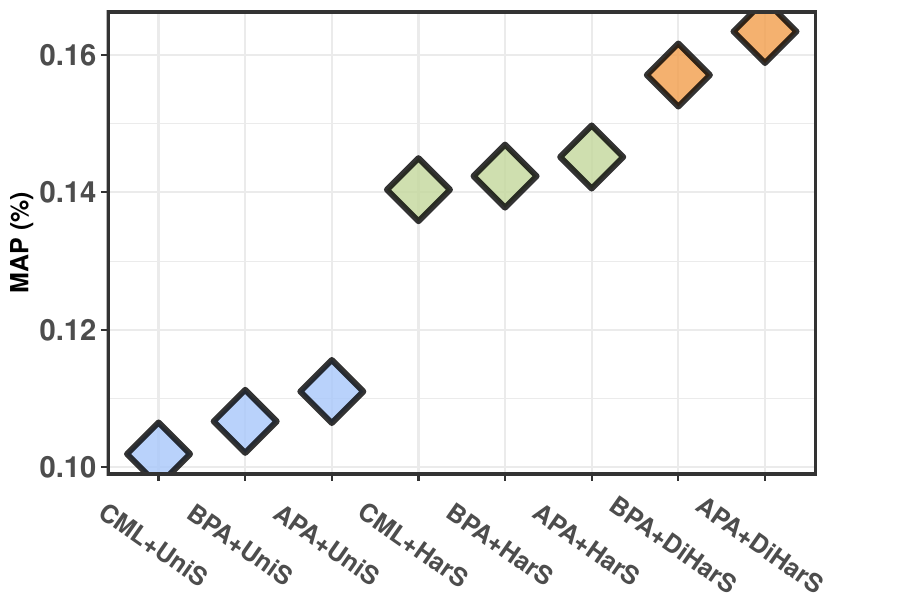}
		}
		\subfigure[Animation]{
			\includegraphics[width=0.231\textwidth]{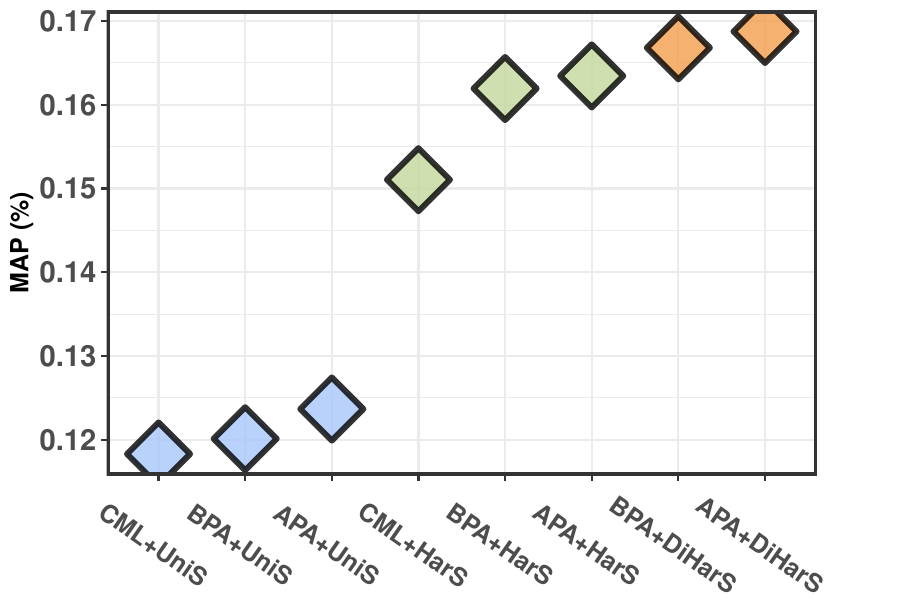}
		}
		\subfigure[Musical]{
			\includegraphics[width=0.231\textwidth]{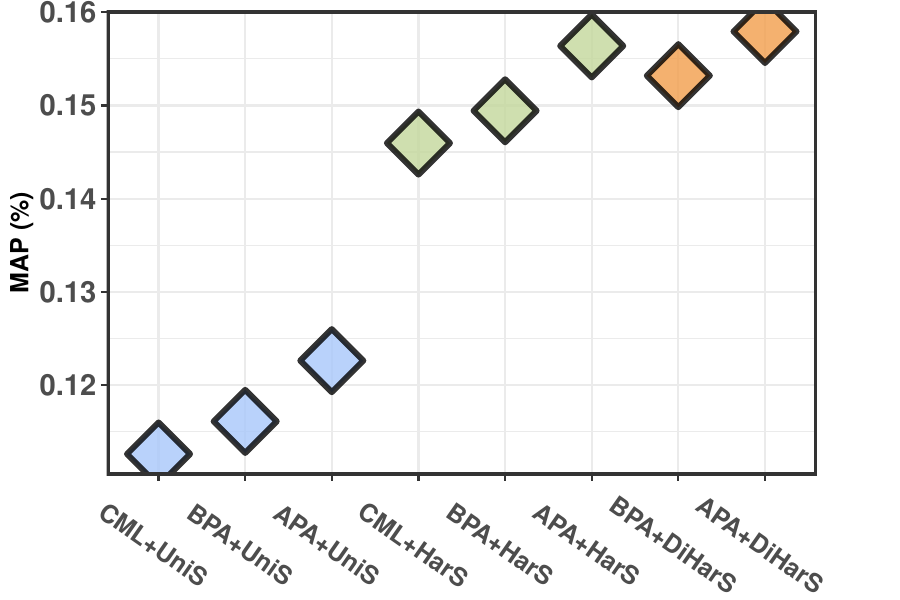}
		}
		\subfigure[Western]{
			\includegraphics[width=0.231\textwidth]{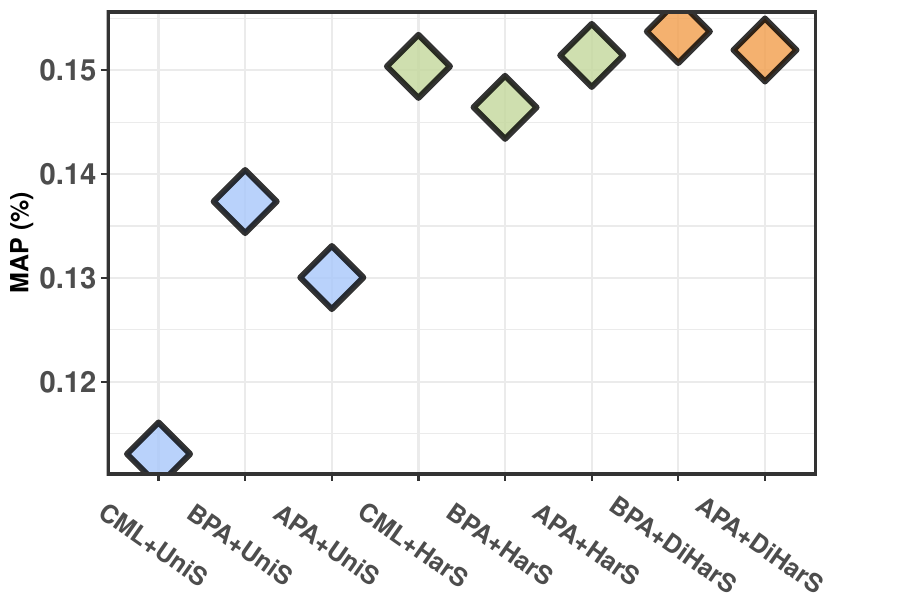}
		}
		\subfigure[Film-Noir]{
			\includegraphics[width=0.231\textwidth]{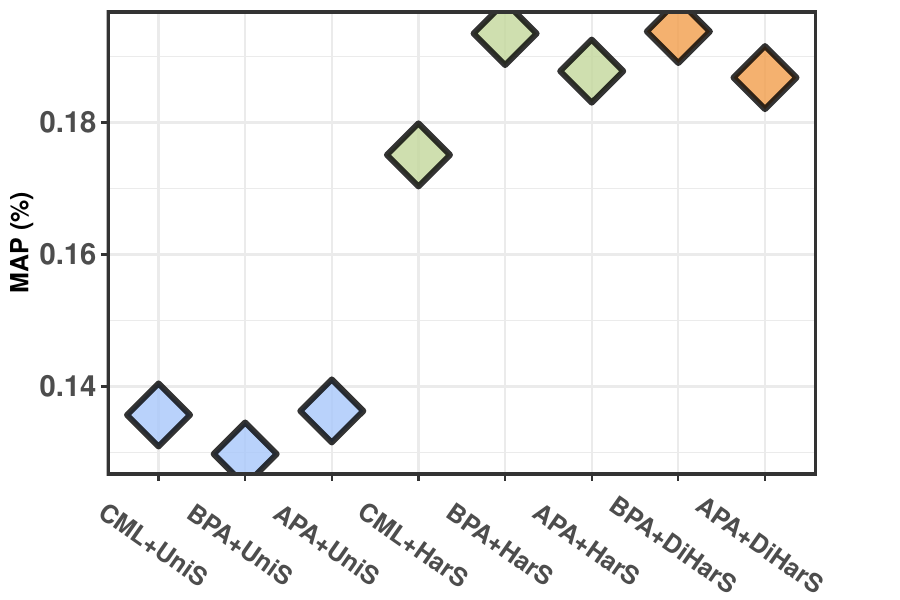}
		}
		\subfigure[Documentary]{
			\includegraphics[width=0.231\textwidth]{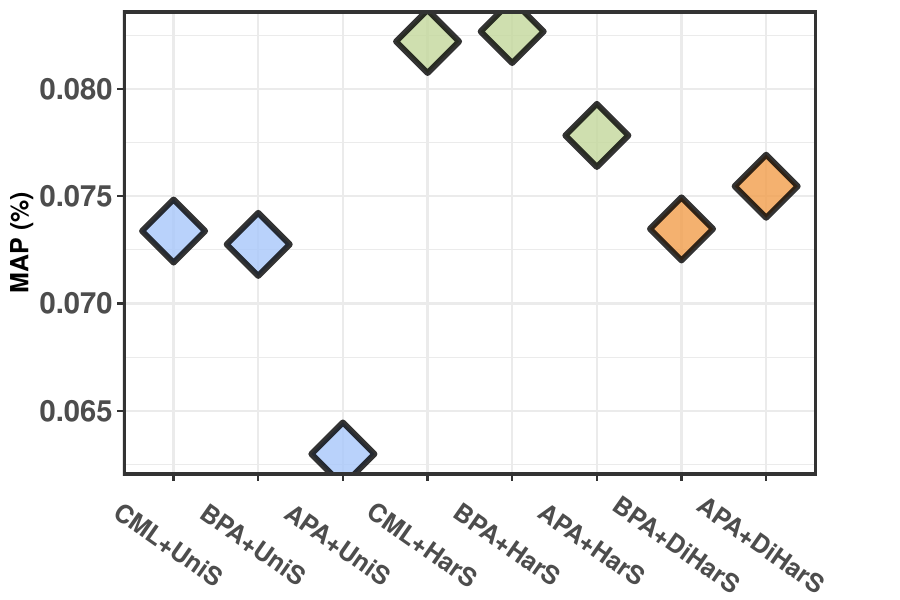}
		}
		\subfigure[IMAX]{
			\includegraphics[width=0.231\textwidth]{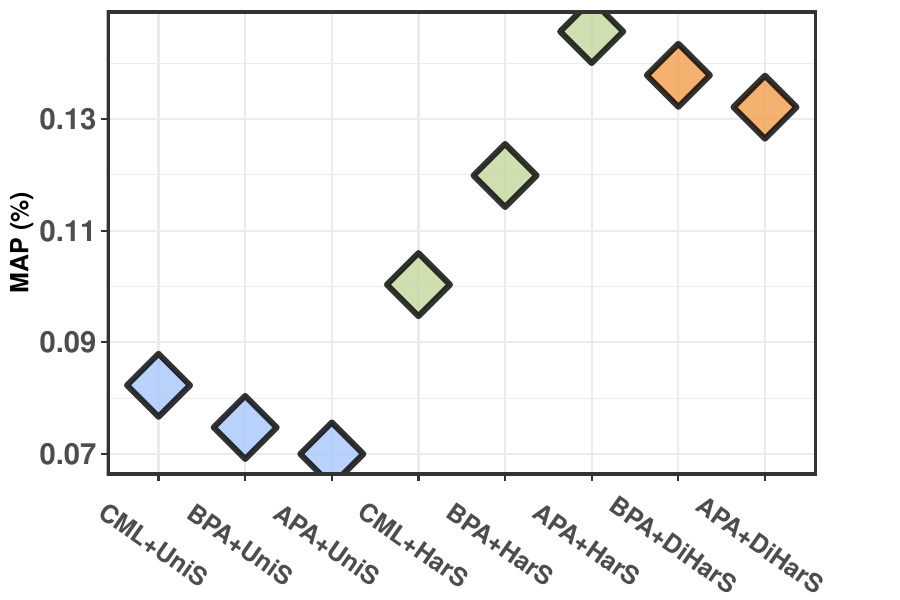}
		}
		\caption{Fine-grained performance over each interest group on MovieLens-10m dataset. }
		\label{supp:per_arrtribute_performance}
	\end{figure*}
	
	
	\begin{figure}[!t]
		\centering
		\subfigure[P@$3$ on Steam-200k]{
			\includegraphics[width=0.231\textwidth]{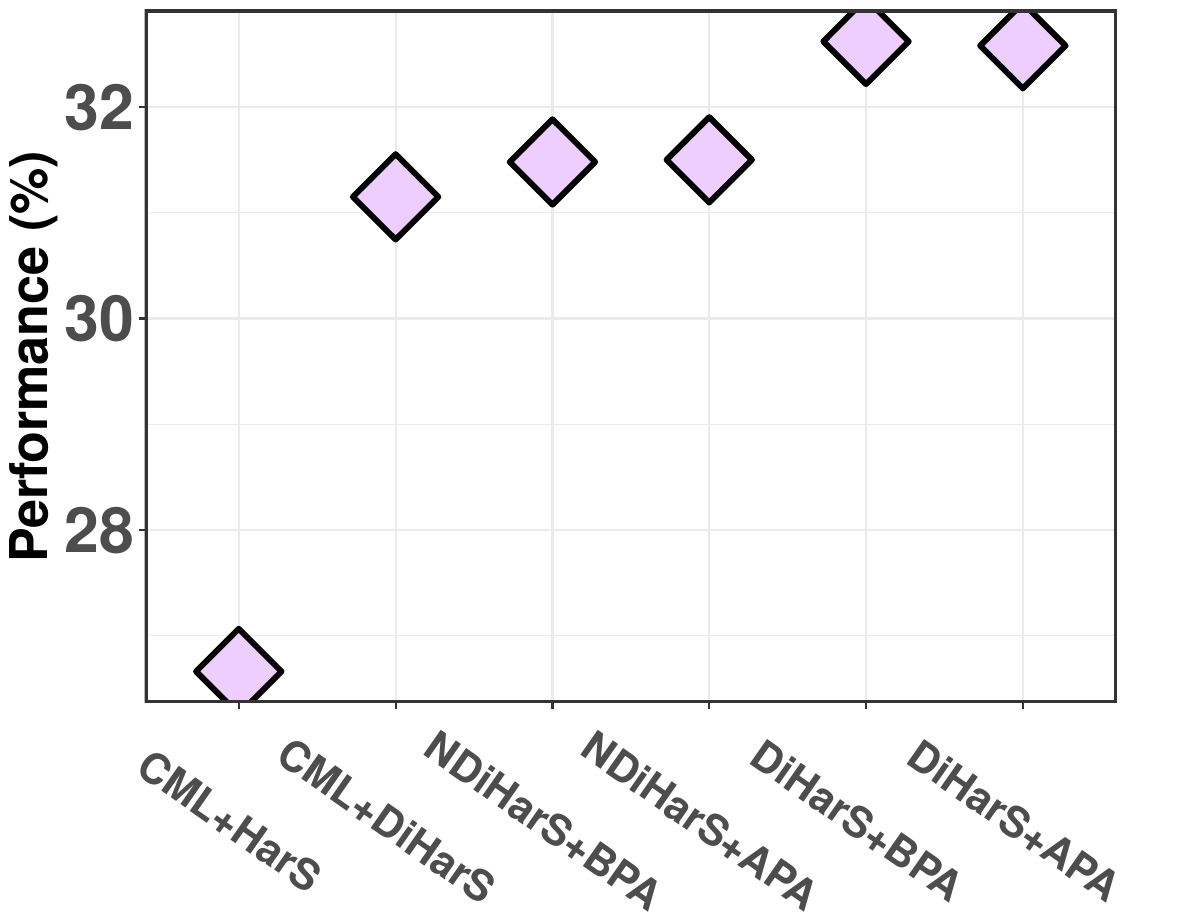}
		}
		\subfigure[MRR@$3$ on Steam-200k]{
			\includegraphics[width=0.231\textwidth]{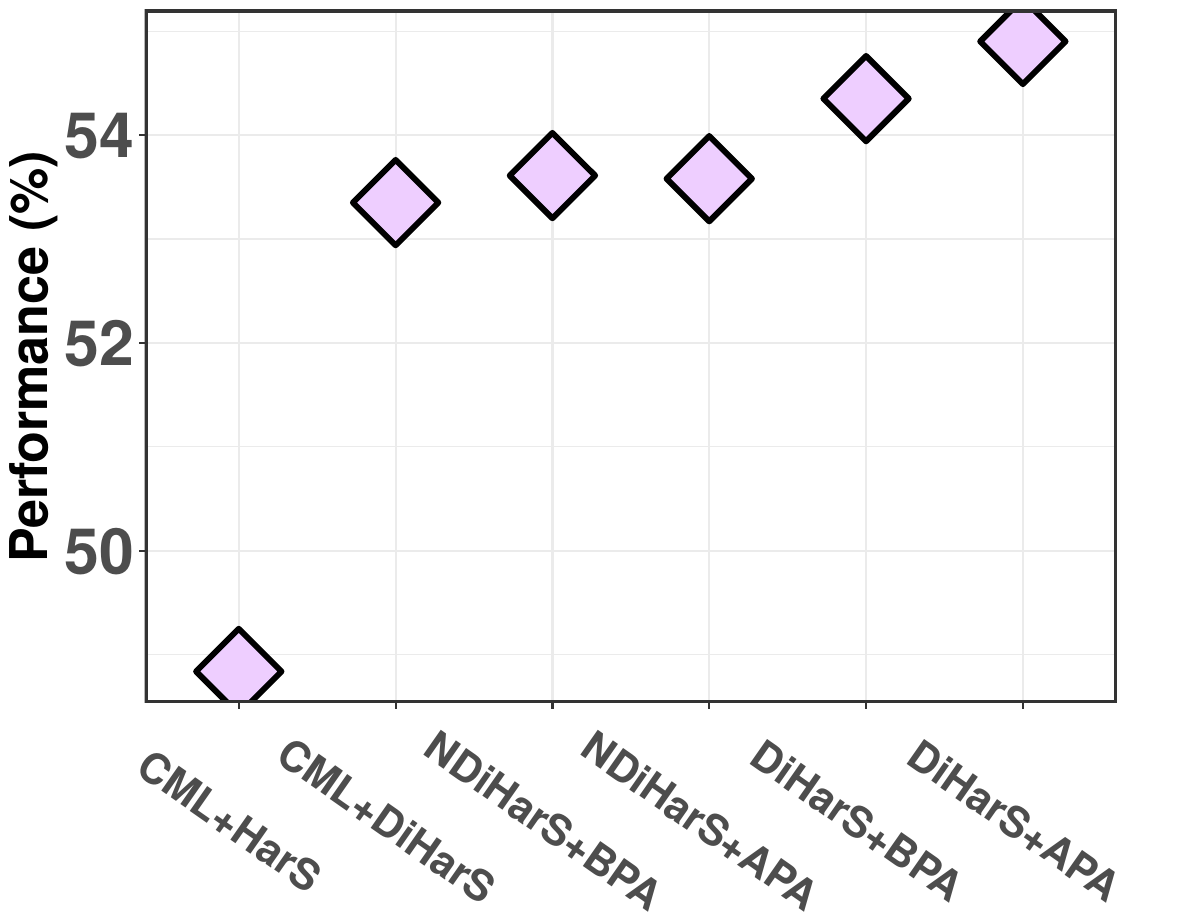}
		}
		\subfigure[P@$3$ on CiteULike]{
			\includegraphics[width=0.231\textwidth]{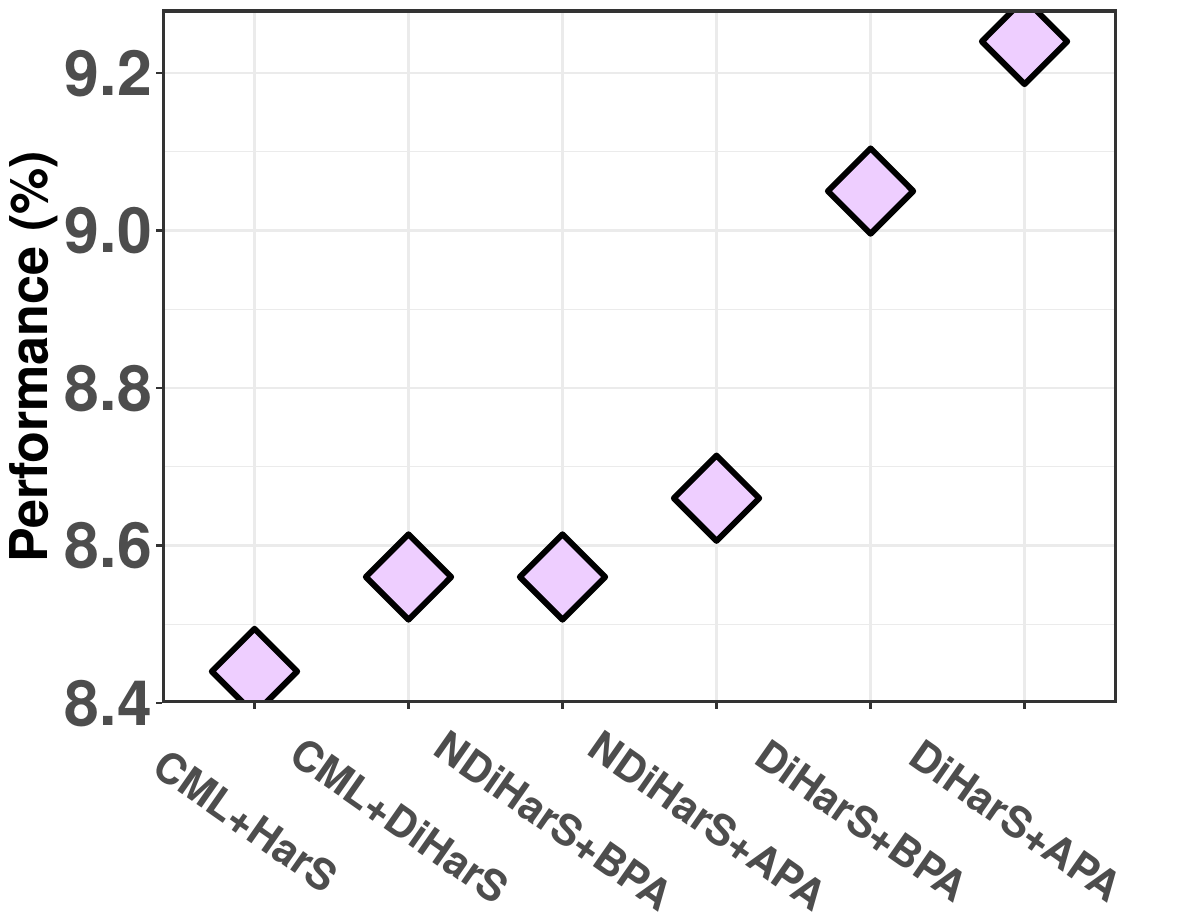}
		}
		\subfigure[MRR@$3$ on CiteULike]{
			\includegraphics[width=0.231\textwidth]{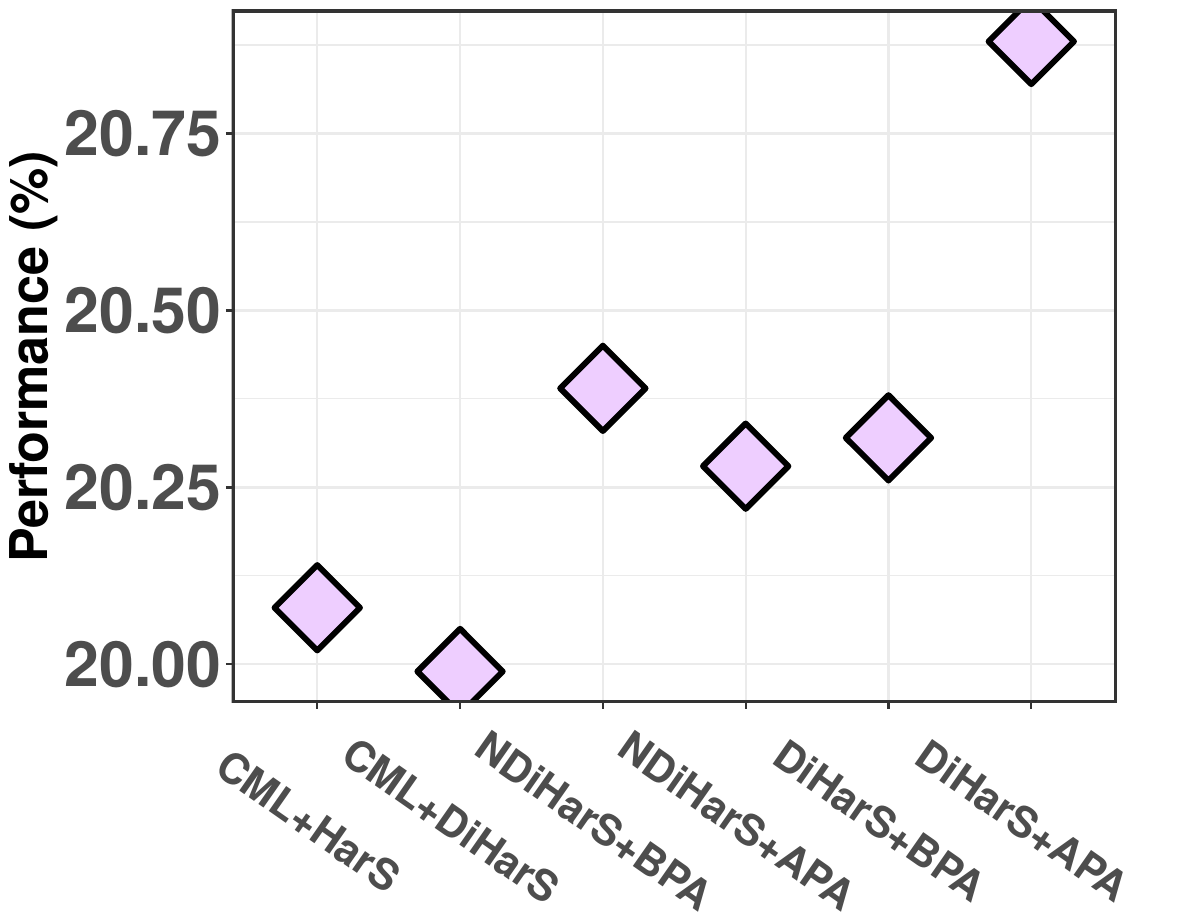}
		}
		\caption{Ablation Performance of DiHarS strategy for CML and DPCML algorithms on Steam-200k and CiteULike datasets.}
		\label{fig:supp_ab_for_DiHarS}
	\end{figure}

\begin{table}[htbp]
	\centering
	\caption{The $MaxDiv@N$ performance comparison of CML-based algorithms on Steam-200k and MovieLens-1M datasets. Here \textbf{a higher value implies more diverse} recommendation results.}
	  \begin{tabular}{c|c|cccc}
	  \toprule
	  \multicolumn{6}{c}{Steam-200k} \\
	  \midrule
	  Type & Method & MaxDiv@3 & MaxDiv@5 & MaxDiv@10 & MaxDiv@20 \\
	  \midrule
	  \multirow{6}[2]{*}{UniS} & CML & \cellcolor[rgb]{1.0,  1.0,  1.0} 1.354  & \cellcolor[rgb]{1.0,  1.0,  1.0} 4.750 & \cellcolor[rgb]{1.0,  1.0,  1.0} 23.520 & \cellcolor[rgb]{1.0,  1.0,  1.0} 117.927  \\
		& CML+DD & \cellcolor[rgb]{ .996,  .984,  .984} 1.719  & \cellcolor[rgb]{1.0,  1.0,  1.0} 5.844  & \cellcolor[rgb]{1.0,  1.0,  1.0} 29.392  & \cellcolor[rgb]{ 1,  .988,  .988} 144.962  \\
		& DPCML+BPA & \cellcolor[rgb]{ .996,  .976,  .98} 1.822  & \cellcolor[rgb]{ 1,  .988,  .988} 6.713  & \cellcolor[rgb]{ 1,  .988,  .988} 34.727  & \cellcolor[rgb]{ .996,  .973,  .973} 179.065  \\
		& DPCML+APA & \cellcolor[rgb]{ .996,  .98,  .98} 1.791  & \cellcolor[rgb]{ 1,  .988,  .992} 6.672  & \cellcolor[rgb]{ .996,  .984,  .984} 35.871  & \cellcolor[rgb]{ .992,  .965,  .969} 189.182  \\
		& DPCML+BPA w/o DCRS  & \cellcolor[rgb]{ 1,  .988,  .988} 1.643  & \cellcolor[rgb]{1.0,  1.0,  1.0} 5.857  & \cellcolor[rgb]{1.0,  1.0,  1.0} 30.425  & \cellcolor[rgb]{ .996,  .984,  .984} 155.193  \\
		& DPCML+APA w/o DCRS & \cellcolor[rgb]{ 1,  .988,  .988} 1.602  & \cellcolor[rgb]{1.0,  1.0,  1.0} 5.814  & \cellcolor[rgb]{1.0,  1.0,  1.0} 29.881  & \cellcolor[rgb]{ .996,  .984,  .984} 151.516  \\
	  \midrule
	  \multirow{6}[2]{*}{HarS} & CML & \cellcolor[rgb]{ .996,  .98,  .98} 1.752  & \cellcolor[rgb]{ 1,  .988,  .988} 6.809  & \cellcolor[rgb]{ .996,  .973,  .976} 40.378  & \cellcolor[rgb]{ .988,  .941,  .945} 236.794  \\
		& CML+DD & \cellcolor[rgb]{ .992,  .957,  .961} 2.214  & \cellcolor[rgb]{ .992,  .969,  .973} 8.089  & \cellcolor[rgb]{ .988,  .953,  .957} 48.142  & \cellcolor[rgb]{ .98,  .914,  .918} 294.911  \\
		& DPCML+BPA & \cellcolor[rgb]{ .98,  .918,  .922} 2.977  & \cellcolor[rgb]{ .98,  .922,  .925} 11.472  & \cellcolor[rgb]{ .98,  .906,  .91} 65.952  & \cellcolor[rgb]{ .973,  .875,  .882} 369.876  \\
		& DPCML+APA & \cellcolor[rgb]{ .984,  .933,  .937} 2.661  & \cellcolor[rgb]{ .984,  .937,  .941} 10.314  & \cellcolor[rgb]{ .984,  .925,  .929} 59.003  & \cellcolor[rgb]{ .976,  .894,  .902} 329.847  \\
		& DPCML+BPA w/o DCRS  & \cellcolor[rgb]{ .98,  .918,  .922} 2.958  & \cellcolor[rgb]{ .984,  .922,  .925} 11.398  & \cellcolor[rgb]{ .98,  .91,  .914} 65.398  & \cellcolor[rgb]{ .973,  .878,  .882} 365.458  \\
		& DPCML+APA w/o DCRS & \cellcolor[rgb]{ .98,  .914,  .918} 3.008  & \cellcolor[rgb]{ .984,  .925,  .929} 11.250  & \cellcolor[rgb]{ .98,  .91,  .914} 64.940  & \cellcolor[rgb]{ .973,  .878,  .882} 364.131  \\
	  \midrule
	  \multirow{4}[2]{*}{DiHarS} & DPCML+BPA & \cellcolor[rgb]{ .941,  .761,  .773} 5.898  & \cellcolor[rgb]{ .941,  .761,  .773} 22.593  & \cellcolor[rgb]{ .941,  .761,  .773} 121.477  & \cellcolor[rgb]{ .941,  .761,  .773} 592.671  \\
		& DPCML+APA & \cellcolor[rgb]{ .957,  .812,  .824} 4.935  & \cellcolor[rgb]{ .953,  .808,  .816} 19.417  & \cellcolor[rgb]{ .949,  .796,  .804} 109.376  & \cellcolor[rgb]{ .949,  .78,  .792} 554.719  \\
		& DPCML+BPA w/o DCRS  & \cellcolor[rgb]{ .945,  .769,  .78} 5.779  & \cellcolor[rgb]{ .945,  .769,  .78} 22.084  & \cellcolor[rgb]{ .945,  .769,  .78} 118.889  & \cellcolor[rgb]{ .945,  .765,  .776} 585.161  \\
		& DPCML+APA w/o DCRS & \cellcolor[rgb]{ .957,  .816,  .827} 4.856  & \cellcolor[rgb]{ .957,  .816,  .827} 18.815  & \cellcolor[rgb]{ .953,  .8,  .808} 107.190  & \cellcolor[rgb]{ .949,  .784,  .796} 548.343  \\
	  \midrule
	  \multicolumn{6}{c}{MovieLens-1M} \\
	  \midrule
	  Type & Method & MaxDiv@3 & MaxDiv@5 & MaxDiv@10 & MaxDiv@20 \\
	  \midrule
	  \multirow{6}[2]{*}{UniS} & CML & \cellcolor[rgb]{ 1,  .988,  .984} 1.739  & \cellcolor[rgb]{ 1,  .996,  .988} 6.142  & \cellcolor[rgb]{ 1,  1,  .996} 30.127  & \cellcolor[rgb]{1.0,  1.0,  1.0} 140.095  \\
		& CML+DD & \cellcolor[rgb]{ 1,  .976,  .961} 1.864  & \cellcolor[rgb]{ 1,  .988,  .976} 6.444  & \cellcolor[rgb]{ 1,  .996,  .988} 31.080  & \cellcolor[rgb]{ 1,  1,  .996} 143.439  \\
		& DPCML+BPA & \cellcolor[rgb]{ 1,  .988,  .976} 1.775  & \cellcolor[rgb]{ 1,  .992,  .984} 6.294  & \cellcolor[rgb]{ 1,  .992,  .988} 31.426  & \cellcolor[rgb]{ 1,  .992,  .988} 150.733  \\
		& DPCML+APA & \cellcolor[rgb]{ 1,  .988,  .98} 1.751  & \cellcolor[rgb]{ 1,  .992,  .984} 6.254  & \cellcolor[rgb]{ 1,  .992,  .988} 31.280  & \cellcolor[rgb]{ 1,  .996,  .988} 148.985  \\
		& DPCML+BPA w/o DCRS  & \cellcolor[rgb]{1.0,  1.0,  1.0} 1.623  & \cellcolor[rgb]{1.0,  1.0,  1.0} 5.857  & \cellcolor[rgb]{1.0,  1.0,  1.0} 29.500 & \cellcolor[rgb]{1.0,  1.0,  1.0} 140.057  \\
		& DPCML+APA w/o DCRS & \cellcolor[rgb]{ 1,  .992,  .988} 1.703  & \cellcolor[rgb]{ 1,  .996,  .992} 6.116  & \cellcolor[rgb]{ 1,  .996,  .992} 30.598  & \cellcolor[rgb]{ 1,  .996,  .992} 145.893  \\
	  \midrule
	  \multirow{6}[2]{*}{HarS} & CML & \cellcolor[rgb]{ 1,  .918,  .867} 2.443  & \cellcolor[rgb]{ 1,  .922,  .875} 8.826  & \cellcolor[rgb]{ 1,  .925,  .875} 46.390 & \cellcolor[rgb]{ 1,  .914,  .863} 244.078  \\
		& CML+DD & \cellcolor[rgb]{ 1,  .894,  .827} 2.685  & \cellcolor[rgb]{ 1,  .906,  .847} 9.484  & \cellcolor[rgb]{ 1,  .914,  .859} 48.593  & \cellcolor[rgb]{ 1,  .902,  .839} 258.683  \\
		& DPCML+BPA & \cellcolor[rgb]{ 1,  .847,  .749} 3.144  & \cellcolor[rgb]{ 1,  .851,  .761} 11.498  & \cellcolor[rgb]{ 1,  .859,  .769} 60.696  & \cellcolor[rgb]{ 1,  .855,  .769} 313.086  \\
		& DPCML+APA & \cellcolor[rgb]{ 1,  .847,  .753} 3.123  & \cellcolor[rgb]{ 1,  .851,  .761} 11.477  & \cellcolor[rgb]{ 1,  .855,  .769} 60.975  & \cellcolor[rgb]{ 1,  .851,  .761} 317.433  \\
		& DPCML+BPA w/o DCRS  & \cellcolor[rgb]{ 1,  .878,  .804} 2.827  & \cellcolor[rgb]{ 1,  .878,  .808} 10.423  & \cellcolor[rgb]{ 1,  .882,  .808} 55.612  & \cellcolor[rgb]{ 1,  .875,  .796} 292.089  \\
		& DPCML+APA w/o DCRS & \cellcolor[rgb]{ 1,  .863,  .776} 2.989  & \cellcolor[rgb]{ 1,  .859,  .773} 11.238  & \cellcolor[rgb]{ 1,  .847,  .753} 62.926  & \cellcolor[rgb]{ 1,  .827,  .725} 345.167  \\
	  \midrule
	  \multirow{4}[2]{*}{DiHarS} & DPCML+BPA & \cellcolor[rgb]{ 1,  .788,  .659} 3.690  & \cellcolor[rgb]{ 1,  .792,  .667} 13.671  & \cellcolor[rgb]{ 1,  .804,  .682} 72.367  & \cellcolor[rgb]{ 1,  .812,  .698} 365.536  \\
		& DPCML+APA & \cellcolor[rgb]{ .996,  .78,  .647} 3.761  & \cellcolor[rgb]{ .996,  .78,  .647} 14.084  & \cellcolor[rgb]{ .996,  .78,  .647} 76.888  & \cellcolor[rgb]{ .996,  .78,  .647} 400.408  \\
		& DPCML+BPA w/o DCRS  & \cellcolor[rgb]{ 1,  .788,  .659} 3.713  & \cellcolor[rgb]{ 1,  .792,  .663} 13.776  & \cellcolor[rgb]{ 1,  .8,  .678} 72.969  & \cellcolor[rgb]{ 1,  .808,  .69} 369.048  \\
		& DPCML+APA w/o DCRS & \cellcolor[rgb]{ 1,  .875,  .796} 2.861  & \cellcolor[rgb]{ 1,  .867,  .788} 10.861  & \cellcolor[rgb]{ 1,  .859,  .773} 60.536  & \cellcolor[rgb]{ 1,  .847,  .749} 325.842  \\
	  \bottomrule
	  \end{tabular}%
	\label{tab:diversity}%
  \end{table}%
  
\begin{table}[htbp]
	\centering
	\caption{The $ILS@N$ and $Coverage@N$ performance of CML-based algorithms on Steam-200k and MovieLens-1M datasets. Here \textbf{a higher value implies more diverse} recommendation results.}
	  \begin{tabular}{c|c|cccc}
	  \toprule
	  \multicolumn{6}{c}{Steam-200k} \\
	  \midrule
	  Type & Method & ILS@5 & ILS@20 & Coverage@5 & Coverage@20 \\
	  \midrule
	  \multirow{4}[2]{*}{UniS} & CML & \cellcolor[rgb]{1.0,  1.0,  1.0} 2.375  & \cellcolor[rgb]{1.0,  1.0,  1.0} 58.964  & \cellcolor[rgb]{ .906,  .957,  .878} 0.148  & \cellcolor[rgb]{ .925,  .965,  .906} 0.275  \\
		& CML+DD & \cellcolor[rgb]{ .988,  .996,  .98} 2.922  & \cellcolor[rgb]{ .988,  .996,  .984} 72.481  & \cellcolor[rgb]{ .98,  .992,  .976} 0.095  & \cellcolor[rgb]{1.0,  1.0,  1.0} 0.093  \\
		& DPCML+BPA & \cellcolor[rgb]{ .976,  .988,  .965} 3.357  & \cellcolor[rgb]{ .969,  .988,  .961} 89.533  & \cellcolor[rgb]{ .871,  .937,  .831} 0.173  & \cellcolor[rgb]{ .894,  .949,  .863} 0.357  \\
		& DPCML+APA & \cellcolor[rgb]{ .976,  .988,  .969} 3.336  & \cellcolor[rgb]{ .965,  .984,  .953} 94.591  & \cellcolor[rgb]{ .867,  .933,  .824} 0.178  & \cellcolor[rgb]{ .882,  .945,  .847} 0.380  \\
	  \midrule
	  \multirow{4}[2]{*}{HarS} & CML & \cellcolor[rgb]{ .973,  .988,  .965} 3.405  & \cellcolor[rgb]{ .941,  .973,  .922} 118.397  & \cellcolor[rgb]{ .933,  .969,  .914} 0.128  & \cellcolor[rgb]{ .918,  .961,  .89} 0.299  \\
		& CML+DD & \cellcolor[rgb]{ .957,  .98,  .941} 4.044  & \cellcolor[rgb]{ .91,  .957,  .882} 147.455  & \cellcolor[rgb]{1.0,  1.0,  1.0} 0.080  & \cellcolor[rgb]{ .996,  1,  .996} 0.105  \\
		& DPCML+BPA & \cellcolor[rgb]{ .91,  .957,  .882} 5.736  & \cellcolor[rgb]{ .871,  .937,  .831} 184.938  & \cellcolor[rgb]{ .906,  .953,  .875} 0.149  & \cellcolor[rgb]{ .847,  .925,  .796} 0.476  \\
		& DPCML+APA & \cellcolor[rgb]{ .925,  .965,  .902} 5.157  & \cellcolor[rgb]{ .89,  .949,  .859} 164.924  & \cellcolor[rgb]{ .906,  .957,  .878} 0.147  & \cellcolor[rgb]{ .871,  .937,  .831} 0.413  \\
	  \midrule
	  \multirow{2}[2]{*}{DiHarS} & DPCML+BPA & \cellcolor[rgb]{ .753,  .878,  .678} 11.297  & \cellcolor[rgb]{ .753,  .878,  .678} 296.331  & \cellcolor[rgb]{ .847,  .925,  .804} 0.189  & \cellcolor[rgb]{ .753,  .878,  .678} 0.696  \\
		& DPCML+APA & \cellcolor[rgb]{ .8,  .902,  .737} 9.708  & \cellcolor[rgb]{ .776,  .89,  .706} 277.359  & \cellcolor[rgb]{ .753,  .878,  .678} 0.256  & \cellcolor[rgb]{ .773,  .89,  .702} 0.657  \\
	  \midrule
	  \multicolumn{6}{c}{MovieLens-1M} \\
	  \midrule
	  Type & Method & ILS@5 & ILS@20 & Coverage@5 & Coverage@20 \\
	  \midrule
	  \multirow{4}[2]{*}{UniS} & CML & \cellcolor[rgb]{ 1,  1,  .996} 3.071  & \cellcolor[rgb]{ 1,  1,  .996} 70.047  & \cellcolor[rgb]{ 1,  .949,  .8} 0.245  & \cellcolor[rgb]{ 1,  .973,  .886} 0.348  \\
		& CML+DD & \cellcolor[rgb]{ 1,  1,  .996} 3.222  & \cellcolor[rgb]{ 1,  1,  .996} 71.720  & \cellcolor[rgb]{ 1,  .957,  .827} 0.237  & \cellcolor[rgb]{ 1,  .992,  .965} 0.251  \\
		& DPCML+BPA & \cellcolor[rgb]{ 1,  1,  .996} 3.147  & \cellcolor[rgb]{ 1,  1,  .992} 75.366  & \cellcolor[rgb]{ 1,  .984,  .925} 0.201  & \cellcolor[rgb]{ 1,  .984,  .933} 0.291  \\
		& DPCML+APA & \cellcolor[rgb]{ 1,  1,  .996} 3.127  & \cellcolor[rgb]{ 1,  1,  .996} 74.492  & \cellcolor[rgb]{ 1,  .984,  .929} 0.200  & \cellcolor[rgb]{ 1,  .984,  .929} 0.292  \\
	  \midrule
	  \multirow{4}[2]{*}{HarS} & CML & \cellcolor[rgb]{ 1,  .984,  .933} 4.413  & \cellcolor[rgb]{ 1,  .98,  .922} 122.039  & \cellcolor[rgb]{ 1,  1,  .996} 0.175  & \cellcolor[rgb]{ 1,  .984,  .929} 0.294  \\
		& CML+DD & \cellcolor[rgb]{ 1,  .98,  .918} 4.742  & \cellcolor[rgb]{ 1,  .98,  .91} 129.342  & \cellcolor[rgb]{ 1,  1,  .996} 0.177  & \cellcolor[rgb]{ 1,  1,  .996} 0.207  \\
		& DPCML+BPA & \cellcolor[rgb]{ 1,  .969,  .867} 5.749  & \cellcolor[rgb]{ 1,  .969,  .871} 156.543  & \cellcolor[rgb]{ 1,  1,  .992} 0.179  & \cellcolor[rgb]{ 1,  .976,  .91} 0.318  \\
		& DPCML+APA & \cellcolor[rgb]{ 1,  .969,  .867} 5.739  & \cellcolor[rgb]{ 1,  .969,  .867} 158.717  & \cellcolor[rgb]{ 1,  .996,  .98} 0.182  & \cellcolor[rgb]{ 1,  .976,  .902} 0.325  \\
	  \midrule
	  \multirow{2}[2]{*}{DiHarS} & DPCML+BPA & \cellcolor[rgb]{ 1,  .953,  .812} 6.835  & \cellcolor[rgb]{ 1,  .957,  .827} 182.768  & \cellcolor[rgb]{ 1,  .98,  .918} 0.205  & \cellcolor[rgb]{ 1,  .949,  .8} 0.446  \\
		& DPCML+APA & \cellcolor[rgb]{ 1,  .949,  .8} 7.042  & \cellcolor[rgb]{ 1,  .949,  .8} 200.204  & \cellcolor[rgb]{ 1,  1,  .996} 0.176  & \cellcolor[rgb]{ 1,  .957,  .827} 0.418  \\
	  \bottomrule
	  \end{tabular}%
	\label{tab:ILS_and_Coverage}%
  \end{table}%
  
	\begin{figure*}[!t]
		\centering
		\subfigure[N=$3$ on Steam-200k]{
			\includegraphics[width=0.22\textwidth]{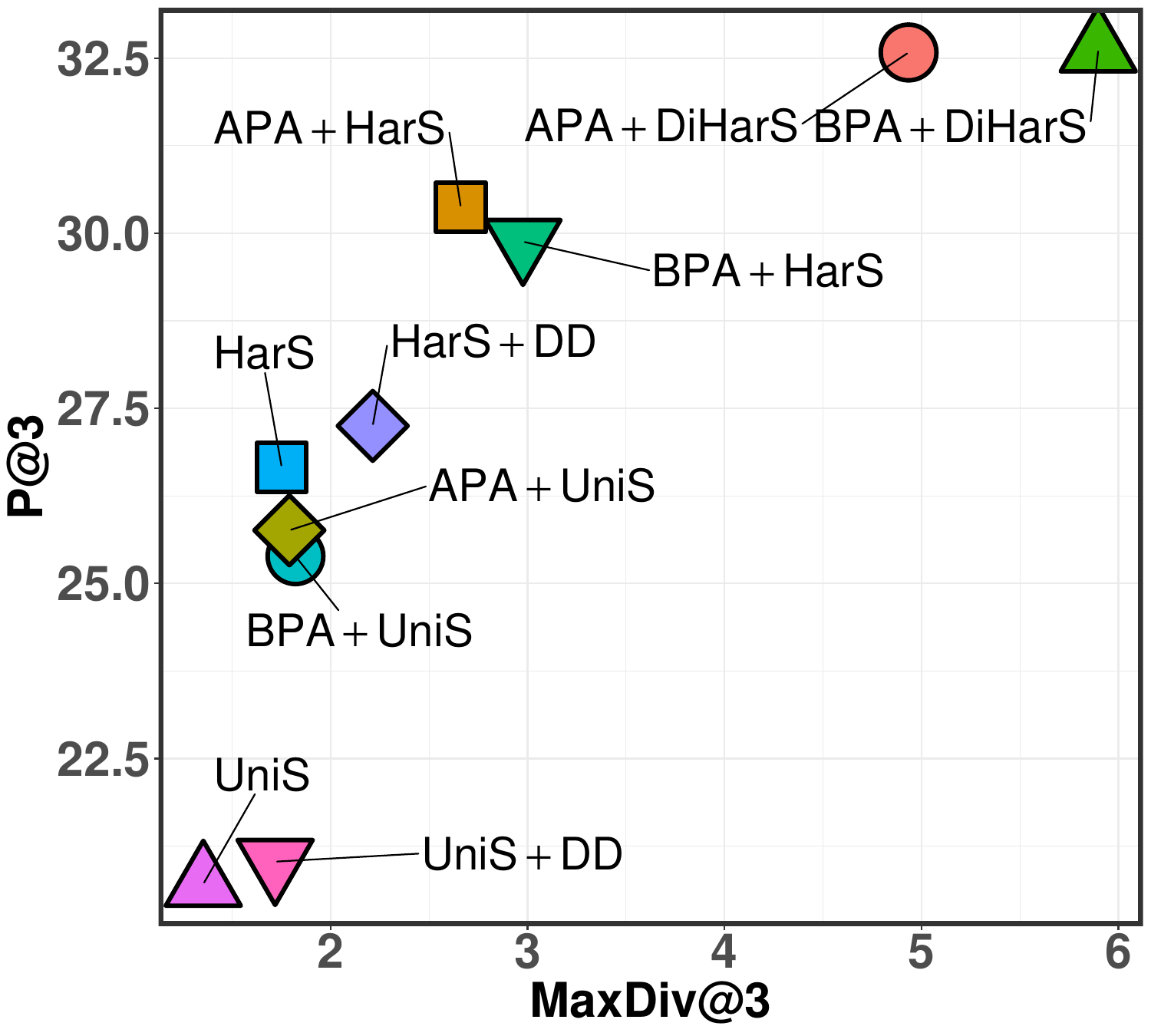}
		}
		\subfigure[N=$5$ on Steam-200k]{
			\includegraphics[width=0.22\textwidth]{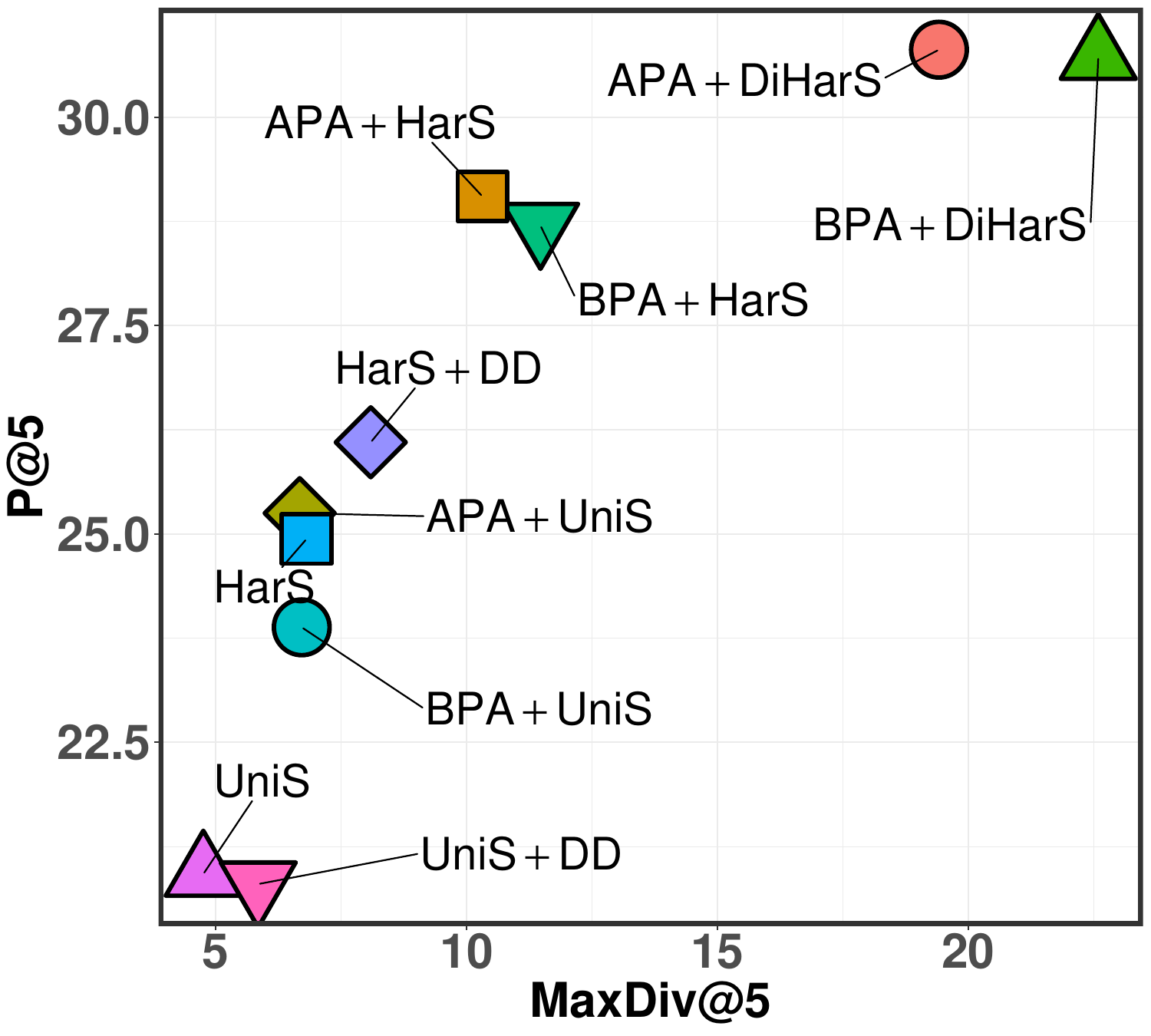}
		}
		\subfigure[N=$5$ on Steam-200k]{
			\includegraphics[width=0.22\textwidth]{img/diverfication/200k_i5.pdf}
		}
		\subfigure[N=$5$ on Steam-200k]{
			\includegraphics[width=0.22\textwidth]{img/diverfication/200k_c5.pdf}
		}
		
		\subfigure[N=$3$ on MovieLens-1M]{
			\includegraphics[width=0.22\textwidth]{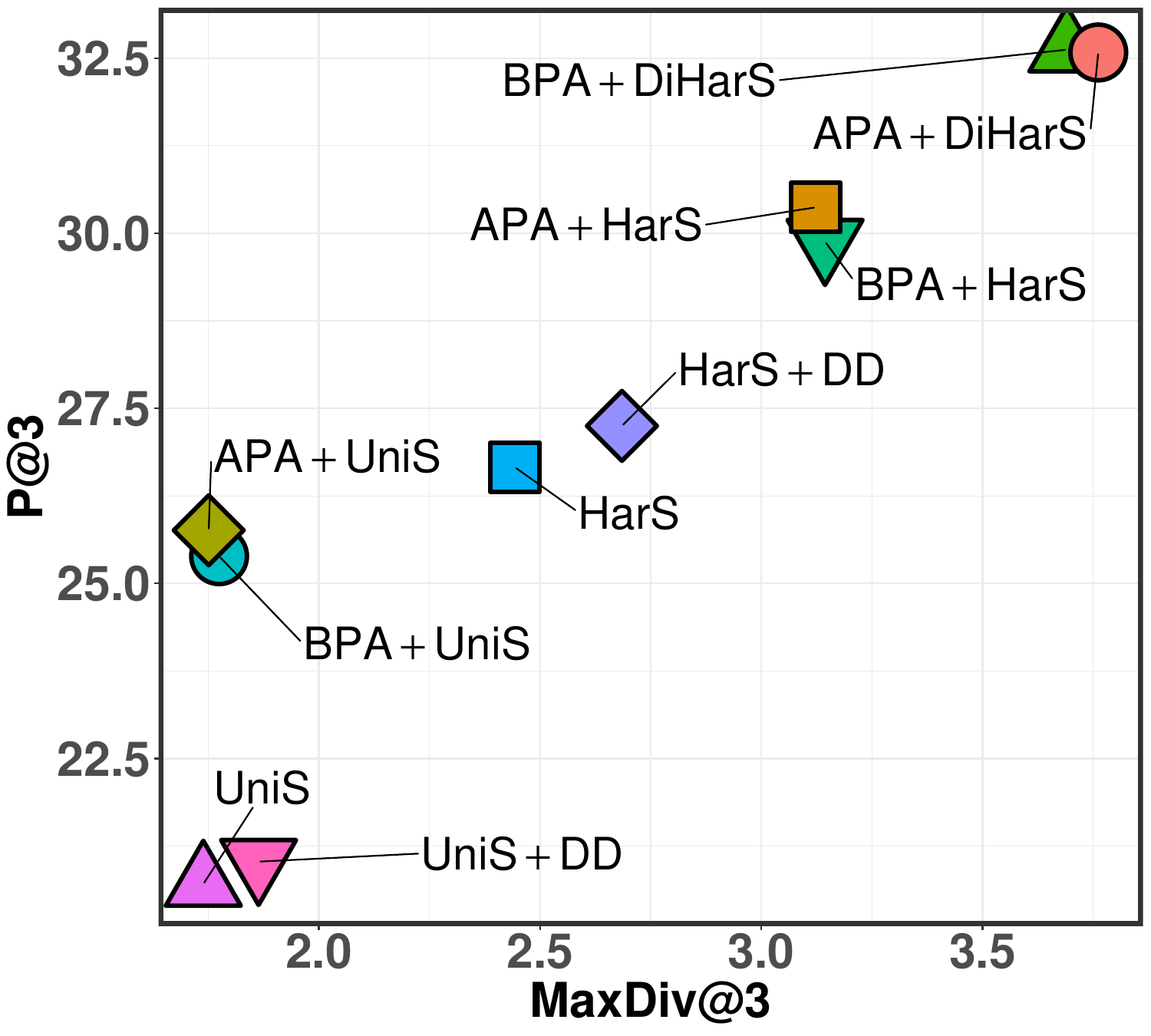}
		}
		\subfigure[N=$5$ on MovieLens-1M]{
			\includegraphics[width=0.22\textwidth]{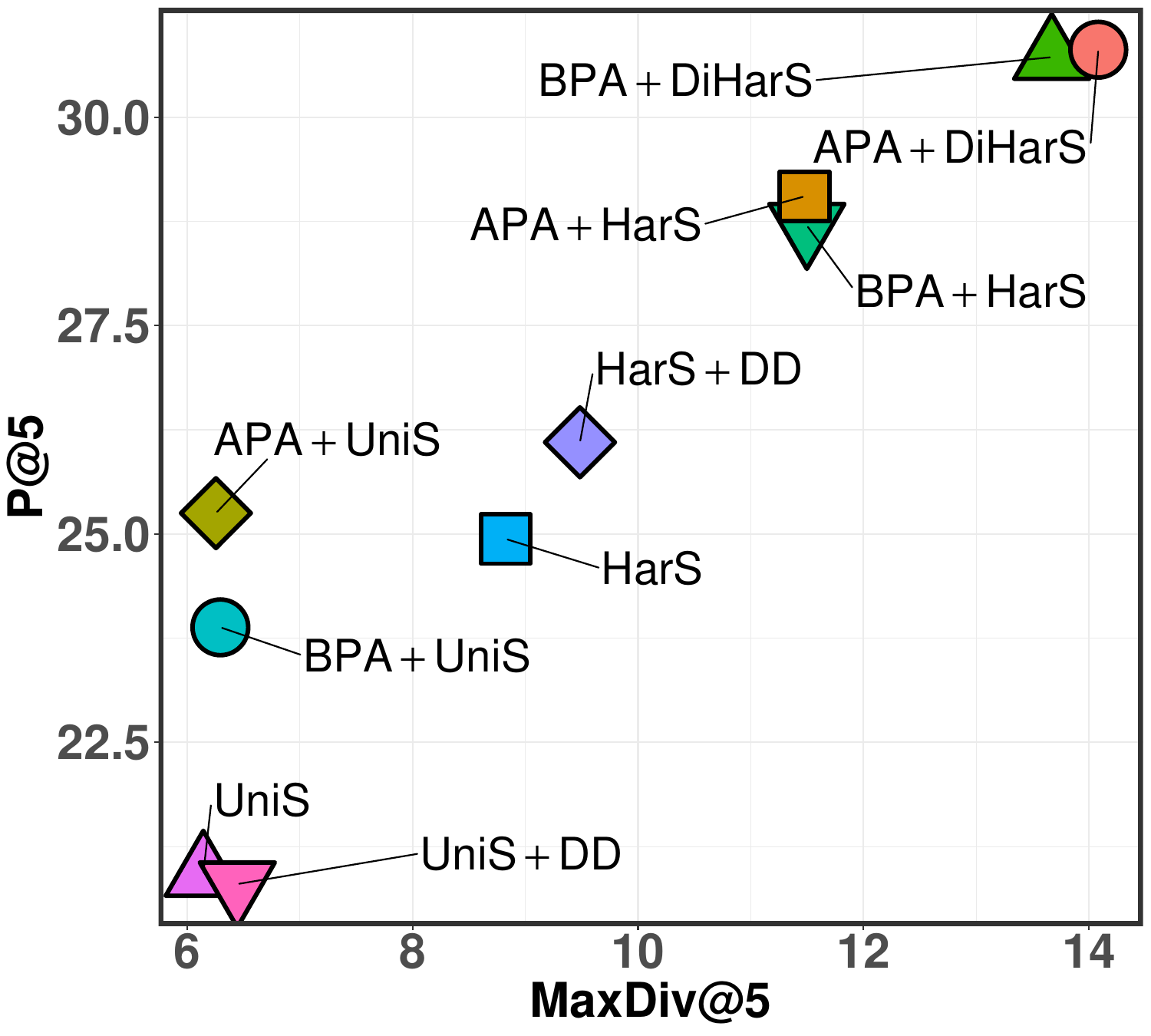}
		}
		\subfigure[N=$5$ on MovieLens-1M]{
			\includegraphics[width=0.22\textwidth]{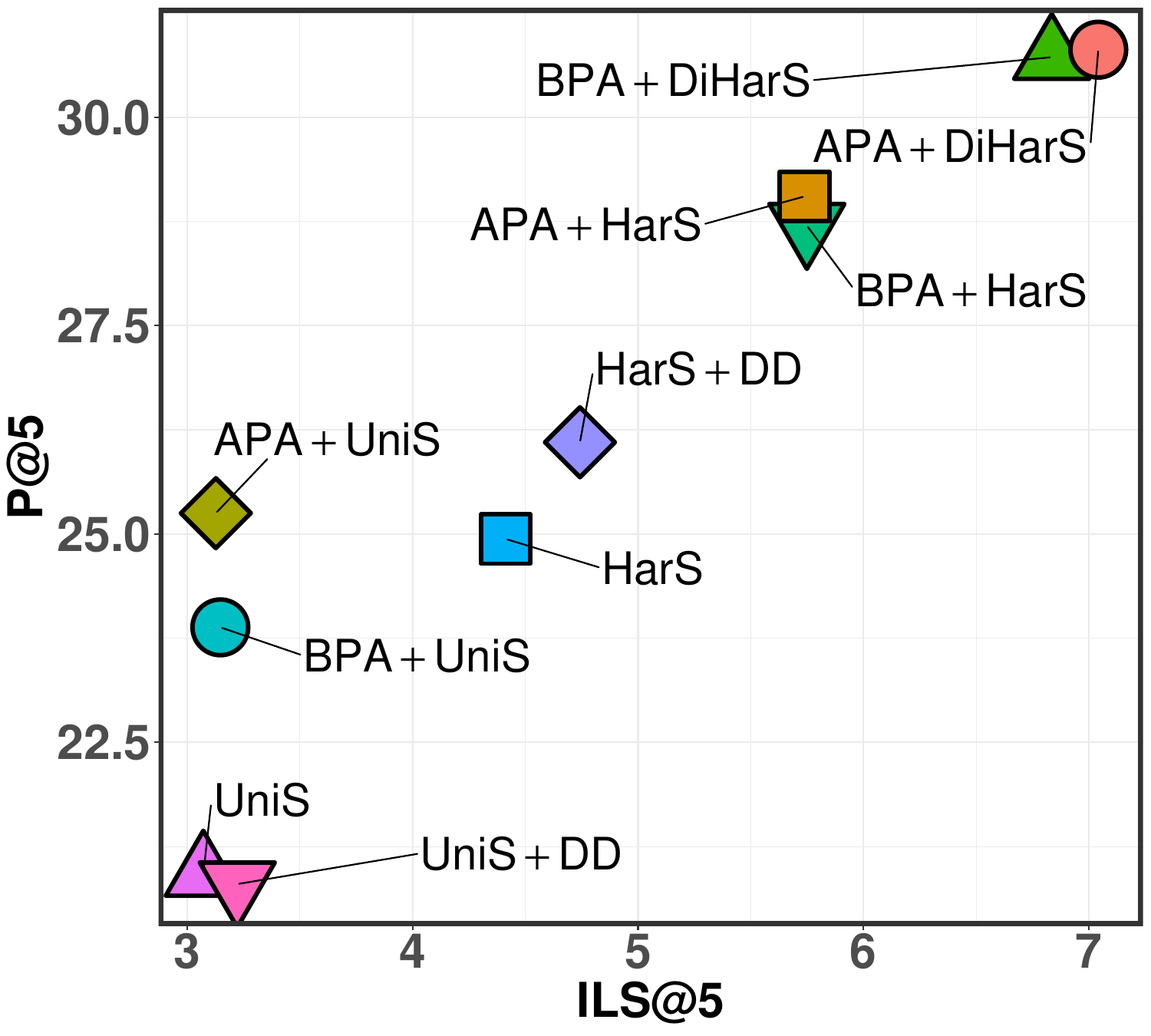}
		}
		\subfigure[N=$5$ on MovieLens-1M]{
			\includegraphics[width=0.22\textwidth]{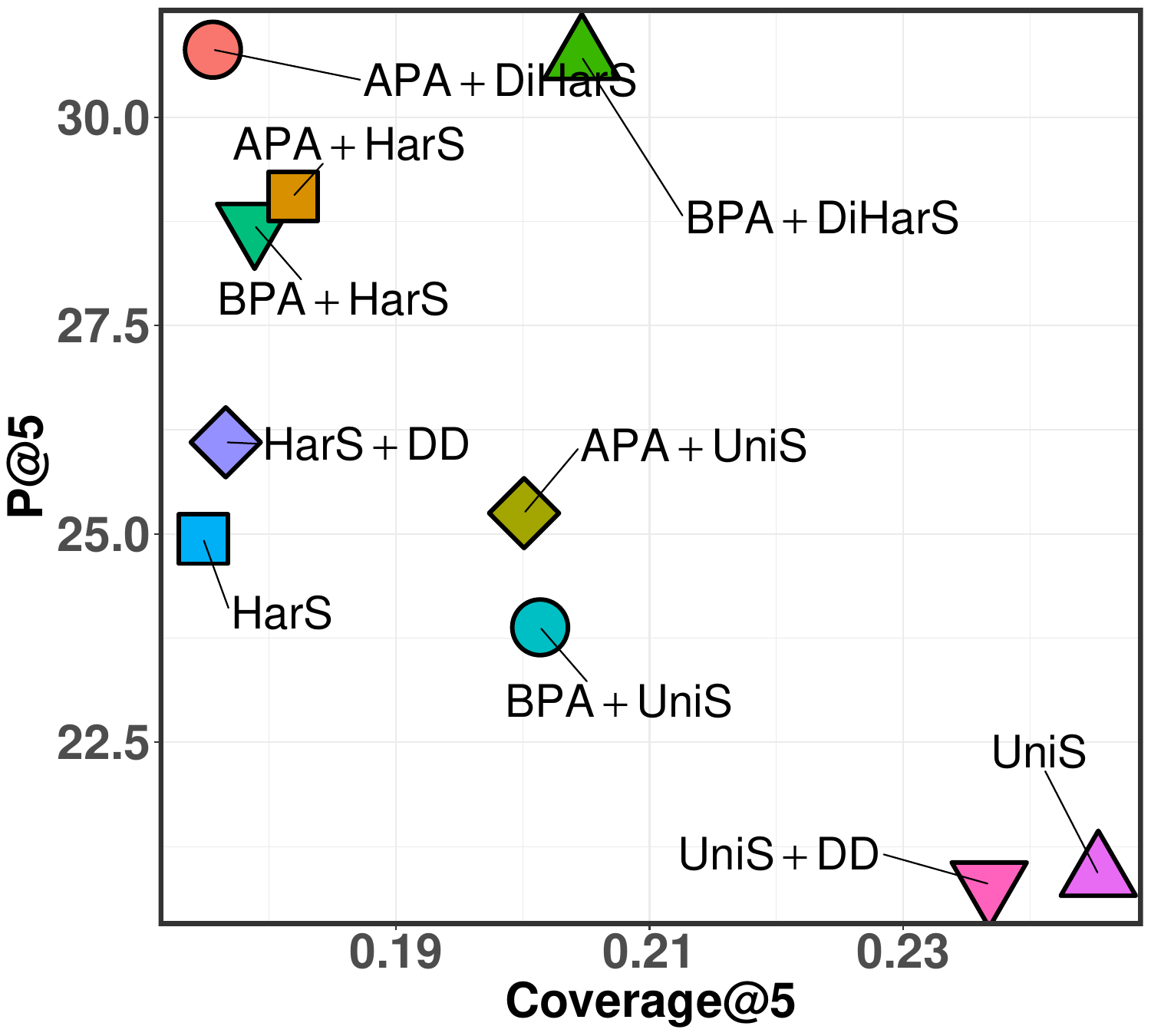}
		}
		\caption{The diversity comparison alongside recommendation performance at $N= \{3, 5\}$ on Steam-200k and MovieLens-1M datasets.}
		\label{fig:supp_div}
	\end{figure*}
	
\begin{table}[htbp]
	\centering
	\caption{Performance comparisons on MovieLens-1M and Steam-200k datasets against other diversity-promoting algorithms. The best performance is highlighted in bold.}
	  \begin{tabular}{c|c|c|cccccccc}
	  \toprule
		& Type & Method & P@3 & R@3 & NDCG@3 & P@5 & R@5 & NDCG@5 & MAP & MRR \\
	  \midrule
	  \multirow{10}[6]{*}{MovieLens-1M} & \multirow{4}[2]{*}{Two-Stage} & UniS+DD & \cellcolor[rgb]{ .961,  .98,  .992} 17.41  & \cellcolor[rgb]{ .996,  1,  1} 3.66  & \cellcolor[rgb]{ .965,  .98,  .996} 17.78  & \cellcolor[rgb]{ .925,  .961,  .988} 18.21  & \cellcolor[rgb]{ .988,  .992,  1} 5.55  & \cellcolor[rgb]{ .937,  .965,  .988} 18.67  & \cellcolor[rgb]{ .941,  .969,  .988} 12.38  & \cellcolor[rgb]{ .949,  .973,  .992} 35.59  \\
		&   & UniS+PD & \cellcolor[rgb]{ .961,  .976,  .992} 17.54  & \cellcolor[rgb]{ .996,  1,  1} 3.71  & \cellcolor[rgb]{ .961,  .98,  .992} 17.88  & \cellcolor[rgb]{ .925,  .957,  .988} 18.33  & \cellcolor[rgb]{ .984,  .992,  1} 5.60  & \cellcolor[rgb]{ .933,  .965,  .988} 18.79  & \cellcolor[rgb]{ .941,  .969,  .988} 12.40  & \cellcolor[rgb]{ .945,  .973,  .992} 35.79  \\
		&   & HarS+DD & \cellcolor[rgb]{ .835,  .91,  .969} 24.89  & \cellcolor[rgb]{ .957,  .976,  .992} 5.87  & \cellcolor[rgb]{ .843,  .914,  .969} 25.38  & \cellcolor[rgb]{ .812,  .894,  .965} 24.90  & \cellcolor[rgb]{ .914,  .953,  .984} 8.26  & \cellcolor[rgb]{ .82,  .902,  .965} 25.77  & \cellcolor[rgb]{ .882,  .933,  .976} 15.78  & \cellcolor[rgb]{ .831,  .906,  .969} 45.11  \\
		&   & HarS+PD & \cellcolor[rgb]{ .835,  .91,  .969} 24.89  & \cellcolor[rgb]{ .957,  .976,  .992} 5.87  & \cellcolor[rgb]{ .843,  .914,  .969} 25.38  & \cellcolor[rgb]{ .812,  .894,  .965} 24.91  & \cellcolor[rgb]{ .914,  .953,  .984} 8.26  & \cellcolor[rgb]{ .82,  .902,  .965} 25.77  & \cellcolor[rgb]{ .882,  .933,  .976} 15.74  & \cellcolor[rgb]{ .831,  .906,  .969} 45.14  \\
  \cmidrule{2-11}      & \multirow{5}[2]{*}{One-Stage} & PRD & \cellcolor[rgb]{ .976,  .988,  .996} 16.47  & \cellcolor[rgb]{ .992,  .996,  1} 4.01  & \cellcolor[rgb]{ .98,  .992,  .996} 16.54  & \cellcolor[rgb]{ .949,  .973,  .992} 17.00  & \cellcolor[rgb]{ .98,  .988,  .996} 5.77  & \cellcolor[rgb]{ .961,  .98,  .992} 17.19  & \cellcolor[rgb]{ .937,  .965,  .988} 12.63  & \cellcolor[rgb]{ .969,  .98,  .996} 34.12  \\
		&   & RecNet & \cellcolor[rgb]{ .941,  .969,  .988} 18.58  & \cellcolor[rgb]{ .992,  .996,  1} 3.96  & \cellcolor[rgb]{ .949,  .973,  .992} 18.76  & \cellcolor[rgb]{ .906,  .949,  .984} 19.33  & \cellcolor[rgb]{ .976,  .988,  .996} 5.96  & \cellcolor[rgb]{ .918,  .957,  .984} 19.70  & \cellcolor[rgb]{ .941,  .969,  .988} 12.38  & \cellcolor[rgb]{ .941,  .969,  .988} 36.27  \\
		&   & DP-RecNet & \cellcolor[rgb]{ .945,  .969,  .992} 18.46  & \cellcolor[rgb]{ .988,  .996,  1} 4.09  & \cellcolor[rgb]{ .949,  .973,  .992} 18.75  & \cellcolor[rgb]{ .914,  .953,  .984} 19.03  & \cellcolor[rgb]{ .973,  .984,  .996} 6.02  & \cellcolor[rgb]{ .922,  .957,  .984} 19.51  & \cellcolor[rgb]{ .941,  .969,  .988} 12.29  & \cellcolor[rgb]{ .937,  .965,  .988} 36.45  \\
		&   & IDCF & \cellcolor[rgb]{ .851,  .918,  .973} 24.12  & \cellcolor[rgb]{ .961,  .98,  .992} 5.56  & \cellcolor[rgb]{ .855,  .918,  .973} 24.67  & \cellcolor[rgb]{ .824,  .902,  .965} 24.05  & \cellcolor[rgb]{ .925,  .961,  .988} 7.79  & \cellcolor[rgb]{ .835,  .906,  .969} 24.96  & \cellcolor[rgb]{ .89,  .937,  .98} 15.37  & \cellcolor[rgb]{ .839,  .91,  .969} 44.29  \\
		&   & GraphDiv & \cellcolor[rgb]{ .953,  .976,  .992} 17.85  & \cellcolor[rgb]{1.0, 1.0,  1.0} 3.40  & \cellcolor[rgb]{ .949,  .973,  .992} 18.71  & \cellcolor[rgb]{ .933,  .965,  .988} 17.88  &\cellcolor[rgb]{1.0, 1.0,  1.0}  4.97  & \cellcolor[rgb]{ .929,  .961,  .988} 19.02  & \cellcolor[rgb]{1.0, 1.0,  1.0} 8.89  & \cellcolor[rgb]{ .941,  .969,  .988} 36.13  \\
  \cmidrule{2-11}      & Ours & APA+DiHarS & \cellcolor[rgb]{ .82,  .898,  .965} \textbf{25.98} & \cellcolor[rgb]{ .949,  .973,  .992} \textbf{6.16} & \cellcolor[rgb]{ .824,  .902,  .965} \textbf{26.71} & \cellcolor[rgb]{ .796,  .886,  .961} \textbf{25.74} & \cellcolor[rgb]{ .902,  .945,  .98} \textbf{8.65} & \cellcolor[rgb]{ .804,  .89,  .961} \textbf{26.90} & \cellcolor[rgb]{ .882,  .933,  .976} \textbf{15.82} & \cellcolor[rgb]{ .808,  .894,  .965} \textbf{46.92} \\
	  \midrule
	  \multirow{10}[6]{*}{Steam-200k} & \multirow{4}[2]{*}{Two-Stage} & UniS+DD & \cellcolor[rgb]{ .902,  .945,  .98} 21.03  & \cellcolor[rgb]{ .839,  .91,  .969} 12.04  & \cellcolor[rgb]{ .902,  .945,  .98} 21.66  & \cellcolor[rgb]{ .882,  .933,  .976} 20.80  & \cellcolor[rgb]{ .859,  .922,  .973} 10.27  & \cellcolor[rgb]{ .89,  .937,  .98} 21.61  & \cellcolor[rgb]{ .827,  .906,  .969} 18.92  & \cellcolor[rgb]{ .89,  .941,  .98} 40.13  \\
		&   & UniS+PD & \cellcolor[rgb]{ .902,  .945,  .98} 20.89  & \cellcolor[rgb]{ .839,  .91,  .969} 12.04  & \cellcolor[rgb]{ .902,  .945,  .98} 21.56  & \cellcolor[rgb]{ .878,  .933,  .976} 20.89  & \cellcolor[rgb]{ .859,  .922,  .973} 10.34  & \cellcolor[rgb]{ .886,  .937,  .98} 21.62  & \cellcolor[rgb]{ .827,  .906,  .969} 18.92  & \cellcolor[rgb]{ .89,  .941,  .98} 40.19  \\
		&   & HarS+DD & \cellcolor[rgb]{ .796,  .886,  .961} 27.25  & \cellcolor[rgb]{ .765,  .871,  .953} 15.99  & \cellcolor[rgb]{ .796,  .886,  .961} 28.48  & \cellcolor[rgb]{ .788,  .882,  .961} 26.10  & \cellcolor[rgb]{ .773,  .875,  .957} 13.50  & \cellcolor[rgb]{ .788,  .882,  .961} 27.65  & \cellcolor[rgb]{ .745,  .859,  .949} 23.61  & \cellcolor[rgb]{ .776,  .875,  .957} 49.45  \\
		&   & HarS+PD & \cellcolor[rgb]{ .808,  .894,  .965} 26.70  & \cellcolor[rgb]{ .769,  .871,  .957} 15.76  & \cellcolor[rgb]{ .804,  .89,  .961} 27.96  & \cellcolor[rgb]{ .808,  .894,  .965} 24.97  & \cellcolor[rgb]{ .792,  .886,  .961} 12.80  & \cellcolor[rgb]{ .808,  .894,  .965} 26.66  & \cellcolor[rgb]{ .753,  .863,  .953} 23.26  & \cellcolor[rgb]{ .784,  .878,  .957} 48.85  \\
  \cmidrule{2-11}      & \multirow{5}[2]{*}{One-Stage} & PRD & \cellcolor[rgb]{ .933,  .965,  .988} 19.01  & \cellcolor[rgb]{ .875,  .929,  .976} 10.27  & \cellcolor[rgb]{ .933,  .965,  .988} 19.56  & \cellcolor[rgb]{ .886,  .937,  .98} 20.57  & \cellcolor[rgb]{ .867,  .925,  .976} 10.02  & \cellcolor[rgb]{ .89,  .941,  .98} 21.49  & \cellcolor[rgb]{ .871,  .929,  .976} 16.52  & \cellcolor[rgb]{ .918,  .957,  .984} 38.02  \\
		&   & RecNet & \cellcolor[rgb]{ .965,  .98,  .996} 17.20  & \cellcolor[rgb]{ .882,  .937,  .976} 9.75  & \cellcolor[rgb]{ .961,  .98,  .992} 17.93  & \cellcolor[rgb]{ .949,  .973,  .992} 16.91  & \cellcolor[rgb]{ .914,  .953,  .984} 8.31  & \cellcolor[rgb]{ .949,  .973,  .992} 17.83  & \cellcolor[rgb]{ .898,  .945,  .98} 14.83  & \cellcolor[rgb]{ .957,  .976,  .992} 34.80  \\
		&   & DP-RecNet & \cellcolor[rgb]{ .992,  .996,  1} 15.59  & \cellcolor[rgb]{ .886,  .937,  .98} 9.57  & \cellcolor[rgb]{ .988,  .996,  1} 16.12  & \cellcolor[rgb]{1.0, 1.0,  1.0} 13.88  & \cellcolor[rgb]{ .941,  .969,  .988} 7.31  & \cellcolor[rgb]{1.0, 1.0,  1.0} 14.64  & \cellcolor[rgb]{ .898,  .941,  .98} 14.94  & \cellcolor[rgb]{ .996,  1,  1} 31.85  \\
		&   & IDCF & \cellcolor[rgb]{ .843,  .914,  .969} 24.45  & \cellcolor[rgb]{ .804,  .89,  .961} 13.92  & \cellcolor[rgb]{ .843,  .914,  .969} 25.41  & \cellcolor[rgb]{ .824,  .902,  .965} 24.11  & \cellcolor[rgb]{ .816,  .898,  .965} 11.94  & \cellcolor[rgb]{ .827,  .902,  .969} 25.38  & \cellcolor[rgb]{ .788,  .882,  .961} 21.12  & \cellcolor[rgb]{ .827,  .906,  .969} 45.29  \\
		&   & GraphDiv & \cellcolor[rgb]{1.0, 1.0,  1.0} 15.01  & \cellcolor[rgb]{ .918,  .953,  .984} 7.89  & \cellcolor[rgb]{1.0, 1.0,  1.0} 15.29  & \cellcolor[rgb]{ .965,  .98,  .996} 15.92  & \cellcolor[rgb]{ .922,  .957,  .984} 7.98  & \cellcolor[rgb]{ .965,  .98,  .996} 16.88  & \cellcolor[rgb]{ .969,  .984,  .996} 10.84  & \cellcolor[rgb]{1.0, 1.0,  1.0} 31.31  \\
  \cmidrule{2-11}      & Ours & APA+DiHarS & \cellcolor[rgb]{ .706,  .835,  .941} \textbf{32.58} & \cellcolor[rgb]{ .706,  .835,  .941} \textbf{19.09} & \cellcolor[rgb]{ .706,  .835,  .941} \textbf{33.98} & \cellcolor[rgb]{ .706,  .835,  .941} \textbf{30.81} & \cellcolor[rgb]{ .706,  .835,  .941} \textbf{15.99} & \cellcolor[rgb]{ .706,  .835,  .941} \textbf{32.68} & \cellcolor[rgb]{ .706,  .835,  .941} \textbf{25.78} & \cellcolor[rgb]{ .706,  .835,  .941} \textbf{54.90} \\
	  \bottomrule
	  \end{tabular}%
	\label{tab:diversity_comp_supp}%
  \end{table}%

	\subsection{Diversity-promoting Performance Comparison}\label{marjor:supp_C.5.1}
	\subsubsection{Compared to other Diversity-promoting Methods}\label{marjor:supp_C.6.1}
	In this section, we compare our proposed DPMCL-based algorithms with other diversity methods in the recommendation system. Note that this paper aims to boost recommendation diversity \textbf{using collaborative data only}. In light of this, we adopt the following competitive baselines that can work well without requiring any external information:

	\noindent \textbf{(1)} Two-stage methods, i.e., post-processing approaches for promoting diversity:
	\begin{itemize}[leftmargin=*]
		\item The Bounded Greedy (BG) selection \cite{DBLP:conf/iccbr/SmythM01, DBLP:conf/ah/BridgeK06} is one of the most effective re-ranking techniques to improve RS diversity. Briefly, the top-$N$ recommendations are generated as follows: (a) picking up the most relevant $L$ ($L > N$) items preferred by a target user $u_i$ and (b) selecting $N$ items with maximum quality among $L$ items in a greedy fashion. Specifically, step (a) is achieved by the general recommendation model (such as the latent-based model). Step (b) is conducted iteratively, where one item with the highest quality relative to so far recommendation candidate $I^{u_i}$ will be added at a time. Here item $v_j$'s quality could be regarded as the average dissimilarities between $v_j$ and the items already included in $I^{u_i}$:
		\begin{equation}\nonumber
			\mathbb{Q}(v_j, I^{u_i}) := (1 - \omega) \cdot Sim(u_i, v_j) + \omega \cdot RelDiv(v_j, I^{u_i}),
		\end{equation}
		where 
		\[
			RelDiv(v_j, I^{u_i}) := \left\{ 
				\begin{aligned} 
					1.0 \ \ \ & if \ \ I^{u_i} = \emptyset,   \\
					\frac{\sum_{v_k \in I^{u_i}} Dis(v_j, v_k)}{|I^{u_i}|} \ \ \ & otherwise,
				\end{aligned}
			\right.
		\]
		$Sim(\cdot, \cdot)$/$Dis(\cdot, \cdot)$ denotes the similar/dissimilar measure function and $\omega \in [0, 1]$ is a trade-off weight.
	\end{itemize}
	
	We refer interested readers to the literature \cite{DBLP:conf/iccbr/SmythM01, DBLP:conf/ah/BridgeK06} for more algorithm details. 
	The key of BG is how to determine the similarity/dissimilarity functions. In this paper, we consider the re-ranking strategy on top of two CML-based methods, i.e., \textbf{UniS} and \textbf{HarS}. Therefore, $Sim(\cdot, \cdot)$ could be directly reflected by the inverse Euclidean distance. Furthermore, we attempt two different ways to measure the dissimilarities $Dis(\cdot, \cdot)$ between items. One is the Euclidean distance between items, where a higher value represents a more significant dissimilarity, denoted as the  \textbf{Distance-based Diversity (DD)} strategy. Another one is the \textbf{Popularity-based Diversity (PD)} strategy \cite{DBLP:conf/compsac/PremchaiswadiPJP13,DBLP:conf/recsys/VargasC11}, where items with different popularity levels (i.e., a larger popularity gap) are expected to be recommended for diversification maximization.
	
	\noindent \textbf{(2)} One-stage methods, i.e., optimizing relevance and diversification jointly during training:
	\begin{itemize}[leftmargin=*]
		\item \textbf{Personalized Ranking with Diversity (PRD)} \footnote{\url{https://github.com/guoguibing/librec}} \cite{DBLP:conf/recsys/Hurley13} incorporates the diversity goal into RankSGD \cite{DBLP:journals/jmlr/JahrerT12a}, which aims to recommend relevant items to users while ranking diverse items closely together as much as possible.
		\item \textbf{RecNet} \footnote{\url{https://github.com/baichuan/Neural_Bayesian_Personalized_Ranking}} and \textbf{Diversity-Promoting RecNet (DP-RecNet)} The original RecNet \cite{DBLP:journals/corr/TrofimovSHLMA17} is a generic learning-to-rank framework for implicit feedback. It designs a novel neural network to simultaneously learn representations of users and items in an embedded space and the users' preferences without any contextual information. On top of RecNet, a Diversity-Promoting RecNet \cite{DBLP:conf/recsys/SidanaLA18} is proposed, which explicitly adopts a Kullback-Leibler (KL)-based loss to regulate the diversity within the list of items recommended to each user during training. 
		\item \textbf{Item-Diversity-based Collaborative Filtering (IDCF)} \cite{DBLP:conf/uic/YangFW18} proposes a general variance regularization method for MF-based CF models to improve recommendation diversification. Under implicit feedback-based recommendations, we adopt one of the effective MF-based methods BPR \cite{DBLP:conf/uai/RendleFGS09} as the backbone and further leverage IDCF to promote diversity.
		\item \textbf{Graph Convolutional Network (GCN) based Accuracy-Diversity Trade-off (GCN-AccDiv)}\footnote{\url{https://github.com/esilezz/accdiv-via-graphconv}} \cite{DBLP:journals/ipm/IsufiPH21} involves two GCN modules, namely, the accuracy-oriented RS model and the diversity-oriented RS model. The former component is to learn representations of users and items from the nearest neighbor graph, while the latter is to strike a balance between accuracy and diversity based on the furthest neighbor graph constructed by $k$ users whose preferences are the most dissimilar to the target user.
	\end{itemize}

	\noindent\textbf{Setups.} For two-stage methods, $\omega$ is tuned from $0.1$ to $0.9$ with $0.1$ step margin. In terms of one-stage approaches, all the parameter adjustment strategies strictly follow the corresponding original paper. 

	\noindent \textbf{Performance Comparison.} Empirical results are summarized in Tab.\ref{tab:diversity_comp_supp}. We observe that although the two-stage paradigm could boost the recommendation performance, its improvement is extremely limited. A possible cause is that separately handling relevance and diversity could not strike a reasonable balance between them. Besides, neural-network-based algorithms (such as DP-RecNet and GCN-AccDiv) show relatively unsatisfactory performance due to the data sparsity. By contrast, our APA+DiHarS method could consistently perform better than all newly added algorithms, which suggests its superiority against current diversity-promoting aspects.
		\subsubsection{More Evidence for Recommendation Diversity \label{pami:supp_4.2}}
	Besides performance evaluations, recommendation diversity \cite{DBLP:journals/tbd/XieLLZ00L22,DBLP:conf/coling/RazaBN22} is another significant concern. In this sense, we test the diversity performance with a series of widely adopted diversity metrics, including 
	\begin{itemize}
		\item \textit{Max-sum Diversification (MaxDiv)} \cite{10.1145/2213556.2213580} measures the recommendation diversification by considering item-side similarity, where a high value implies that the recommendation results are relatively diverse:
		\[
			\textit{MaxDiv}@N = \frac{1}{|\mathcal{U}|}\sum_{u_i \in \mathcal{U}} \sum_{\substack{v_i, v_j \in \mathcal{I}^N_{u_i}, \\ v_i \neq v_j}} s(v_i, v_j),
			\]
			where $s(v_i, v_j)=\|\boldsymbol{g}_{v_i} - \boldsymbol{g}_{v_j}\|^2$ is the square of Euclidean distance between item $v_i$ and $v_j$; $\mathcal{I}^N_{u_i}$ is the top-$N$ recommendation items for user $u_i$.
		\item \textit{Intra-List Similarity (ILS)} \cite{DBLP:conf/www/ZieglerMKL05} shows the average diversity of a list recommended to all users, which is  permutation-insensitivity:
		\[
			\textit{ILS}@N = \frac{1}{|\mathcal{U}|}\sum_{u_i \in \mathcal{U}} \frac{\sum_{\substack{v_i \in \mathcal{I}^N_{u_i}}} \sum_{\substack{v_j \in \mathcal{I}^N_{u_i}, v_i \neq v_j}} Sim(v_i, v_j)}{2},
		\]
		where $Sim(v_i, v_j)$ is the custom-defined criterion \cite{DBLP:conf/www/ZieglerMKL05}. This paper employs $s(v_i, v_j)$ as our criterion since it is the direct and unique standard for CML-based methods to make recommendations, where \textbf{a high value indicates a more diverse result.}
		\item \textit{Coverage} \cite{DBLP:journals/tiis/KaminskasB17} (a.k.a ``aggregate diversity'' \cite{DBLP:journals/tkde/AdomaviciusK12} or simply ``diversity'' \cite{DBLP:conf/recsys/Shi13}) reflects the holistic diversity of an algorithm, which is usually expressed as the degree of available items presented to users, i.e.,:
		\[
			\textit{Coverage}@N = \frac{|\bigcup _{u_i \in \mcu}\mathcal{I}^N_{u_i}|}{|\mathcal{I}|}.
		\] 
		
		Generally speaking, a higher value represents that users can access a broader range of items, improving the potential for diverse recommendations.
	\end{itemize}

	\noindent \textbf{Results.} Since developing CML-based diversity-promoting methods is our goal, we consider the following approaches: a) traditional CML optimized by \textbf{UniS} and \textbf{HarS}. b) traditional CML (UniS and HarS) with \textbf{Distance-based Diversity (DD)} promoting. c) Our proposed DPCML framework with both BPA and APA strategies. 
	Here we also consider three negative sampling tricks, denoted as \textbf{BPA+UniS}, \textbf{BPA+HarS}, \textbf{BPA+DiHarS}, \textbf{APA+UniS}, \textbf{APA+HarS} and \textbf{APA+DiHarS}, respectively. Besides, we also consider DPCML without (w/o) DCRS. The experiments are conducted on the Steam-200k and MovieLens-1M datasets with $N \in \{3, 5, 10, 20\}$ for $\textit{MaxDiv}@N$ and $N \in \{5, 20\}$ for $\textit{ILS}@N$ and $\textit{Coverage}@N$. The empirical results are provided in Fig.\ref{fig:supp_div}, Tab.\ref{tab:diversity} and Tab.\ref{tab:ILS_and_Coverage}. From these results, we can conclude: a) Within the same negative sampling strategy, DPCML could achieve better diversity in most cases, even CML using the reranking trick \textbf{DD}. b) More significantly, our proposed DiHarS strategy could further boost recommendation diversity. This suggests the effectiveness of promoting recommendation diversity. c) Even without the regularization term, DPCML still outperforms CML. Most importantly, equipped with DCRS, DPCML could achieve better diversification results against w/o DCRS in most cases. The above experiments suggest that DPCML could perform better than traditional CML in recommendation accuracy and diversity.
	
\subsection{More Evidence of Quantitative Analysis} \label{ext_qa}
\subsubsection{Ablation Study for DiHarS Framework \label{pami:supp_ab_for_DiHarS}}
To show the effectiveness of our proposed DiHarS algorithm, we investigate the performance of different DiHarS variants. At first, we consider the usage of DiHarS for the CML framework (i.e., \textbf{CML+DiHarS}) and regard the HarS approach (\textbf{CML+HarS}) as the benchmark. Furthermore, we also consider the non-differentiable version of DiHarS (short for \textbf{NDiHarS}), i.e., directly using the sort operation to achieve the sparse sample selections in (\ref{ref_eq28}). Compared with the traditional HarS fixing $U\equiv1$ in (\ref{cml_sampling_hars}), the major difference of NDiHarS is the number of $U=\lfloor n_{i}^- \cdot \beta_i \rfloor \ge 1$ determined by the FPR range $\beta_i$ in Thm.\ref{pami:thm2}. For DiHarS and NDiHarS, we conduct experiments for DPCML with both BPA (i.e., \textbf{BPA+NDiHarS} and \textbf{BPA+DiHarS}) and APA (i.e., \textbf{APA+NDiHarS} and \textbf{APA+DiHarS}) strategies. The hyper-parameter setups stay the same as DiHarS. The empirical results are presented in Fig.\ref{fig:supp_ab_for_DiHarS}. Our proposed DiHarS could outperform its sort-based counterpart (i.e., NDiHarS-driven methods) significantly because the non-differentiable loss function might be challenging to optimize. Besides, we can observe that applying DiHarS to the standard CML could also perform better than the conventional HarS trick in most cases. These results consistently provide evidence for the superiority of our proposed DiHarS. 
	\begin{table}[htbp]
		\centering
		\caption{Sensitivity analysis for DPCML with the proposed BPA strategy and the UniS sampling method ($C=5$) on the Steam-200k dataset. }
		\begin{tabular}{c|cccccccc}
			\toprule
			$\eta$ & P@3 & R@3 & NDCG@3 & P@5 & R@5 & NDCG@5 & MAP & MRR \\
			\midrule
			1 & 25.04 & 14.65 & 26.01 & 24.60 & 12.55 & 25.81 & 21.65 & 45.55 \\
			3 & 24.67 & 14.43 & 25.50 & 23.88 & 12.25 & 24.96 & 21.56 & 44.73 \\
			5 & 25.24 & 14.91 & 26.65 & 23.80 & 12.17 & 25.34 & 22.17 & 47.23 \\
			10 & 25.39 & 14.84 & 26.56 & 23.88 & 12.11 & 25.25 & 22.26 & 46.79 \\
			20 & 24.60 & 14.34 & 25.79 & 24.03 & 12.05 & 25.17 & 21.87 & 46.20 \\
			30 & 25.23 & 14.69 & 26.19 & 24.25 & 12.08 & 25.58 & 21.94 & 46.00 \\
			\bottomrule
		\end{tabular}%
		\label{tab:sen_dpcml1}%
	\end{table}%
	
	\begin{table}[htbp]
		\centering
		\caption{Sensitivity analysis for for DPCML with the proposed BPA strategy and the HarS sampling ($C=5$) on the Steam-200k dataset.}
		\begin{tabular}{c|cccccccc}
			\toprule
			$\eta$ & P@3 & R@3 & NDCG@3 & P@5 & R@5 & NDCG@5 & MAP & MRR \\
			\midrule
			1 & 28.55 & 16.35 & 29.92 & 27.82 & 13.94 & 29.65 & 22.90 & 50.57 \\
			3 & 28.68 & 16.32 & 29.96 & 27.71 & 13.90 & 29.59 & 23.13 & 50.19 \\
			5 & 29.34 & 16.82 & 30.45 & 27.98 & 13.95 & 29.75 & 23.42 & 50.62 \\
			10 & 29.88 & 17.13 & 31.22 & 28.70 & 14.51 & 30.56 & 24.10 & 51.95 \\
			20 & 29.81 & 17.12 & 31.08 & 29.11 & 14.65 & 30.77 & 24.35 & 51.90 \\
			30 & 29.43 & 16.99 & 30.67 & 28.96 & 14.53 & 30.56 & 24.50 & 51.36 \\
			\bottomrule
		\end{tabular}%
		\label{tab:sen_dpcml2}%
	\end{table}%
	
	\begin{figure}[!t]
		\centering
		\subfigure[DPCML+BPA+UniS]{
			\includegraphics[width=0.31\textwidth]{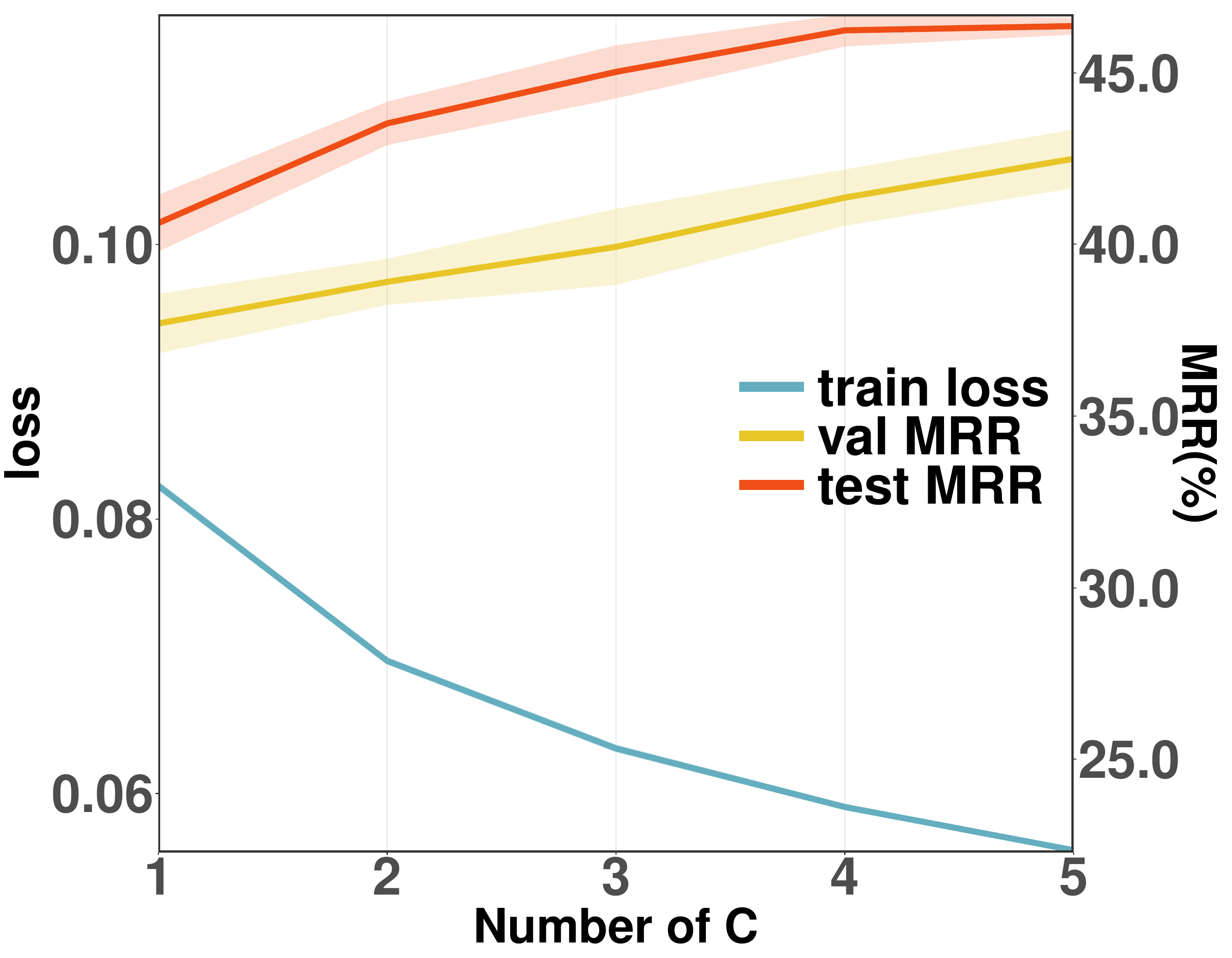}
			\label{DPCML1}
		}
		\subfigure[DPCML+BPA+HarS]{
			\includegraphics[width=0.31\textwidth]{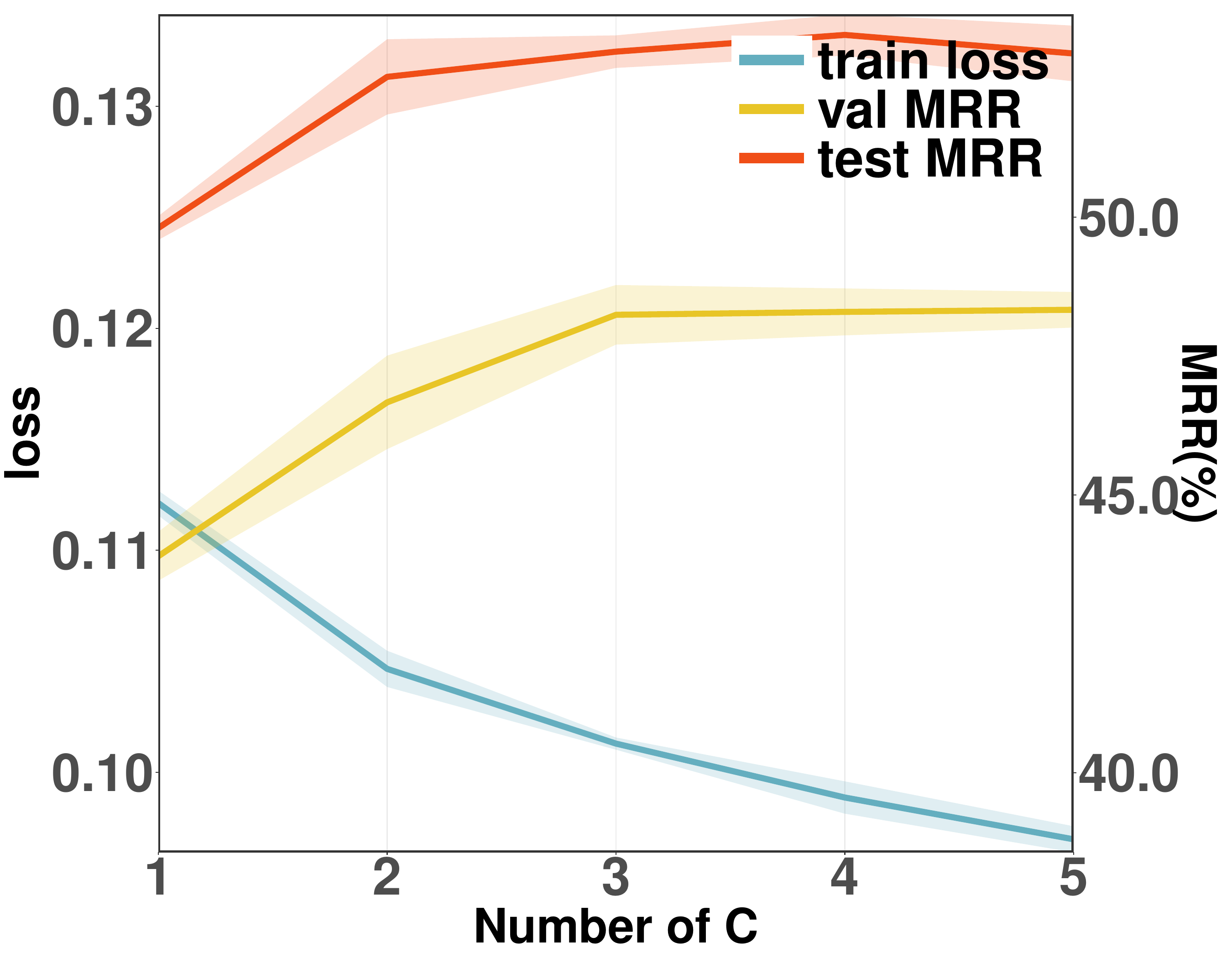}
			\label{DPCML2}
		}
		\subfigure[DPCML+BPA+DiHarS]{
			\includegraphics[width=0.31\textwidth]{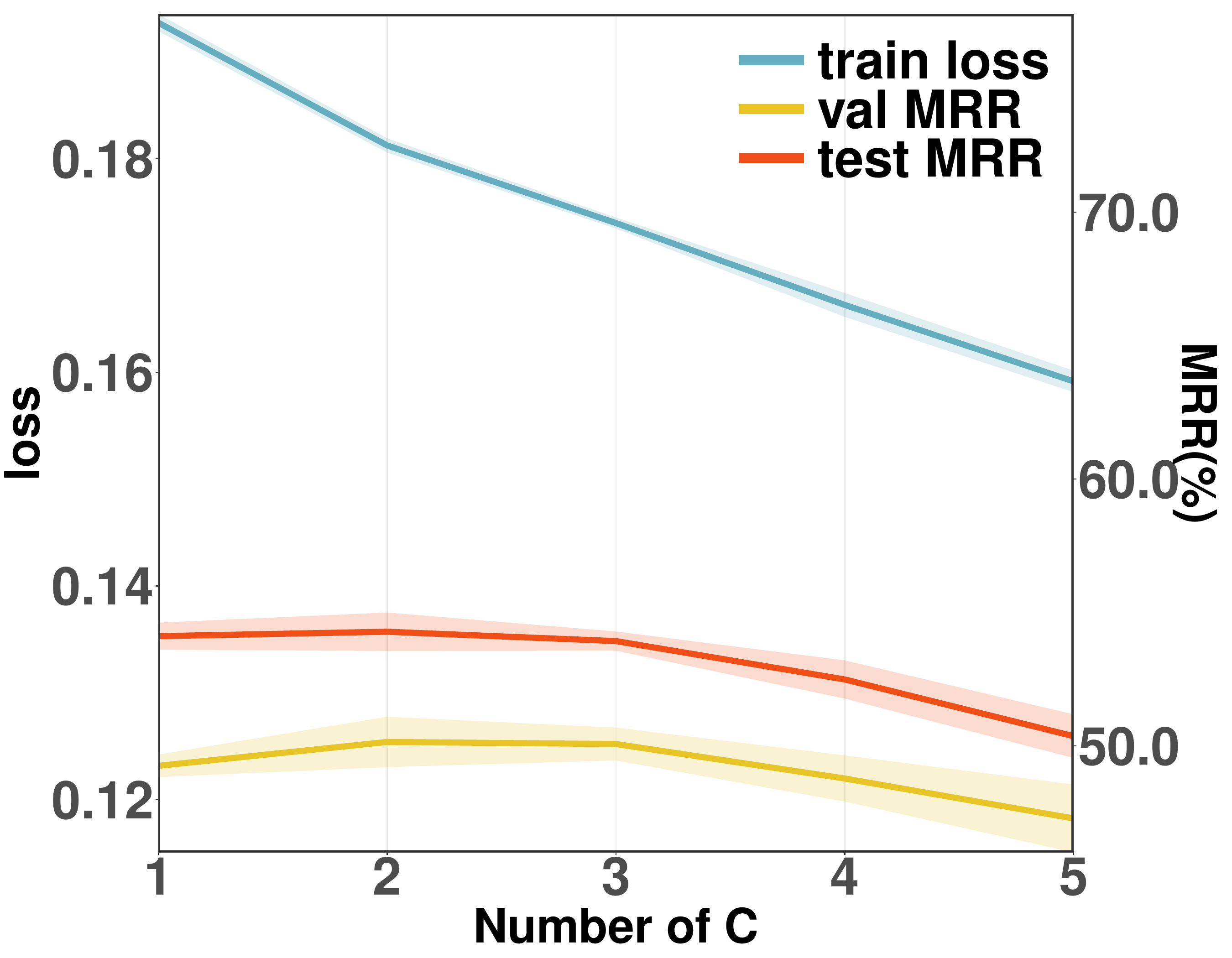}
			\label{DiHarS}
		}
		\caption{Empirical justification of Thm.\ref{them1} on the Steam-200k dataset. Here we report the qualitative performance of DPCML with the BPA strategy and consider three difference negative sampling tricks.}
		\label{just_thm1}
	\end{figure}

	\subsubsection{Empirical Justification of Corol.\ref{cor1}} \label{pami:supp_sec.7.6.4}
	To demonstrate the validity of Corol.\ref{cor1}, we conduct empirical studies on the Steam-200k dataset. Note that we merely consider our proposed DPCML with the BPA strategy here since the APA strategy would make users' vector numbers dynamic and thus be challenging to analyze directly. Expressly, we set $C \in \{1, 2, 3, 4, 5\}$ and record the results of train loss, validation (val) and test MRR metrics. Moreover, to ensure a fair comparison, all experiments are repeated $5$ times with $5$ different random seeds. The empirical results are shown in Fig.\ref{just_thm1}, where the shades represent the variance among $5$ experiments. According to these results, we can observe that, with the increase of $C$, the empirical risk (i.e., training loss) of DPCML ($C>1$) learning with any of three sampling strategies could be significantly smaller than the corresponding CML ($C=1$) counterpart. Furthermore, DPCML could substantially improve the recommendation performance on the validation/test set. Therefore, the above empirical results consistently present that our proposed DPCML framework could induce a smaller generalization error than the traditional CML paradigm, empirically suggesting the correctness of Corol.\ref{cor1}.
	
	\subsubsection{Sensitivity analysis of $\eta$}\label{major:ab_eta}
	We investigate the sensitivity of $\eta \in \{0, 1, 3, 5, 10, 20, 30\}$ for recommendation results on the Steam-200k dataset. The experimental results are listed in Tab.\ref{tab:sen_dpcml1} and Tab.\ref{tab:sen_dpcml2} for DPCML1 and DPCML2, respectively. We can conclude that a proper $\eta$ (roughly $10$) could significantly improve the performance, suggesting the essential role of the proposed diversity control regularization scheme.
	

	\subsubsection{Ablation Studies of Diversity Control Regularization Scheme (DCRS) \label{ab_stu}}
\begin{figure*}[]
		\centering
		\subfigure[P@$3$]{
			\includegraphics[width=0.23\columnwidth]{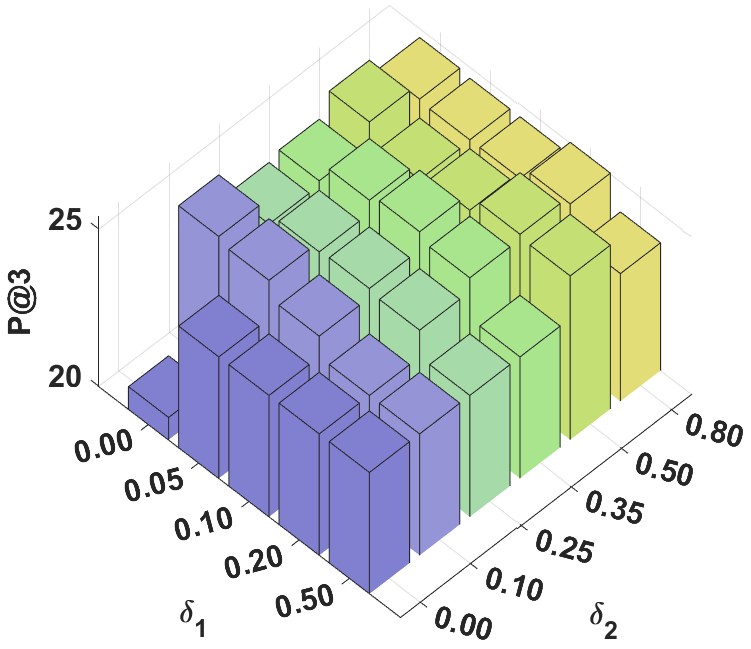}
		}
		\subfigure[R@$3$]{
			\includegraphics[width=0.23\columnwidth]{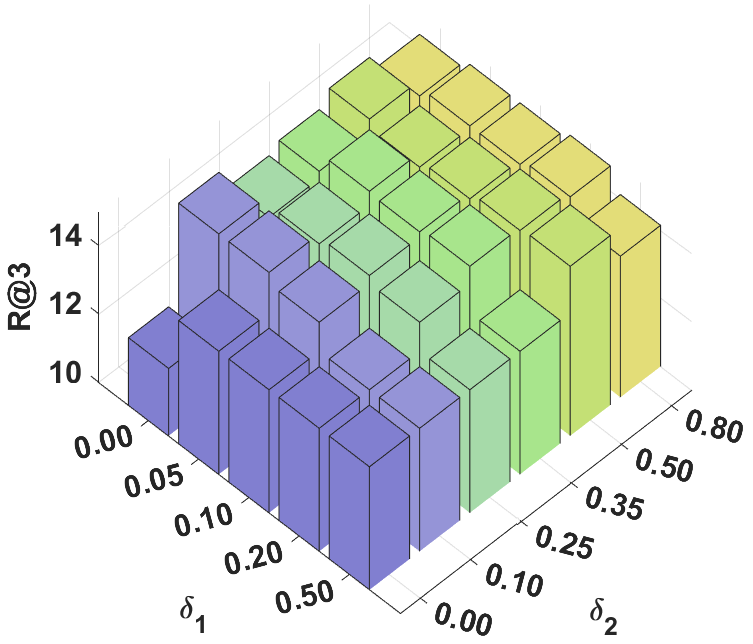}
		}
		\subfigure[NDCG@$3$]{
			\includegraphics[width=0.23\columnwidth]{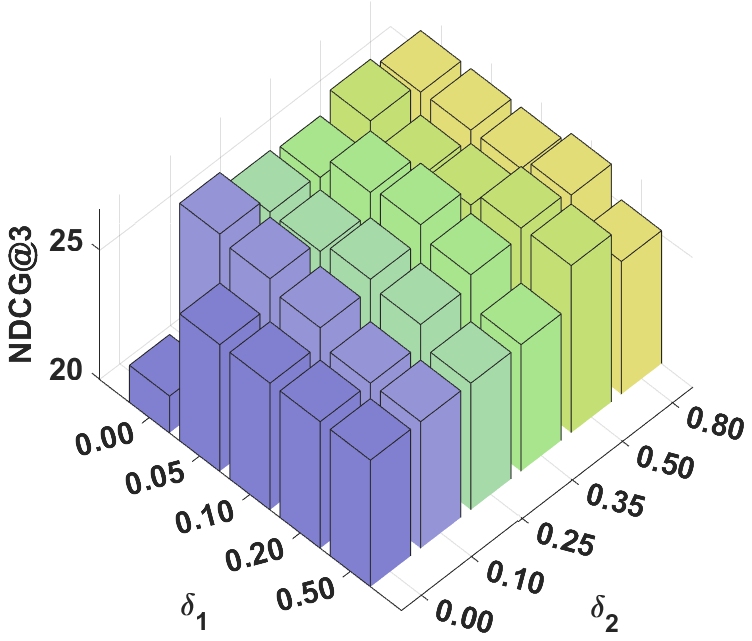}
		}
		\subfigure[P@$5$]{
			\includegraphics[width=0.23\columnwidth]{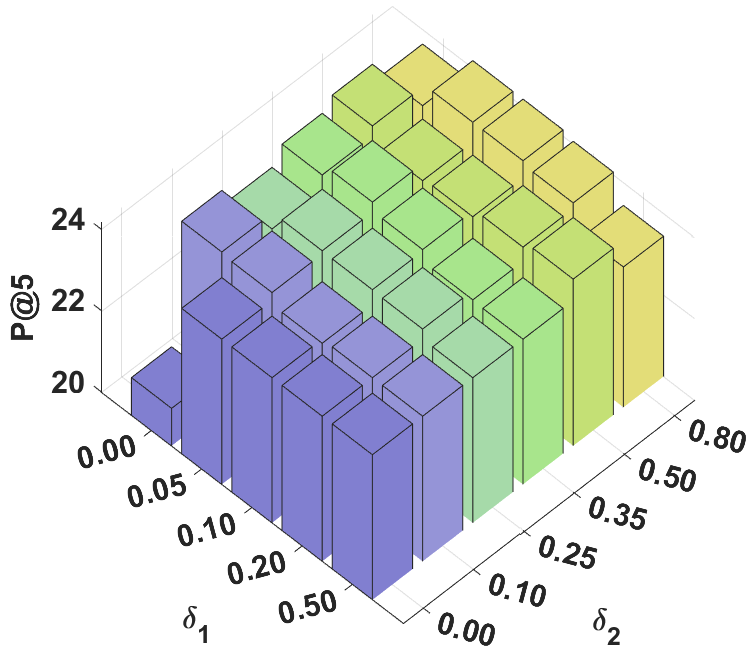}
		}
		\subfigure[R@$5$]{
			\includegraphics[width=0.23\columnwidth]{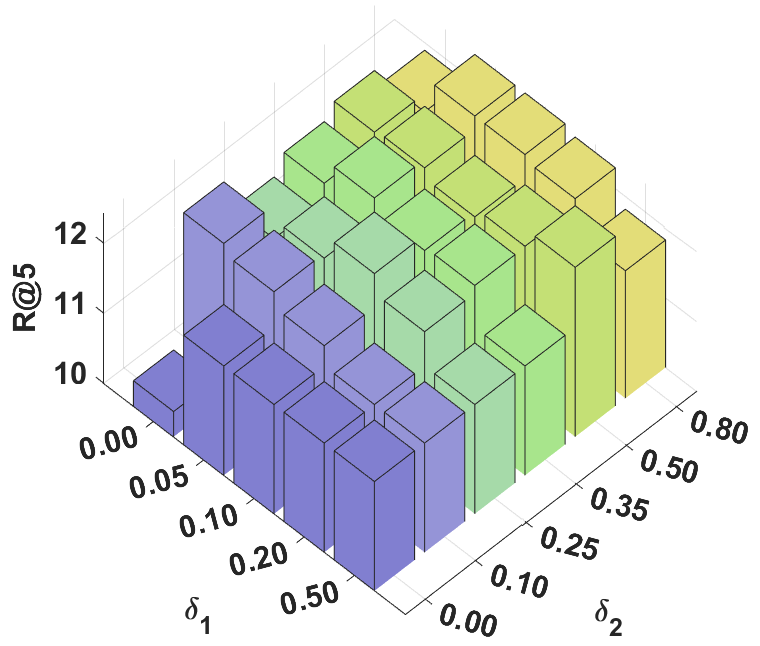}
		}
		\subfigure[NDCG@$5$]{
			\includegraphics[width=0.23\columnwidth]{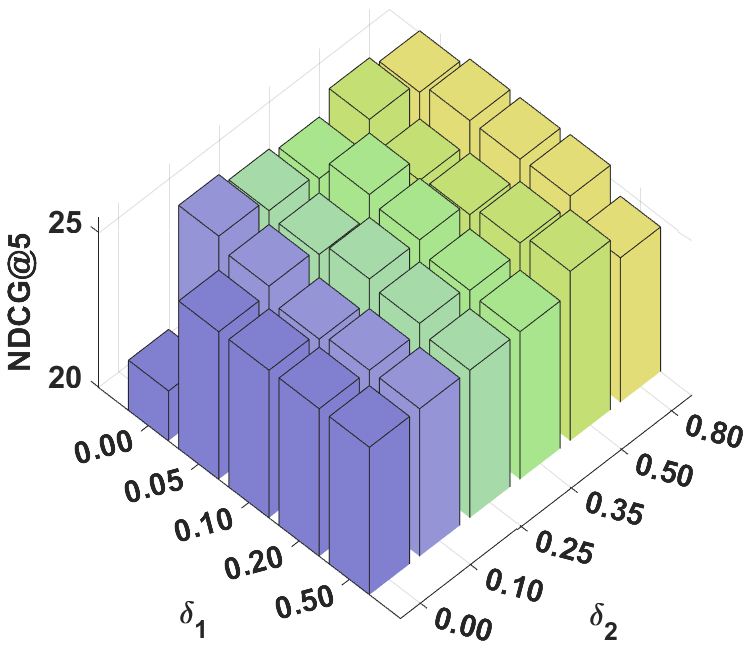}
		}
		\subfigure[MAP]{
			\includegraphics[width=0.23\columnwidth]{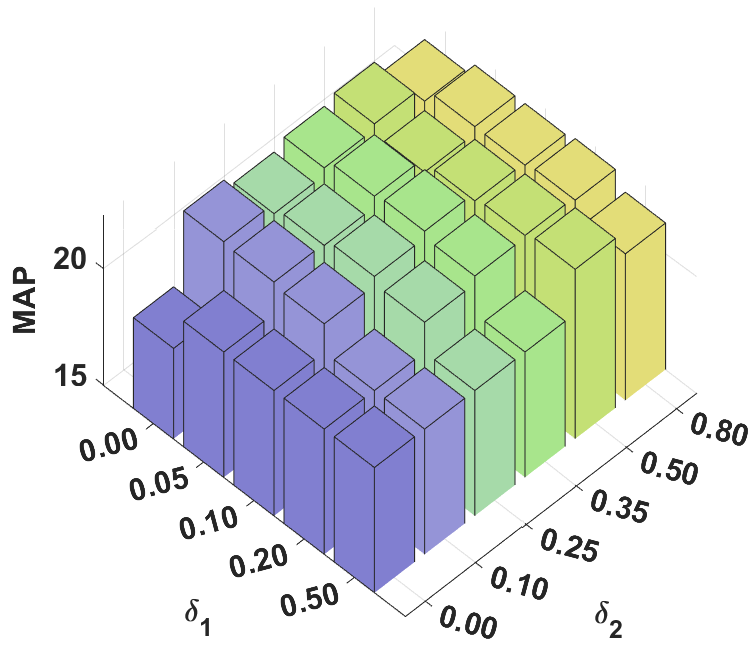}
		}
		\subfigure[MRR]{
			\includegraphics[width=0.23\columnwidth]{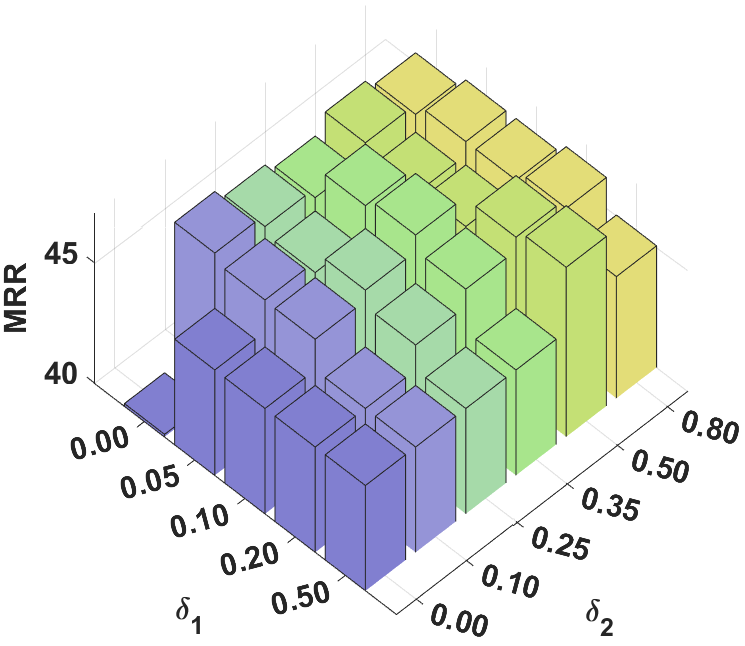}
		}
		\caption{Sensitivity against $\delta_1$ and $\delta_2$ for \textbf{BPA+UniS} on Steam-200k. The $x$- and $y$-axis stand for the value of $\delta_1$ and $\delta_2$ respectively, and the $z$-axis shows the performance.}
		\label{sensitivity_supp1}
	\end{figure*}

\begin{figure*}[]
		\centering
		
		\subfigure[P@$3$]{
			\includegraphics[width=0.23\columnwidth]{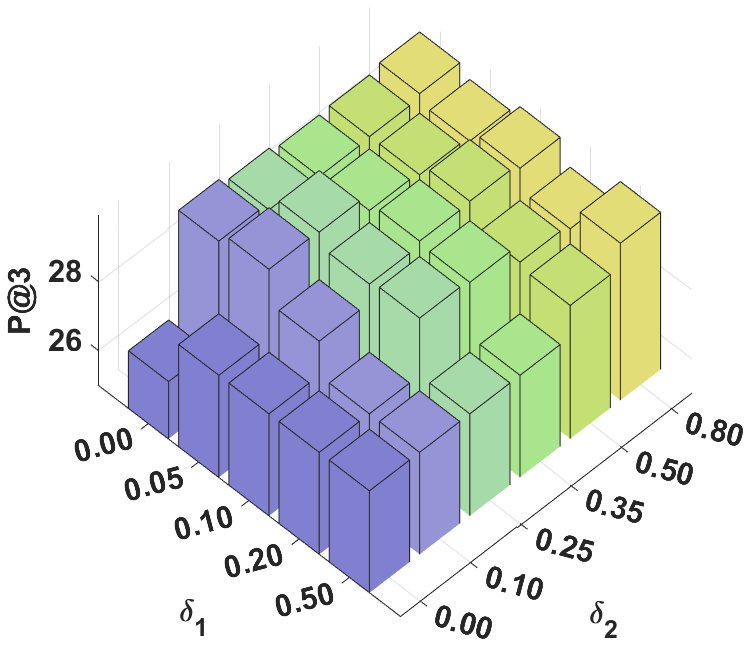}
		}
		\subfigure[R@$3$]{
			\includegraphics[width=0.23\columnwidth]{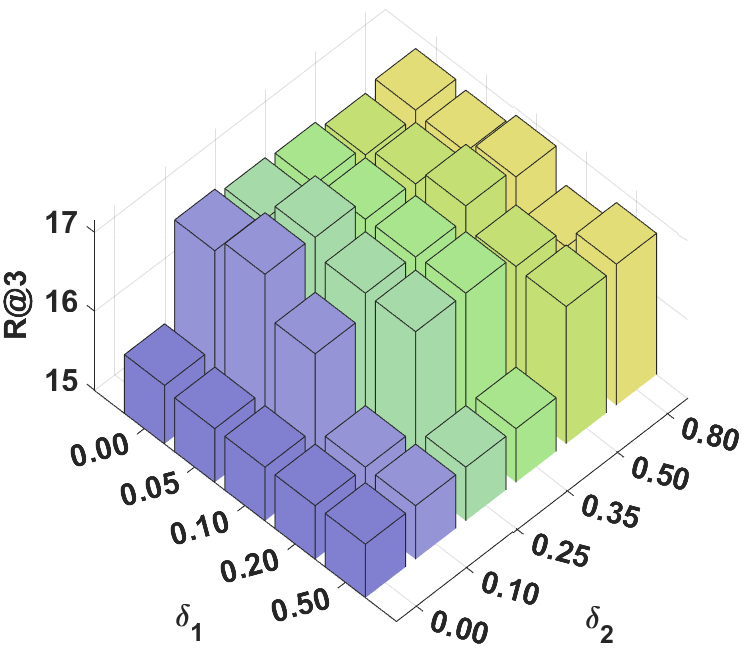}
		}
		\subfigure[NDCG@$3$]{
			\includegraphics[width=0.23\columnwidth]{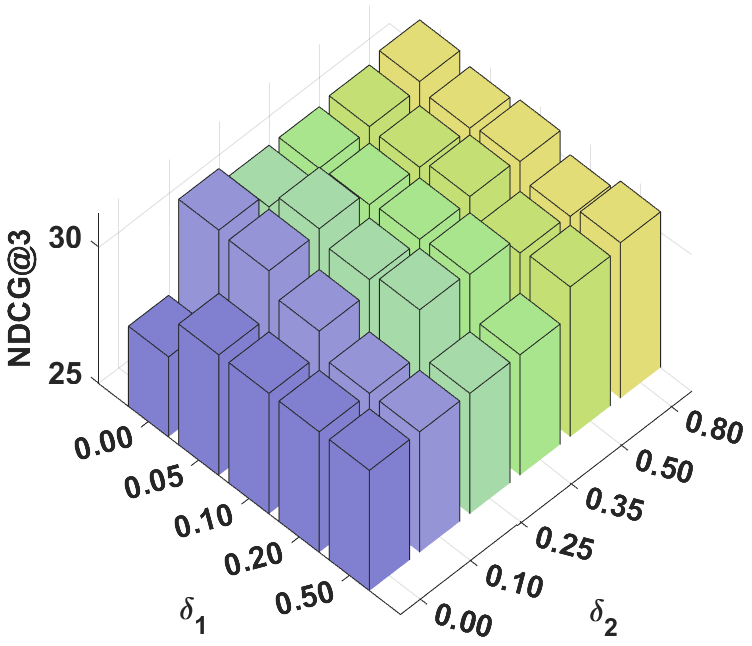}
		}
		\subfigure[P@$5$]{
			\includegraphics[width=0.23\columnwidth]{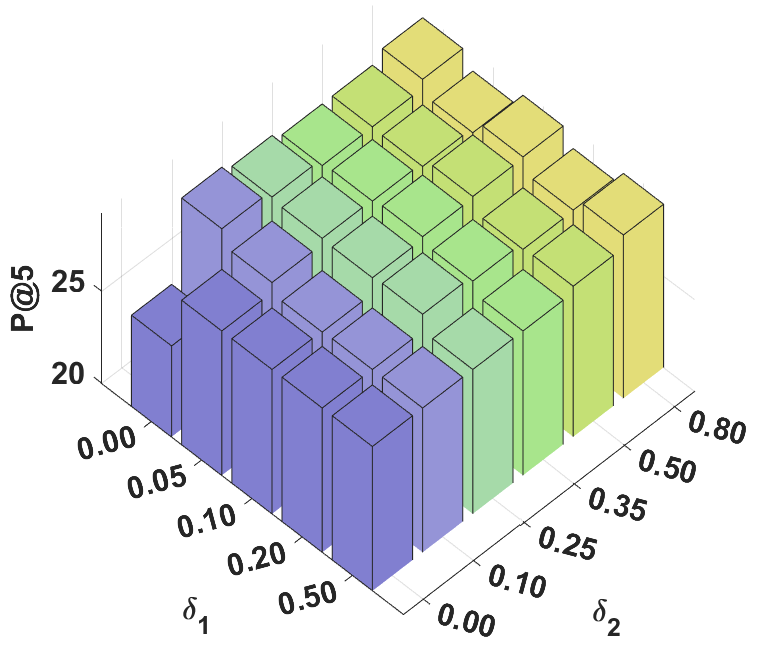}
		}
		\subfigure[R@$5$]{
			\includegraphics[width=0.23\columnwidth]{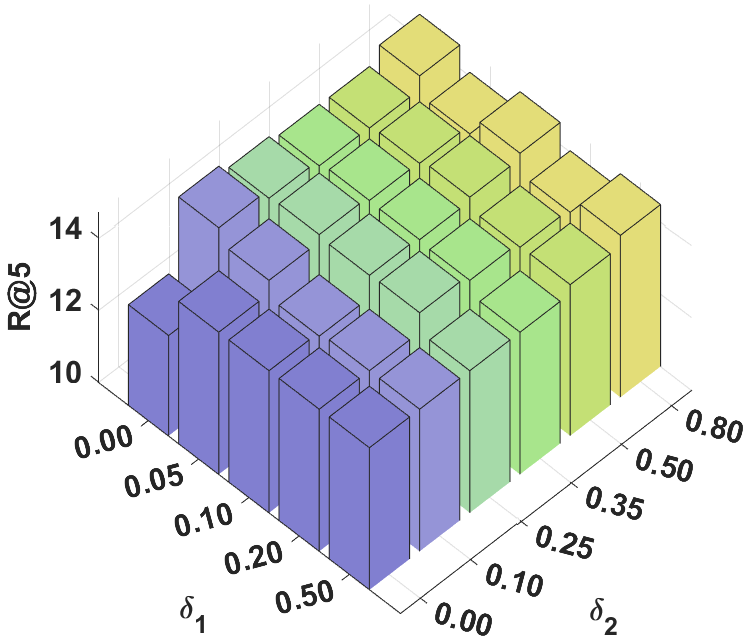}
		}
		\subfigure[NDCG@$5$]{
			\includegraphics[width=0.23\columnwidth]{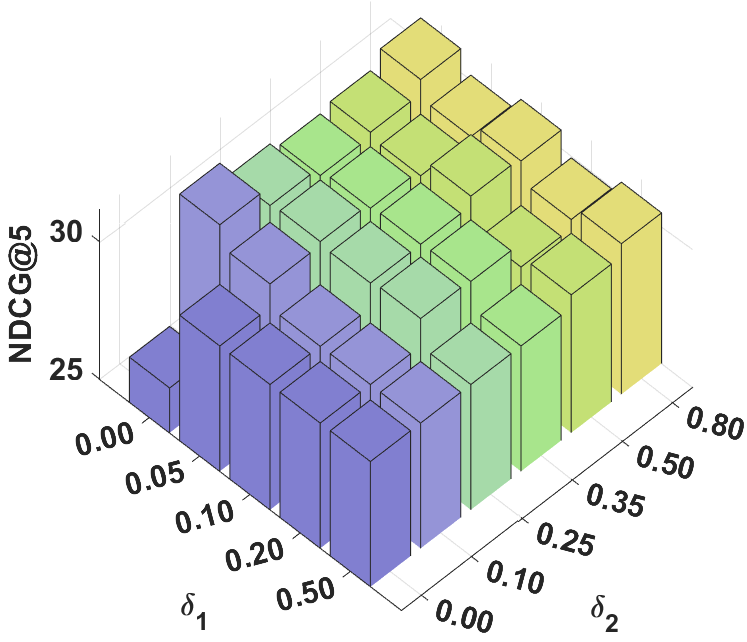}
		}
		\subfigure[MAP]{
			\includegraphics[width=0.23\columnwidth]{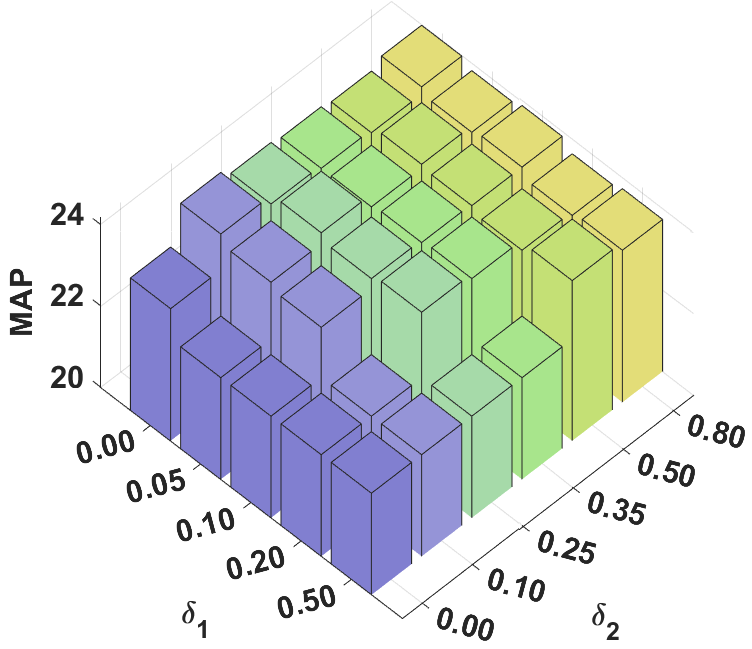}
		}
		\subfigure[MRR]{
			\includegraphics[width=0.23\columnwidth]{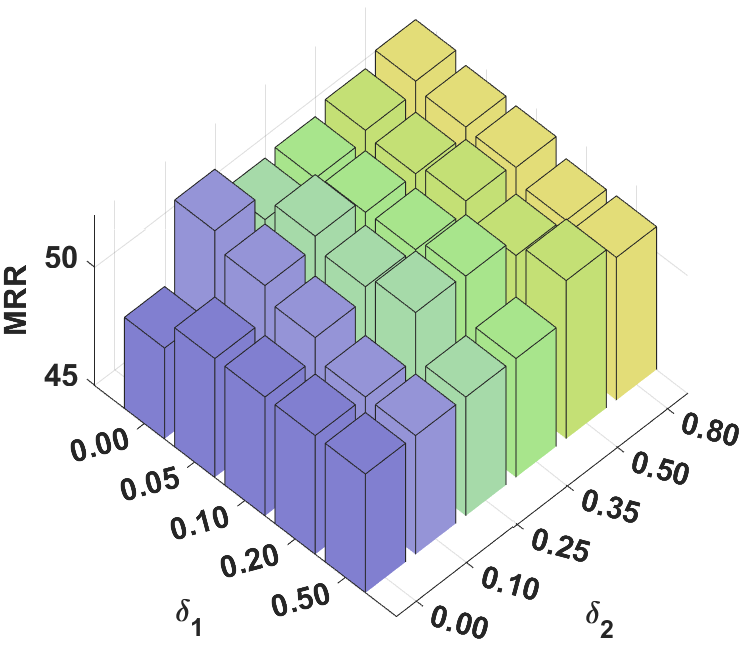}
		}
		\caption{Sensitivity against $\delta_1$ and $\delta_2$ for \textbf{BPA+HarS} on Steam-200k. The $x$- and $y$-axis stand for the value of $\delta_1$ and $\delta_2$ respectively, and the $z$-axis shows the performance.}
		\label{sensitivity_supp2}
	\end{figure*}

	First, we analyze the influence of two main hyper-parameters in DCRS, $\delta_1$ and $\delta_2$. We illustrate a $3$D-barplot based on the results of the grid search on Steam-200k. The results are presented in Fig.\ref{sensitivity_supp1} and Fig.\ref{sensitivity_supp2}. For a clear comparison, $\delta_1 = \delta_2 = 0$ represents the performance of the standard single-vector counterparts and $\delta_1 > \delta_2$ indicates the results of DPCML removing the diversity control regularization scheme. Moreover, we set the trade-off coefficient $\eta=10$ and the representation number $C=5$ here. From these results, we can observe that the proposed regularization scheme could significantly boost performance on all metrics, which demonstrates the effectiveness of the DCRS term. In addition, one can see that there would induce different performances with different diversity values. This suggests that controlling a proper diversity of the embeddings for the same user is essential to accommodate their preferences better.

\begin{figure*}[!t]
		\centering
		\subfigure[BPA+UniS (P@3)]{
			\includegraphics[width=0.18\textwidth]{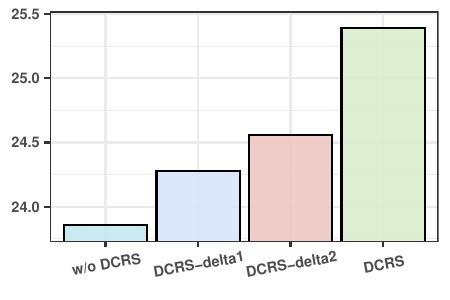}
		}
		\subfigure[BPA+UniS (R@3)]{
			\includegraphics[width=0.18\textwidth]{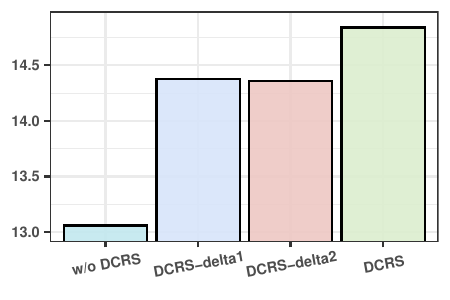}
		}
		\subfigure[BPA+UniS (NDCG@3)]{
			\includegraphics[width=0.18\textwidth]{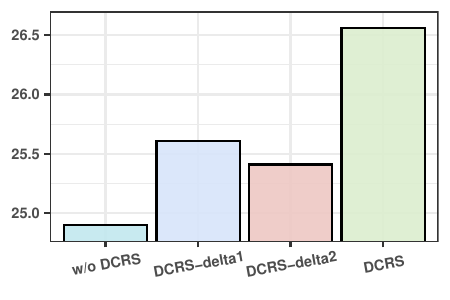}
		}
		\subfigure[BPA+UniS (MAP)]{
			\includegraphics[width=0.18\textwidth]{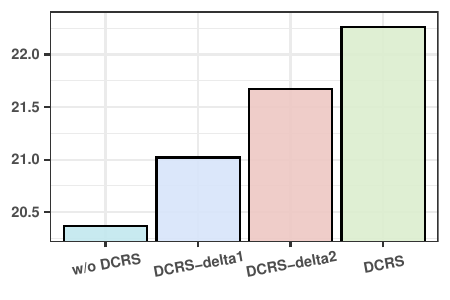}
		}
		\subfigure[BPA+UniS (MRR)]{
			\includegraphics[width=0.18\textwidth]{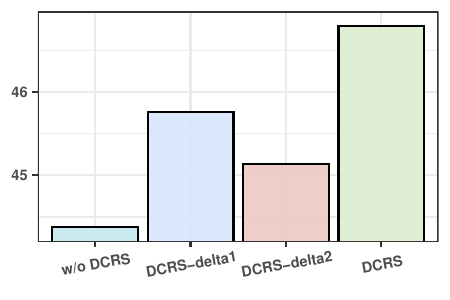}
		}
		\subfigure[BPA+HarS (P@3)]{
			\includegraphics[width=0.18\textwidth]{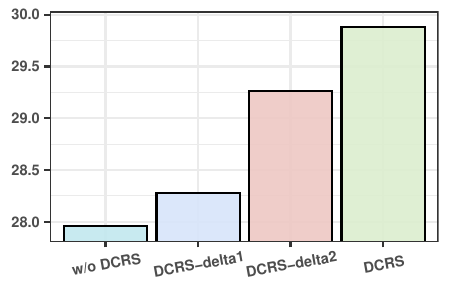}
		}
		\subfigure[BPA+HarS (R@3)]{
			\includegraphics[width=0.18\textwidth]{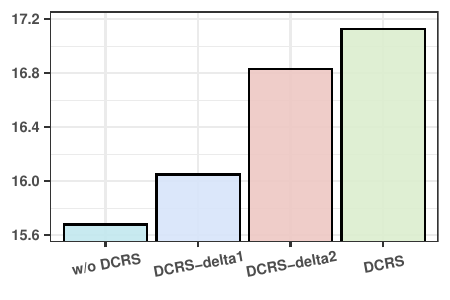}
		}
		\subfigure[BPA+HarS (NDCG@3)]{
			\includegraphics[width=0.18\textwidth]{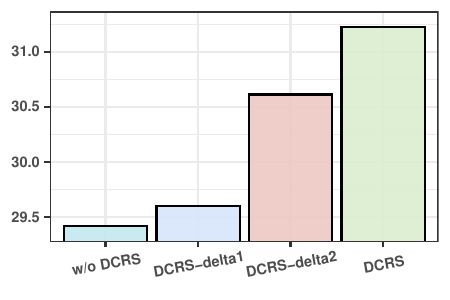}
		}
		\subfigure[BPA+HarS (MAP)]{
			\includegraphics[width=0.18\textwidth]{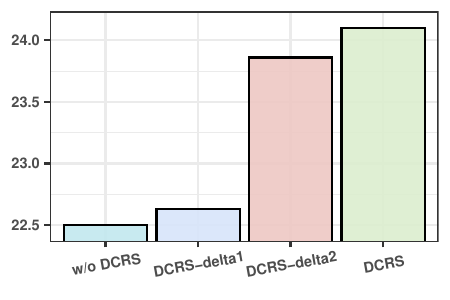}
		}
		\subfigure[BPA+HarS (MRR)]{
			\includegraphics[width=0.18\textwidth]{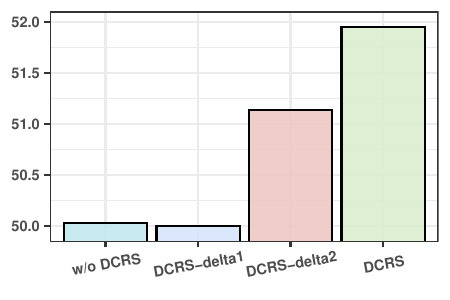}
		}
		\caption{Ablation studies of DCRS on Steam-200k datasets. Please refer to Appendix.\ref{ab_stu} for the detailed performance.}
		\label{ab_DCRS}
	\end{figure*}

	Furthermore, we compare its performance with the following three variants of DCRS:
	\begin{itemize}[leftmargin=*]
		\item $\textbf{w/o DCRS}$: This is a variant of our method where no regularization is adopted at all. 
		\item $\textbf{DCRS}-\delta_1$: This is a variant of our method where the punishment on a \textbf{large} diversity is \textbf{removed}. In other words, we will use the following regularization term:
		$$
		\psi_{\boldsymbol{g}}(u_i) = \max(0, \delta_1 - \delta_{\boldsymbol{g}, u_i}).
		$$ 
		\item $\textbf{DCRS}-\delta_2$: This is a variant of our method where the punishment on a \textbf{small} diversity is \textbf{removed}. In other words, we will use the following regularization term:
		$$
		\psi_{\boldsymbol{g}}(u_i) = \max(0, \delta_{\boldsymbol{g}, u_i} - \delta_2).
		$$
	\end{itemize}
	The empirical results on the Steam-200k dataset are provided in Fig.\ref{ab_DCRS}, and we also present the detailed performance in Tab.\ref{tab:DPCML1} and Tab.\ref{tab:DPCML2}. In most cases, only employing one of the two terms of DCRS could still improve the recommendation performance. However, none of them could outperform our proposed method. This strengthens the effectiveness of our proposed regularization scheme.
	
	
	\begin{figure*}[!t]
		\centering
		\subfigure[MovieLens-1M]{
			\includegraphics[width=0.23\textwidth]{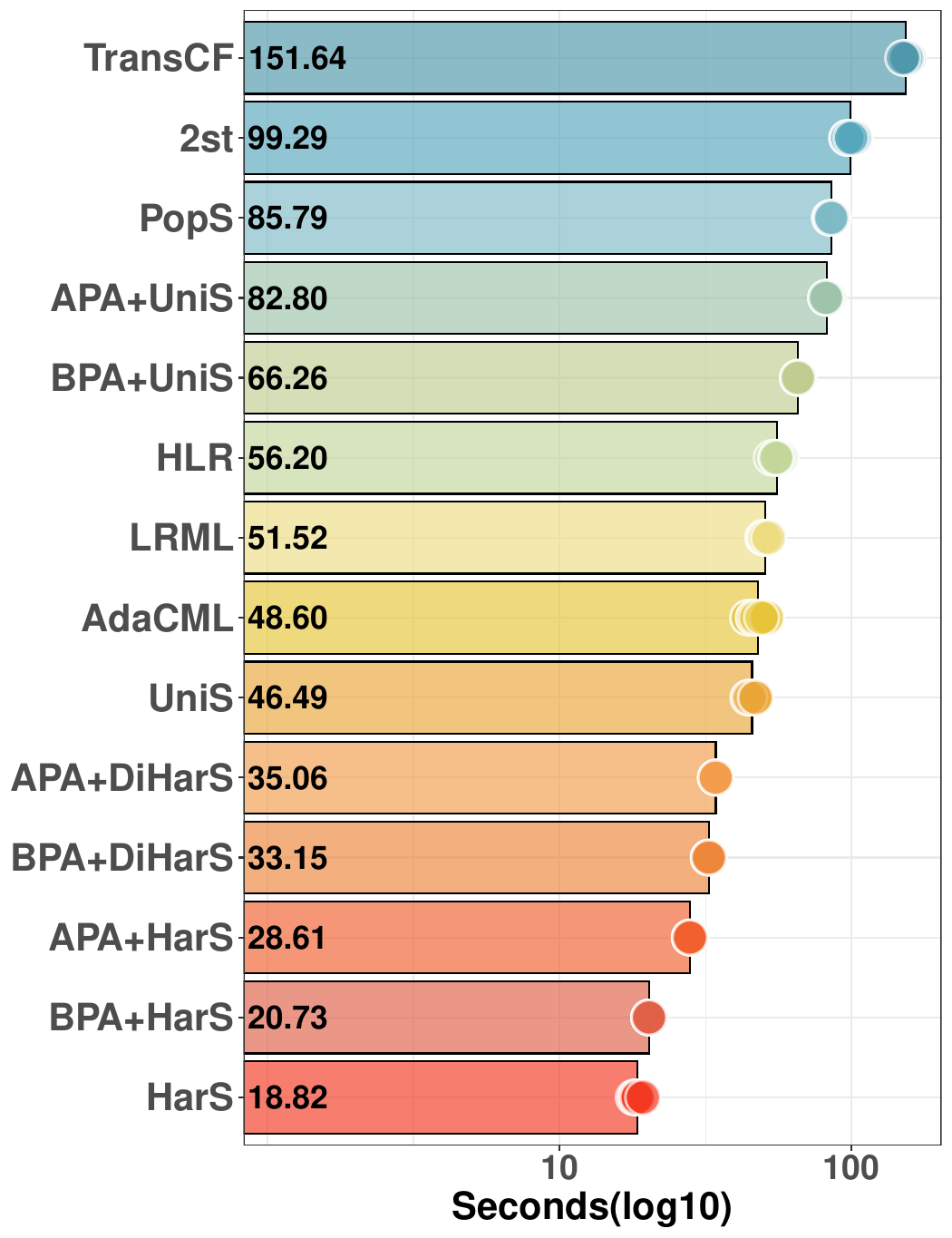}
			\label{ab.sub.ml-1m}
		}
		\subfigure[Steam-200k]{
			\includegraphics[width=0.23\textwidth]{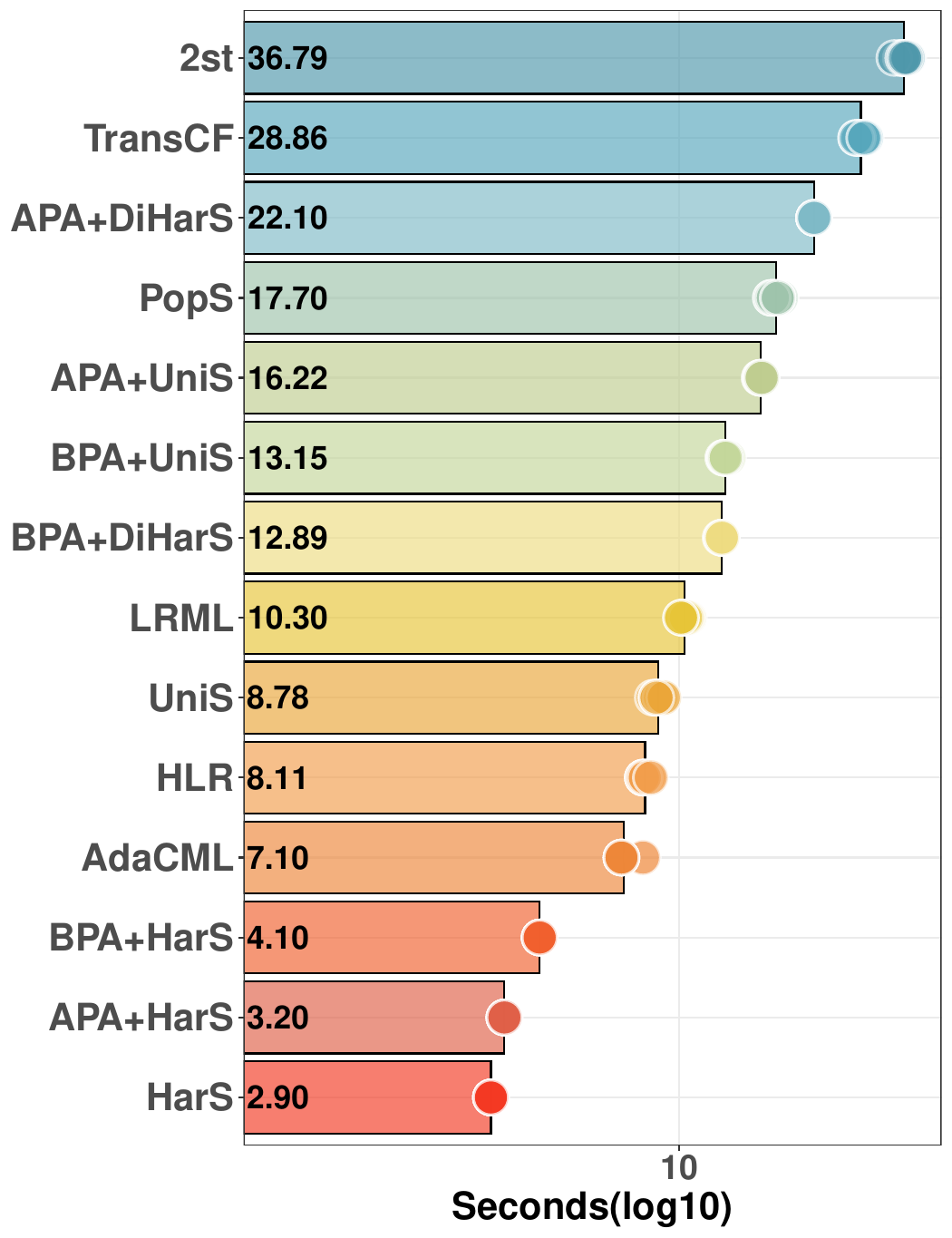}
			\label{ab.sub.200k}
		}
		\subfigure[CiteULike]{
			\includegraphics[width=0.23\textwidth]{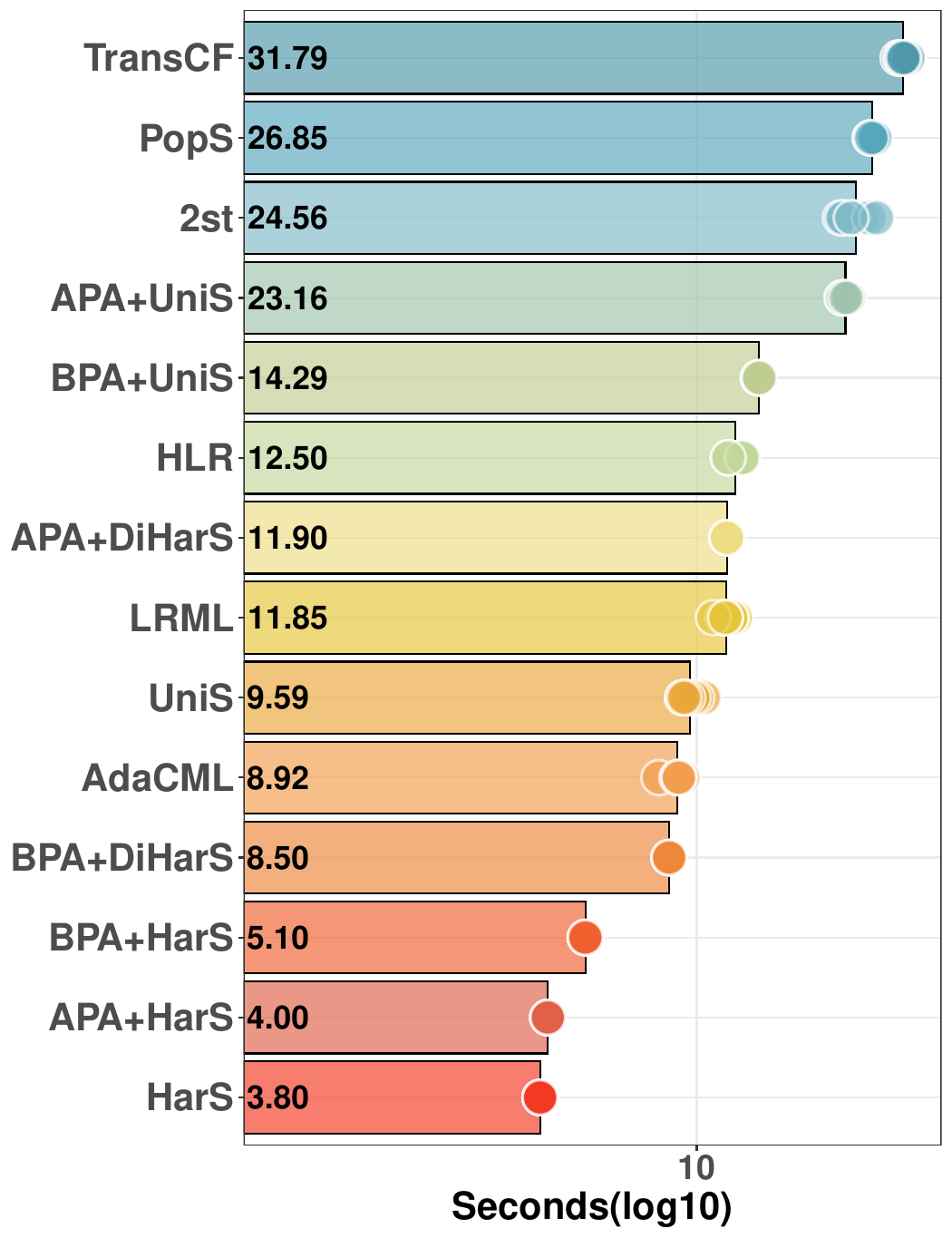}
			\label{ab.sub.citeulike}
		}
		\subfigure[MovieLens-10M]{
			\includegraphics[width=0.23\textwidth]{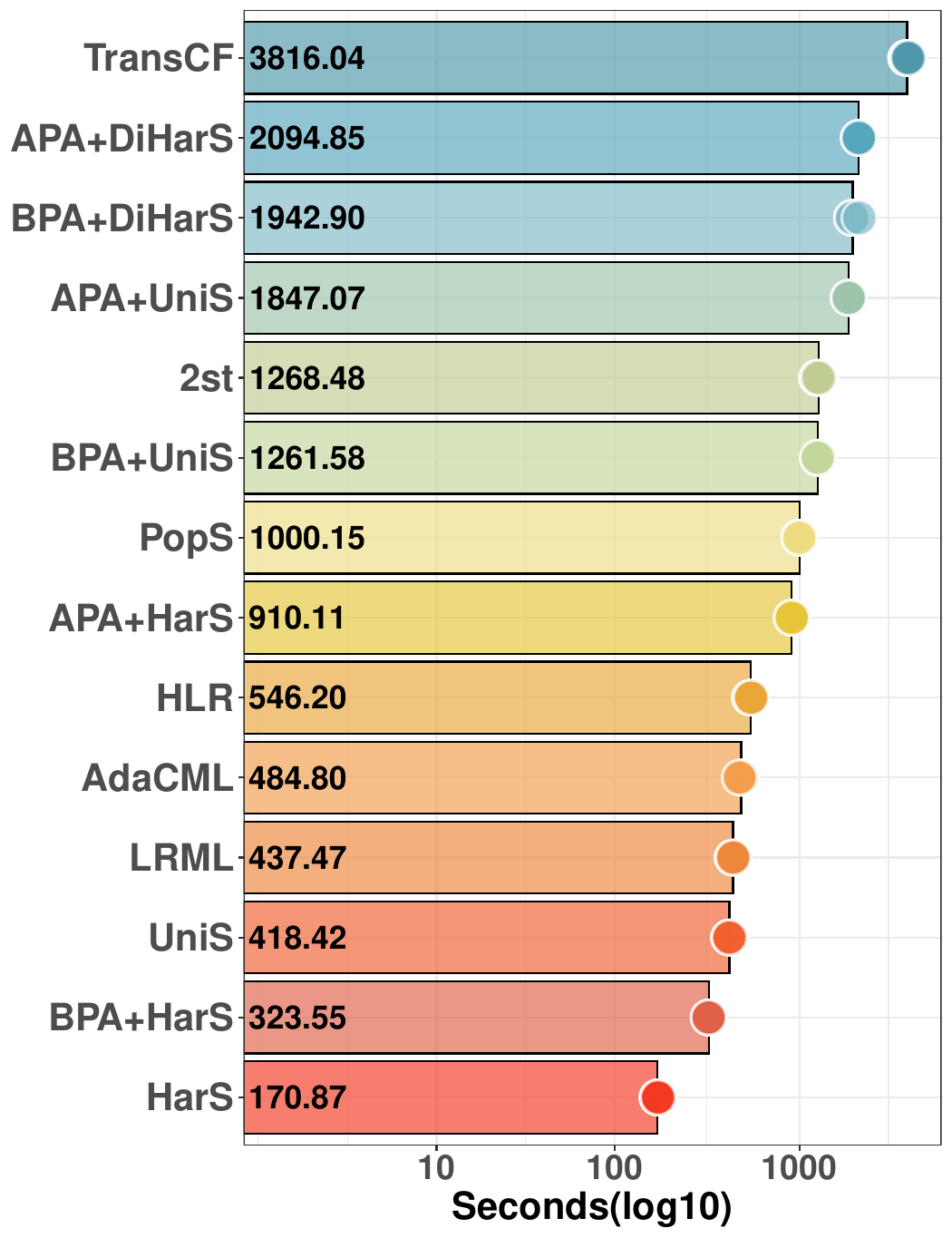}
			\label{ab.sub.ml-10m}
		}
		\caption{Training efficiency comparison among CML-based competitors.}
		\label{app:runtime}
	\end{figure*}

	\begin{table*}[!t]
		\centering
		\caption{Inference efficiency (unit: seconds) comparison among CML-based competitors.}
		  \begin{tabular}{c|ccccccc}
		  \toprule
		  Dataset & CML & LRML & TransCF & AdaCML & HLR & DPCML+BPA & DPCML+APA \\
		  \midrule
		  MovieLens-1M & 0.06  & 0.22  & 1.00  & 0.85  & 4.18  & 0.16  & 0.33  \\
		  Steam-200k & 0.05  & 0.20  & 0.33  & 0.67  & 3.34  & 0.13  & 0.22  \\
		  CiteULike-T & 0.34  & 1.05  & 1.56  & 3.73  & 19.81  & 1.01  & 2.03  \\
		  MovieLens-10M & 1.80  & 5.77  & 91.41  & 24.54  & 121.27  & 5.17  & 10.40  \\
		  \bottomrule
		  \end{tabular}%
		\label{app:inference}%
	  \end{table*}
	
	\begin{table}[htbp]
		\centering
		\caption{Ablation studies of \textbf{BPA+UniS} on Steam-200k dataset.}
		\begin{tabular}{c|cccccccc}
			\toprule
			Method & P@3 & R@3 & NDCG@3 & P@5 & R@5 & NDCG@5 & MAP & MRR \\
			\midrule
			w/o DCRS & 23.86 & 13.06 & 24.90 & 23.57 & 11.56 & 24.77 & 20.37 & 44.38 \\
			DCRS-$\delta_1$ & 24.28 & 14.38 & 25.61 & 22.48 & 11.35 & 24.13 & 21.02 & 45.76 \\
			DCRS-$\delta_2$ & 24.56 & 14.36 & 25.41 & 23.82 & 11.97 & 24.74 & 21.67 & 45.14 \\
			DCRS & \textbf{25.39} & \textbf{14.84} & \textbf{26.56} & \textbf{23.88} & \textbf{12.11} & \textbf{25.25} & \textbf{22.26} & \textbf{46.79} \\
			\bottomrule
		\end{tabular}%
		\label{tab:DPCML1}%
	\end{table}%
	
	\begin{table}[htbp]
		\centering
		\caption{Ablation studies of \textbf{BPA+HarS} on Steam-200k dataset.}
		\begin{tabular}{c|cccccccc}
			\toprule
			Method & P@3 & R@3 & NDCG@3 & P@5 & R@5 & NDCG@5 & MAP & MRR \\
			\midrule
			w/o DCRS & 27.96 & 15.68 & 29.42 & 27.85 & 13.94 & 29.56 & 22.50 & 50.03 \\
			DCRS-$\delta_1$ & 28.28 & 16.05 & 29.60 & 27.25 & 13.75 & 29.17 & 22.63 & 50.00 \\
			DCRS-$\delta_2$ & 29.26 & 16.83 & 30.61 & 28.47 & 14.28 & 30.16 & 23.86 & 51.14 \\
			DCRS & \textbf{29.88} & \textbf{17.13} & \textbf{31.22} & \textbf{28.70} & \textbf{14.51} & \textbf{30.56} & \textbf{24.10} & \textbf{51.95} \\
			\bottomrule
		\end{tabular}%
		\label{tab:DPCML2}%
	\end{table}%

	\subsubsection{Training \& Inference Efficiency}\label{major:supp_efficiency}
	We investigate the training and inference overheads of DPCML-based algorithms against all conventional CML-based approaches. Specifically, in terms of training comparisons, we consider DPCML under both BPA and APA strategies and three negative sampling techniques. Here, every method is executed for $10$ epochs, and the average running time across $10$ epochs is finally reported at the bottom of the box plot. Moreover, considering that adopting different negative sampling strategies will only ease the training optimization burdens, we consider the inference overhead comparisons among several CML architectures, including vanilla CML, LRML, TransCF, AdaCML, HLR, DPCML+BPA and DPCML+APA. We run $10$ times for each method to estimate the inference expenses precisely and report the average efficiency. The training and inference efficiency on MovieLens-1M, Steam-200k, CiteULike, and MovieLens-10M datasets are summarized in Fig.\ref{app:runtime} and Tab.\ref{app:inference}, respectively. Unsurprisingly, DPCML-based algorithms can achieve the best efficiency neither in the training nor inference phase since the multi-vector representation strategies inevitably lead to some additional overheads. However, the efficiency of DPCML is satisfactory and competitive in general. For example, in terms of training efficiency, DPCML-based methods could outperform 2st and TransCF in most cases. In addition, we notice that a series of sophisticated CML-based algorithms (say, TransCF, AdaCML, and HLR) demonstrate poor performance during the inference phase because they involve heavy relation computations between users and items. By contrast, DPCML-based variants are still competitive and predict faster than those sophisticated CML architectures. Overall, we can conclude that our proposed DPCML framework could offer promising recommendation performance within acceptable efficiencies. In the future, we will pay more attention to further accelerating DPCML without hurting recommendation accuracy. 
	
	\subsubsection{Effectiveness of DCRS for Joint Accessibility Model}\label{supp:C.7.5}
	To see this, we attempt to apply the proposed diversity control regularization scheme (DCRS) for M2F \cite{DBLP:conf/recsys/WestonWY13,DBLP:conf/eaamo/GuoKJG21}. In addition, we further explore the effectiveness of DCRS for the general framework of joint accessibility (GFJA, Eq.(\ref{eq3123}) in the main paper). Here we also conduct a grid search to choose the best performance of M2F with DCRS on the Steam-200k and MovieLens-1M datasets, where the parameters space stays the same as DPCML. The experimental results are summarized in Tab.\ref{tab:reg_for_MF}. From the above results, we can draw the following observations: 1) The proposed DCRS does not work well for MF-based models. A possible reason here is that the metric space of MF-based and CML-based methods are intrinsically different. MF adopts the inner-product space while CML adopts the Euclidean space. In this paper, we merely consider the DCRS for Euclidean space. The corresponding strategy for the inner-product space is left as future work. 2) In most metrics, GFJA+DCRS could outperform GFJA significantly, which supports the advantages of our proposed DCRS. 3) Compared with M2F, the performance gain of GFJA is sharp on both datasets. This suggests the superiority of our proposed method against the current multi-vector-based competitors.

	\begin{table}[htbp]
		\centering
		\caption{Performance comparison of joint accessibility model equipped with DCRS on the Steam-200k and MovieLens-1M datasets.}
		\begin{tabular}{c|cccccccc}
			\toprule
			\multicolumn{9}{c}{Steam-200k} \\
			\midrule
			Method & P@3 & R@3 & NDCG@3 & P@5 & R@5 & NDCG@5 & MAP & MRR \\
			\midrule
			M2F & 11.33 & 5.69 & 11.95 & 11.44 & 5.73 & 12.98 & 6.43 & 25.05 \\
			M2F+DCRS & 10.92 & 5.58 & 11.49 & 10.89 & 5.48 & 12.37 & 6.25 & 24.26 \\
			GFJA & 21.53 & \textbf{12.60} & 22.52 & 20.37 & \textbf{10.16} & 21.49 & 19.32 & 40.69 \\
			GFJA+DCRS & \textbf{21.63} & 12.40 & \textbf{22.72} & \textbf{20.38} & 9.98 & \textbf{21.74} & \textbf{19.53} & \textbf{40.92} \\
			\midrule
			\multicolumn{9}{c}{MovieLens-1M} \\
			\midrule
			M2F & 8.61 & 1.84 & 9.36 & 7.60 & 2.30 & 8.67 & 2.95 & 20.40 \\
			M2F+DCRS & 7.59 & 1.49 & 8.16 & 7.10 & 2.02 & 7.92 & 2.53 & 18.51 \\
			GFJA & 15.79 & 3.19 & 16.11 & 16.02 & 4.77 & 16.66 & 11.04 & 32.54 \\
			GFJA+DCRS & \textbf{16.71} & \textbf{3.54} & \textbf{16.94} & \textbf{17.24} & \textbf{5.27} & \textbf{17.71} & \textbf{11.75} & \textbf{33.87} \\
			\bottomrule
		\end{tabular}%
		\label{tab:reg_for_MF}%
	\end{table}%


\begin{table}[htbp]
	\centering
	\caption{Basic Information of the RecSys dataset, where $\cdot/\cdot$ reports the number of interactions for cold start users and items.}
	  \begin{tabular}{c|cc|cc}
	  \toprule
	  \multicolumn{5}{c}{RecSys} \\
	  \midrule
	  \multirow{2}[4]{*}{Dataset} & \multicolumn{2}{c|}{Subset 1} & \multicolumn{2}{c}{Subset 2} \\
  \cmidrule{2-5}      & Warm Start & Cold Start & Warm Start & Cold Start \\
	  \midrule
	  \#Users & 2,799 & 2,116 & 20,134 & 3,610 \\
	  \#Items & 12,612 & 1,310 & 42,214 & 6,104 \\
	  \#Ratings & 94,016 & 15,336/1,520 & 639,742 & 34,562/9,125 \\
	  \%Density & 0.2663\% & - & 0.0753\% & - \\
	  \bottomrule
	  \end{tabular}%
	\label{tab:recsys}%
  \end{table}%

\begin{table}[htbp]
	\centering
	\caption{Parameter size (\textbf{MB}) comparisons on MovieLens-1M, Steam-200k, CiteULike and MovieLens-10M datasets. Here we consider BPA-based DPCML ($C=3$).}
	  \begin{tabular}{c|cccc}
	  \toprule
	  Method & MovieLens-1M & Steam-200k & CiteULike & MovieLens-10M \\
	  \midrule
	  CML+HarS & 3.81  & 3.38  & 11.90  & 30.21  \\
	  DPCML+HarS & 8.41  & 6.25  & 15.88  & 82.98  \\
	  DPCML+DiHarS & 9.72  & 6.51  & 16.16  & 94.32  \\
	  \bottomrule
	  \end{tabular}%
	\label{tab:param_size}%
  \end{table}%
  
	\subsubsection{Parameter Size of $\bmga$}
	As introduced in Thm.\ref{pami:thm3}, $\bmga$ represents all learnable $\gamma_{ij}$ for each $(u_i, v_j^+)$ where the parameter size is $\sum\limits_{u_i \in \mcu} n_i^+$, and $n_i^+$ is the number of the observed actions of user $u_i$. During the implementations, we express $\boldsymbol{\gamma}$ as a $|\mcu| \times |\mci|$ sparse matrix where only the index $(i, j)$ corresponding to $(u_i, v_j^+)$ is used and the other positions are fixed as 0. In terms of the large-scale recommendation scenario, $\boldsymbol{\gamma}$ could be efficiently implemented by the sparse tensor operations in the current deep learning library (such as \texttt{torch.sparse}). To see this, we investigate the parameter size of DiHarS and HarS on MovieLens-1M, Steam-200k, CiteULike and MovieLens-10M datasets. The results are shown in Tab.\ref{tab:param_size}. Although DiHarS would inevitably bring a heavier memory burden than HarS, it is still acceptable in a practical system because \textbf{the interactions of a user are usually very limited} (i.e., $n_i^+ \lll |\mathcal{I}|$).

	\subsection{Potential Challenges and Solutions of DPCML \label{major:coldstartC.8}}
	Although DPCML has demonstrated superiority from theoretical and empirical aspects, there are still two significant limitations in practice. On the one hand, DPCML merely learns users' and items' representations from the one-hot encoding transformations in Sec.\ref{sec.3.3}. Nonetheless, \textcolor{blue}{\textbf{(L1)}} the usage of other semantic information is not explored, which is usually non-negligible to user preferences. On the other hand, \textcolor{blue}{\textbf{(L2)}} the latent representation assignment strategies inherently depend on sufficient user-item interaction records, which will lose efficacy when no interest records are available for some users (i.e., cold start users). 

	Note that \textcolor{blue}{\textbf{(L1)}} and \textcolor{blue}{\textbf{(L2)}} widely exist for most latent collaborative filtering models. Many efforts have been devoted to addressing these issues \cite{DBLP:conf/kdd/LiuBXG022,DBLP:journals/corr/abs-1907-08674}. Typically, \cite{DBLP:conf/nips/VolkovsYP17} proposes a simple but effective framework named DropoutNet (DN), which can be applied to any inner-product-based methods, such as MF-based algorithms. We refer interested readers to the original paper \cite{DBLP:conf/nips/VolkovsYP17} for more details due to space limitations. 
	
	To enhance the scalability of DPCML, we explore extending the CML-based framework into the DN framework. To the best of our knowledge, this is the early trial to address these problems along the CML research line. Without loss of generality, DPCML with the BPA strategy is considered. 
	
	We follow the notations introduced in Sec.\ref{major:sec3.1} and further let $\mathcal{C}_{u_i}$ be the content features (say occupation, gender and age) for user $u_i$ and $\mathcal{C}_{v_j}$ represents the content information (such as tags, prices and visual features) for item $v_j$. Following the roadmaps of DN, the first step is to separately transform both sparse preference and content information into dense features and then unify them as latent embedding in a new metric space. Specifically, we have
	\begin{equation}\label{major:eq32}
		\tilde{\bm{g}}^{c}_{u_i} = h_{\mcu}^{c}([\bm{g}^{c}_{u_i}, \Phi_{\mcu}(\mathcal{C}_{u_i})]), \ \ \forall c \in [C], u_i \in \mcu,
	\end{equation}
	\begin{equation}\label{major:eq323}
		\tilde{\bm{g}}_{v_h} = h_{\mathcal{I}}([\bm{g}_{v_j}, \Phi_{\mathcal{I}}(\mathcal{C}_{v_j})]), \ \ \forall v_j \in \mathcal{I},
	\end{equation}
	where $[\cdot, \cdot]$ means the concatenate operation for two vectors, $\bm{g}^{c}_{u_i}$ and $\bm{g}_{v_j}$ are the same as (\ref{pami:eq7}), $\Phi_{\mcu}$ and $\Phi_{\mathcal{I}}$ can be any DNN-based feature extractor toward different content inputs, $h_{\mcu}^{c}$ and $h_{\mathcal{I}}$ are the final fusion models. 
	
	(\ref{major:eq32}) and (\ref{major:eq323}) allow us to exploit side information adequately to assist DPCML in pursuing more expressive representations, which overcomes \textcolor{blue}{\textbf{(L1)}}. However, the preference embedding is generally absent for the cold start user or item. To alleviate this issue, given the preference and content inputs, the critical recipe of DN borrows the idea of Dropout \cite{DBLP:journals/jmlr/SrivastavaHKSS14}, which randomly samples a fraction of users and items and masks their corresponding preference inputs as $\bm{0}$. Concretely, for each ``dropouted'' user or item, we have 
	\begin{equation}\label{major:eq33}
		\tilde{\bm{g}}^{c}_{u_i} = h_{\mcu}^{c}([\bm{0}, \Phi_{\mcu}(\mathcal{C}_{u_i})]), \ \ \forall c \in [C],
	\end{equation}
	\begin{equation}\label{major:eq34}
		\tilde{\bm{g}}_{v_j} = h_{\mathcal{I}}([\bm{0}, \Phi_{\mathcal{I}}(\mathcal{C}_{v_j})]), \ \ \forall v_j \in \mathcal{I}.
	\end{equation}

	After incorporating all inputs into a set of multiple vectors in the new latent space, we still leverage the minimum item-user Euclidean distance as the relevance score:
	\begin{equation}\label{major:eq3}
		\tilde{s}(u_i, v_j) = \min\limits_{c \in [C]} \|\tilde{\boldsymbol{g}}_{u_i}^c - \tilde{\boldsymbol{g}}_{v_j}\|^2, \forall v_j \in \mci.
	\end{equation}

	Finally, we train DPCML+DN with a similar objective in Sec.\ref{major:sec4.2.4}, where different user representations will be activated to fit diverse preference groups. By doing so, owning to the dropout effect, DPCML could be capable of producing promising representations for both users and items to accommodate diverse interests even if the latent input is not provided, which solves the \textcolor{blue}{\textbf{(L2)}}. 
	
	\begin{table}[!t]
		\centering
		\caption{Performance comparisons for both warmstart and coldstart cases on ResSys dataset.}
		  \begin{tabular}{c|c|ccc|ccc|ccc}
		  \toprule
		  \multirow{2}[4]{*}{Type} & \multirow{2}[4]{*}{Method} & \multicolumn{3}{c|}{WarmStart} & \multicolumn{3}{c|}{ColdStart User} & \multicolumn{3}{c}{ColdStart Item} \\
	  \cmidrule{3-11}      &   & \multicolumn{1}{c|}{P@3} & \multicolumn{1}{c|}{R@3} & N@3 & \multicolumn{1}{c|}{P@3} & \multicolumn{1}{c|}{R@3} & N@3 & \multicolumn{1}{c|}{P@3} & \multicolumn{1}{c|}{R@3} & N@3 \\
		  \midrule
		  \multicolumn{11}{c}{Subset 1} \\
		  \midrule
		  \multirow{3}[2]{*}{Joint-Training} & MGMF+DN & \cellcolor[rgb]{ .996,  .953,  .957} 11.55  & \cellcolor[rgb]{ .996,  .941,  .945} 4.56  & \cellcolor[rgb]{ 1,  .973,  .976} 11.56  & \cellcolor[rgb]{ .996,  .953,  .957} 2.47  & \cellcolor[rgb]{ .996,  .953,  .957} 1.07  & \cellcolor[rgb]{ .996,  .953,  .957} 2.26  & \cellcolor[rgb]{ .996,  .953,  .957} 9.05  & \cellcolor[rgb]{ .996,  .953,  .957} 3.26  & \cellcolor[rgb]{ .996,  .953,  .957} 9.16  \\
			& CML+DN & \cellcolor[rgb]{ .996,  .933,  .941} 11.63  & \cellcolor[rgb]{ .996,  .925,  .933} 4.59  & \cellcolor[rgb]{ 1,  .976,  .976} 11.56  & \cellcolor[rgb]{ .996,  .961,  .965} 3.56  & \cellcolor[rgb]{ .996,  .961,  .965} 1.44  & \cellcolor[rgb]{ .996,  .957,  .961} 3.42  & \cellcolor[rgb]{ .98,  .8,  .824} 13.69  & \cellcolor[rgb]{ .984,  .808,  .831} 4.62  & \cellcolor[rgb]{ .984,  .804,  .827} 13.37  \\
			& DPCML+DN & \cellcolor[rgb]{ .98,  .784,  .804} \underline{12.27}  & \cellcolor[rgb]{ .984,  .804,  .827} \underline{4.89}  & \cellcolor[rgb]{ .988,  .851,  .867} \underline{12.21}  & \cellcolor[rgb]{ .984,  .816,  .839} \underline{7.26}  & \cellcolor[rgb]{ .98,  .776,  .8} \underline{3.15}  & \cellcolor[rgb]{ .984,  .816,  .835} \underline{7.08}  & \cellcolor[rgb]{ .98,  .776,  .804} \underline{14.79}  & \cellcolor[rgb]{ .976,  .761,  .788} \textbf{5.27} & \cellcolor[rgb]{ .98,  .776,  .804} 14.57  \\
		  \midrule
		  \multirow{3}[2]{*}{Pre-Training} & MGMF+DN & \cellcolor[rgb]{ .996,  .953,  .957} 11.34  & \cellcolor[rgb]{ .996,  .953,  .957} 4.41  & \cellcolor[rgb]{ .996,  .953,  .957} 11.42  & \cellcolor[rgb]{ .988,  .851,  .871} 6.33  & \cellcolor[rgb]{ .984,  .812,  .831} 2.83  & \cellcolor[rgb]{ .984,  .831,  .851} 6.62  & \cellcolor[rgb]{ .996,  .953,  .957} 4.64  & \cellcolor[rgb]{ .996,  .953,  .957} 1.81  & \cellcolor[rgb]{ .996,  .953,  .957} 4.37  \\
			& CML+DN & \cellcolor[rgb]{ .988,  .875,  .886} 11.87  & \cellcolor[rgb]{ .992,  .898,  .91} 4.66  & \cellcolor[rgb]{ .992,  .91,  .922} 11.90  & \cellcolor[rgb]{ .988,  .847,  .863} 6.49  & \cellcolor[rgb]{ .984,  .831,  .851} 2.64  & \cellcolor[rgb]{ .988,  .867,  .882} 5.74  & \cellcolor[rgb]{ .98,  .78,  .808} 14.57  & \cellcolor[rgb]{ .98,  .784,  .808} 4.98  & \cellcolor[rgb]{ .976,  .761,  .788} \textbf{15.27} \\
			& DPCML+DN & \cellcolor[rgb]{ .969,  .702,  .729} \textbf{12.60} & \cellcolor[rgb]{ .976,  .761,  .788} \textbf{4.99} & \cellcolor[rgb]{ .976,  .761,  .788} \textbf{12.67} & \cellcolor[rgb]{ .976,  .761,  .788} \textbf{8.65} & \cellcolor[rgb]{ .976,  .761,  .788} \textbf{3.27} & \cellcolor[rgb]{ .976,  .761,  .788} \textbf{8.41} & \cellcolor[rgb]{ .976,  .761,  .788} \textbf{15.45} & \cellcolor[rgb]{ .98,  .773,  .796} \underline{5.15}  & \cellcolor[rgb]{ .98,  .769,  .796} \underline{15.04}  \\
		  \midrule
		  \multicolumn{11}{c}{Subset 2} \\
		  \midrule
		  \multirow{3}[2]{*}{Joint-Training} & MGMF+DN & \cellcolor[rgb]{ .965,  .984,  .996} 27.50  & \cellcolor[rgb]{ .965,  .984,  .996} 12.76  & \cellcolor[rgb]{ .969,  .984,  .996} 27.76  & \cellcolor[rgb]{ .965,  .984,  .996} 10.84  & \cellcolor[rgb]{ .965,  .984,  .996} 3.41  & \cellcolor[rgb]{ .965,  .984,  .996} 7.65  & \cellcolor[rgb]{ .965,  .984,  .996} 6.56  & \cellcolor[rgb]{ .965,  .984,  .996} 3.84  & \cellcolor[rgb]{ .965,  .984,  .996} 9.13  \\
			& CML+DN & \cellcolor[rgb]{ .933,  .969,  .992} 27.78  & \cellcolor[rgb]{ .902,  .949,  .988} 12.88  & \cellcolor[rgb]{ .945,  .973,  .992} 27.86  & \cellcolor[rgb]{ .659,  .827,  .949} \underline{23.44}  & \cellcolor[rgb]{ .616,  .804,  .941} \textbf{7.38} & \cellcolor[rgb]{ .71,  .855,  .957} \underline{21.81}  & \cellcolor[rgb]{ .945,  .973,  .992} 8.51  & \cellcolor[rgb]{ .961,  .98,  .996} 4.66  & \cellcolor[rgb]{ .996,  1,  1} 9.34  \\
			& DPCML+DN & \cellcolor[rgb]{ .965,  .984,  .996} 27.52  & \cellcolor[rgb]{ .965,  .984,  .996} 12.69  & \cellcolor[rgb]{ .965,  .984,  .996} 27.63  & \cellcolor[rgb]{ .616,  .804,  .941} \textbf{24.90} & \cellcolor[rgb]{ .624,  .808,  .945} \underline{7.34}  & \cellcolor[rgb]{ .616,  .804,  .941} \textbf{26.31} & \cellcolor[rgb]{ .698,  .847,  .957} 17.31  & \cellcolor[rgb]{ .745,  .871,  .965} 9.08  & \cellcolor[rgb]{ .718,  .855,  .957} 17.34  \\
		  \midrule
		  \multirow{3}[2]{*}{Pre-Training} & MGMF+DN & \cellcolor[rgb]{ .886,  .941,  .984} 27.98  & \cellcolor[rgb]{ .847,  .922,  .976} 12.98  & \cellcolor[rgb]{ .882,  .941,  .984} 28.11  & \cellcolor[rgb]{ .961,  .98,  .996} 12.35  & \cellcolor[rgb]{ .953,  .976,  .996} 3.93  & \cellcolor[rgb]{ .973,  .988,  .996} 9.03  & \cellcolor[rgb]{ .788,  .894,  .969} 14.14  & \cellcolor[rgb]{ .796,  .898,  .969} 8.08  & \cellcolor[rgb]{ .773,  .886,  .969} 15.71  \\
			& CML+DN & \cellcolor[rgb]{ .753,  .875,  .965} \underline{28.51}  & \cellcolor[rgb]{ .765,  .882,  .965} \underline{13.13}  & \cellcolor[rgb]{ .773,  .886,  .969} \underline{28.55}  & \cellcolor[rgb]{ .953,  .976,  .996} 12.70  & \cellcolor[rgb]{ .953,  .976,  .996} 3.91  & \cellcolor[rgb]{ .902,  .953,  .988} 12.43  & \cellcolor[rgb]{ .631,  .812,  .945} \underline{19.65}  & \cellcolor[rgb]{ .686,  .839,  .953} \underline{10.33}  & \cellcolor[rgb]{ .62,  .808,  .945} \underline{20.17}  \\
			& DPCML+DN & \cellcolor[rgb]{ .616,  .804,  .941} \textbf{29.07} & \cellcolor[rgb]{ .616,  .804,  .941} \textbf{13.40} & \cellcolor[rgb]{ .616,  .804,  .941} \textbf{29.17} & \cellcolor[rgb]{ .89,  .945,  .984} 14.88  & \cellcolor[rgb]{ .953,  .976,  .996} 3.90  & \cellcolor[rgb]{ .843,  .922,  .976} 15.43  & \cellcolor[rgb]{ .616,  .804,  .941} \textbf{20.18} & \cellcolor[rgb]{ .616,  .804,  .941} \textbf{11.73} & \cellcolor[rgb]{ .616,  .804,  .941} \textbf{20.23} \\
		  \bottomrule
		  \end{tabular}%
		\label{tab:cold}%
	  \end{table}%

	\noindent\textbf{Clarifications.} Our DPCML upgraded by DN directly follows the model architecture in \cite{DBLP:conf/nips/VolkovsYP17}, but makes some technical changes. First, the original DN is developed for MF-based methods, while we consider the CML with Euclidean space. Also, it merely assigns unique embedding for each user in the new unified space, leading to the preference bias more or less, as discussed in Sec.\ref{Sec3.2}. In stark contrast, DPCML+DN still introduces multiple representations for each user, enjoying similar benefits to the original DPCML. Moreover, the significant difference is that we do not use the least square loss to recover the score gap between the latent preference and the dropouted version \cite{DBLP:conf/nips/VolkovsYP17}. Instead, we use a similar way to (\ref{cml_eq}) to train the DPCML+DN after unifying the latent and content information as the new representations because we merely care about the relative preference of users toward different items rather than the specific preference value. 

  
	\subsubsection{Experiment Setups}\label{major:coldstartc8.1.1}
	\noindent\textbf{Dataset.} The experiments are conducted on the RecSys dataset, which is a part of the data in the ACM RecSys 2017 Challenge \cite{DBLP:conf/recsys/AbelDEK17}. Specifically, it contains both user and item content information, such as education, work experience for users and location, title/tags for items. Here, we directly adopt the released 1-of-n features in DN \cite{DBLP:conf/nips/VolkovsYP17} \footnote{\url{https://github.com/layer6ai-labs/DropoutNet}}, where the user feature is $831$ dimensions, and the item feature is $2738$ dimensions. In addition, we remove duplicate actions by reserving the latest user-item interactions and also delete users with interaction lengths less than $25$ to ensure a reasonable dataset sparsity. Finally, we consider two different scales of RecSys to simulate distinct deployed scenarios. The statistical information for warm and cold start cases is summarized in Tab.\ref{tab:recsys}. 
	
	\noindent\textbf{Competitors.} To show the effectiveness of DPCML, we also apply DN to \textbf{MGMF} and UniS-based \textbf{CML}. The main distinction between these methods lies in the input latent model, where the \textbf{MGMF} directly follows the DN paper \cite{DBLP:conf/nips/VolkovsYP17} and CML fixes $C=1$ in (\ref{major:eq3}). The others are handled similarly to DPCML. Moreover, the latent model should be trained first in the study \cite{DBLP:conf/nips/VolkovsYP17}. In this paper, \textbf{we consider two cases}: (1) the latent model is trained together with the DN model from scratch (denoted as \textbf{Joint-Training (JT)}), and (2) the latent model is pre-trained and then fixed when training the DN model (denoted as \textbf{Pre-Training (PT)}).

	\noindent\textbf{Implementation Details.} As suggested by \cite{DBLP:conf/recsys/CovingtonAS16,DBLP:conf/nips/VolkovsYP17}, a pyramid architecture of neural networks is employed with the batch norm and tanh activation functions. Following the literature \cite{DBLP:conf/recsys/CovingtonAS16,DBLP:conf/nips/VolkovsYP17},
	we directly apply $h_{\mcu}^{c}$ and $h_{\mathcal{I}}$ to the concatenated joint preference-content inputs in the RecSys dataset. More complicated architectures will be explored in our future study. To be specific, $h_{\mcu}^{c}$ is implemented by $(d + 831) \rightarrow 800 \rightarrow 400 \rightarrow 200 \rightarrow d$ and $h_{\mathcal{I}}$ is implemented by
	$(d + 2738) \rightarrow 800 \rightarrow 400 \rightarrow 200 \rightarrow d$, where $d=100$ is the dimensional of the latent embedding. During training, the ``dropout'' rate for the preference inputs of users and items in each mini-batch is fixed at $0.5$. Furthermore, we consider the performances of (DP)CML-based approaches optimized by the UniS. The other implementations follow the same as Sec.\ref{app:train}. Finally, during the test phase, both preference and content information of the warm start users/items will be fed into DPCML+DN, while merely the content input is known for these cold start ones. 
	\subsubsection{Overall Performance} \label{major:C.8.2}
	The detailed performance results are reported in Tab.\ref{tab:cold}. From these results, we can observe that our proposed DPCML+DN could tackle \textcolor{blue}{\textbf{(L1)}} and \textcolor{blue}{\textbf{(L2)}} well. It significantly outperforms the competitors in cold start cases while achieving competitive or even better performance in most warm start cases. This supports the potential of DPCML and deserves more research attention in the future.

  
\end{document}